\documentclass[a4paper,12pt,times,authoryear,print,index]{PhDThesisPSnPDF}

\input{preamble}
\title{EXPLORING THE ORIGIN AND DYNAMICS OF SOLAR MAGNETIC FIELDS}

\subtitle{}

\degreetitle{Doctor of Philosophy in Science}

\author{Soumitra Hazra}

\dept{Department of Physical Sciences\\
and\\
Centre of Excellence in Space Sciences India}
\university{Indian Institute of Science Education and Research Kolkata}
\crest{\includegraphics[width=0.25\textwidth]{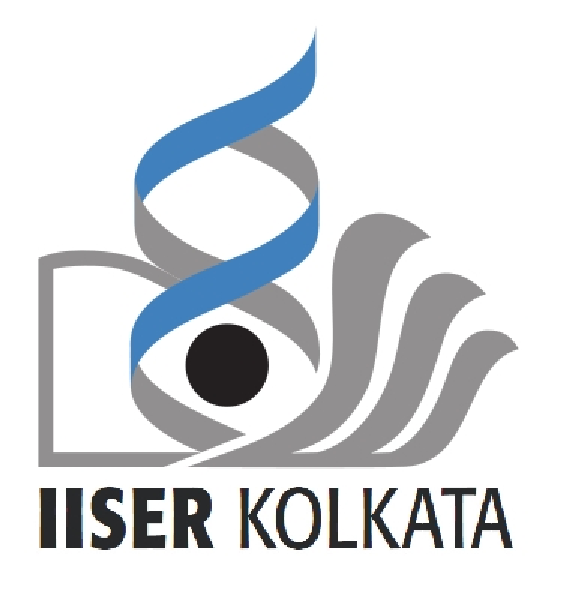}}

\degreedate{2015} 

\subject{LaTeX} \keywords{{LaTeX} {PhD Thesis} {Science} {Indian Institute of Science Education and Research, Kolkata}}

\begin{document}

\frontmatter

\begin{titlepage}
  \maketitle
\end{titlepage}

\begin{dedication} 
\centering

\LARGE{\it Dedicated to My Parents...}

\end{dedication}


\begin{declaration}

This thesis is a presentation of my original research work. Wherever contributions of others
are involved, every effort is made to indicate this clearly, with due reference to the literature and
acknowledgments of collaborative research and discussions.
The work has not been submitted earlier either in entirety or in parts for a degree or
diploma at this or any other Institution or University. Some chapters of this thesis have either been published
or are in the process of being published.
This work was done under the guidance of Dr. Dibyendu Nandi, at the Indian Institute of Science
Education and Research Kolkata (IISER Kolkata).
\vspace{1cm} ~~\\

\end{declaration}

\begin{certificate}

This is to certify that this thesis entitled 
\emph{\bf {"Exploring the Origin and Dynamics of Solar Magnetic Fields"}} -- which is being submitted by Soumitra Hazra (who registered on 11th August, 2009 for a PhD Degree with registration number {\textbf{09RS028}} at the Indian Institute of Science Education and Research Kolkata) -- is 
based upon his own research work under my supervision and that neither this 
thesis nor any part of it has been submitted for any degree or any other academic award anywhere else.
\vspace{2cm} ~~\\
Dr. Dibyendu Nandi \\ 
Associate Professor\\
{Department of Physical Sciences and Center of Excellence in Space Sciences India}\\
Indian Institute of Science Education and Research Kolkata ~~~\\
West Bengal 741246\\
India

\end{certificate}
\begin{acknowledgements}      

This is a special moment for me and I wish to express my sincere gratitude to my thesis supervisor Dibyendu Nandi, who introduced me to the exciting world of research in Astrophysics. This thesis work would never have been possible without his help and constant support during this period. He also taught me that one of the major components of scientific research is to be able to communicate your scientific ideas with others. I have learnt a lot of science as well as the methodolgy of tackling a difficult problem through discussions with him. I also thank him for his support and encouragement to attend different conferences which enabled me to interact with other scientists.\\

I would like to thank B Ravindra for giving me the opportunity to visit his institute and work with him. His constant help, encouragement and discussion helped me grasp the skills of satelite data anyalisis quickly. I have completed an important project with him "The Relationship between Solar Coronal X-Ray Brightness and Active Region Magnetic Fields: A Study with High Resolution Hinode Observations". I am also extremly thankful to Aveek da, Andres, Piet, Dario, Bidya da, Dipankar da, Shravan Hanasoge, Paul Rajaguru, Durgesh Tripathi, Nandita Srivastava, Dhrubaditya Mitra and Prasad Subramanian for their helpful discussions and readiness to answer my numerious queries. I also want to thank Arnab Rai Choudhuri for his two books "Physics of Fluids and Plasma' and "Astrophysics for Physicist" which I used as a text book during my Ph.D period.\\

I am also greatful to all the faculty members and colleagues of the Department of Physical Sciences, IISER Kolkata for their help and support. I want to thank Barun (nada), Dyuti, choto and baro Nandan (Das \& Roy), Abhinav, Subhrajit (kaka), Gopal, Tanmoy, Sudipta, Rupak, Richarj, Sumi, Harkirat, Sanhita, Anirudha, Abhishek, Arghya da (mama), Basu da, Chandan da, Rabi da, Vivek da, Priyam da for making my initial life at IISER Kolkata fun and enjoyable. It gives me pleasure to thank my CESSI lab mates Mayukh, Prantika, Avyarthana, Rakesh, Sushant, Tamoghna, Athira, Sanchita, Prasenjit, Abhinna, Chandu da for their support. A special word of thanks to Mayukh and Prantika for their readiness to help me with MATLAB related issues. Mayukh has also collaborated with me in a paper. I want to thank Sanjib Ghosh for his initial help in my first work. I also thank Madhusudan, Arun Babu, Wageesh, Sushant Bishoi, Grijesh Gupta, Krishna da, Vemareddy for discussion and help. It is also my pleasure to thank Ankan, Debmalya (motu), Chiranjeeb, Soumya (gambat), Diptesh, Deepak, Radhe, Soumen, Rafikul and all those who made my life memorable and enjoyable at IISER Kolkata. I also want to thank Prasanta da and Soma di for their help and support during my stay at Montana.\\

It is also my pleasure to thank my masters and bachelors degree friends Monalisa, Arup, Amit, Amaresh, Arnab, Nikhil, Sarengi, Suman, Prithwish da, Asim da, Samaresh da and my childhood friends Palash, Dilip, Jagneswar, Gopi, Bubun, Bulbul da, Bappaditya, Sushanta, Arup Samanta, Dipak da, Abhi, Piu for their help and support. I also want to thank my maternal uncles and aunts (mama and mamima), paternal uncles and aunts for their encouragement and support. \\ 

Finally, I would like to thank my parents and sister for their constant patience and support. They have always shared my joys and sorrows and have given their unconditional love. Thanks to all of my family members and friends for their understanding and supporting me.

\end{acknowledgements}


\begin{abstract}
The Sun is a magnetically active star and is the source of the solar wind, electromagnetic radiation and energetic particles which affect the heliosphere and the
Earth's atmosphere. The magnetic field of the Sun is responsible for most of the dynamic activity of the Sun. This thesis research seeks to understand solar magnetic field generation and the role that magnetic fields play in the dynamics of the solar atmosphere.  Specifically, this thesis focuses on two themes: in the first part, we study the origin and behaviour of solar magnetic fields using magnetohydrodynamic dynamo theory and modelling, and in the second part, utilizing observations and data analysis we study two major problems in solar physics, namely, the coronal heating problem and initiation mechanisms of solar flares.\\

The magnetic field of the Sun, whose evolution is evident in the 11 year cycle of sunspots, is created within the Sun through complex interactions between internal plasma flows and fields. It is widely believed that magnetic fields of not just the Sun, but all astronomical bodies are produced by this hydromagnetic dynamo process. To explain the origin of the solar cycle, Eugene Parker first proposed the idea of flux recycling between the toroidal (which is in the azimuthal i.e., $\phi$-direction) and poloidal (which is in r-$\theta$ plane) field components . Currently, flux transport dynamo models based on the Babcock-Leighton mechanism for poloidal field generation appears as a promising candidate for explaining different aspects of the solar cycle. In this scenario, the toroidal field is produced within the convection zone due to stretching of the poloidal component by strong differential rotation while the poloidal field is produced at the solar surface due to decay and dispersal of tilted bipolar sunspot pairs. Flux transport mechanisms such as diffusion, meridional flow and turbulent pumping shares the role of communicator between these two largely separated source layers. In chapter 1, we describe observations of solar magnetic fields and development of solar dynamo theory to motivate our work.\\

An outstanding issue related to the solar cycle is extreme fluctuations, specifically the occurrence of extended periods of reduced or no activity, known as grand minima episodes. The origin of such episodes have eluded a consistent theory. The Maunder minima was such an episode during 1645-1715 AD when there was almost no sunspots observed on the Sun. There is also observational evidence of many such episodes in the past. The crucial fact is that the solar cycle has recovered from these episodes every time and regained normal activity levels. It is expected that during grand minima phases, the Babcock-Leighton mechanism would not be able to produce poloidal field as this mechanism relies on the presence of sunspots on the solar surface. This leads to the following fundamental question: How does the solar cycle recover every time from these episodes? We address this question through diverse means. In chapter 2 of this thesis, we develop a mathematical, low order time delay dynamo model (based on delay differential equations) removing all spatial dependence terms from the magnetic induction equation and mimic flux transport through the introduction of finite time delays in the system. By introducing fluctuations in the Babcock-Leighton source term of this low order dynamo model, we, for the first time explicitly demonstrate that a solar cycle model based on the Babcock-Leighton mechanism alone can not recover from a grand minima. We find that an additional poloidal field generation mechanism effective on weak magnetic field is necessary for recovery of the sunspot cycle from grand minima like episodes.\\

Modeling the Babcock-Leighton mechanism in a correct way to capture the observed surface dynamics is a challenging task. Two different approaches, mainly near-surface alpha-coefficient formulation and the double ring formalism has been followed to model the Babcock-Leighton mechanism for poloidal field generation. Earlier it has been shown that the second approach, i.e., the double ring formalism is more successful in explaining observational results compared to the near-surface alpha-coefficient formulation. Inspired by this, we develop a 2.5D kinematic solar dynamo model where we simulate the Babcock-Leighton mechanism via the double ring algorithm. In chapter 3 of this thesis, we utilize this state-of-the spatially extended model in a solar like geometry to validate our findings on entry and exit from Maunder minima like episodes. We find that stochastic fluctuations in the Babcock-Leighton mechanism is a possible candidate for triggering entry into grand minima phases. However, the Babcock-Leighton mechanism alone is not able to recover the solar cycle from a grand minimum. An additional mean field $\alpha$-effect effective on week magnetic field is necessary for self-consistent recovery. Thus, this result puts the earlier findings based on a low order time delay model on firmed grounds. This spatially extended model also allows one to explore latitudinal asymmetry and hemispheric coupling of the sunspot cycle. Based on simulations with this model, we find that stochastic fluctuation in both poloidal field sources makes hemispheric coupling weak, thus introducing asymmetry in sunspot eruptions in the Northern and Southern hemispheres. The phase locking between the two hemispheres is thus impacted resulting in a switching of the parity of the solutions. Particularly we find that parity shifts in the sunspot cycle is more likely to occur when solar activity in one hemisphere strongly dominates over the other hemisphere for a period of time significantly longer than the sunspot cycle timescale. While direct observations over the last fifty years have shown that the solar magnetic cycle exhibits dipolar (odd) parity, our results suggest that the solar cycle has a significant probability to reside in the quadrapolar (even) parity state. Our findings may open the pathway for predicting parity flip in the Sun. \\

Meridional circulation is an essential ingredient in present day kinematic solar dynamo models. Earlier studies based on theoretical considerations and numerical simulations have suggested that a deep equatorward meridional flow near the base of the solar convection zone is plausibly essential in explaining the observed equatorward migration of the sunspot belt. However there is no observational evidence of such a deep meridional flow as it is difficult to probe such deep layers. Some recent observational studies, on the other hand,  indicate that the meridional flow could be shallow or more complex than previously assumed. In chapter 4 of this thesis, we explore whether flux transport dynamos could function with such a shallow meridional flow and discuss the consequences that this scenario would have on our traditional understanding of magnetic field dynamics in the solar interior. We demonstrate that dynamo models of the solar cycle can produce solar-like solutions with a shallow meridional flow if the effects of turbulent pumping of magnetic flux is taken into account.\\

In the second part of this thesis, we utilize satellite observations to explore the dynamics of magnetic fields in the solar atmosphere. The solar corona, the outer atmosphere of the Sun is very hot compared to the solar surface and can reach millions of degrees. There is controversy regarding the physical processes that heat the solar outer atmosphere to such high temperatures. Such high temperatures result in X-ray emission from the solar corona. In chapter 5, we discuss the observational techniques and the current theoretical understanding developed over time, necessary to explain coronal dynamics. In chapter 6 of this thesis, we explore the relationship between coronal X-ray brightness and sunspot magnetic fields using high resolution observations from the Solar Optical Telescope and X-Ray Telescope onboard the Hinode satellite (a joint JAXA-NASA space mission). We find that the total magnetic flux within active regions sunspot structures is the primary determinant of solar coronal X-ray luminosity suggesting that magnetic flux is the fundamental quantity that determines coronal heating.  This result sets important constraints on theories of solar and stellar coronal heating.

Solar flares are highly energetic eruptions from the Sun that hurl out magnetize plasma and emit high energy radiation thereby impacting space weather and space- and some ground-based technologies. Thus predictions of solar flare is necessary for developing advance warning systems. Previous studies indicate that the non-potentiality of the magnetic field is closely related with solar flare productivity. However traditional measures of magnetic field non-potentiality based on the force-free parameter have been questioned and studies based on the force-free parameter have given diverging results. In chapter 7 of this thesis, a recently developed, flux-tube fitting technique is utilized to measure the non-potentiality (twist) of solar magnetic fields and test whether the kink-instability mechanism (following magnetic helicity conservation) can be a plausible initiation mechanism for solar flares. We demonstrate that those sunspot magnetic field structures in which the twist exceeds the threshold for kink instability are more prone to generate solar flares. This finding may lead to more accurate solar flare prediction schemes based on the kink instability mechanism. \\

The chapters that follow, outline the outcome of this thesis research, are written in the form of research publications; as such they are self-contained with independent introductions and conclusions. 

{\bf Research publications emnating out of this thesis work:}

\begin{itemize}
\item{A stochastically forced time delay solar dynamo model: self-consistent
recovery from a Maunder-like grand minimum necessitates a mean-field alpha
effect.
Hazra, S., Passos, D. \& Nandy, D. { \bf ApJ},  789, 5 (2014)} \\
(Dario Passos was a collaborator in exploring this idea using a different, spatially extended dynamo model the results of which are published in Passos, D, Nandy, D, Hazra, S \& Lopes, I. 2014, A\&A, 563, A18.)
\item{Double ring algorithm of solar active region eruptions within the framework of kinematic dynamo model.
Hazra, S. \& Nandy, D., {\it Bulletin of
Astronomical Society of India, ASI conference series}, vol 10, p 117-121 (2013)}
\item{ The Relationship between Solar Coronal X-Ray Brightness and Active Region Magnetic Fields: A Study Using High Resolution Hinode Observations.
Hazra, S., Nandy, D. \& Ravindra, B. {\bf Solar Physics}, 290, 771 (2015)}\\
(Ravindra Belur was a collaborator in this work in which he provided the knowhow on the software tools necessary to reduce the observational data.)
\item{ A New Paradigm of Magnetic Field Dynamics at the Basis of the Sunspot Cycle.
Hazra, S. \& Nandy, D. in preparation}
\item{Strong Hemispheric Asymmetry can Trigger Parity Changes in the Sunspot Cycle. 
Hazra, S. \& Nandy, D. in preparation}
\item{ Exploring the Relationship between Kink Instability and Solar Flare: A Study Using High Resolution Hinode Observations.
Panja, M., Hazra, S. \& Nandy, D. in preparation}\\
(Mayukh Panja was an undergraduate summer student who contributed to the observational analysis under the supervision of Soumitra Hazra and Dibyendu Nandy.)
\end{itemize}

\end{abstract}


\tableofcontents

\listoffigures

\listoftables


\printnomencl


\mainmatter
\chapter{Introduction to the Solar Magnetic Cycle}

\section{The Sun: Interior and Atmosphere}

\subsection{Solar Interior}
The Sun is a completely gaseous body consisting mostly of hydrogen and helium. Solar interior consists of three regions, namely, Core, Radiative zone and Convection zone. Inside the core, energy is generated via nuclear fusion by converting hydrogen into helium. When we go outwards from the core to the surface, density decreases gradually. Energy is transported via radiation in the inner 70 \% of the Sun and by convection in the outer 30 \% of the Sun. In the interface between these two regions, a strong radial shear in the rotation (known as the tachocline) exists between 0.675 and 0.725 solar radius (Charbonneau et al. 1999). 
\subsection{Solar Atmosphere}
The region above the solar photosphere is known as solar atmosphere. A small region above the photosphere where temperature rises from 6000K to 20000K is known as the chromosphere. The very low density region above the chromosphere where temperature is of the order of $10^6$ K is called  the solar corona. The solar corona is only visible at the time of total solar eclipse. The layer which separates the chromosphere and the corona is known as the transition region. The rapid rise in temperature from the chromosphere to corona can not be of thermal origin as this would violate the second law of thermodynamics. This problem is known as the coronal heating problem. Different layers of the Sun (both interior and atmosphere) are shown in Fig.~1.1. 

\begin{figure}[t!]
\centering
\includegraphics*[width=0.75\linewidth]{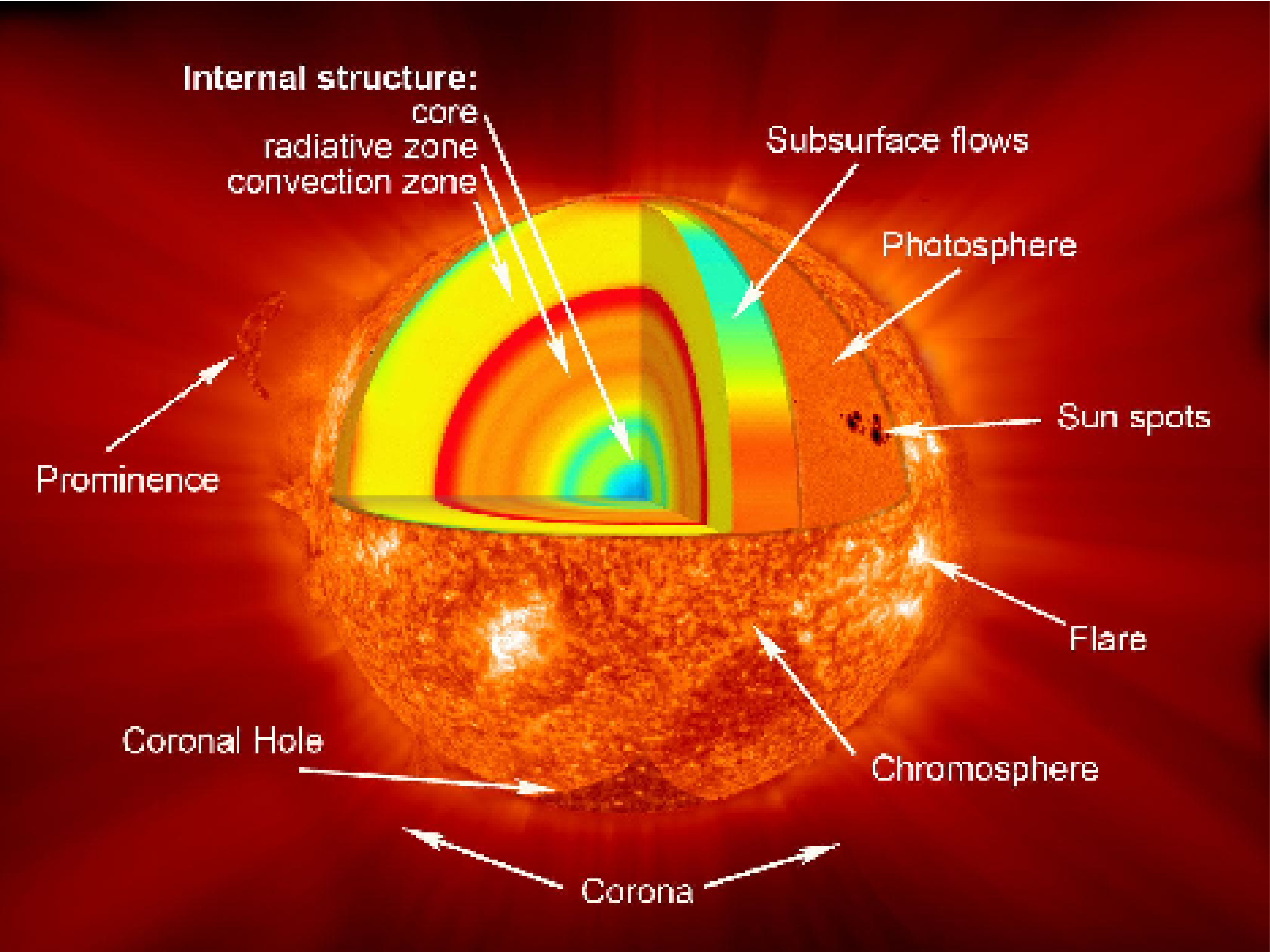}
  \caption{A cartoon image of the structure of the Sun.}

\end{figure}

\section{Discovery of the Solar Magnetic Cycle}
 Sunspots are transient dark spots on the solar surface. Large number of naked eye sunspot observations were reported by Chinese astronomers around 27 BC. There are almost no records of sunspot observations in Europe before 17th century. In 807 AD, a large sunspot was seen for more than 8 days and it was simply interpreted as a planetary transit. The invention of the telescope in the early seventeenth century brought about a revolution in the field of astronomy. This invention made possible the detailed observational study of sunspots and astronomers like Galileo Galilei, Thomas Harriot, Johanes and David Fabricius, and Christoph Scheiner were quick to harness the immense potential of the telescope. In 1611, Johanes Fabricius was the first to publish a description of sunspots in his book "De Maculis in Sole Observatis" (On the spots observed in the Sun). 
 
  Very few sunspots were observed during the second part of seventeenth century. Later it was revealed that this was not due to lack of observations. Surprisingly for a long period of time there were almost no sunspots on the solar surface. This period between 1645 to 1715 is known as the Maunder minimum. Absence of aurorae and lack of a bright solar corona during solar eclipses was also noted during this period.

 \begin{figure}[t!]
\centering
\includegraphics*[width=0.75\linewidth]{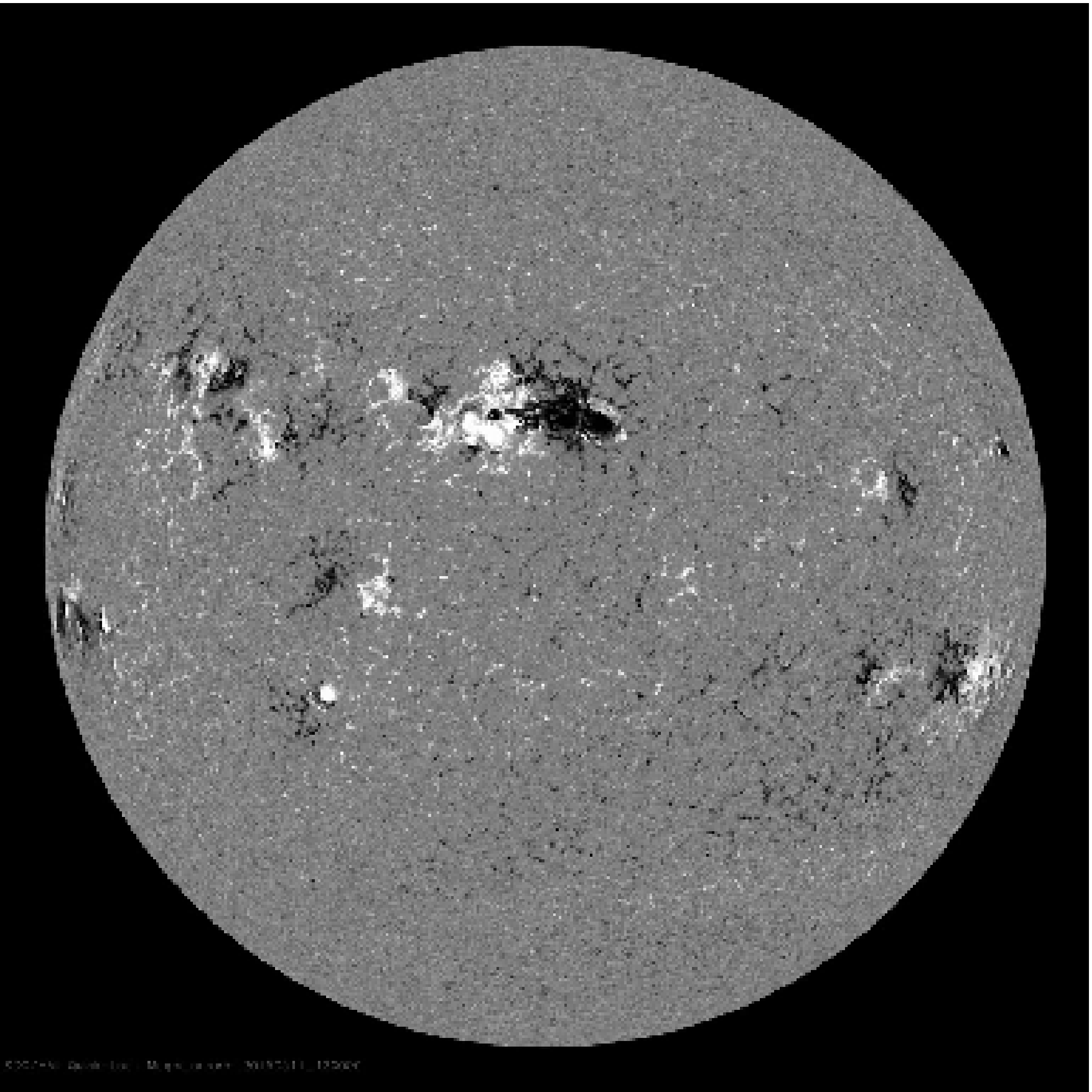}
   \caption{SDO-HMI magnetogram image recorded on May 11, 2015 showing bipolar sunspot pairs within active region structures. In the image, white signifies positive polarity while black signifies negative polarity sunspots.}
\end{figure}

 After analyzing more than two decades of sunspot observations, Samuel Schwabe (1844) discovered the cyclic rise and fall of sunspot numbers with a periodicity of 11 years. This is now well known as the solar cycle (see top panel of Fig.~1.3). Just after the discovery of the solar cycle, in 1852 four astronomers pointed out that the period of changes of geomagnetic activity at the Earth was identical with the periodicity of the solar cycle which provided a significant clue about possible Sun-Earth connections. Later, Carrington (1858) noted that sunspots first appear at mid-latitudes and then appear at lower and lower latitudes (closer to the equator) as the cycle progresses. In 1904, Edward and Annie Maunder introduced a new way of visualizing this characteristic by plotting sunspot emergence latitude with time (popularly known as butterfly diagram, bottom panel of Fig.~1.3). Till then there was no evidence about the interconnection between sunspots and the magnetic field. The first evidence about this interconnection came from George Hale (1908), who identified sunspots as strong magnetic regions on the solar surface (by observing the Zeeman splitting of the sunspot spectra). In 1919, Hale and his coworkers discovered most of the properties of sunspot groups (also known as Active Regions, Fig.~1.2):
 
$\bullet$ Sunspots frequently appear in pairs at the surface and the relative orientation of the magnetic field of most active regions is opposite across the equator.\\

$\bullet$ The polarity of active regions change from one to another solar cycle but remains same in a hemisphere for a given solar cycle.\\ 

$\bullet$ The line joining the bipolar sunspot pairs are tilted with respect to the east-west direction (leading spot along the direction of rotation is closer to the equator than the following spot) and this tilt angle increases with latitude. This is known as Joy's law. \\
\begin{figure}[t!]
\centering
\includegraphics*[width=\linewidth]{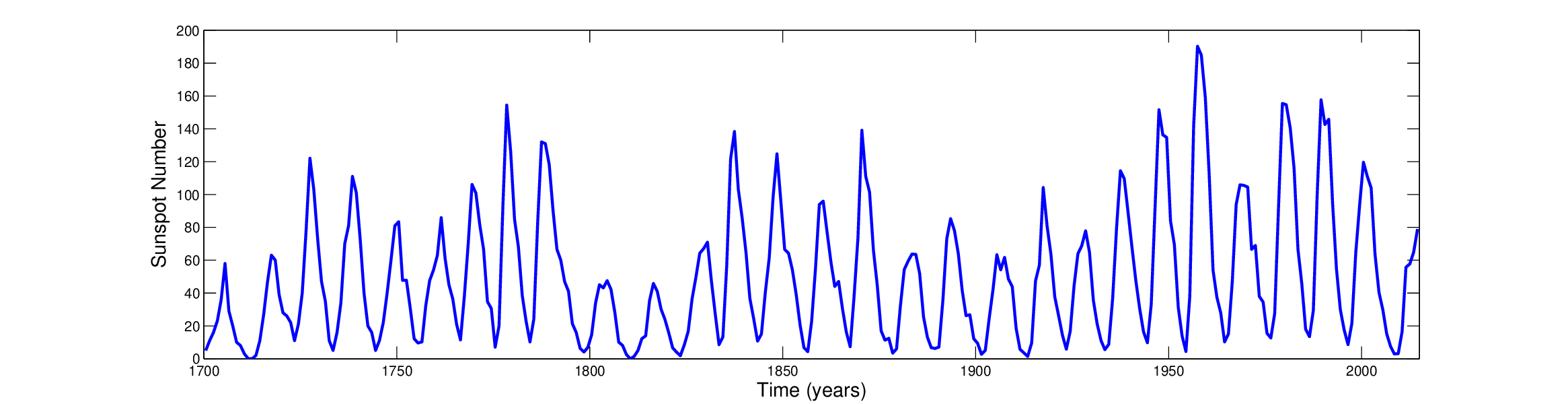}
\includegraphics*[width= \linewidth]{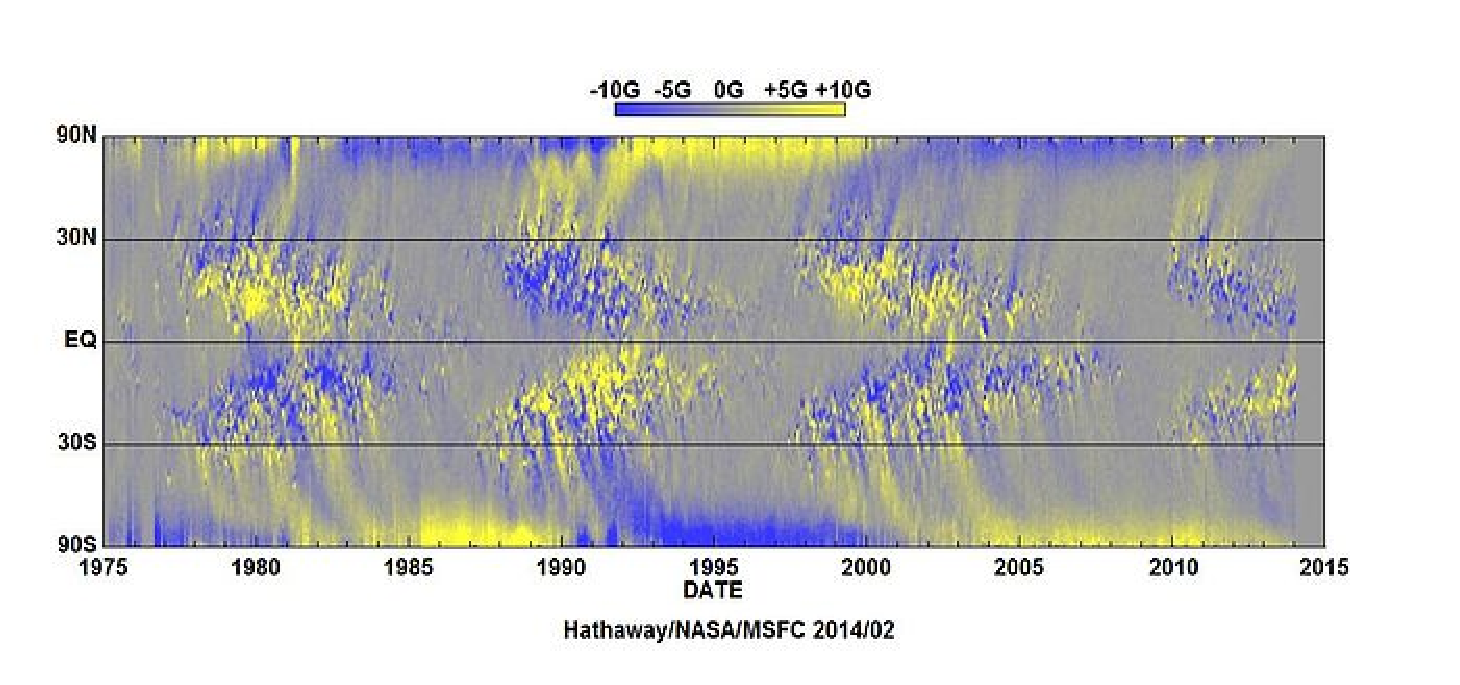}
   \caption{Top panel: Plot of international sunspot number as a function of time in years. Bottom panel: Latitude vs time plot from recent high resolution observations. Background shows weak, diffuse radial field on the photosphere. This plot is widely known as the butterfly diagram. Image credit: Hathaway/NASA/MSFC.}
\end{figure}

Harold Babcock and his son Horace Babcock (1955) developed the magnetograph and used this magnetograph to study the distribution of magnetic fields on the region outside of sunspots. A very weak field of the order of 10 Gauss was found to be mostly concentrated in the latitude above $55^\circ$, (Babcock, 1959). This weak diffuse magnetic field migrates poleward and changes the sign of the polar field every 11 years. Note that the polar field reverses its polarity when the sunspot number is maximum. We also note that recent high resolution observations find vertically oriented magnetic flux tubes with strong kilo-Gauss magnetic field strength in the polar region (Tsuneta et al. 2008), which might also contribute to the polar field.

There is also another type of photospheric magnetic field structure known as small scale magnetic field.  This mixed polarity small scale magnetic field is very dynamic and does not vary much with the solar cycle. Although the origin of this small scale magnetic field is unknown, some recent studies suggest that a small scale dynamo near the solar surface (local dynamo), may be the source for this small scale magnetic field (Petrovay \& Szakaly 1993; Cattaneo 1999; Danilovic et al. 2010; Lites 2011).

\section{Generation of the Large Scale Solar Magnetic Field}
\subsection{Magnetohydrodynamics}
Matter inside the Sun exists in the ionized (plasma) state. To  explain the magnetic nature of the solar cycle, one has to understand the behaviour of magnetic fields inside electrically conducting fluids -- which is the heart of the subject Magnetohydrodynamics (MHD). The interaction of the plasma velocity field with the magnetic field can be described through the magnetic induction equation:
\begin{equation}
\frac{\partial \mathbf{B}}{\partial t} = \nabla \times (\mathbf{v} \times \mathbf{B}) + \lambda \nabla^2 \mathbf{B},
\end{equation}
where the first term in the right hand side of the equation is the source term and the second term is the diffusion term. Here we assume the situation where the diffusivity ($\lambda$) does not vary with space.

One important input in the induction equation is the velocity ($\mathbf{v}$). To describe a MHD system self-consistently, we also require the Navier-Stokes equation, which describes the evolution of the velocity field.
\begin{equation}
 \rho \frac{\partial \mathbf{v}}{\partial t} + \rho (\mathbf{v} \cdot \nabla)\mathbf{v}= -\nabla p + \mathbf{J} \times \mathbf{B}  + \rho \mathbf{g} + \nabla. \mathbf{\tau},
\end{equation}
where $\mathbf{v}$ is the fluid velocity, $-\nabla p$ is the force due to pressure gradient, $\mathbf{J} \times \mathbf{B}$ is the Lorentz force term and $\mathbf{\tau}$ is the viscous stress tensor. The Lorentz force term is calculated using the solution of induction equation ($B$), which acts as a forcing term in the Navier-Stokes equation. 

Note that $B=0$ is a valid solution of the induction equation, so that no magnetic field generation is possible if we start with zero magnetic field. So there must be some mechanism through which initial seed magnetic fields can be generated and amplified. As we are mainly interested in the generation of large scale solar magnetic fields, it is suffice to assume that we start with a pre-existing seed magnetic field. Dynamo is a process which can amplify this seed magnetic field to produce large scale magnetic fields by converting the kinetic energy of plasma into magnetic energy. So, in order to explore the full dynamical behaviour of the magnetized plasma, we have to solve equations (1.1), (1.2) together with the mass continuity and energy conservation equations:
\begin{equation}
\frac{\partial \rho}{\partial t} +\nabla.(\rho \mathbf{v})=0,
\end{equation}

\begin{equation}
\frac{\partial p}{\partial t} + \mathbf{v}. \nabla p + \gamma p \nabla.\mathbf{v}= - (\gamma-1)L,
\end{equation}
where $L$ is the heat loss rate which consists of the terms due to thermal conduction, ohmic heating etc. and $\gamma$ represents the ratio between specific heats. These equations along with $\nabla.B=0$ and equation of state, comprise the full set of MHD equations.\\

The evolution of magnetic field (equation 1.1) inside the plasma is governed by the competition between induction and diffusion of the magnetic field. If we take the ratio of two terms on the right hand side of the equation then we get the magnetic Reynold's number, $R_m= \frac{VL}{\eta}$ where $V$ is the velocity and $L$ is the spatial length-scale. Now it is obvious that $R_m \gg 1$ for astrophysical systems as length scales (L) are very large. In this case one may approximate the induction equation as:
 \begin{equation}
\frac{\partial \mathbf{B}}{\partial t} \simeq \nabla \times (\mathbf{v} \times \mathbf{B}).
\end{equation}
In this situation (i.e., in the ideal MHD limit), Alfv\'en (1942a) pointed out that magnetic flux is conserved inside the plasma system and moves with the fluid. This theorem is known as Alfv\'en's theorem of flux-freezing. It is well known from early nineteenth century observations that the Sun rotates differentially with the equator rotating faster than the pole. Since the flux is frozen inside the plasma, it allows differential rotation to stretch magnetic field lines along the direction of rotation (i.e., the toroidal or $\phi$ direction). This process is known as the $\Omega$-effect and was first pointed out by Larmor (1919).
 
 Theoretical and numerical magnetoconvection studies performed by Chandrasekhar (1952) and Weiss (1981) suggest that in the presence of magnetic field, convective systems get separated into regions that are free of magnetic field where vigorous convection takes place while magnetic fields are concentrated into thin structures in the form of flux tubes. It is also known that the presence of strong magnetic field makes the magneto-fluid more stable against convection, i.e., convection is suppressed within regions of strong magnetic field due to the tension of magnetic field lines (Thompson 1951; Chandrasekhar 1952). Since convection is suppressed in regions of strong magnetic field, there is less efficient heat transport in these regions. Because of this, sunspots appear darker than the surroundings. To sum up, it is expected that magnetic field exists in the form of flux tubes inside the solar convection zone and strong differential rotation of the Sun stretches these flux tubes in the toroidal i.e., $\phi$-direction.

Let us assume that the gas pressure inside the flux tube is $p_{int}$ and outside it is $p_{ext}$, $B$ is the strength of magnetic field inside the flux tube.
To maintain pressure balance across the surrounding surface of the flux tube:
\begin{equation}
p_{ext}=p_{int}+\frac{B^2}{2 \mu_0}.
\end{equation}
When the flux tubes are in isothermal condition, the above equation implies, $\rho_{ext} \geq \rho_{int}$. If such a situation arises in any part of the flux tube, then this part will experience a buoyancy force. Due to magnetic buoyancy this part rises up against gravity and generates bipolar sunspot pairs on the solar surface (Parker 1955a, 1955b). 

\subsection{Parker's Mean-Field Dynamo}

In spherical geometry, we can write the magnetic field as:
\begin{equation}
\mathbf{B}=B_r \hat{r} +B_\theta \hat{\theta} +B_\phi \hat{\phi}.
\end{equation}
We consider the stellar system as axisymmetric with the rotation axis coinciding with the axis of symmetry. Then $B_r$, $B_\theta$ and $B_\phi$ do not vary with $\phi$. In this situation, we can write the magnetic field as a sum of the toroidal ($\mathbf{B_t}$) and poloidal ($\mathbf{B_p}$) field components.
   \begin{equation}
\mathbf{B}=\mathbf{B_t}+\mathbf{B_p},
\end{equation}
 where $\mathbf{B_t}= B_\phi \hat{\phi}$  and $\mathbf{B_p}=   B_r \hat{r} +B_\theta \hat{\theta}$. \\

In this case, $\mathbf{B_p}$ can be expressed in terms of the vector potential:\\
\begin{equation}
\mathbf{B_p}=\nabla \times A \hat{\phi}.
\end{equation}

\begin{figure}[!t]
\centering
\begin{tabular}{cc}
\includegraphics*[width=0.5\linewidth]{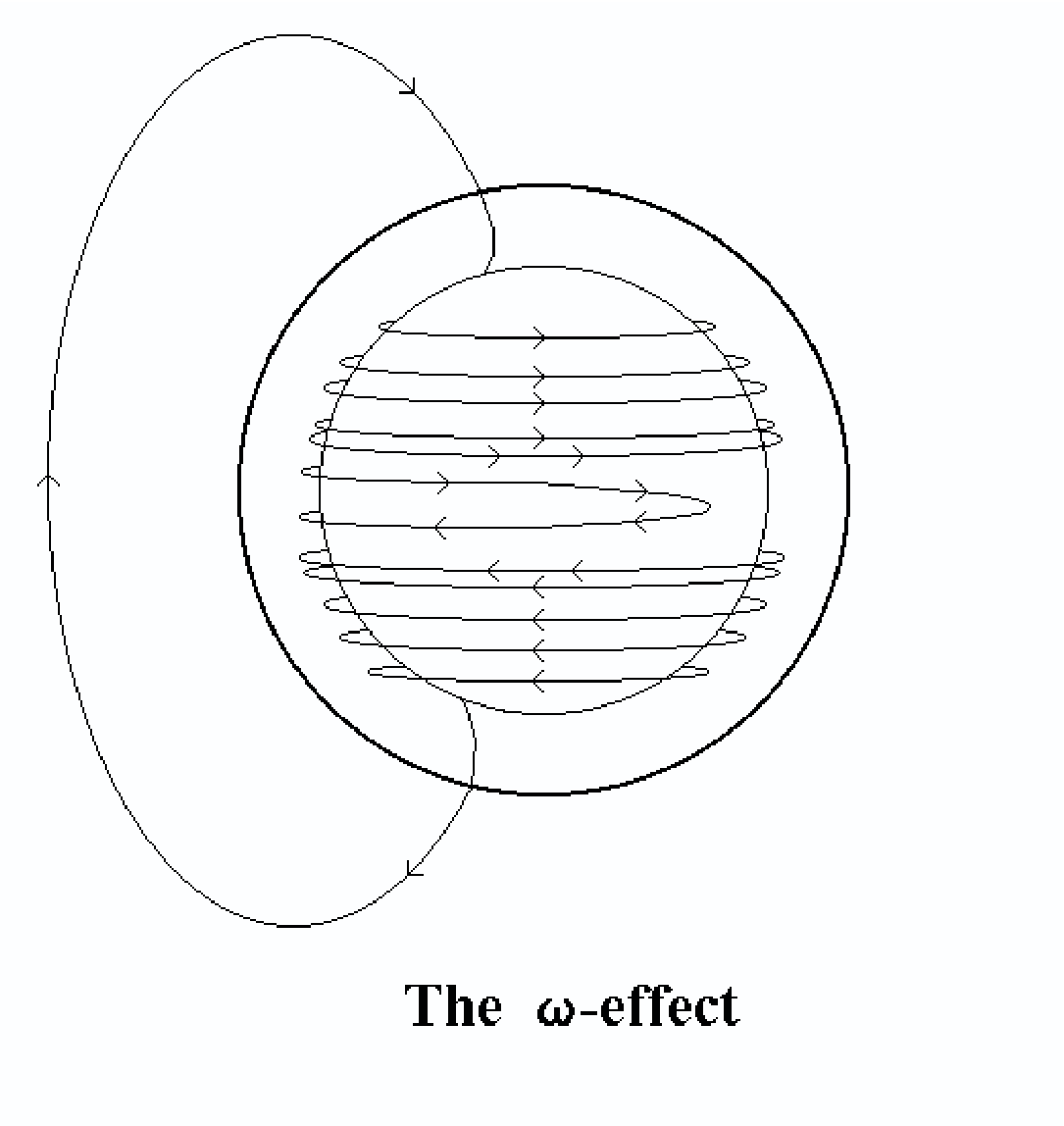} & \includegraphics*[width=0.5\linewidth]{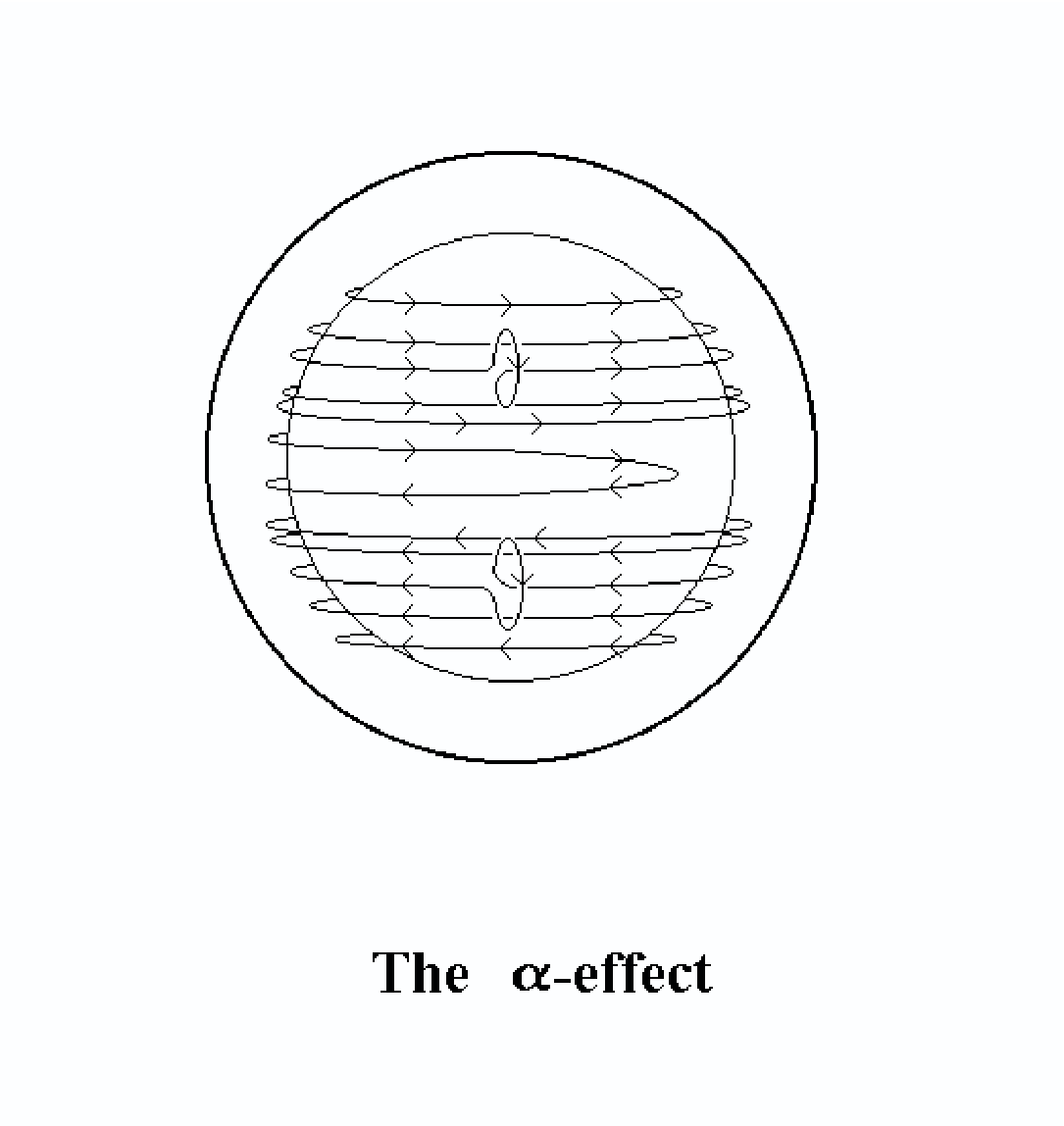}\\
                (a)                                           &                   (b)\\
 \end{tabular}                
\caption{Parker's turbulent dynamo: (a) The $\omega$-effect: Poloidal field lines are stretched by differential rotation in the solar interior and produces the toroidal component of magnetic field.  (b) $\alpha$-effect: At the time of rise through the convection zone, toroidal flux tubes are twisted due to helical turbulence and produces magnetic field components in the poloidal plane. Image credit: Hathaway/NASA/MSFC}
\end{figure}                
 To explain the origin of the solar cycle, Parker (1955b) first proposed the idea of flux recycling between the toroidal and poloidal field components.

The first part of the full dynamo mechanism ($Poloidal \rightarrow Toroidal$) relies on the idea proposed by Larmor (1919). In this process shearing of the large scale poloidal field due to strong differential rotation produces the toroidal field. 

The second part of the dynamo mechanism ($Toroidal \rightarrow Poloidal$) is a debated issue. The first breakthrough in this direction was proposed by Parker (1955b). The idea was that when some part of the toroidal flux tube rises through the convection zone due to magnetic buoyancy, they are subject to helical turbulence which imparts a twist to the rising plasma blobs. As magnetic field is frozen inside the plasma, this helical twist of plasma blobs also imparts a helical twist to the magnetic field. Thus rising toroidal flux tubes are helically twisted out of the plane and produce magnetic field component in the poloidal plane. Although the idea of Parker at that time was largely intuitive, it was put on rigorous mathematical footing after a decade through the development of mean field electrodynamics (Steenbeck, Krause \& R\"adler 1966).

\subsection{Mean Field Electrodynamics}
Turbulence is expected to play an important role in the solar dynamo as the solar convective zone is highly turbulent. As it is not possible to develop a deterministic theory to tackle turbulence, it is necessary to develop a statistical scheme based on average properties of turbulence. 

In a turbulent medium, we can decompose the fluid velocity ($\mathbf{v}$) and magnetic field ($\mathbf{B}$) in terms of mean and fluctuating parts. Thus:

\begin{equation}
  ~~~~~~~~~~~~~~~~~~~~~\mathbf{v}= \mathbf{\bar{v}} + \mathbf{v'}, ~~~~~~~~~~~~~~~~~~  \mathbf{B}= \mathbf{\bar{B}}+ \mathbf{B'}
\end{equation}
  where the term with overline corresponds to the mean and the primed terms correspond to the fluctuating parts. In mean field theory the mean typically denotes ensemble averages over length-scales and time-scales much larger than turbulent eddy length-scale and eddy turn over time scales. Also by definition, $\mathbf{\bar{v'}}=\mathbf{\bar{B'}}=0$, i.e., mean of the fluctuating components are zero. 
  Substituting (1.10) into the magnetic induction equation (1.1) we get:
  
\begin{equation}
 \frac{\partial \mathbf{\bar{B}}}{\partial t}+ \frac{\partial \mathbf{B'}}{\partial t}= \nabla \times (\mathbf{\bar{v}} \times \mathbf{\bar{B}} + \mathbf{v'} \times \mathbf{\bar{B}} + \mathbf{\bar{v}} \times \mathbf{B'} + \mathbf{v'} \times \mathbf{B'}) + \lambda \nabla^2(\mathbf{\bar{B}}+ \mathbf{B'}),
\end{equation}
where diffusivity ($\lambda$) is constant.   
Again averaging equation (1.11) term by term, we get:
  \begin{equation}
 \frac{\partial \mathbf{\bar{B}}}{\partial t}= \nabla \times (\mathbf{\bar{v}} \times \mathbf{\bar{B}}) + \nabla \times \mathbf{\varepsilon} + \lambda \nabla^2(\mathbf{\bar{B}} + \mathbf{B'}),
\end{equation}
where $\mathbf{\varepsilon}= \overline{\mathbf{v'} \times \mathbf{B'}}$ is known as the mean electromotive force which arises because of the correlation between fluctuating components of velocity and magnetic fields (Steenbeck, Krause \& R\"adler 1966; Krause \& R\"adler 1980).
Subtracting (1.12) from (1.11), we get
 \begin{equation}
 \frac{\partial \mathbf{B'}}{\partial t}= \nabla \times (\mathbf{v'} \times \mathbf{\bar{B}} + \mathbf{\bar{v}} \times \mathbf{B'} + \mathbf{v'} \times \mathbf{B'} - \mathbf{\varepsilon}) + \lambda \nabla^2 \mathbf{B'}.
\end{equation}
Let us assume at initial time ($t=0$), fluctuation in the magnetic field is zero. From equation (1.13), it is clear that if there is no fluctuation in the magnetic field ($\mathbf{B'}=0$) then there is a linear relationship between mean electromotive force ($\varepsilon$) and mean magnetic field ($\mathbf{\bar{B}}$). Now if we assume that the spatial scale of the fluctuation is very small compared to the mean magnetic field components, then we can express the mean emf in a Taylor series:
\begin{equation}
 \varepsilon_i= \alpha_{ij} \bar{B_j} + \beta_{ijk} \frac{\partial \bar{B_j}}{\partial x_k}+ \gamma_{ijkl} \frac{\partial^2 \bar{B_j}}{\partial x_k \partial x_l}+ ...+ a_{ij} \frac{\partial \bar{B_j}}{\partial t}+ b_{ijk} \frac{\partial^2 \bar{B_j}}{\partial x_k \partial t}+........
\end{equation}
 Considering spatial derivatives upto the first order, we get:
\begin{equation}
 \varepsilon_i= \alpha_{ij} \bar{B_j} + \beta_{ijk} \frac{\partial \bar{B_j}}{\partial x_k},
\end{equation}
 where the quantities $\alpha_{ij}$ and $\beta_{ijk}$ are pseudo tensors depending on $\bar{v}$ and $v'$. At this point, it is necessary to constrain $\alpha_{ij}$ and $\beta_{ijk}$, which is difficult because of our lack of knowledge about convective turbulence.

 We consider the simplest situation where the mean velocity field vanishes i.e., $\bar{v}=0$ and the turbulent velocity field ($v'$) is steady, homogeneous and isotropic. In that case we can construct $\alpha_{ij}$ and $\beta_{ijk}$ using only isotropic tensors. Thus we can write:
 \begin{equation}
 \alpha_{ij} = \alpha \delta_{ij} , ~~~~~~~\beta_{ijk} = -\beta \epsilon_{ijk},
\end{equation}
 where  $\alpha$ is a pseudo scalar. Thus we get the expression for turbulent emf as:
 \begin{equation}
 \mathbf{\varepsilon} = \alpha \mathbf{\bar{B}} - \beta \nabla \times \mathbf{\bar{B}}.
\end{equation}
 
Let us consider the turbulent medium as isotropic and inhomogeneous, then we can express $\alpha_{ij}$ as a sum of symmetric and antisymmetric components (as we are mainly interested in considering terms upto the first order, we do not express $\beta_{ijk}$ as a sum of symmetric and antisymmetric components). 
\begin{equation}
 \alpha_{ij} = \alpha^{(S)}_{ij} +\alpha^{(A)}_{ij}= \alpha \delta_{ij} -\epsilon_{ijk} \gamma_k .
\end{equation}
So, $\alpha^{(A)}_{ij} \bar{B_j}= (\mathbf{\gamma} \times \mathbf{B})_i$ ; Thus the expression for turbulent emf for isotropic inhomogeneous medium becomes:
\begin{equation}
 \mathbf{\varepsilon} = \alpha \mathbf{\bar{B}} -\mathbf{\gamma} \times \mathbf{\bar{B}} -\beta \nabla \times \mathbf{\bar{B}},
\end{equation}
 where,
\begin{equation}
 \alpha = -\frac{1}{3} \overline{\mathbf{v'} \cdot (\nabla \times \mathbf{v'})} \tau ,
\end{equation}
\begin{equation}
 \beta=\frac{1}{3} \overline{\mathbf{v'} \cdot \mathbf{v'}} \tau .
\end{equation}

We see from equation (1.20) that $\alpha$ is proportional to the helical motion in the turbulent medium, thus it represents the average helical motion inside the turbulent convective zone. The term $\tau$ indicates the correlation time for turbulence. The term $\beta$ has the same dimension of diffusivity but its origin is turbulence, therefore this term is known as turbulent diffusivity. The term $\mathbf{\gamma} \times \mathbf{\bar{B}}$ represents the advection of average field ($\mathbf{\bar{B}}$) with an effective pumping velocity $\mathbf{\gamma}$. Since the term ($\mathbf{\gamma}$) creates inhomogeneity in a homogeneous medium this term is known as turbulent pumping. Given that all physical quantities like pressure, temperature etc. inside the solar convection zone are strongly dependent on the radial coordinate, one can treat solar convection as anisotropic and inhomogeneous in the radial coordinate only. Due to this highly stratified nature of convection, there is an asymmetry between upward and downward flows (Hurlburt, Toomre \& Massaguer 1984). This asymmetry between upward and downward flow causes turbulent pumping. Similarly, gradient in density produces density pumping, topological asymmetry produces topological pumping. 

Let us substitute equation (1.17) in equation (1.12), then we get:
 \begin{equation}
 \frac{\partial \mathbf{\bar{B}}}{\partial t}= \nabla \times (\mathbf{\bar{v}} \times \mathbf{\bar{B}}) + \nabla \times (\alpha \mathbf{\bar{B}})+ \eta \nabla^2 \mathbf{\bar{B}},
\end{equation}
where $\eta=\lambda+\beta$, is the net magnetic diffusivity. Equation (1.22) represents the evolution of magnetic field in a homogeneous, isotropic turbulent medium. The first term in the right hand side (RHS) of the equation represents the advection of the magnetic field and the toroidal field generation process due to shearing, the second term represents the poloidal field generation process due to helical motions present in the turbulent medium and the last term in the RHS represents turbulent diffusion.

\subsection{Flux Tube Dynamics and the Babcock-Leighton Mechanism for Poloidal Field Generation}
\begin{figure}[t!]
\centering
\includegraphics*[width=0.75\linewidth]{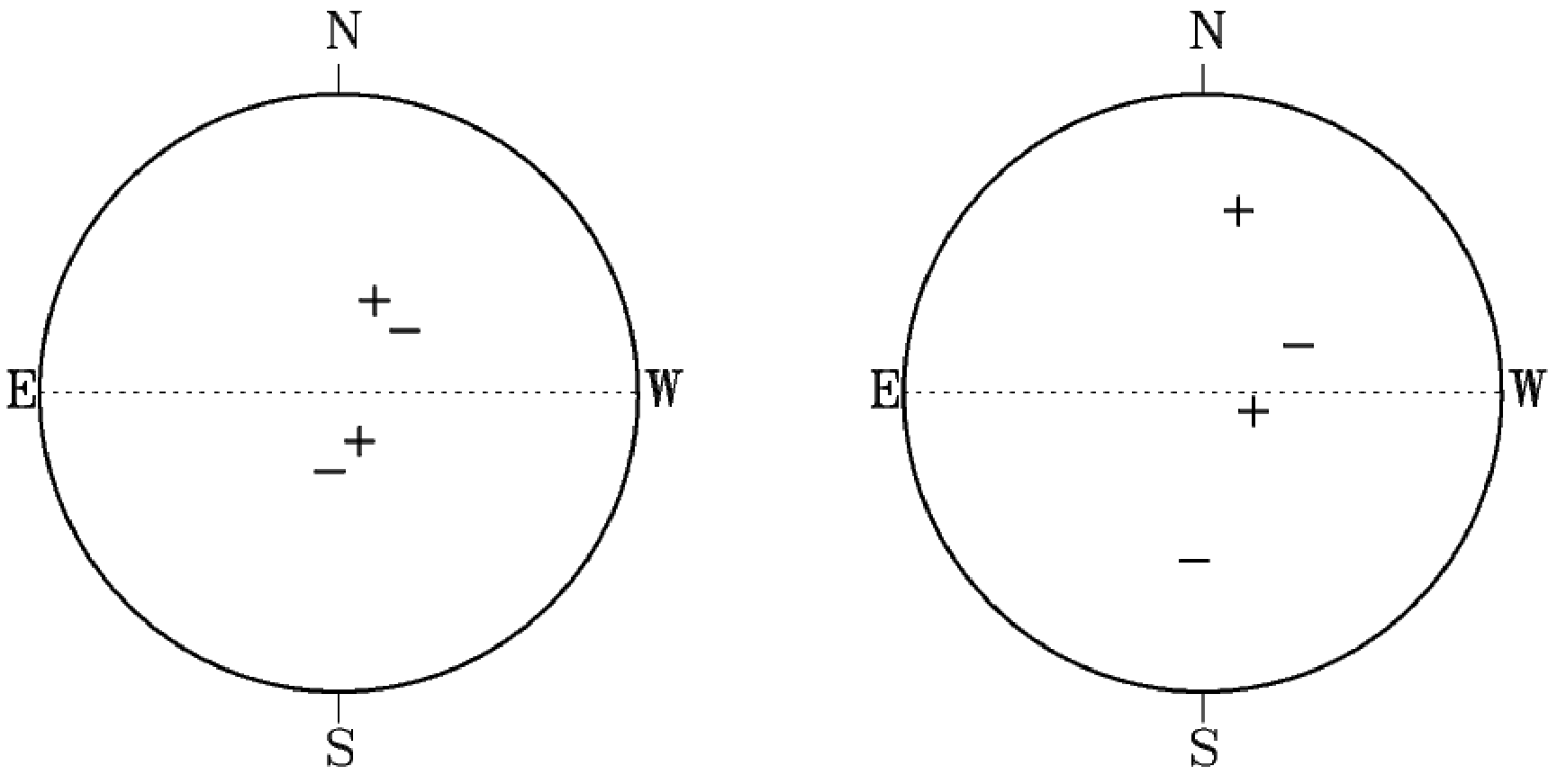}
  \caption{A cartoon image of the Babcock-Leighton mechanism: (a) Newly emerged bipolar magnetic regions with opposite leading/ following polarity patterns obeying Hale's polarity law. (b) Decaying bipolar magnetic regions, Trailing polarity goes to higher latitude while leading components reconnect across the equator. Image credit: Paul Charbonneau}
\end{figure}

Can the toroidal field generation take place throughout the full convection zone of the Sun? It was soon understood that the toroidal field generation due to shearing of the poloidal field is not possible throughout the full convection zone because of the destabilizing effect of magnetic buoyancy (Parker 1975; Moreno-Insertis 1983). Subsequently, dynamo theorists favored the thin overshoot layer at the base of the convection zone as the ideal place for amplification and storage of the magnetic field (Spiegel \& Weiss 1980; van Ballegooijen 1982; DeLuca \& Gilman 1986; Choudhuri 1990). After the helioseismic discovery of the tachocline with a strong radial gradient in rotation at the base of the convection zone, it is thought that the toroidal field generation and storage takes place in that layer.
 
Numerical simulations of buoyant flux tubes suggest that only flux tubes with initial field strength 50-100 KGauss are consistent with the observed tilt and emergence latitude of active regions (Choudhuri \& Gilman 1987; D'Silva \& Choudhuri 1993; Fan, Fisher \& DeLuca 1993; Fan, Fisher \& McClymont 1994; Caligari, Moreno-Insertis \& Sch\"ussler 1995; Fan \& Fisher 1996; Caligari, Sch\"ussler \& Moreno-Insertis 1998; Fan \& Gong 2000); also see D'Silva (1993). Flux tube simulations thus constrain the value of toroidal field at the base of the convection zone. This value is one order of magnitude higher than the equipartition field strength. At this strong field strength, helical turbulence would not be able to impart significant twist as required by the classical mean-field $\alpha$-effect suggested by Parker (1955).

This realization has resulted in adoption of an alternative idea for poloidal field generation. Babcock(1961) and Leighton (1969) proposed that poloidal field can be regenerated at the surface due to the decay and re-distribution of bipolar sunspot flux. This process is known as the Babcock-Leighton mechanism. It is also a well known observational fact that sunspots always appear in pairs at the surface with a systematic tilt with respect to the east-west direction. Because of this tilt angle, when sunspots decay, the flux from leading polarity preferentially diffuses towards equator whereas flux from trailing polarity is advected towards the poles (Fig.~1.5 a,b). As the polarity orientation is opposite in each hemisphere there is a net cancellation of flux across the equator and in the polar region accumulation of the new flux cancels the opposite polarity flux of the previous cycle and creates the new cycle polar field. Observationally it was found that the mean tilt angle of bipolar sunspot regions vary from cycle to cycle and there is a large scatter in the tilt angles (Dasi-Espuig et al. 2010). Since Poloidal field generation in this mechanism is strongly dependent on the tilt angle of the bipolar sunspot pairs, this mechanism itself is a source of irregularity (Choudhuri et al. 2007; Jiang et al. 2007). Recent observational results also lend strong support to this mechanism (Dasi-Espuig et al. 2010; Kitchatinov \& Olemskoy 2011a; Mu\~noz-Jaramillo et al. 2013). In recent years most of the kinematic dynamo models are based on the scenario that -- a) The toroidal field is produced due to strong differential rotation in the convection zone b) The poloidal field is produced near the solar surface due to decay of bipolar sunspot regions.

\subsection{Differential Rotation and Meridional Circulation: Essential Ingredients of Solar Dynamo Modelling}
In his classic paper, Parker (1955) showed that linear dynamo equations support periodically propagating dynamo wave solutions -- which signifies the solar cycle. The direction of such periodic propagating dynamo waves is given by:
 \begin{equation}
 s= \alpha \nabla  \Omega  \times \hat{\phi},
\end{equation}
  where $\Omega$ is the solar differential rotation  which arises mainly because of Reynolds stresses $<v_r v_\theta>$ and $<v_\theta v_\phi>$ (which creates angular momentum flux). To obtain the equatorward propagation of dynamo waves (in keeping with the equatorward migration of the sunspot belt), the following condition must be satisfied:
\begin{equation}
 \alpha \frac{\partial \Omega}{\partial r} < 0.
\end{equation}
This is known as the Parker-Yoshimura sign rule (Parker 1955; Yoshimura 1975). Since at that time the profile of the differential rotation inside the convection zone was unknown, there was full freedom to choose the profile of differential rotation such that results match with observation. When the differential rotation was measured by helioseismology with great accuracy (Thompson et al. 1996; Kosovichev et al. 1997; Schou et al. 1998) it was found that the observed differential rotation profile would give rise to poleward propagating dynamo solutions as there is a negative radial shear at low latitudes. At this point it was necessary to address this problem.

Observations of small magnetic features on the solar surface show that they are carried by surface flows from equator to pole with an estimated speed of 10-20 m/s (Komm, Howard \& Harvey 1993; Latushko 1994; Snodgrass \& Dailey 1996; Hathaway 1996). This axisymmetric poleward flow in the meridional plane is known as meridional circulation. Helioseismic measurements in later time also confirmed these observations and measured these poleward flows more accurately at the top 10 \% of the solar convection zone (Giles et al. 1997; Schou \& Bogart 1998; Braun \& Fan 1998, Gonz\'alez Hern\'andez et al. 1999). Although measurement of the meridional circulation deep in the convection zone is still not possible, it is reasonable to assume that there must be an equatorward return flow somewhere in the convection zone because of mass conservation. The latitudinal structure of the surface meridional flow is well observed (Hathaway \& Rightmire 2010,2011; Hathaway 1996) but the radial structure still remains largely controversial (Zhao et al. 2013). If these two largely segregated source layers ($\alpha$-effect is at the surface while $\Omega$-effect is at the base of the convection zone) are only coupled by diffusion (no meridional circulation is present), then there is poleward propagation of the sunspot belt according to the Parker-Yoshimura sign rule. However if these two layers are coupled by meridional circulation then we can get equatorward propagation of sunspot belts even if the Parker-Yoshimura sign rule is violated (Choudhuri et al. 1995; Durney 1995). After this insight meridional circulation became an essential ingredient in solar dynamo models.

\subsection{Kinematic Babcock-Leighton Dynamo Models} 
 In the kinematic dynamo problem, plasma flows are taken as inputs and it is assumed that the mean flows are not significantly altered by the Lorentz force. Observationally it is found that mean flows do not vary significantly with time thus the kinematic approach may be a good approximation to study the evolution of solar magnetic fields. In the axisymmetric spherical coordinate system, we can represent the magnetic and velocity fields as:

\begin{equation}
 ~~\mathbf{B}(r,\theta,t) = \nabla \times (A(r,\theta,t) \hat{\phi}) + B(r,\theta,t) \hat{\phi},~~~~ \mathbf{v} = r~ \sin (\theta) \Omega \hat{\phi}+ \mathbf{v_p},
\end{equation}
where $B(r,\theta,t)$ and $A(r, \theta,t)$ represents the toroidal magnetic field and vector potential for the poloidal magnetic field, $\Omega$ is the differential rotation and $\mathbf{v_p}$ is the meridional flow. 

Substituting these into the magnetic induction equation, we get the standard $\alpha \omega$ -dynamo equations:
\begin{equation}
   \frac{\partial A}{\partial t} + \frac{1}{s}\left[ \mathbf{v_p} \cdot \nabla (sA) \right] = \eta\left( \nabla^2 - \frac{1}{s^2}  \right)A + S(r,\theta,B),
\end{equation}
\begin{equation}
   \frac{\partial B}{\partial t}  + s\left[ \mathbf{v_p} \cdot \nabla\left(\frac{B}{s} \right) \right] + (\nabla \cdot \mathbf{v_p})B = \eta\left( \nabla^2 - \frac{1}{s^2}  \right)B + s\left(\left[ \nabla \times (A\bf \hat{e}_\phi) \right]\cdot \nabla \Omega\right)   + \frac{1}{s}\frac{\partial (sB)}{\partial r}\frac{\partial \eta}{\partial r},
\end{equation}
where $s= r~\sin (\theta)$. The terms with $\mathbf{v_p}$ in left hand side of both the equations correspond to the advection and deformation of magnetic field by meridional circulation. In the right hand side of both the equations, first term corresponds to the diffusion of magnetic field and second term corresponds to the source term. On the RHS of equation (1.27), the third term corresponds to the advection of magnetic field due to the gradient of turbulent diffusivity. In equation (1.26), $S(r,\theta,B)$ represents the generation of poloidal field due to the Babcock-Leighton 
mechanism. For the research work described in this thesis, we model the Babcock-Leighton mechanism i.e., the source term for poloidal field evolution equation by the methods of double ring proposed by Durney (1997) and  subsequently used by Nandy \& Choudhuri (2001), Mu\~noz-Jaramillo, 
Nandy, Martens \& Yeates (2010) and Nandy, Mu\~noz-Jaramillo \& Martens (2011). In equation (1.26) and (1.27), we have to define three input ingredients so that we can solve these system of equations: magnetic diffusivity, differential rotation and meridional circulation.

\begin{figure}[!t]
\centering
\begin{tabular}{cc}
\includegraphics*[width=0.5\linewidth]{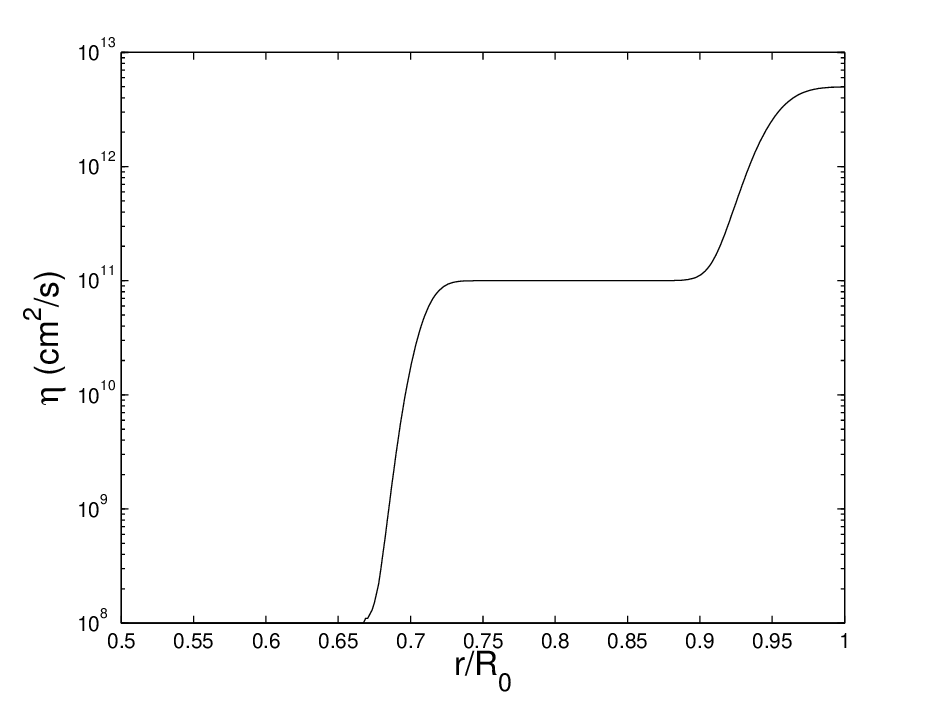} & \includegraphics*[width=0.5\linewidth]{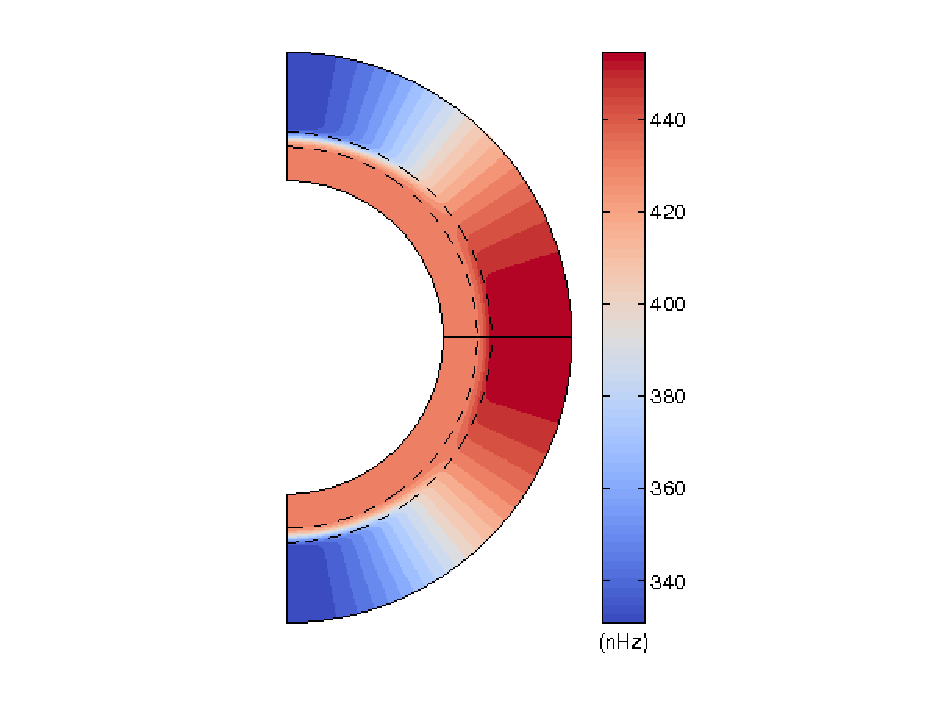}\\
                (a)                                           &                   (b)\\
 \end{tabular}                
\caption{(a) Variation of turbulent magnetic diffusivity diffusivity with radius. (b) Analytical differential rotation profile (in nHz) used in dynamo model. Region between two dashed circular arcs indicates the tachocline.}
\end{figure}
Now we discuss these three essential input ingredients in kinematic solar dynamo models, namely, turbulent diffusivity, differential rotation and meridional circulation. With the turbulent diffusivity term we try to capture the net effect of convective turbulence on the large scale magnetic field. This term also acts as a communicator between two source layers (as two source layers are segregated; the poloidal field generation takes place at the solar surface while the toroidal field generation takes place at the base of the convection zone). Here we use double step diffusivity profile (Dikpati et al. 2002; Chatterjee, Nandy \& Choudhuri 2004; Guerrero \& de Gouveia Dal Pino 2007; Jouve \& Brun 2007, Mu\~noz-Jaramillo, Nandy \& Martens 2009):
\begin{eqnarray}
 \eta(r) = \eta_{bcd} + \frac{\eta_{cz} - \eta_{bcd}}{2}\left( 1 + \operatorname{erf}\left( \frac{r - r_{cz}}{d_{cz}}  \right)
      \right) \nonumber \\
      ~~~~~~~~~~~+ \frac{\eta_{sg} - \eta_{cz} - \eta_{bcd}}{2}\left( 1 + \operatorname{erf}\left( \frac{r - r_{sg}}{d_{sg}},  \right) \right).
\end{eqnarray}
 where $\eta_{bcd}= 10^8$ cm$^2$/s, $\eta_{cz} = 10^{11}$ cm$^2$/s and $\eta_{sg} = 5 \times 10^{12}$ cm$^2$/s corresponds to the diffusivity at the bottom of computational domain, diffusivity in the convection zone and near surface supergranular diffusivity respectively. Transition from one value of diffusivity to another is characterized by $r_{cz} = 0.73R_\odot$, $d_{cz} = 0.015R_\odot$, $r_{sg} = 0.95R_\odot$ and $d_{sg} = 0.015R_\odot$. The typical double step magnetic diffusivity profile is shown in Fig.~1.6 (a). See Mu\~noz-Jaramillo et al. (2011) for a discussion on constraining the diffusivity profile.

One necessary input parameter in solar dynamo models is the differential rotation ($v_{\phi}$; which stretches poloidal field lines in the $\phi$-direction and creates the toroidal field). Here we use the differential rotation profile as prescribed by Mu\~noz-Jaramillo et al. (2009) (see Fig.~1.6 (b)):
 \begin{equation}\label{DRan}
   \begin{array}{cc}
      \Omega_A(r,\theta) = 2\pi\Omega_{c} + \pi\left( 1 + \operatorname{erf}\left( \frac{r - r_{tc}}{d_{tc}}  \right)\right) 
      \left( \Omega_{e} - \Omega_{c} + ( \Omega_{p} - \Omega_{e} )\Omega_S(\theta) \right) ,\\
      \\
      \Omega_S(\theta) = a\cos^2(\theta) + (1-a)\cos^4(\theta), \\
    \end{array}
\end{equation}
 where $r_{tc} = 0.7R_\odot$ i.e. the location of the tachocline, $d_{tc} = 0.025R_\odot$ i.e. half of thickness of the tachocline,
$\Omega_{c} = 432$ nHz i.e. rotation frequency of the core, $\Omega_{e} = 470$ nHz i.e. rotation frequency of the equator, 
 $\Omega_{p} = 330$ nHz i.e. rotation frequency of the pole and $a = 0.483$.
 
 \begin{figure}[!t]
\centering
\begin{tabular}{cc}
\includegraphics*[width=0.5\linewidth]{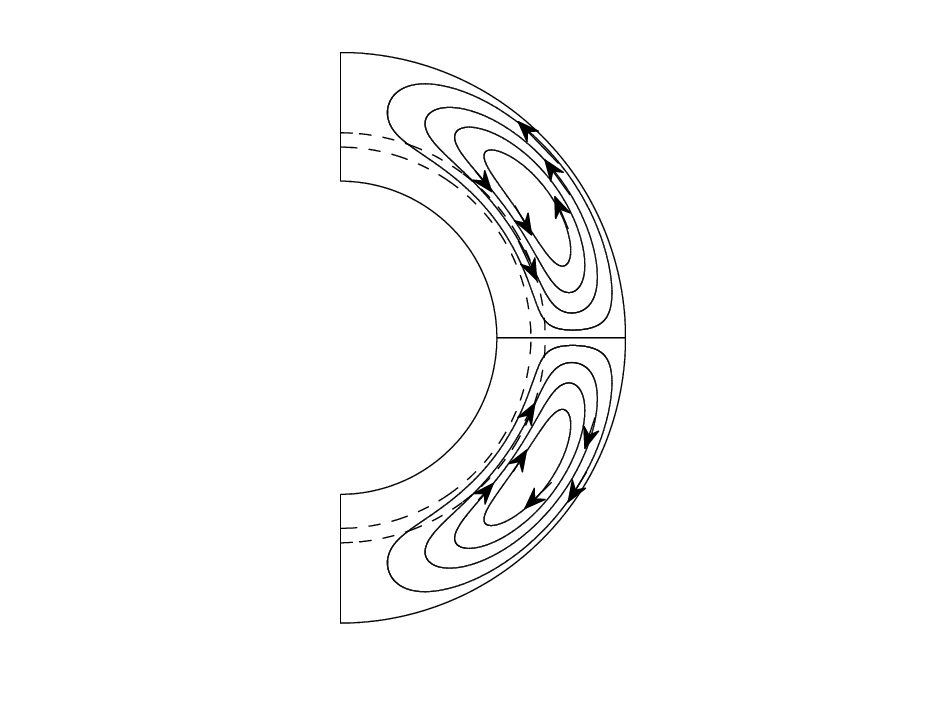} & \includegraphics*[width=0.5\linewidth]{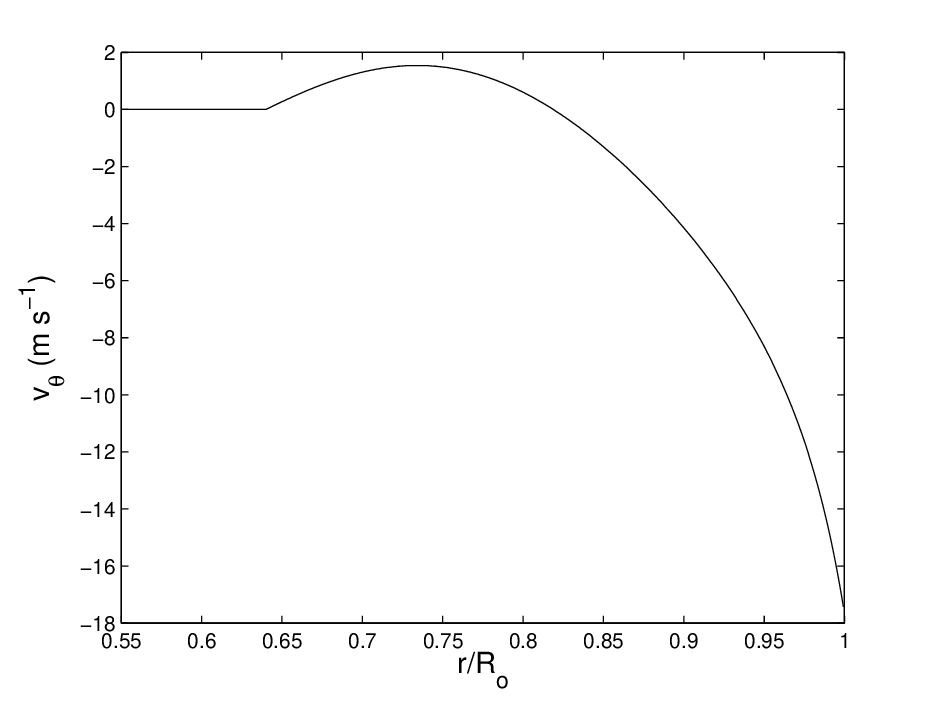}\\
                (a)                                           &                   (b)\\
 \end{tabular}                
\caption{(a) Meridional circulation streamlines used in our model. Region between two dashed
circular arcs indicates the tachocline. (b) Plot of latitudinal velocity ($v_\theta$ in m/s) as a function of $r/R_0$ at $45 ^0$ latitude.}
\label{fig1}
\end{figure}
\begin{figure}[!t]
\centering
\begin{tabular}{cc}
\includegraphics*[width=0.5\linewidth]{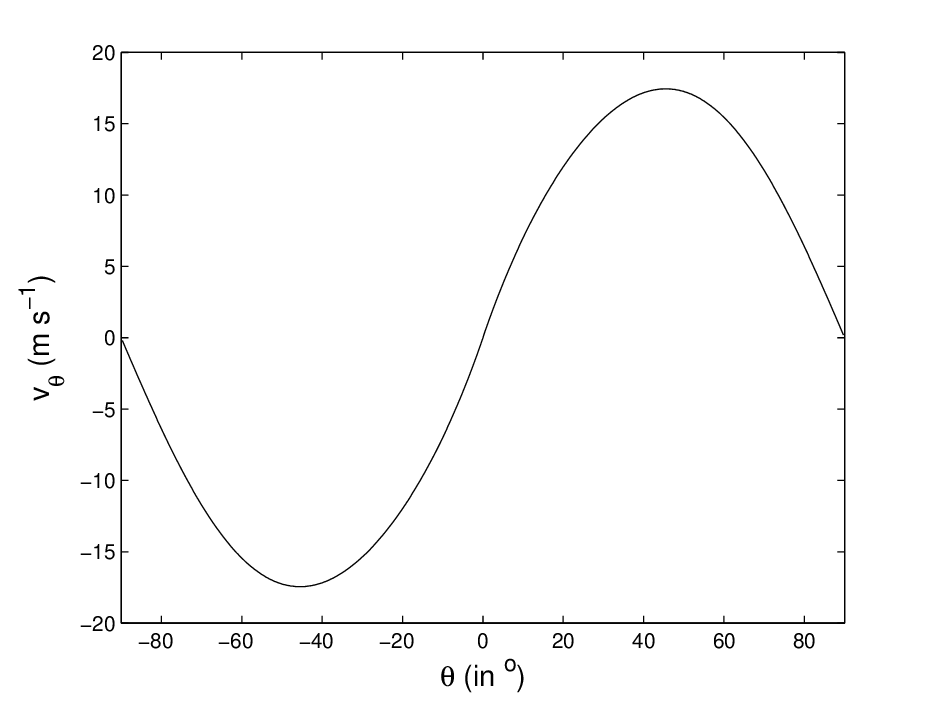} 
 \end{tabular}                
\caption{ Plot of latitudinal velocity ($v_\theta$ in m/s) as a function of latitude ($\theta$) at the solar surface.}
\label{fig1}
\end{figure}
The meridional circulation (i.e., velocity component in the $r-\theta$ plane) can be defined as
 \begin{equation}
    \mathbf{v_p}\left(r,\theta\right) =
    \frac{1}{\rho(r)} \nabla \times \left(\psi(r,\theta)\widehat{\textbf{e}}_{\phi}\right).
\end{equation}
which is estimated from the stream function (defined within $0 \leq \theta \leq \pi/2$, i.e., in the northern hemisphere) as described in Chatterjee, Nandy and Choudhuri (2004):
\begin{eqnarray}
\psi r \sin \theta = \psi_0 (r - R_p) \sin \left[ \frac{\pi (r - R_p)}
{(R_\odot - R_p)} \right] \{ 1 - e^{- \beta_1 \theta^{\epsilon}} \}\nonumber \\
 \{1 - e^{\beta_2 (\theta - \pi/2)} \} e^{-((r -r_0)/\Gamma)^2},       
\end{eqnarray}
where the parameters are defined as follow: $ \beta_1=1.5,~ \beta_2=1.8,~ \epsilon=2.0000001,~ r_0=(R_\odot-R_b)/4,~ \Gamma=3.47 \times 10^8~m$. The term $\psi_0$ determines the the maximum speed of the flow. $R_p$ is the penetration depth of the meridional flow. The meridional circulation profile in the southern hemisphere is generated by a mirror reflection of the velocity profile across the equator. Streamlines of a typical meridional circulation profile are shown in Fig.~1.7 (a). 

\begin{figure}[t!]
\centering
\includegraphics*[width=0.75\linewidth]{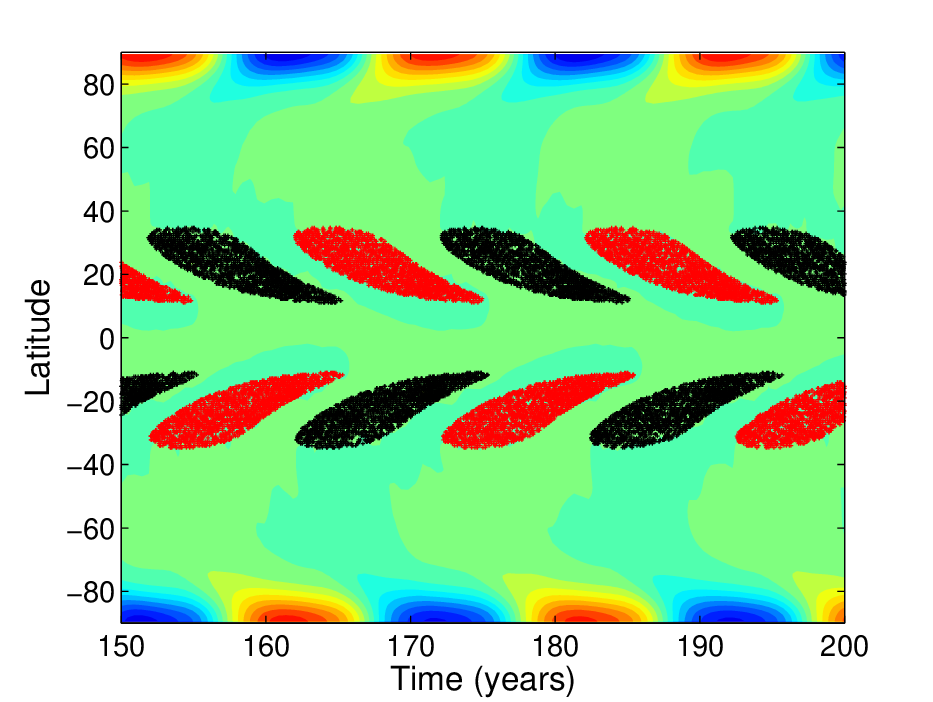}
  \caption{Butterfly diagram obtained from our flux transport dynamo model where background shows weak diffuse radial field on solar surface and eruption latitudes are denoted by symbols black (“+”) and red (“+”), indicating underlying negative and positive toroidal field respectively.}
\end{figure}
 Taking the density stratification in the convection zone as $\rho= C(\frac{R_\odot}{r}-\gamma)^m$ (from standard solar model; Christensen-Dalsgaard et al. 1996), we get the speed of meridional circulation:\\
 \begin{equation}
 v_r=\frac{1}{\rho r^2 \sin \theta} \frac{\partial}{\partial \theta} (\psi r \sin \theta), 
 \end{equation}
 \begin{equation}
   v_\theta=-\frac{1}{\rho r \sin \theta} \frac{\partial}{\partial r} (\psi r \sin \theta),
 \end{equation}
 where, $\gamma=0.95$, $m=3/2$.
 
Some studies find the peak speed of the flow is in the range of 12-20 m/s (Komm, Howard \& Harvey 1993; Braun \& Fan 1998; Gizon \& Rempel 2008). Here we set the peak speed of the meridional circulation at mid latitudes to be 18 m/s. Fig.~1.7 (b) shows the variation of the $\theta$ component of this velocity at mid latitude as a function of solar radius. Fig.~1.8 shows the variation of the $\theta$ component of this velocity as a function of latitude at the solar surface. 
 
 Having defined all the input ingredients we can solve equations (1.26) and (1.27) with appropriate boundary conditions. We perform all of our spatially extended dynamo simulations in a meridional slab $0.55 R_\odot < r < R_\odot$ and within $0 < \theta < \pi$. Since we use axisymmetric equations, both the poloidal and toroidal field vectors need to be zero ($ A=0$ and $B=0$) at the pole ($\theta=0$ and $\theta=\pi$), to avoid any singularity. At the bottom of the computational domain i.e., at $r=0.55 R_\odot$, we assume perfectly conducting material, thus both field components vanish ($ A=0$ and $B=0$) at the bottom boundary. At the top ($r=R_\odot$), we assume that there is only radial component of magnetic field ($B=0$ and $\partial (rA)/\partial r =0$); this is necessary for stress balance between subsurface and coronal magnetic fields (van Ballegooijen \& Mackay 2007). We set $A=0$ and $B \propto \sin(2 \theta)*\sin(\pi*((r-0.55 R_\odot)/(R_\odot-0.55 R_\odot)))$ as initial conditions for our 2.5D dynamo simulation.
 
Fig.~1.9 shows a simulation from our solar dynamo model. From this plot, we find that our dynamo model is able to represent some of the important features of the solar cycle, e.g., periodicity, phase relationship and relative hemispheric orientation of both the toroidal and poloidal field, sunspot eruption latitudes, equatorward migration of sunspot belts and the pole-ward migration of the radial field.
 \begin{figure}[t!]
\centering
\includegraphics*[width=0.75\linewidth]{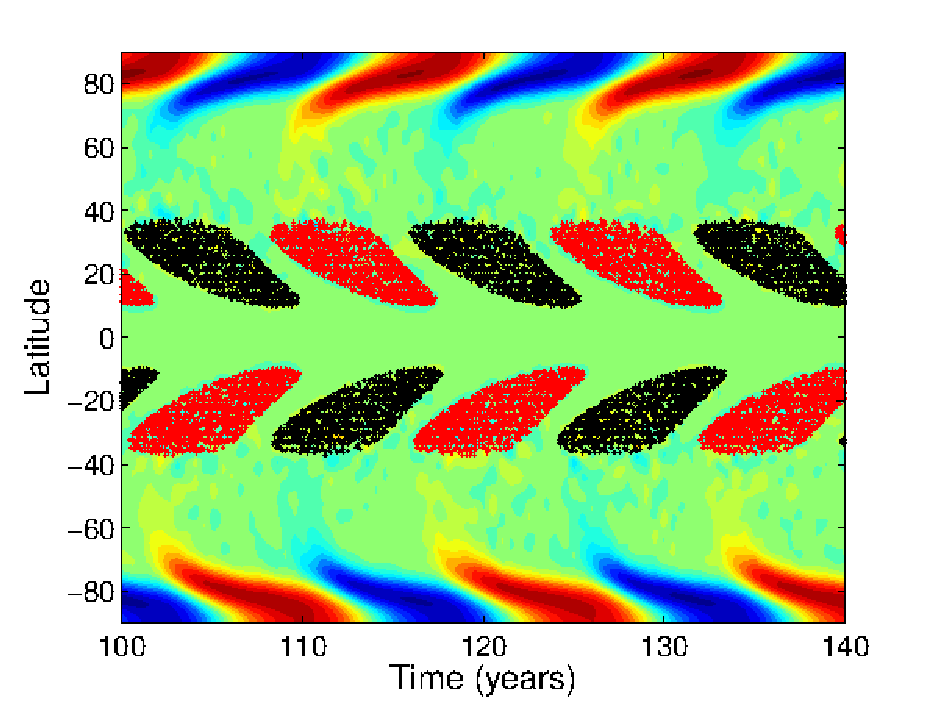}
  \caption{Butterfly diagram obtained from our flux transport dynamo model where background shows weak diffuse radial field on solar surface and eruption latitudes are denoted by symbols black (“+”) and red (“+”), indicating underlying negative and positive toroidal field respectively. This plot is obtained using the meridional flow profile suggested by Mu\~noz-Jaramillo et al. (2009).}
\end{figure} 

We also point out that our model results are robust w.r.t. the assumed profile of internal meridional flow. We demonstrate this by utilizing a different meridional flow profile as suggested by Mu\~noz-Jaramillo et al. (2009) based on fits to the helioseismic data, wherein the stream function and profile is given by:
 \begin{equation}
 \psi(r,\theta) = \frac{v_0}{r} (r-R_p)(r-R_\odot) \left[\sin (\pi \frac{r-R_p}{R_1-R_p})\right]^a \sin ^{q+1} (\theta) \cos \theta,
  \end{equation}
  where, q=1 governs the latitudinal dependence, $R_p = 0.65 R_\odot$ is the penetration depth, a=1.92 and $ R_1= 1.027 R_\odot$ govern the location of the peak of poleward flow and the amplitude and location of equatorward return flow. $v_0= 18$ is used to control the amplitude of meridional flow. We find that the results do not change qualitatively with the alternative meridional flow profile (see Fig.~1.10). 
      
 In the next few chapters of this thesis, we will investigate some important features of the solar cycle using this model. 

\chapter{Exploring Grand Minima Phases with a Low Order, Time Delay Dynamo Model}

Fluctuations in the Sun's magnetic activity, including episodes of grand minima such as the Maunder minimum have important consequences for space and planetary environments. However, the underlying dynamics of such extreme fluctuations remain ill-understood. In this chapter we develop a low order time delay dynamo model removing all space dependent terms. Introduction of time delays capture the physics of magnetic flux transport between spatially segregated dynamo source regions in the solar interior. We follow the Babcock-Leighton approach to treat the poloidal field generation mechanism due to decay and dispersal of tilted bipolar sunspot region. Introducing stochastic fluctuation in Babcock-Leighton source term for poloidal field generation, we demonstrate that the Babcock-Leighton poloidal field source based on dispersal of tilted bipolar sunspot flux, alone, can not recover the sunspot cycle from a grand minimum. We find that an additional poloidal field source effective on weak fields -- e.g., the mean-field $\alpha$-effect driven by helical turbulence -- is necessary for self-consistent recovery of the sunspot cycle from grand minima episodes.

\section {Introduction} 

Sunspots, which are strongly magnetized regions, play a key role in governing the activity of the Sun. The number of sunspots observed on the solar surface waxes and wanes with time generating the 11-year solar cycle. While there is a small variation in this periodicity, fluctuations in the amplitude of the solar cycle are large. Extreme fluctuations are manifest in grand maxima episodes -- when the cycle amplitudes are much higher than normal, and grand minima episodes -- when the cycle amplitudes fall drastically, even leading to the disappearance of sunspots for an extended period of time. The most striking evidence of such a minimum in the recorded history of sunspot numbers is the so-called Maunder minimum between 1645 and 1715 AD (Eddy 1988). The lack of sunspots during this period is statistically well-proven and is not due to the lack of observations -- which covered 68\% of the days during this period  (Hoyt \& Schatten 1996). The occurrence of these solar activity extremes is correlated with temperature records over millennium scale (Usoskin et al. 2005); the solar Maunder minimum coincided with the severest part of the Little Ice Age -- a period of global cooling on Earth.

Over the last decade, solar activity reconstructions based on cosmogenic isotopes and geomagnetic data (Usoskin et al. 2000, 2003, 2007; Miyahara et al. 2004; Steinhilber et al. 2010; Lockwood \& Owens 2011), which are indirect proxies for probing long-term solar activity have brought to the fore various properties of these grand minima episodes. These observations show that there have been many such activity minima in the past; however the solar cycle has recovered every time and regained normal activity levels. There is some evidence for persistent, but very weak amplitude cycles during the Maunder minimum and a slow strengthening of cycle amplitudes to normal levels during the recovery phase. While the general perception was that the onset of the Maunder minimum was sudden, a recent reconstruction based on historical sunspot records has challenged that notion indicating that the onset phase of the minimum may have been gradual (Vaquero et al. 2011).

A magnetohydrodynamic (MHD) dynamo mechanism, involving interactions of plasma flows and magnetic fields drives the solar cycle. Our understanding of the solar dynamo, see e.g., the reviews by  Ossendrijver (Ossendrijver 2003) and Charbonneau (Charbonneau 2010), is based on the generation and recycling of the toroidal and poloidal components of the Sun's magnetic field. The toroidal magnetic field is produced by stretching of poloidal field lines by differential rotation -- a process termed as the $\Omega$-effect (Parker 1955). It is thought this process is concentrated near the base of the solar convection zone (SCZ) -- where the upper part of the tachocline (a region of strong radial gradient in the rotation) and overshoot layer (which is stable to convection) offers an ideal location for toroidal field amplification and storage. Sufficiently strong toroidal flux tubes are magnetically buoyant and erupt radially outwards producing sunspots where they intersect the solar surface.

For the dynamo to function, the poloidal component has to be regenerated back from the toroidal component, a step for which, diverse propositions exist. The first such proposition invoked helical turbulent convection as a means of twisting rising toroidal flux tubes to regenerate the poloidal component (a mechanism traditionally known as the the mean field $\alpha$-effect; Parker 1955). Numerous dynamo models based on the mean-field $\alpha$-effect were constructed and such models enjoyed a long run as the leading contender for explaining the origin of the solar cycle (Charbonneau 2010). However, subsequent simulations of the dynamics of buoyant toroidal flux tubes and observational constraints set by the tilt angle distribution of sunspots pointed out that the toroidal magnetic field at the base of the SCZ must be as high as $10^5$ G (D'Silva \& Choudhuri 1993; Fan et al. 1993; Caligari et al. 1995); such strong toroidal flux tubes being one order of magnitude stronger than the equipartition magnetic field in the SCZ would render the mean field $\alpha$-effect ineffective. This consideration revived interest in an alternative mechanism of poloidal field production based on the flux transport mediated decay and dispersal of tilted bipolar sunspots pairs in the near-surface layers (Babcock 1961; Leighton 1969), hereby, referred to as the Babcock-Leighton mechanism. 
 
In the last couple of decades, multiple dynamo models have been based on this idea (Durney 1997; Dikpati \& Charbonneau 1999; Nandy \& Choudhuri 2002; Chatterjee et al. 2004; Mu\~noz-Jaramillo et al. 2009) and have successfully reproduced many nuances of the solar cycle. Some (Tobias et al. 2006; Bushby \& Tobias 2007; Cattaneo \& Hughes 2009) have criticised the usage of such mean-field dynamo models to predict the solar cycle, however it should be noted that recent studies (Simard et al. 2013; Dube \& Charbonneau 2013) indicate that if input profiles are extracted from three-dimensional full MHD simulations and fed into two-dimensional mean-field dynamo models, they are capable of producing qualitatively similar solutions to those found in the full MHD simulations. The major advantage of the mean-field dynamo framework is that it allows for much faster integration times compared to the full MHD simulations and are therefore computationally efficient as well as physically transparent. Recent observations also lend strong support to the Babcock-Leighton mechanism (Dasi-Espuig et al. 2010; Mu\~noz-Jaramillo et al. 2013) and this is now believed to be the dominant source for the Sun's poloidal field. Surface transport models (Wang et al. 1989; van Ballegooijen et al. 2010) also provide theoretical evidence that this mechanism is in fact operating in the solar surface. Randomness or stochastic fluctuations in the Babcock-Leighton poloidal field generation mechanism is an established method for exploring variability in solar cycle amplitudes (Charbonneau \& Dikpati 2000; Charbonneau et al. 2004, 2005; Passos \& Lopes 2011; Passos et al. 2012, Choudhuri \& Karak 2012) as are deterministic or non-linear feedback mechanisms (Wilmot-Smith et al. 2005; Jouve et al. 2010). Stochastic fluctuations within the dynamo framework are physically motivated from the random buffeting that a rising magnetic flux tube endures during its ascent through the turbulent convection zone and from the observed scatter around the mean (Joy's law) distribution of tilt angles. It is to be noted that similar fluctuations are to be expected in the mean-field $\alpha$ effect as well (Hoyng 1988) and such phenomenon can be explored within the framework of truncated mean-field dynamo models (Yoshimura 1975).

Since the two source layers for toroidal field generation (the $\Omega$ effect) and poloidal field regeneration (the $\alpha$-effect) are spatially segregated in the SCZ, there must be effective communication to complete the dynamo loop.  Magnetic buoyancy efficiently transports toroidal field from the bottom of the convection zone to the solar surface. On the other hand, meridional circulation, turbulent diffusion and turbulent pumping share the role of transporting the poloidal flux from the surface back to the solar interior (Karak \& Nandy 2012) where the toroidal field of the next cycle is generated thus keeping the cycles going. Thus, there is a time delay built into the system due to the finite time required for transporting magnetic fluxes from one source region to another within the SCZ.

Based on delay differential equations and the introduction of randomness on the poloidal field source, here we construct a novel, stochastically forced, non-linear time delay dynamo model for the solar cycle to explore long-term solar activity variations. We particularly focus our investigations on the recovery from grand minima phases and demonstrate that the Babcock-Leighton mechanism alone -- which is believed to be the dominant source for the poloidal field -- cannot restart the solar cycle once it settles into a prolonged grand minimum. The presence of an additional poloidal field source capable of working on weak magnetic fields, such as the mean field $\alpha$-effect is necessary for recovering the solar cycle.\\

\section{Stochastically Forced, Non-Linear, Time Delay Solar Dynamo Model}

The model is an extension of the low order time delay dynamo equations previously explored by Wilmot-Smith et al. (2006). This model was derived considering only the source and dissipative mechanisms in the dynamo process. All space dependent terms were removed and instead the physical effect of flux transport through space was captured through the explicit introduction of time delays in the system of equations.

The time delay dynamo equations are given by
 \begin{eqnarray}
    \frac{dB_\phi(t)}{dt}&=& \frac{\omega}{L} A(t-T_0)-\frac{B_\phi(t)}{\tau}\ \\
    \frac{dA(t)}{dt}&=&\alpha_0 {f_1} (B_\phi(t-T_1)) B_\phi(t-T_1) -\frac{A(t)}{\tau} \, ,
 \end{eqnarray}
where $B_\phi$ represents toroidal field strength and $A$ represents poloidal field strength. The evolution of each magnetic component is due to the interplay of the source and dissipative terms in the system. In the toroidal field evolution equation $\omega$ is the difference in rotation rate over the depth of the SCZ and $L$ is the depth of SCZ. Thus $\omega/L$ corresponds to the average shear in the differential rotation. The dissipative term is governed by turbulent diffusion, characterized by the diffusion time scale ($\tau$). The parameter $T_0$ is the time delay for the conversion of poloidal field into toroidal field and is justified by the finite time that the meridional circulation or turbulent pumping takes to transport the poloidal magnetic flux from the surface layers to the tachocline. $T_1$ is the time delay for the conversion of toroidal field into poloidal field and accounts for the buoyant rise time of toroidal flux tubes through the SCZ. 
\begin{figure}[t!]
 \centering
\includegraphics*[width=0.75\linewidth]{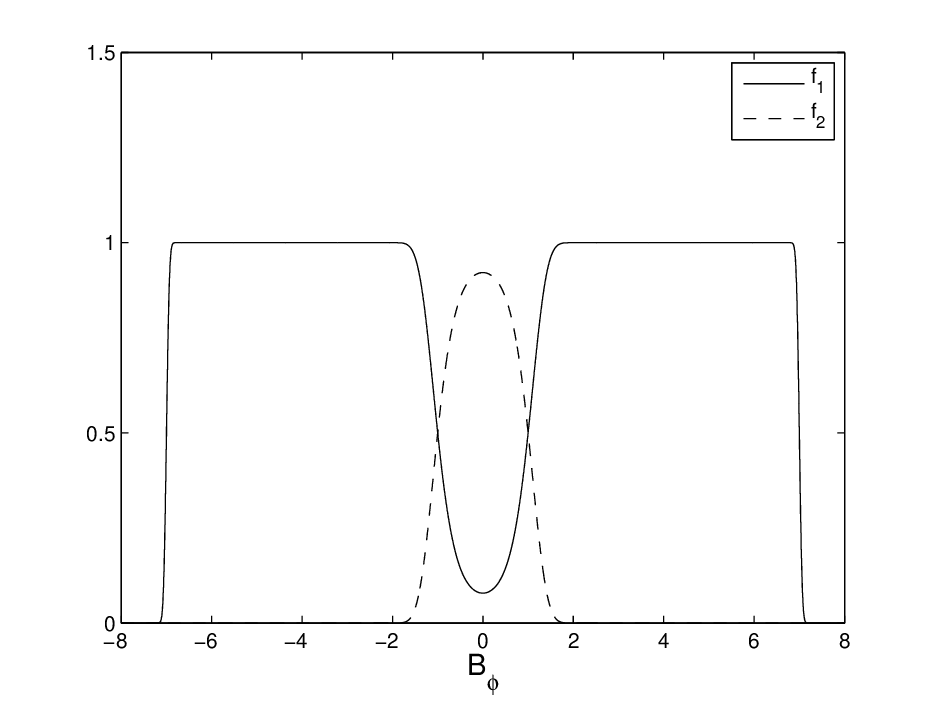}
  \caption{Profile of the quenching function $f_1$ for the Babcock-Leighton $\alpha$ and $f_2$ for the weak, mean field $\alpha$-effect (described later in the text). The plot of $f_1$ corresponds to parameters $B_{min}=1$ and $B_{max} =7$ and $f_2$ corresponds to $B_{eq}=1$ (all in arbitrary code units).}
  \label{fig1}
\end{figure}
The meridional circulation timescale is about 10 yr for a peak flow speed 20 m$s^{-1}$ (from mid-latitudes at near-surface layers  to mid-latitudes above the convection zone base; see Yeates et al. 2008 for detailed calculation of meridional circulation time scale). Another dominant flux transport mechanism for downward transport of magnetic field could be turbulent flux pumping with a timescale of about one yr (with a relatively high pumping speed of 5 m$s^{-1}$). The buoyant rise time of flux tubes from the SCZ base to surface is about three months (assuming the rise timescale is of the order of Alfv\'enic time scale, which is also a general agreement with simulations; see also Fan et al. 1993). As the magnetic buoyancy time scale is much shorter compared to the meridional circulation (or turbulent diffusion or flux pumping) timescale, we assume $T_1<<T_0$. Since it is not clear which is the most dominant flux transport mechanism - meridional circulation or turbulent pumping, we explore our model in two different regimes of operation to test for robustness. In one setup, we consider $T_0 = 4 T_1$ (if $T_0$ corresponds to pumping time scale) and $T_0= 40 T_1$ (if $T_0$ corresponds to meridional circulation time scale). This model setup mimics spatial separation between two source layers in the Sun's convection zone and the role of magnetic flux transport between them and is therefore physically motivated. On the other hand, due to its nature, this model is amenable to long time-integration without being computationally expensive.

To account for quenching of the Babcock-Leighton poloidal source $\alpha$, we take a general form of $\alpha$, i.e $\alpha=\alpha_0 {f_1}$, where $\alpha_0$ is the amplitude of the $\alpha$ effect and $f_1$ is the quenching factor approximated here by a nonlinear function
\begin{eqnarray}
     f_1=\frac{[1+\operatorname{erf}(B^2_{\phi}(t-T_1)-B^2_{min})]}{2}
     \times \frac{[1-\operatorname{erf}(B^2_{\phi}(t-T_1)-B^2_{max})]}{2}.
\end{eqnarray}
Figure 2.1 depicts this quenching function, constructed with the motivation that only flux tubes with field strength above $B_{min}$  (and not below) can buoyantly rise up to the solar surface and contribute to the Babcock-Leighton poloidal field source, i.e., sunspots (Parker 1955) and that flux tubes stronger than $B_{max}$ erupt without any tilt therefore quenching the poloidal source (D'Silva \& Choudhuri 1993; Fan et al. 1993). Accounting for these lower and upper operating thresholds for the Babcock-Leighton poloidal source is fundamentally important for the dynamics.
\begin{figure}[t!]
 \centering
\includegraphics*[width=0.75\linewidth]{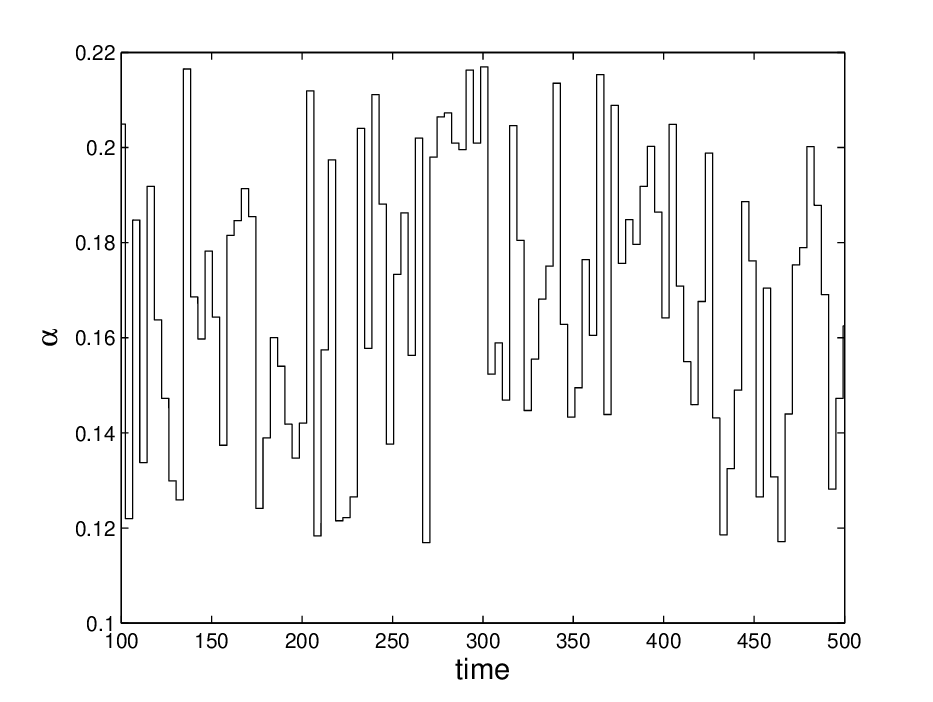}
  \caption{Stochastic fluctuations in time in the poloidal source term $\alpha$ at a level of 30\% ($\delta=30$) with a correlation time ($\tau_{cor}=4$) using our random number generating programme.}
\label{fig2}
\end{figure}
Our aim here is to explore the impact of stochastic fluctuations in this time delay solar dynamo model. For $\alpha=\alpha_0$, we get a strictly periodic solution. In order to introduce stochastic fluctuations, we redefine $\alpha$ as
\begin{equation}
    \alpha=\alpha_0\,[1+\frac{\delta}{100} \sigma(t,\tau_{cor})],
\end{equation}
where $\sigma(t,\tau_{cor})$ is a uniform random function lying in the range [+1,-1], changing values at a coherence time, $\tau_{cor}$. Statistical fluctuations are characterized by $\delta$ and $\tau_{cor}$, which correspond to percentile level of fluctuation and coherence time correspondingly. Figure 2.2 shows a typical $\alpha $ fluctuation generated by our random number generation program. Stochastic variations in the Babcock-Leighton $\alpha$ coefficient are natural because they arise from the cumulative effect of a finite number of discrete flux emergences, i.e., active region eruptions, all with various degrees of tilt randomly scattered  around a mean Joy's law distribution.

In this system the dynamo number ($N_D= \alpha_0 \omega \tau^2/L$) is defined as the ratio between the source and dissipative terms, which is a measure of the efficiency of the dynamo mechanism. The product of source terms is $|\alpha_0 \omega/L|$ while that of the dissipative terms is $1/\tau^2$. In terms of physical parameters, the expected diffusion time scale ($L^2/\eta$) in the SCZ is 13.8 yr for a typical diffusivity of $10^{12} cm^2 s^{-1}$ implying that the dissipative term ($1/\tau^2$) is of the order of $10^{-18} s^{-2}$. Now, if we take the value of $\omega$ as the difference in rotation rate across the SCZ in nHz (as measured; for details see Howe 2009) , L as the length of SCZ and $\alpha_0$ as 1 m $s^{-1}$ then the source term $|\alpha_0 \omega/L|$, is of the same order as the dissipative term and the dynamo number can be made higher than unity by slightly adjusting the $\alpha$ coefficient. In fact, if the tachocline is considered as the interface across which flux transport is occurring, then the dynamo number becomes even greater as the radial differential rotation is about the same while the length scale reduces further. In this model we always take the value of $|\alpha_0 \omega/L|$ (source term) to be greater than $1/\tau^2$ (decay term), and set the magnitude of $|\omega/L|$ and $|\alpha_0|$ in a way such that the strength of toroidal field is greater than the strength of poloidal field (as suggested by observations). In summary, keeping all of the other physically motivated parameters fixed, the dynamo number can be varied by adjusting the value of $\alpha_0$. Since $B_{min}$ corresponds to the equipartition field strength (on the order of $10^4$ Gauss) above which magnetic flux tubes become buoyant while $B_{max}$ is on the order of $10^5$ Gauss (above which flux tubes emerge without any tilt, thus shutting off the Babcock-Leighton source; D'Silva \& Choudhuri 1993), we take the ratio of $B_{max}/B_{min}$ as 7 for all of our calculation. Here we explore our low order time delay model in two parameter space regimes to test for robustness. In the first case we fix the parameters as $ \tau=15, B_{min}=1, B_{max}=7, T_0=4 T_1, T_1=0.5$ and $\omega/L=-0.34$ while in the second case we take $ \tau=25, B_{min}=1, B_{max}=7, T_0=40 T_1, T_1=0.5$ and $\omega/L=-0.102$.  Initial conditions are taken to be $(B_{min}+B_{max})/2$ for both A and $B_\phi$.  Our choice of parameters ensures that in both cases the diffusive timescale is much higher than flux transport timescales ($\tau > T_0 + T_1$). The simulations are robust over a range of negative $N_D$ values; for a detailed parameter space study of the underlying model without stochastic fluctuations, please refer to Wilmot-Smith et al. (2006). Below, we present the results of our stochastically forced dynamo simulations focusing on entry and exit from grand minima episodes.

\section{Results and Discussions}

We first perform simulations without the lower operating threshold in the Babcock-Leighton $\alpha$-effect (setting ${B_{min}} = 0$) in Eqn.~2.3. A majority of Babcock-Leighton dynamo models, including many that have explored the dynamics of grand minima do not use this lower operating threshold. As already known (Charbonneau \& Dikpati 2000; Choudhuri \& Karak 2012) we find that the Babcock-Leighton dynamo with this setup generate cycles of varying amplitudes, including episodes of higher than average activity levels (grand maxima) and occasional episodes of very low amplitude cycles reminiscent of Maunder-like grand minima (Fig.~2.3, upper panel; Fig.~2.4, upper panel). When we do switch on the lower operating threshold, however, we find that the Babcock-Leighton dynamo is unable to recover once it settles into a grand minimum
\begin{figure}[t!]
\centering
\includegraphics*[width=\linewidth]{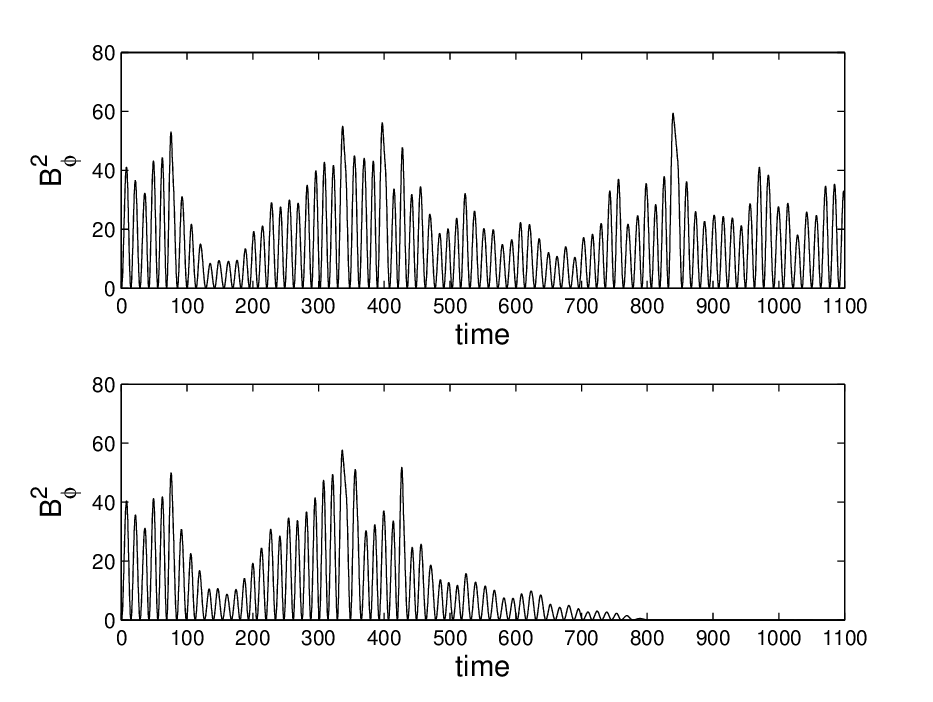}
  \caption{(a) Time evolution of the magnetic energy proxy without considering the lower operating threshold in the quenching function ($B_{min}=0$); (b) Same as above but with a finite lower operating threshold ($B_{min}=1$). The solar dynamo never recovers in the latter case once it settles into a grand minima. All other parameters are fixed at $ \tau=15, B_{max}=7, T_0=2, T_1=0.5, \omega/L=-0.34$ and $\alpha_0=0.17$}
\label{fig3}
\end{figure}
(Fig.~2.3, lower panel; Fig.~2.4, lower panel). This striking result can be explained invoking the underlying physics of the solar cycle. When a series of poloidal field fluctuations lead to a decline in the toroidal field amplitude below the threshold necessary for magnetic buoyancy to operate (with a consequent failure of sunspots to form), the Babcock-Leighton poloidal field source which relies on bipolar sunspot eruptions completely switches off resulting in a catastrophic quenching of the solar cycle. Earlier simulations, which did not include the lower quenching missed out on this physics because even very weak magnetic fields, which in reality could never have produced sunspots, continued to (unphysically) contribute to poloidal field creation. Earlier, it has been shown that the lower threshold due to magnetic buoyancy plays a crucial amplitude limiting role in the Babcock-Leighton solar cycle (Nandy 2002) and this study indicates that this should be accounted for in all Babcock-Leighton solar dynamo models.
\begin{figure}[t!]
\centering
\includegraphics*[width=\linewidth]{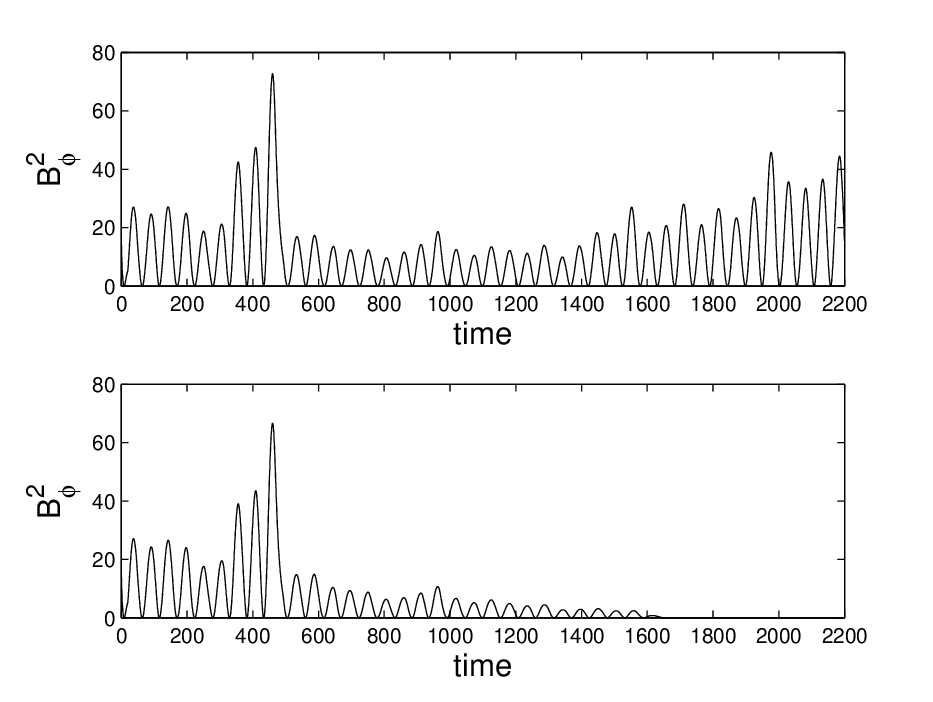}
  \caption{(a) Time evolution of the magnetic energy proxy without considering the lower operating threshold in the quenching function ($B_{min}=0$); (b) Same as above but with a finite lower operating threshold ($B_{min}=1$). The solar dynamo never recovers in the latter case once it settles into a grand minima. All other parameters are fixed at $ \tau=25, B_{max}=7, T_0=20, T_1=0.5, \omega/L=-0.102$ and $\alpha_0=0.051$}
\label{fig3}
\end{figure}
To circumvent this problem faced by the stochastically forced Babcock-Leighton dynamo, we explicitly test an idea (Nandy 2012) for the recovery of the solar cycle based on an additional poloidal source effective on weak toroidal fields. Since the tachocline is the seat of strong toroidal field, any weak field $\alpha$ which is effective only on sub-equipartition strength field will get quenched there. Thus, this $\alpha$-effect must reside above the base of the SCZ (Parker 1993) in a layer away from the strongest toroidal fields. Motivated by this, we devise a new system of dynamo equations governed by
\begin{eqnarray}
    \frac{dB_\phi(t)}{dt} &=& \frac{\omega}{L} A(t-T_0)-\frac{B_\phi(t)}{\tau},\\
   \frac{dA(t)}{dt} &=& \alpha_0 f_1(B_\phi(t-T_1)) B_\phi(t-T_1) 
    + \alpha_{mf} f_2(B_\phi(t-T_2)) B_\phi(t-T_2) -\frac{A(t)}{\tau}, \nonumber\\ 
\end{eqnarray}
where $f_2$, the quenching function for the weak field poloidal source $\alpha_{mf}$ is shown in Fig.~2.1 and is parameterized by
\begin{equation}
f_2 = \frac{\operatorname{erfc}(B^2_\phi(t - T_2)-B^2_{eq})}{2}.
\end{equation}
Taking $B_{eq}=1$ ensures that the weak field source term gets quenched at or below the lower operating threshold for the Babcock-Leighton $\alpha$ and the former, therefore, can be interpreted to be the mean field $\alpha$ effect. In equation~2.6, the time delay $T_2$ is the time necessary for the toroidal field to enter the source region where the additional, weak-field $\alpha$ effect is located.
\begin{figure}[t!]
\centering
\includegraphics*[width=\linewidth]{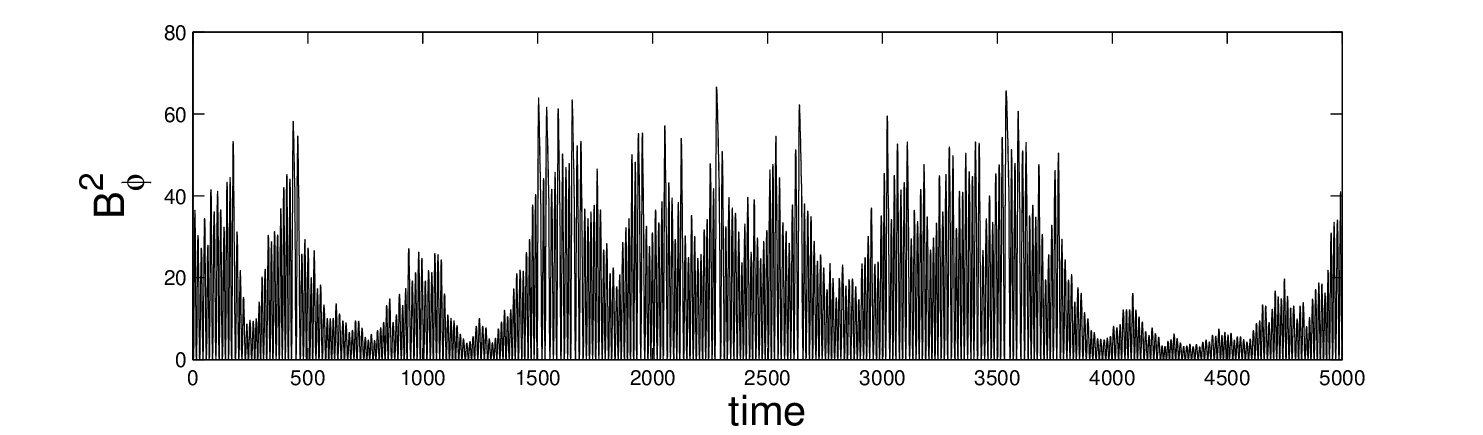}
  \caption{Time series of the magnetic energy ($B_\phi^2$) with both Babcock-Leighton and a weak (mean-field like) $\alpha$ effect for 30\% fluctuation in $\alpha$, $\tau$=15, $T_0$=2, $T_1$=0.5, $T_2$=0.25, $B_{min}$ = $B_{eq}$ =1, $B_{max}$=7, $\omega/L$=$-0.34$, $\alpha_0$=0.17 and $\alpha_{mf}$=0.20. This long-term simulation depicts the model's ability to recover from grand minima episodes.}
\label{fig4}
\end{figure}
\begin{figure}[t!]
\centering
\includegraphics*[width=\linewidth]{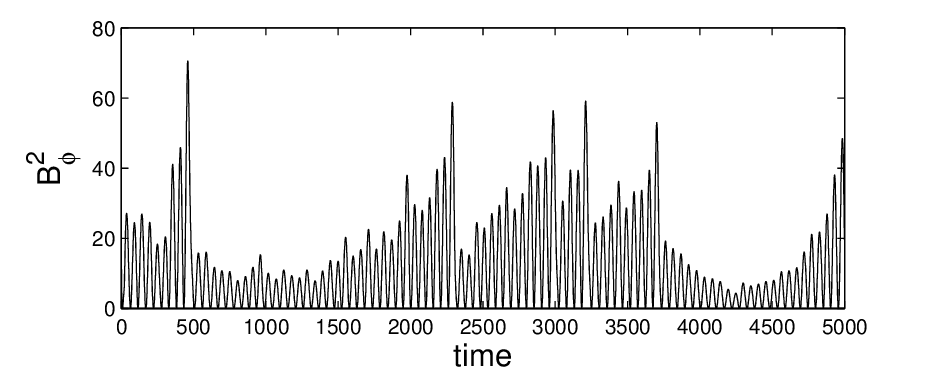}
  \caption{ Time series of the magnetic energy ($B_\phi^2$) with both Babcock-Leighton and a weak (mean-field like) $\alpha$ effect for 50\% fluctuation in $\alpha$, $\tau$=25, $T_0$=20, $T_1$=0.5, $T_2$=0.25, $B_{min}$ = $B_{eq}$ =1, $B_{max}$=7, $\omega/L$=$-0.102$, $\alpha_0$=0.051 and $\alpha_{mf}$=0.04. This long-term simulation depicts the model's ability to recover from grand minima episodes.}
\label{fig4}
\end{figure}

\begin{figure}[t!]
\centering
\includegraphics*[width=\linewidth]{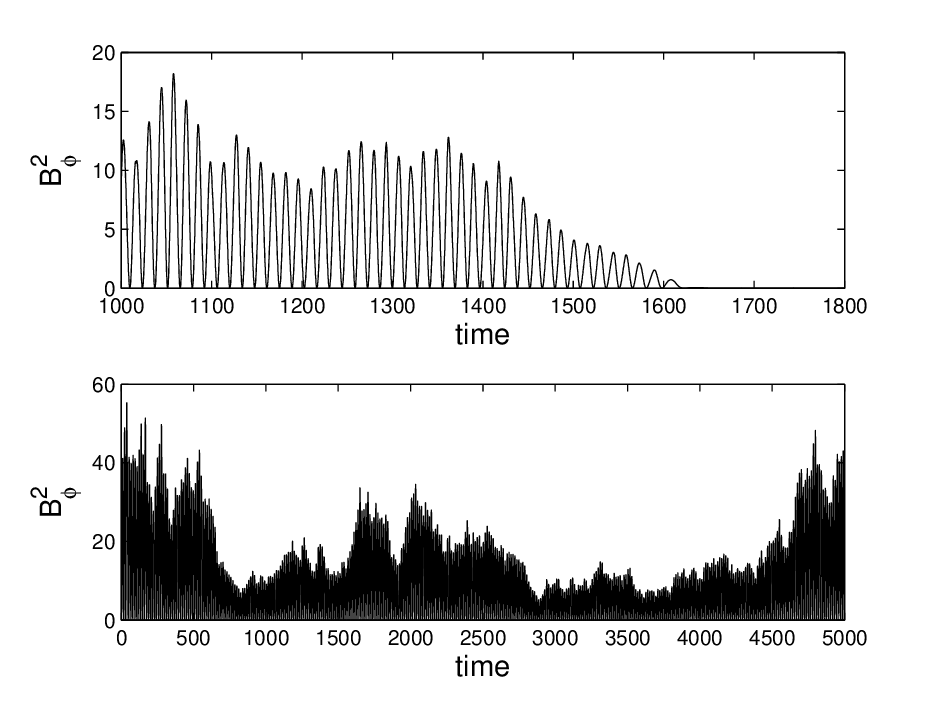}
  \caption{Top panel: Time evolution of the magnetic energy proxy with a finite lower operating threshold ($B_{min}=1$) and 30 \% fluctuation in time delay ($T_0$). The solar dynamo never recovers in the latter case once it settles into a grand minima. All other parameters are fixed at $ \tau=15, B_{max}=7, T_0=2, T_1=0.5, \omega/L=-0.34$ and $\alpha_0=0.17$. Bottom panel: Same as above but with both Babcock-Leighton and a weak (mean-field like) $\alpha$ effect for 30\% fluctuation in time delay ($T_0$), $\tau$=15, $T_0$=2, $T_1$=0.5, $T_2$=0.25, $B_{min}$ = $B_{eq}$ =1, $B_{max}$=7, $\omega/L$=$-0.34$, $\alpha_0$=0.17 and $\alpha_{mf}$=0.20. This long-term simulation depicts the model's ability to recover from grand minima episodes.}
\label{fig4}
\end{figure}  
If $T_2=0$, i.e. the generation layer of the additional $\alpha$ effect is coincident with the $\Omega$ effect (layer) then we find that the stochastically forced dynamo again fails to recover from a grand minimum. This is reminiscent of the original motivation behind the introduction of the interface dynamo idea with spatially segregated source regions (Parker 1993). However, if $T_2$ is finite and $T_1 > T_2$ (i.e., there is some segregation between the $\Omega$ effect toroidal source, the additional weak-field $\alpha$ and the Babcock-Leighton $\alpha$), we find that the solar cycle can recover from grand minima like episodes in a robust manner. Figures 2.5 and 2.6 depict such solutions (for two different sets of parameter), where we explicitly demonstrate self-consistent entry and exit from grand minima like episodes. We note that this recovery of the solar cycle from grand minima like episodes is possible with or without fluctuations in the additional, weak field poloidal source term $\alpha_{mf}$.

\section{Conclusions}

In summary, we have constructed a new model of the solar dynamo for exploring solar cycle fluctuations based on a system of stochastically forced, non-linear, delay differential equations. Utilizing this model for long-term simulations we have explicitly demonstrated that the currently favored mechanism for solar poloidal field production, the Babcock-Leighton mechanism, alone, cannot recover the solar cycle from a grand minimum. We have also demonstrated that an additional, mean field like $\alpha$-effect capable of working on weaker fields is necessary for self-consistent entry and exit of the solar cycle from grand minima episodes. We have demonstrated that our results and conclusions hold over two very diverse regimes of parameter choices. Although we have utilized fluctuations in the poloidal source in our study, we do not claim that this is the only possible source of fluctuations in the solar cycle. We also point out that our model results are robust if we run simulations considering fluctuations in time delay (Fig.~2.7). We note that simulations motivated from this current study and based on a spatially extended dynamo model in a solar-like geometry supports the results from this mathematical time delay model (Passos et al. 2014). Taken together, these strengthen the conclusion that a mean field like $\alpha$-effect effective on weak toroidal fields must be functional in the Sun's convection zone and that this is vitally important for the solar cycle, even if the dominant contribution to the poloidal field comes from the Babcock-Leighton mechanism during normal activity phases.
\newcommand{\etal}{{\it et al.}}

\newcommand{\be}{\begin{equation}}
\newcommand{\ee}{\end{equation}}
\newcommand{\bea}{\begin{eqnarray}}
\newcommand{\eea}{\end{eqnarray}}
\newcommand{\ba}{\begin{array}}
\newcommand{\ea}{\end{array}}
\newcommand{\bit}{\begin{itemize}}
\newcommand{\eit}{\end{itemize}}
\newcommand{\ben}{\begin{enumerate}}
\newcommand{\een}{\end{enumerate}}

\chapter{Strong Hemispheric Asymmetry can Trigger Parity Changes in the Sunspot Cycle}

Although sunspots have been systematically observed on the Sun's surface over the last four centuries, their magnetic properties have been revealed and documented only since the early 1900s. Sunspots typically appear in pairs of opposite magnetic polarity which have a systematic orientation. This polarity orientation is opposite across the equator -- a trend that has persisted over the last century since observations of sunspot magnetic fields exist. Taken together with the configuration of the global poloidal field of the Sun -- that governs the heliospheric open flux and cosmic ray flux at Earth -- this phenomena is consistent with the dipolar parity state of an underlying magnetohydrodynamic dynamo mechanism. Although, hemispheric asymmetry in the emergence of sunspots is observed in the Sun, a parity shift has never been observed. We simulate hemispheric asymmetry through introduction of random fluctuations in a computational dynamo model of the solar cycle and demonstrate that changes in parity are indeed possible over long time-scales. In particular, we find that a parity shift in the underlying nature of the sunspot cycle is more likely to occur when sunspot activity dominates in any one hemisphere for a time which is significantly longer compared to the sunspot cycle period. Our simulations suggest that the sunspot cycle may have resided in quadrupolar parity states in the distant past, and provides a possible pathway for predicting parity flips in the future.

\section{Introduction}
Samuel Heinrich Schwabe discovered the 11 year solar cycle in 1843, but detailed observations about the dipolar nature of solar magnetic fields exist only for last hundred years (Hale et al. 1919). One may pose the question whether solar magnetic fields have always been in the dipolar state?\\

To investigate this issue we use an axisymmetric kinematic flux transport solar dynamo model which involves the generation and recycling of the toroidal and poloidal field (Parker 1955). In this model, the toroidal field is produced by stretching of poloidal field lines at the base of the convection zone due to strong differential rotation (Parker 1955) and the poloidal field is generated through a combination of mean field $\alpha$-effect due to helical turbulence in the solar convection zone (Parker 1955) and the Babcock-Leighton mechanism due to decay and dispersal of tilted bipolar sunspot region at the near-surface layers (Babcock 1961; Leighton 1969). The kinematic flux transport dynamo model based on the Babcock-Leighton mechanism for poloidal field generation has been successful in explaining different observational aspects of the solar cycle (Dikpati \& Charbonneau 1999; Nandy \& Choudhuri 2002; Chatterjee et al. 2004; Goel \& Choudhuri 2009; Nandy et al. 2011; Karak \& Nandy 2012; DeRosa et al. 2012). Recent observations also lend strong support to the Babcock-Leighton mechanism as a primary source for poloidal field generation (Dasi-Espuig et al. 2010; Mu\~noz-Jaramillo et al. 2013).\\

It is widely thought that stochastic fluctuations in the poloidal field generation mechanism is the primary source for irregularity in the solar cycle (Hoyng 1988; Choudhuri 1992; Charbonneau \& Dikpati 2000; Charbonneau et al. 2004). In the Babcock-Leighton framework, poloidal field generation depends on the tilt angle of bipolar sunspot pairs, which is imparted by the action of Coriolis force on buoyantly rising toroidal flux tubes from the base of the solar convection zone. Observational scatter of tilt angles around the mean given by Joy's law may be produced by turbulent buffeting that a rising flux tube encounters during its journey through the convection zone (Longcope \& Choudhuri 2002). Thus the Babcock-Leighton mechanism for poloidal field generation is not a deterministic process but a random one (Choudhuri et al. 2007). Another major source in solar cycle irregularity is fluctuations in the meridional circulation (Lopes \& Passos 2009; Karak 2010).\\

On the one hand, two different types of symmetries are obtained, in general, in solutions of the dynamo equations. The global magnetic field is of dipolar nature (dipolar or odd parity) if the toroidal field is antisymmetric across the equator; conversely, if the toroidal field is symmetric across the equator then the global field is of quadrupolar nature (quadrupolar or even parity). Some previous studies have found solutions  that are of quadrupolar nature using low diffusivity in their kinematic dynamo models. It has been suggested that an additional alpha effect at the base of the convection zone is necessary to produce the observed dipolar parity (Dikpati \& Gilman 2001; Bonanno et al. 2002). However, other studies suggest that strong hemispheric coupling by higher diffusivity is necessary for generation of the global dipolar magnetic field without the presence of an additional alpha effect at the base of the convection zone (Chatterjee et al. 2004; Chatterjee \& Choudhuri 2006; Hotta \& Yokoyama 2010). These past studies have been inspired with the primary aim of ensuring dipolar solutions to the dynamo equations with the notion that the solar dynamo has always persisted in the dipolar parity state with antisymmetric toroidal fields across the equator.

On the other hand, unequal solar activity in northern and southern hemispheres (known as hemispheric asymmetry) is well documented (Waldmeier 1955, 1971; Chowdhury et al. 2013; McClintock \& Norton 2013). Observational evidence of strong hemispheric asymmetry exists during the onset of grand-minima like episodes (Sokoloff \& Nesme-Reibes 1994). Theoretical and observational studies also suggest that hemispheric polar field at the minimum of the solar cycle can be used as a precursor to predict the amplitude of the next cycle (Schaten et al. 1978; Schaten 2005; Jiang et al. 2007; Karak \& Nandy 2012; Mu\~noz-Jaramillo et al. 2013). Thus possibly, the hemispheric asymmetry of the polar field at solar minima may be responsible for the hemispheric asymmetry in the next cycle too. Feeding the data of the polar flux of previous cycles in kinematic solar dynamo models, some studies are able to explain hemispheric asymmetry like phenomenon in the current cycle (Goel \& Choudhuri, 2009). Details about hemispheric coupling and hemispheric asymmetry can be found in a review paper by Norton et al. (2014).

To explore hemispheric asymmetry and parity issues and their inter-relationship, we introduce stochastic fluctuations in the Babcock-Leighton poloidal field source and find that stochastic fluctuations can trigger the solar cycle into grand minima like episodes. As proposed earlier, we confirm that an additional $\alpha$-effect is necessary for cycle recovery. In the next step, we introduce stochastic fluctuations in both the Babcock-Leighton mechanism and the additional mean field $\alpha$-effect and find dynamo solutions can self-consistently change parity. The above result begs the question whether it is possible to predict parity flips in the Sun. We find that parity flips in the sunspot cycle tend to occur when solar activity in one hemisphere strongly dominates over the other hemisphere for a period of time significantly longer than the sunspot cycle timescale. However, strong domination of activity in one hemisphere does not necessarily always guarantee a parity change.
\section{Model}
Our model is based on $\alpha\Omega$ dynamo equations in the axisymmetric spherical formulation wherein the dynamo equations are:
\begin{equation}
   \frac{\partial A}{\partial t} + \frac{1}{s}\left[ \mathbf{v_p} \cdot \nabla (sA) \right] = \eta\left( \nabla^2 - \frac{1}{s^2}  \right)A + S (r, \theta, B),
\end{equation}\\
\begin{equation}
   \frac{\partial B}{\partial t}  + s\left[ \mathbf{v_p} \cdot \nabla\left(\frac{B}{s} \right) \right] + (\nabla \cdot \mathbf{v_p})B = \eta\left( \nabla^2 - \frac{1}{s^2}  \right)B + s\left(\left[ \nabla \times (A \bf \hat{e}_\phi) \right]\cdot \nabla \Omega\right)   + \frac{1}{s}\frac{\partial (sB)}{\partial r}\frac{\partial \eta}{\partial r},
\end{equation}\\
where, $B (r, \theta)$ (i.e. $B_\phi$) and $A (r, \theta)$ are the toroidal and vector potential for the poloidal components of the magnetic field respectively. Here $\Omega$ is the differential rotation, $\mathbf{v_p}$ is the meridional flow, $\eta$ is the turbulent magnetic diffusivity and $s = r\sin(\theta)$. 

Here, we use a two step radially dependent magnetic diffusivity profile as described in \citet{munoz09}. In our case, the diffusivity at the bottom of convection zone is $10^8 cm^2/s$, diffusivity in the convection zone is $10^{11} cm^2/s$ and supergranular diffusivity is $5 \times 10^{12} cm^2/s$. $r_{cz} = 0.73R_\odot$, $d_{cz} = 0.025R_\odot$, $r_{sg} = 0.95R_\odot$ and $d_{sg} = 0.015R_\odot$ describes the transition from one diffusivity value to another. We also use the same differential rotation profile as described in \citet{munoz09}.
We generate the meridional circulation profile ($\mathbf{v_p}$) for a compressible flow inside the convection zone by using a stream function along with mass conservation constraint:
 \begin{equation}
    \mathbf{v_p}\left(r,\theta\right) =
    \frac{1}{\rho(r)} \nabla \times \left(\psi(r,\theta)\widehat{\textbf{e}}_{\phi}\right).
\end{equation}
which is estimated from the stream function (defined within $0 \leq \theta \leq \pi/2$, i.e., in the northern hemisphere) as described in Chatterjee, Nandy and Choudhuri (2004):
\begin{eqnarray}
\psi r \sin \theta = \psi_0 (r - R_p) \sin \left[ \frac{\pi (r - R_p)}
{(R_\odot - R_p)} \right] \{ 1 - e^{- \beta_1 \theta^{\epsilon}} \}
 \{1 - e^{\beta_2 (\theta - \pi/2)} \} e^{-((r -r_0)/\Gamma)^2},
\end{eqnarray}
where $\psi_0$ is the factor which determines the maximum speed of the flow. We use the following parameter values $ \beta_1=1.5, \beta_2=1.8, \epsilon=2.0000001, r_0=(R_\odot-R_b)/4, \Gamma=3.47 \times 10^8~ m, \gamma=0.95, m=3/2$. Here $R_p=0.64R_\odot$ is the penetration depth of the meridional flow. The meridional circulation profile in the southern hemisphere is generated by a mirror reflection of the velocity profile across the equator. In this work, we use the surface value of meridional circulation as $17~m s^{-1}$.

Recent observations and theory both indicate that the Babcock-Leighton mechanism for poloidal field creation plays an important role in the solar cycle. In Equation (3.1), $S(r, \theta, B)$ represents the Babcock-Leighton mechanism. But modelling this mechanism to capture correctly the underlying physics is a challenging task. Two different approaches exist in the kinematic dynamo literature to model Babcock-Leighton mechanism-one is alpha-coefficient formulation and another is double ring approach proposed by Durney (1995). Nandy \& Choudhuri (2001) has shown that these two different approaches for modelling Babcock-Leighton mechanism produce qualitatively similar result. However Mu\~noz-Jaramillo et al. (2010) has shown that double ring approach is more successful in capturing the observed surface dynamics compared to the alpha-coefficient formulation.  Motivated by this, here we model the Babcock-Leighton mechanism (i.e., poloidal field source term) by the emergence and flux dispersal of double-rings structures.
\subsection{Modelling Active Regions as Double Rings and Recreating the poloidal field}
 We model the Babcock-Leighton mechanism by the methods of double ring proposed by Durney (1997) and subsequently used by some other groups (Nandy \& Choudhuri 2001; Mu\~noz-Jaramillo et al. 2010; Nandy et al. 2011). In this algorithm we define the $\phi$ component of potential vector $A$ corresponding to active region as:
  \begin{equation}\label{Eq_AR}
    A_{ar}(r,\theta,t)= K_1 A(\Phi,t)F(r)G(\theta),
\end{equation}
where $K_1$ is a constant which ensures super-critical solutions and strength of ring doublet is defined by  $A(\Phi,t)$. $\Phi$ is basically the magnetic flux. We define $F(r)$  as:
\begin{equation}
    F(r)= \left\{\begin{array}{cc}
            0 & r<R_\odot-R_{ar}\\
            \frac{1}{r}\sin^2\left[\frac{\pi}{2 R_{ar}}(r - (R_\odot-R_{ar}))\right] & r\geq R_\odot-R_{ar}
          \end{array}\right.,
\end{equation}
where $R_\odot$ is the solar radius and $R_{ar}=0.15R_\odot$. $R_\odot-R_{ar}$ is the radial extent of the active regions, i.e., how deep they extend from the surface. This extent (surface to $0.85 R_\odot$) is motivated from results indicating that active region flux tube disconnection happens around this depth in the convection zone (Longcope \& Choudhuri 2002). We define $G(\theta)$ in integral form as:
\begin{equation}
    G(\theta) = \frac{1}{\sin{\theta}}\int_0^{\theta}[B_{-}(\theta')+B_{+}(\theta')]\sin(\theta')d\theta',
\end{equation}
where $B_{+}$ ($B_{-}$) defines the strength of positive (negative) ring:
 \begin{equation}\label{Eq_AR_Dp}
    B_{\pm}(\theta)= \left\{\begin{array}{cc}
                     0 & \theta<\theta_{ar}\mp\frac{\chi}{2}-\frac{\Lambda}{2}\\
                     \pm\frac{1}{\sin(\theta)}\left[1+\cos\left(\frac{2\pi}{\Lambda}(\theta-\theta_{ar}\pm\frac{\chi}{2})\right)\right] & \theta_{ar}\mp\frac{\chi}{2}-\frac{\Lambda}{2} \leq \theta < \theta_{ar}\mp\frac{\chi}{2}+\frac{\Lambda}{2}\\
                     0 & \theta \geq \theta_{ar}\mp\frac{\chi}{2}+\frac{\Lambda}{2}
               \end{array}\right..
\end{equation}
Here $\theta_{ar}$ is emergence co-latitude, $\Lambda$ is the diameter of each polarity of the double ring  and $\chi = \arcsin[\sin(\gamma)\sin(\Delta_{ar})]$ is the latitudinal distance between the centers, where the angular distance between polarity centers $\Delta_{ar}=6^o$ and the AR tilt angle is $\gamma$.  We set $\Lambda$, i.e. diameter of ring doublet as $6^o$. Figure 3.1 illustrates the variation of field strength of double ring bipolar pair (namely, $B_{+}$ in red colour and $B_{-}$ in blue colour) with colatitude. The top panel of figure 3.2(a) represents the axisymmetric signature of double rings in both northern and southern hemisphere.
\begin{figure}[!htb]
\centering
\begin{tabular}{cc}
\includegraphics[scale=0.50]{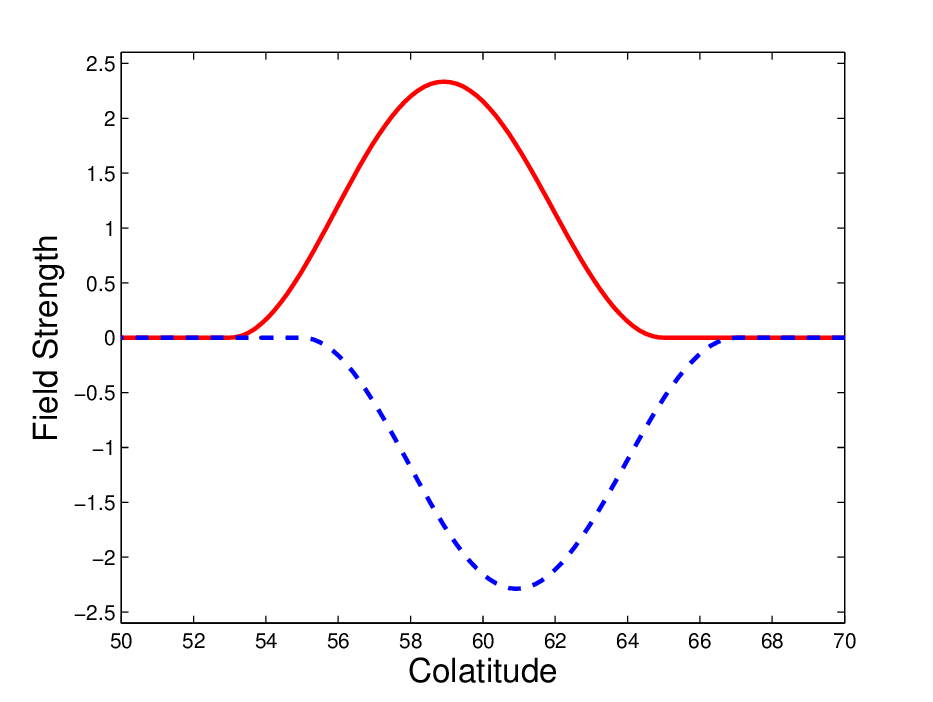} 
 \end{tabular}                
\caption{Diagram illustrating the quantities which define the latitudinal dependence of a double-ring bipolar pair. Variation of strengths for positive ($B_{+}$) and negative ($B_{-}$) ring with colatitude is shown in red and blue colour respectively.}
\end{figure}

Now to recreate the poloidal field, first we check where the toroidal field is higher than the buoyancy threshold at the bottom of convection zone in both northern and southern hemispheres. Then we choose one of the latitudes randomly from both the hemispheres simultaneously at a certain interval of time, using a non uniform probability distribution function such that randomly chosen latitudes remain within the observed active latitudes. The probability distribution function is made to drop steadily to zero between 30$^o$ (-30$^o$) and 40$^o$ (-40$^o$) in the northern (southern) hemisphere. Second, we calculate the magnetic flux of this toroidal ring.
Then we find tilt of corresponding active region, using the expression given in Fan, Fisher \& McClymont (1994\nocite{fan-fisher-mcclymont94})
\begin{equation}
   \gamma \propto \Phi_0^{1/4}B_0^{-5/4}\sin(\lambda),
\end{equation}
where $B_0$ is the local field strength, $\Phi_0$ is the flux associated with the toroidal ring and $\lambda$ is the emergence latitude. We set the constant such that tilt angle lies between  3$^o$ and  12$^o$. \\ Third, we remove a chunk of magnetic field with same angular size as the emerging active region from this toroidal ring and calculate the magnetic energy of the new partial toroidal ring. Then we fix the value of toroidal field such that the energy of the full toroidal ring filled with new magnetic field strength is the same as the magnetic field strength for the partial toroidal ring. This exercise also generated the strength of the ring doublet given by $A(\Phi)$. Finally, we place the ring duplets with these calculated properties at the near-surface layer at the latitudes where they erupt, thus defining the source term $S(r, \theta, B)$.

\section{Results and Discussion}
\begin{figure}[htb!]
  \begin{center}
\begin{tabular}{cc}
\includegraphics[width=15cm]{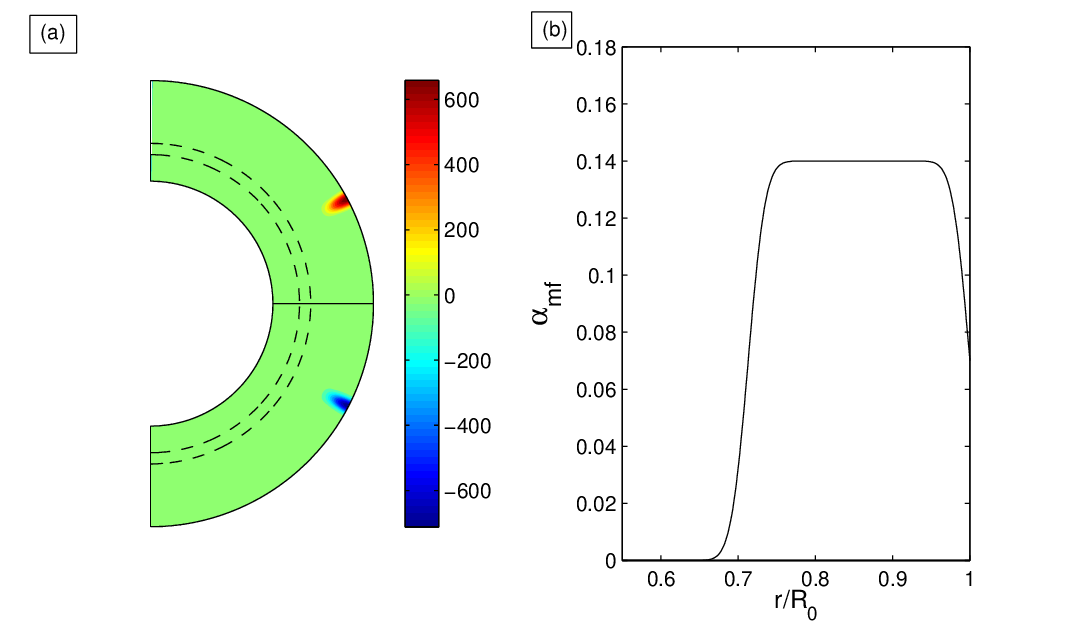} \\
\includegraphics[width=15cm]{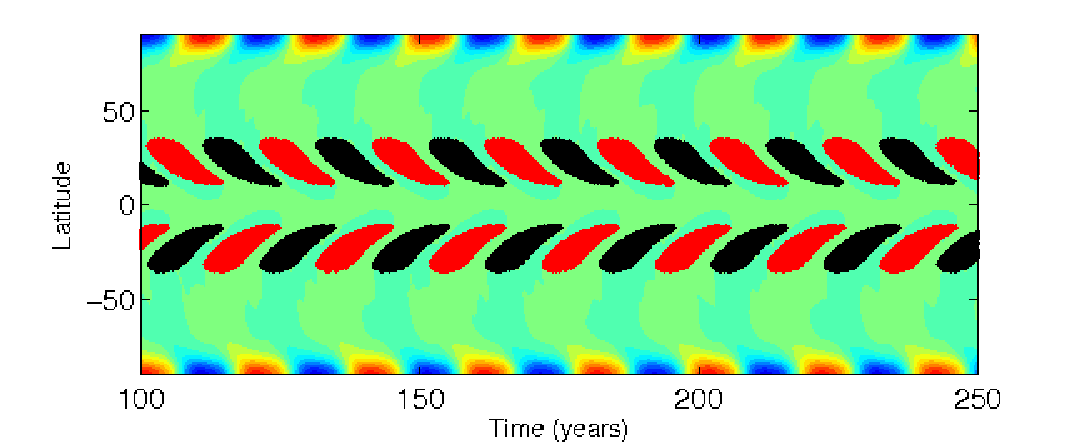} 
\end{tabular}
\end{center}  
\caption{Top panel: (a) Babcock-Leighton mechanism modeled by double-ring algorithm: Poloidal field line contour of double-rings in both northern and southern hemisphere. (b) Radial profile of mean field $\alpha$-coefficient used to model the additional poloidal field generation mechanism. Bottom panel:  Representative butterfly diagram from our solar dynamo model with double ring algorithm, without fluctuation in Babcock-Leighton mechanism.  Here background is the weak diffuse radial field on solar surface and eruption latitudes are denoted by symbols black (“+”) and red (“+”), indicating underlying negative and positive toroidal field respectively.}
\end{figure}
\begin{figure*}[htb!]
        \centering
        \includegraphics[width=15 cm]{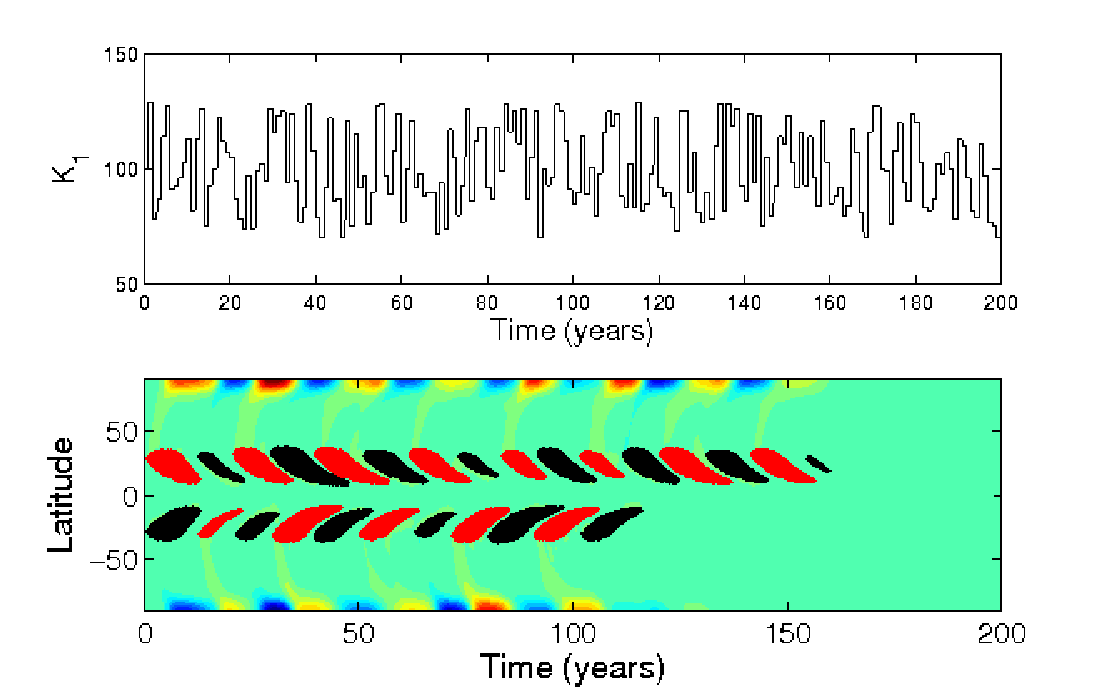}
        \includegraphics[width=15 cm]{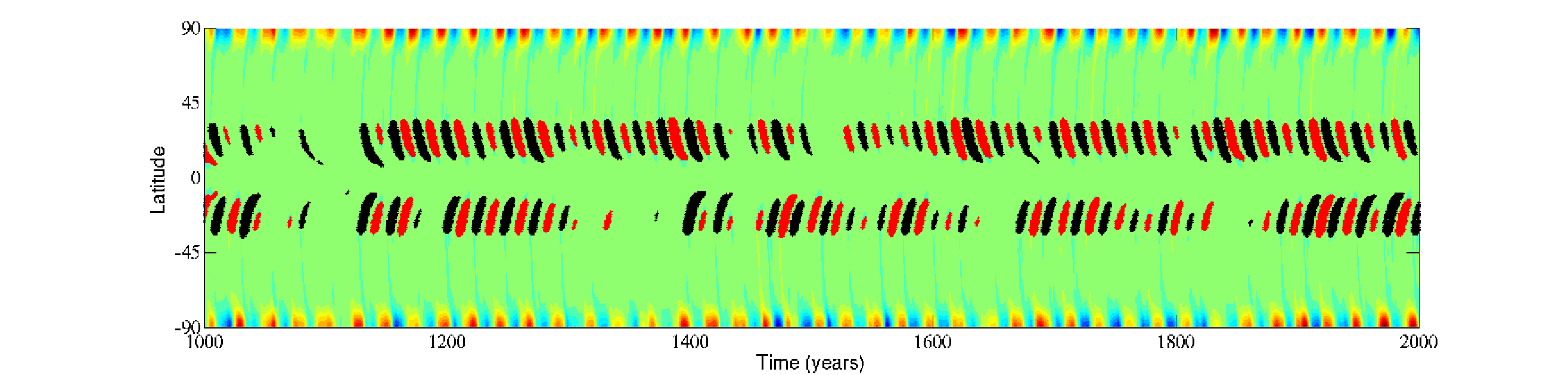}
        \caption{The top panel shows typical figure of stochastic fluctuations in time in the Babcock-Leighton source term constant K1 at a level of 30 \% fluctuation with a correlation time of 1 year using our random number generating program. Middle panel shows simulated butterfly diagram at the base of the convection zone after introducing fluctuation in Babcock-Leighton source term without the presence of additional mean field $\alpha$ source term. Bottom panel shows simulated butterfly diagram at the base of the convection zone when both Babcock-Leighton source term and mean field $\alpha$ effect is present. In last case we introduce 75 \% fluctuation in Babcock-Leighton mechanism and 150 \% fluctuation in mean field $\alpha$.}
        \label{fig:3}
\end{figure*}

To explore the parity issue with our dynamo simulations we define the parity $P(t)$ following the prescription of Chatterjee, Nandy and Choudhuri (2004):\\
 \begin{equation}
 P(t) = \frac{\int_{t-T/2}^{t+T/2}(B_N(t') - \overline{B}_N)(B_S(t') -
  \overline{B}_S)dt'}
{\sqrt{\int_{t-T/2}^{t+T/2}(B_N(t') - \overline{B}_N)^2 dt'}
\sqrt{\int_{t-T/2}^{t+T/2}(B_S(t') - \overline{B}_S)^2 dt'}},
 \end{equation}
where $B_N$ and $B_S$ are the amplitudes of the toroidal field at $25^{\circ}$ latitude in both northern and southern hemispheres at the base of the solar convection zone. We take the averages of $B_N$ and $B_S$ over a dynamo period (i.e. $\overline{B}_N$ and $\overline{B}_S$ ) as zero. The value of parity function should be +1 for quadrupolar parity and -1 for dipolar parity. In the first scenario, we run dynamo simulations without fluctuations, considering only the Babcock-Leighton mechanism as a poloidal field generation process. We find the parity of the solutions are always dipolar (see bottom panels of Fig.~3.2 for representative solution).

The Babcock-Leighton mechanism is not a deterministic process but a random one. This random nature arises due to scatter in tilt angles (an observed fact) of bipolar sunspot pairs whose underlying flux tubes are subject to turbulent buffeting during their ascent through the turbulent convection zone (Longcope \& Choudhuri 2002). Motivated by this fact, we introduce stochastic fluctuations in the Babcock-Leighton mechanism by redefining the coefficient $K_1$ (which governs the strength of the ring doublet) as:
\begin{equation}
    K_1= K_0[1+\frac{\delta}{100} \sigma(t,\tau_{cor})],
\end{equation}
where for $K_1=K_0$, we get supercritical solutions. $\sigma(t, \tau_{cor})$ represents a uniform random function with values between -1 and +1 which changes value at intervals of coherence time $\tau_{cor}$. $\delta$ and $\tau_{cor}$ represents percentile level of fluctuation and coherence time, respectively. Although it is difficult to estimate the actual level of fluctuation within the solar convection zone, our use of fluctuation levels is motivated by the fluctuations present in the observed polar field data from Wilcox Solar Observatory, the eddy velocity distributions present in 3D MHD simulations of turbulent solar convection (Racine et al. 2011; Passos et al. 2012) and some earlier results (Charbonneau \& Dikpati 2000; Dasi-Espuig et al. 2010). Our choice of coherence time ($\tau_{cor}$) is motivated by the rise time of flux tube through the turbulent convection zone (of the order of three months, assuming the rise time scale is equal to the Alfv\'enic time scale; see Caligari et al. 1995) and the time taken by the surface flows to redistribute the sunspots (on the order of several months to a year).

\begin{figure}[h]
\includegraphics[height=6cm, width=15cm]{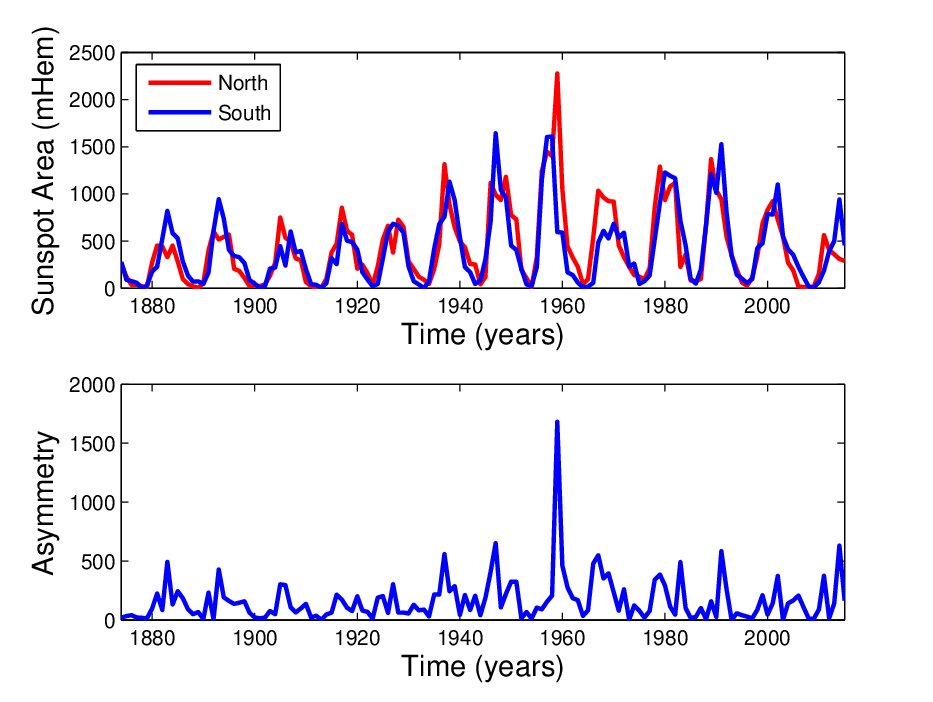}\\
\includegraphics[height=6cm, width=15cm]{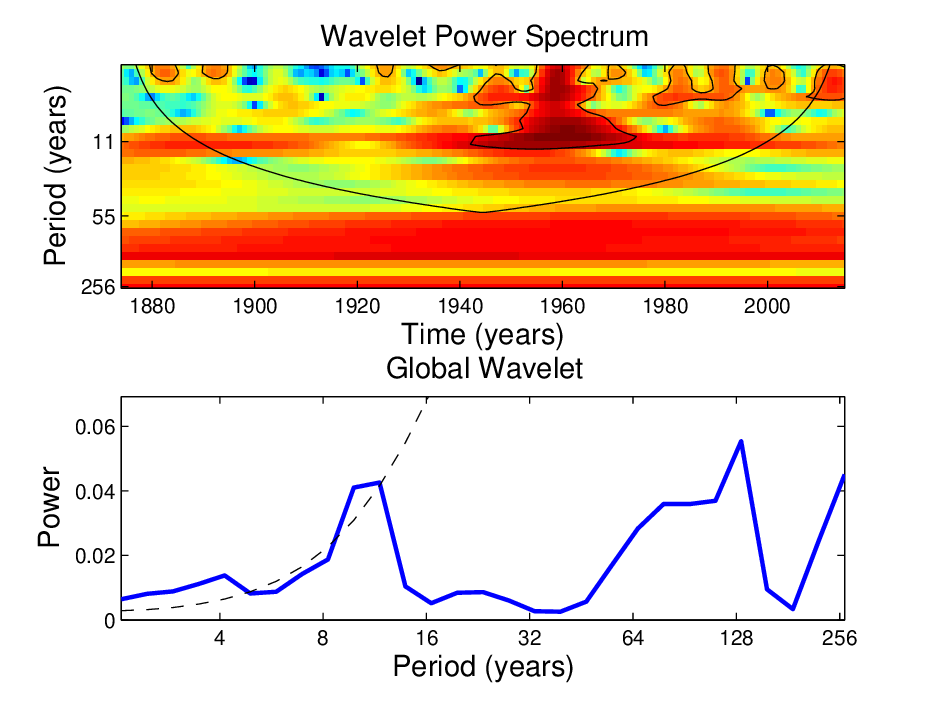}
\caption{First panel shows the time series of yearly averaged sunspot area by hemisphere, the second panel is the time series of yearly averaged  absolute asymmetry generated from observed sunspot area data series, the third panel is the wavelet power spectrum of absolute asymmetry time series and fourth panel shows the global wavelet analysis of absolute asymmetry. Both wavelet power spectrum and global wavelet analysis shows a clear signature of 11 year periodicity in the absolute asymmetry data generated from observation.}
 \end{figure}
 \begin{figure}[h]
\includegraphics[height=3cm, width=15cm]{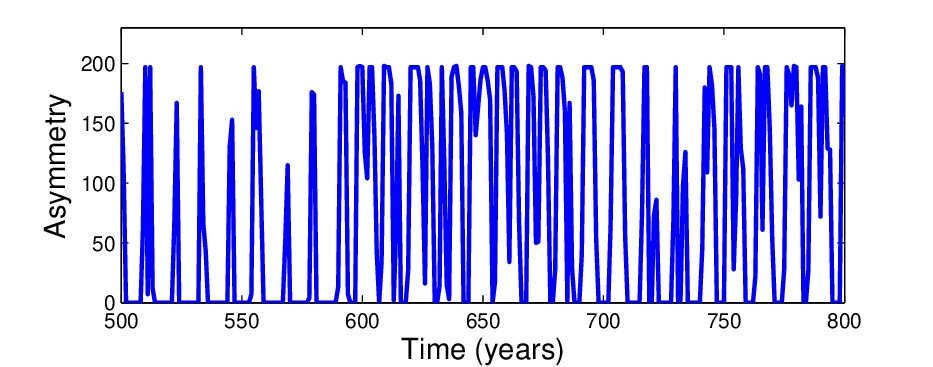}\\
\includegraphics[height=6cm, width=15cm]{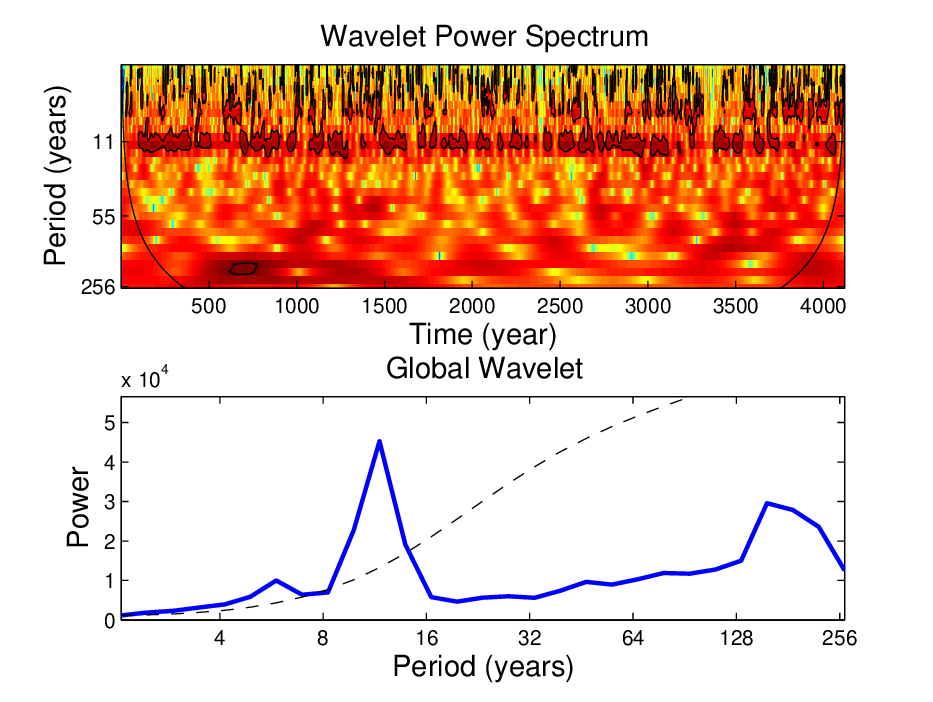}
 \caption{Top panel shows the time series of yearly averaged absolute asymmetry generated from our kinematic dynamo simulation with stochastic fluctuation. In this case we take 60 \% fluctuation in the Babcock-Leighton mechanism and 50 \% fluctuation in mean field $\alpha$-effect. Middle panel and bottom panel shows the wavelet power spectrum and global wavelet analysis of this absolute asymmetry time series, respectively. Both wavelet power spectrum and global wavelet analysis shows a clear signature of 11 year periodicity in the absolute asymmetry data generated from the simulations.}
 \end{figure}
\begin{figure*}[!htb]
        \centering
        \includegraphics[width=15 cm]{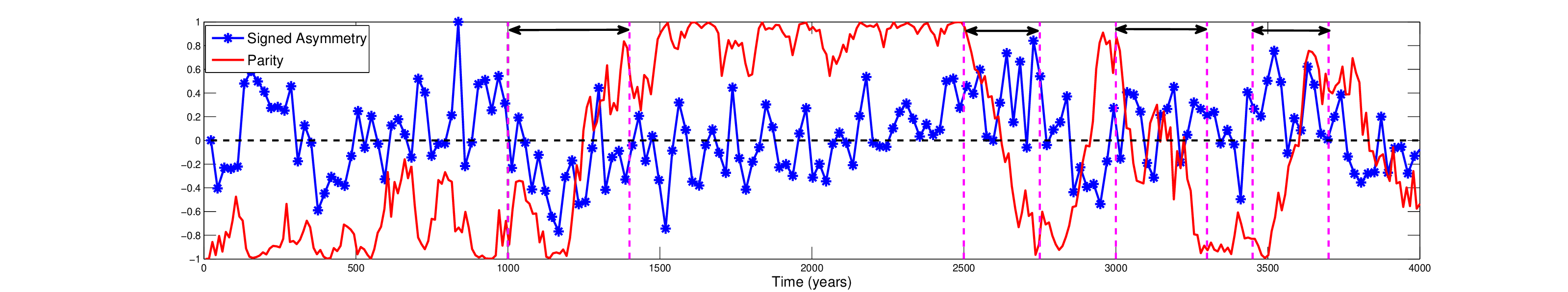}
        \includegraphics[height=3cm, width=15 cm]{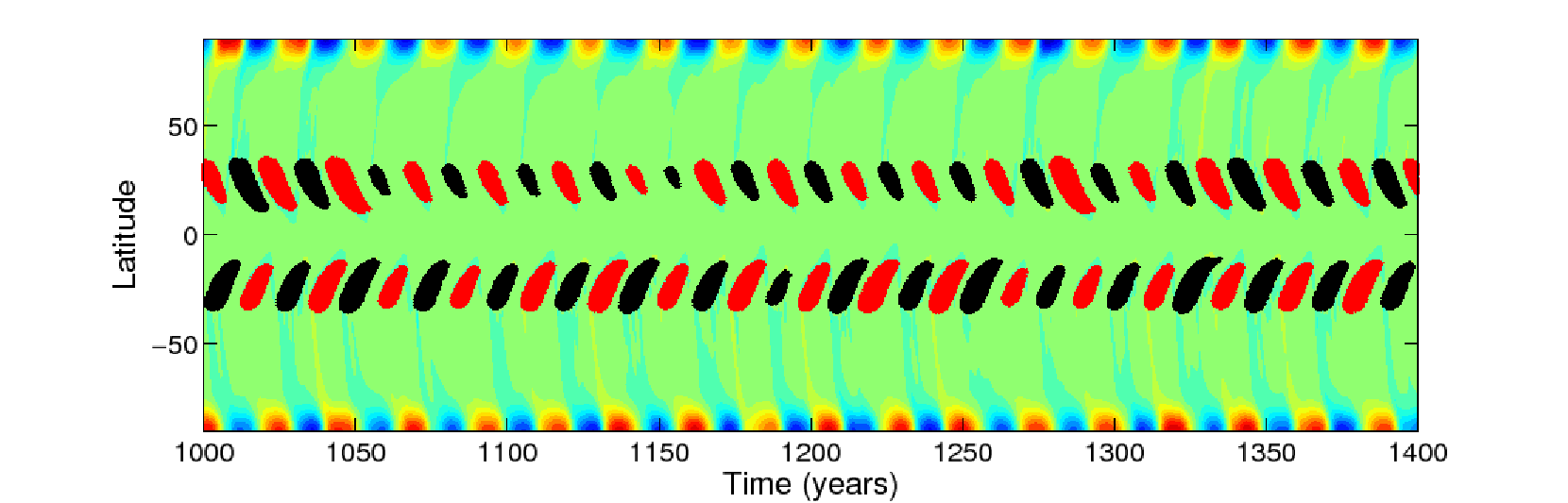}
        \includegraphics[height=3cm, width=15 cm]{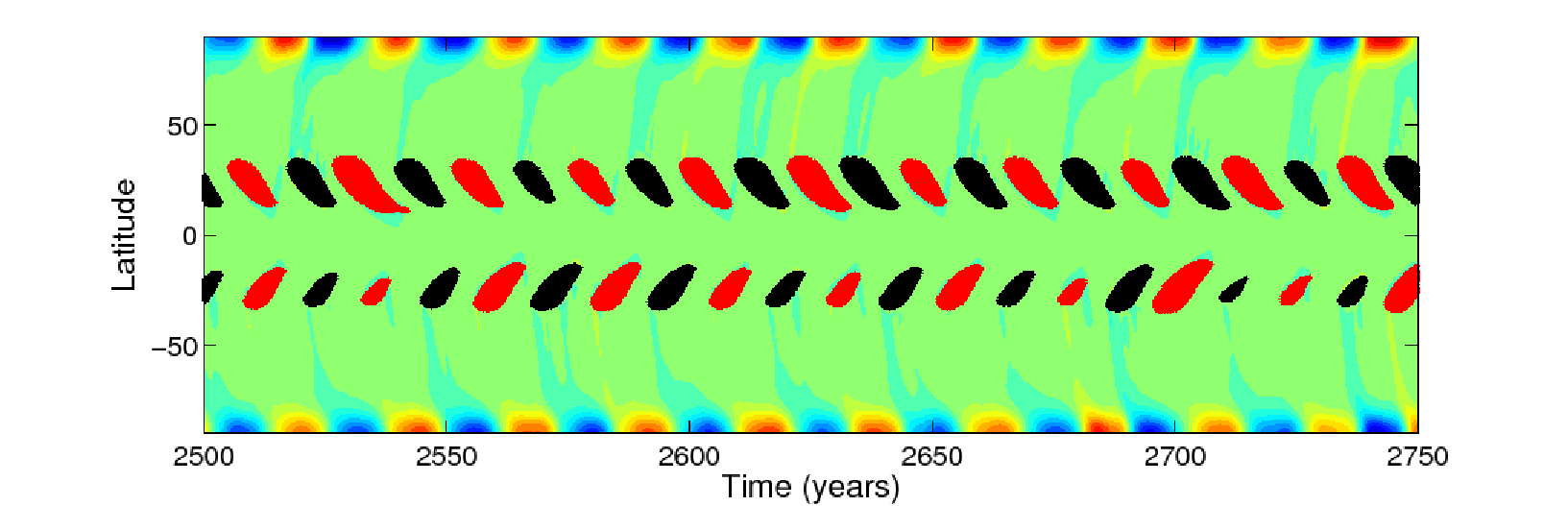}
        \includegraphics[height=3cm, width=15 cm]{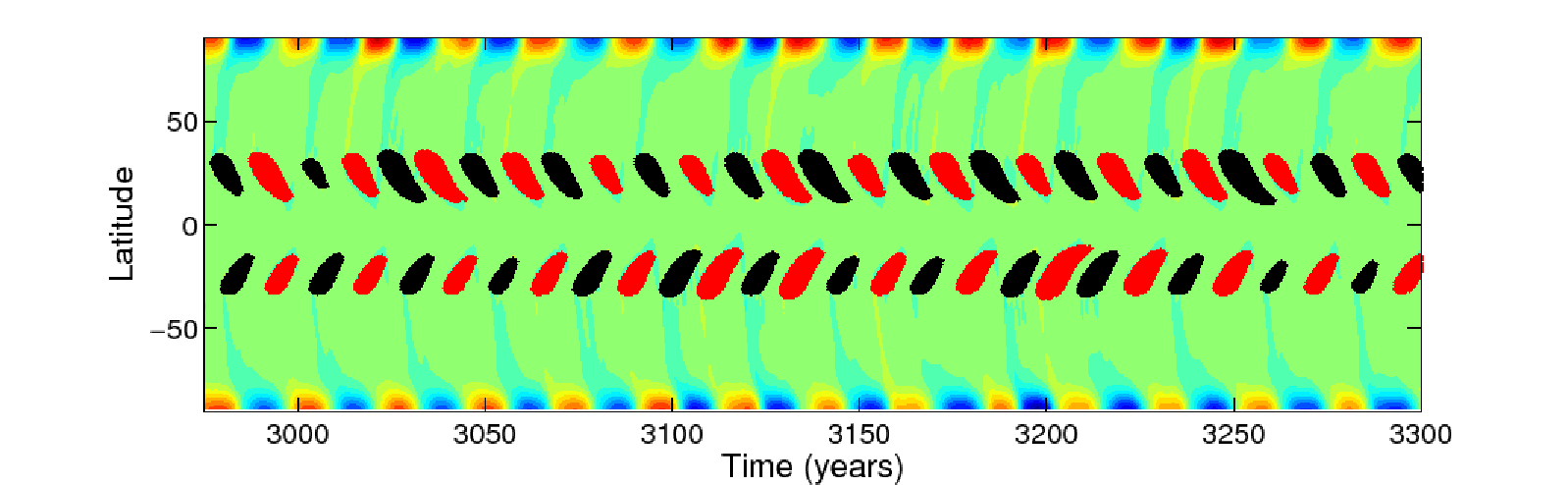}
        \includegraphics[height=3cm, width=15 cm]{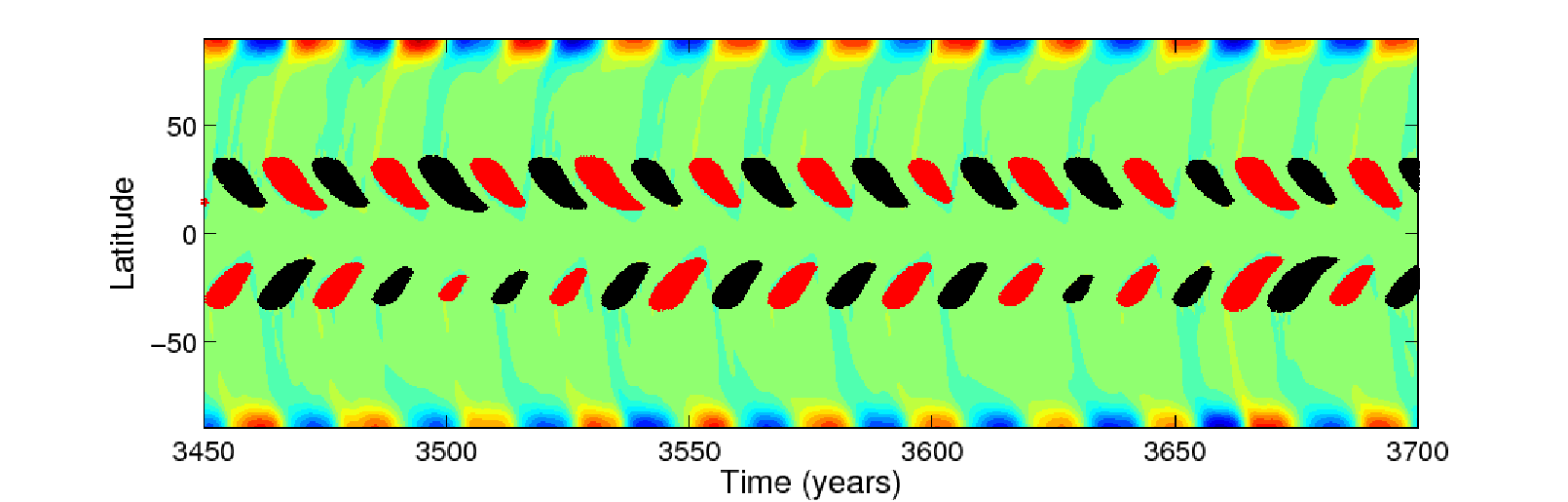}
        \caption{First panel shows the evolution of parity (red colour) and 22 year averaged normalized signed asymmetry (blue color) obtained from our simulations. Second, third, fourth and fifth panels are simulated butterfly diagrams for different time intervals where parity change takes
place. Selected time intervals are shown in top panel by double arrow. All these plots indicate that a change in solar parity takes place only when sunspot activity in one hemisphere dominates over the other for a sufficiently large period of time. This simulations corresponds to 60\% fluctuations in Babcock-Leighton mechanism and 50 \% fluctuations in mean field $\alpha$.}
        \label{fig:3}
\end{figure*}
Introduction of random fluctuations in the Babcock-Leighton mechanism results in the decay and loss of the solar cycle after some time depending on the level of fluctuations. The middle panel of figure 3.3 depicts such a solution where we introduce $50\%$ level of fluctuation with a coherence time of 1 year. From observations of solar cycle and the analysis of cosmogenic isotopes it is well-known that the Sun has gone through several grand minima like episodes but has recovered always. Now we introduce an additional poloidal field generation mechanism operating on weak magnetic fields (which is below the threshold magnetic field strength necessary for sunspot eruption, i.e., akin to the mean-field poloidal source).

We define this additional $\alpha$-effect as:
 \begin{eqnarray}
    \alpha_{mf}= \alpha_{0mf} \frac{ \cos \theta }{4} \left[1+\textrm{erf}
    \left( \frac{r-r_1}{d_1}\right)\right]
    \left[1-\textrm{erf}\left( \frac{r-r_2}{d_2}\right)\right] \nonumber \\
    \times \frac{1}{1+\left(\frac{B_\phi}{B_{up}}\right)^2} ~~~~
   \end{eqnarray}
where $\alpha_{0mf}$ controls the amplitude of this additional mean-field $\alpha$-effect, $r_1=0.71 R_\odot$, $r_2=R_\odot$, $d_1=d_2=0.25 R_\odot$, and $B_{up}= 10^4~G$ i.e. the upper threshold. The function $\frac{1}{1+\left(\frac{B_\phi}{B_{up}}\right)^2}$ ensures that this additional $\alpha$ effect is only effective on weak magnetic field strengths (below the upper threshold $B_{up}$) and the value of $r_1$ and $r_2$ confirms that this additional mechanism takes place inside the bulk of the convection zone (see top panel of figure 3.2(b) for radial profile of the mean field $\alpha$-coefficient). In a model set up this way, poloidal field generation takes place due to the combined effect of the Babcock-Leighton mechanism and a mechanism effective on weak magnetic fields. Simulations with stochastic fluctuation in both the Babcock-Leighton mechanism and an additional mean field $\alpha$-effect show that the cycle recovers from grand minima-like episodes and regains normal activity levels (see bottom panel of figure 3.3 for representative solution with fluctuations in both the Babcock-Leighton and mean field $\alpha$). This result confirms our earlier results based on somewhat different dynamo models (Hazra et al. 2014; Passos et al. 2014).

Introducing fluctuations in both the Babcock-Leighton source ($K_{ar}$) and mean field poloidal source terms ($\alpha_{mf}$), we find the parity of dynamo solutions change (see top panels of figure 3.6 and 3.7). Earlier studies indicate that the dipolar parity of dynamo solutions is associated with strong hemispheric coupling -- which can be obtained either by increasing diffusivity or by introducing an additional mean field $\alpha$ effect (distributed through the convection zone, or tachocline) (Dikpati \& Gilman 2001; Chatterjee et al. 2004). These models do not consider stochastic fluctuation in their simulation. Introducing stochastic fluctuations in dynamo simulations we always find solutions of single periodicity in both hemispheres even if we introduce very large fluctuations in poloidal field source terms (both BL and mean field $\alpha$ effect).

What is the cause of parity change in our model? One possible reason is the different levels of fluctuations in poloidal field source terms associated with northern and southern hemispheres. Stochastic fluctuations or randomness in the poloidal source is plausibly at the heart of hemispheric asymmetry (Hoyng 1988). Thus we think that there may be a relationship between hemispheric asymmetry and parity change. To investigate the relationship between parity and hemispheric asymmetry, we have to define hemispheric asymmetry in the context of our simulations. In our kinematic flux transport dynamo model, we model the Babcock-Leighton mechanism by double ring algorithm. We believe that double ring algorithm is a more realistic way to capture the essence of the Babcock-Leighton mechanism as well as sunspots. For this work, we take the difference between double ring eruptions in the northern and southern hemispheres as a measure of hemispheric asymmetry. We call this difference the signed asymmetry and only the magnitude of this difference as asymmetry for the rest of the chapter.

\begin{figure*}[!htb]
        \centering
        \includegraphics[width=15 cm]{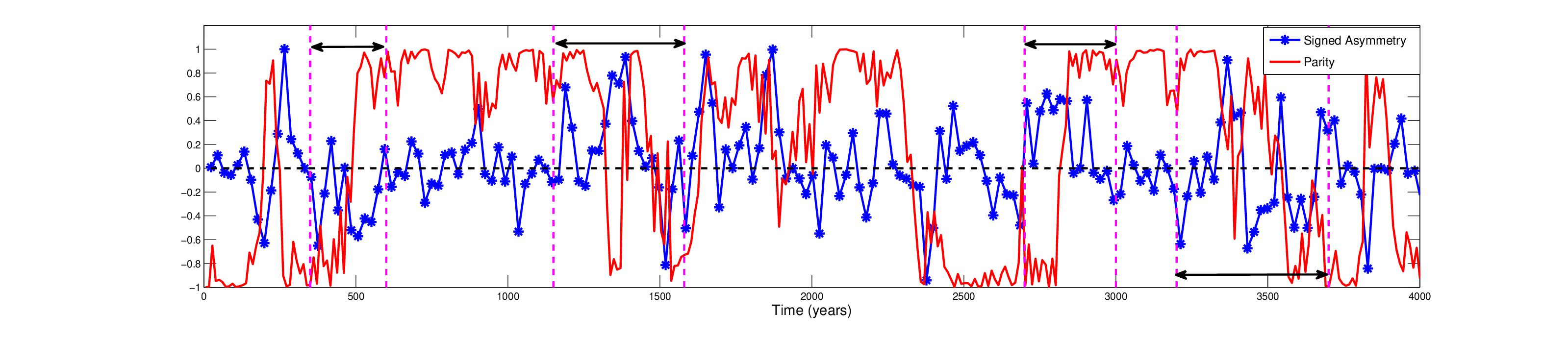}
        \includegraphics[height=3cm, width=15 cm]{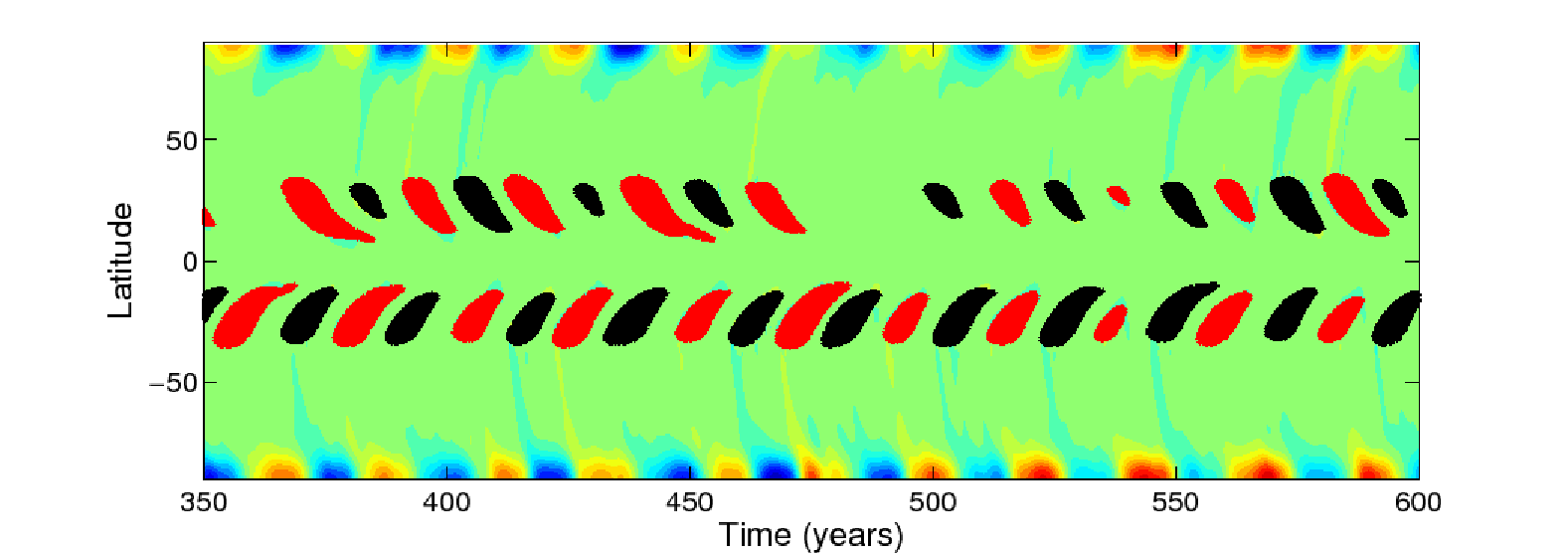}
        \includegraphics[height=3cm, width=15 cm]{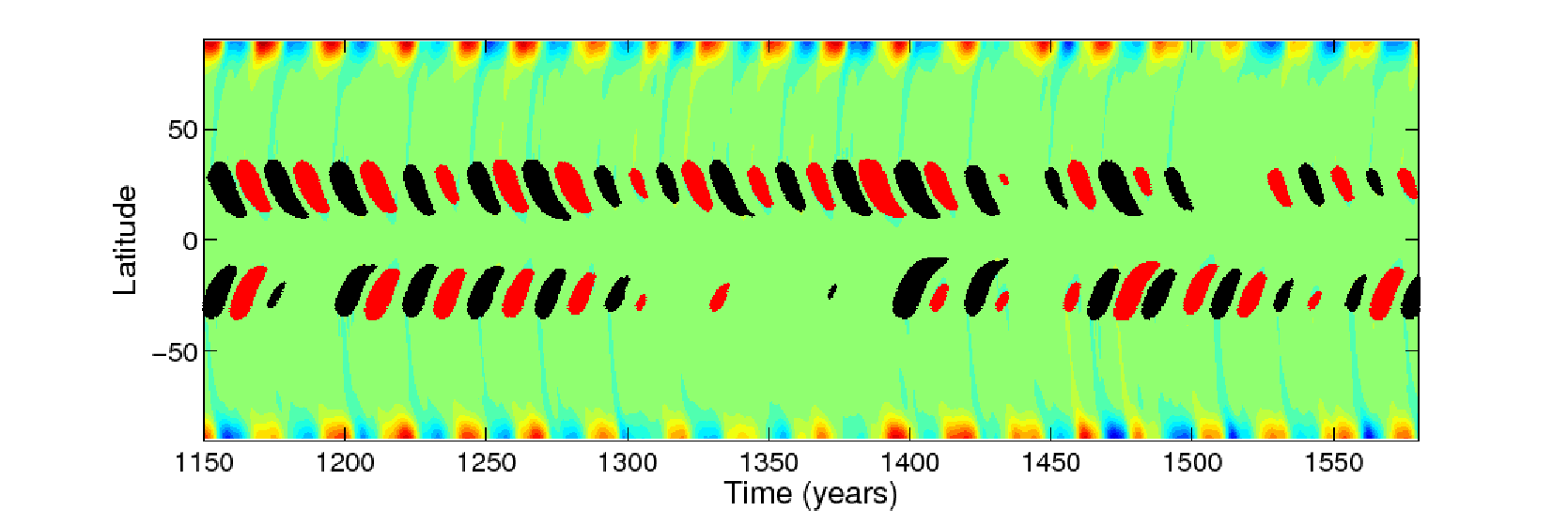}
        \includegraphics[height=3cm,width=15 cm]{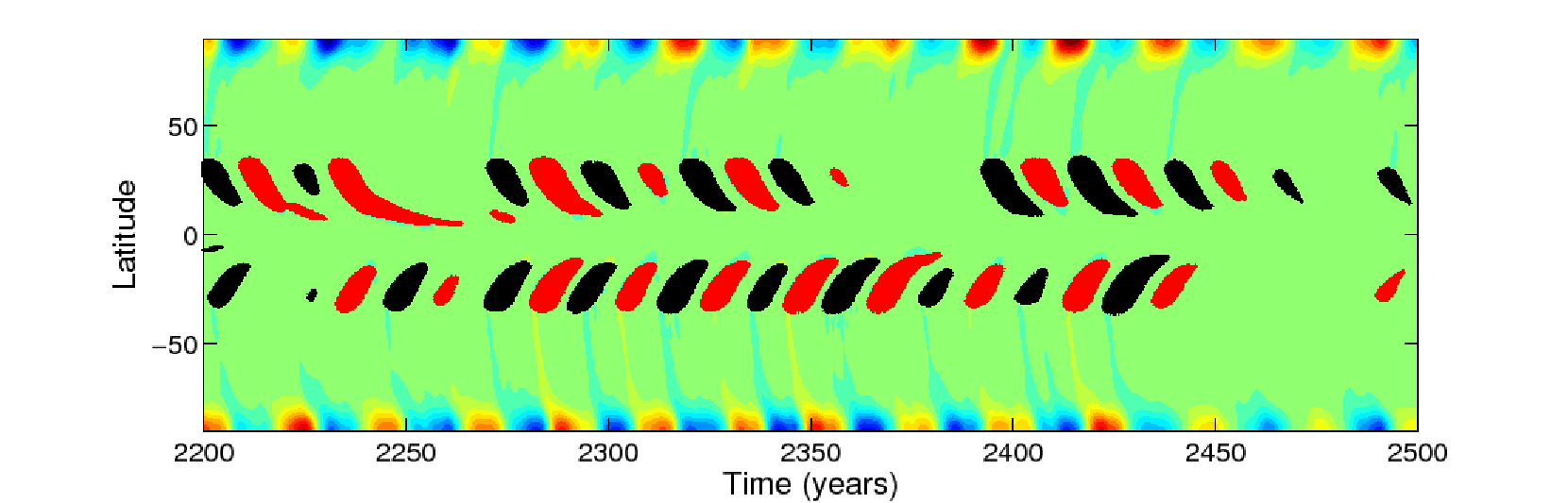}
        \includegraphics[height=3cm,width=15 cm]{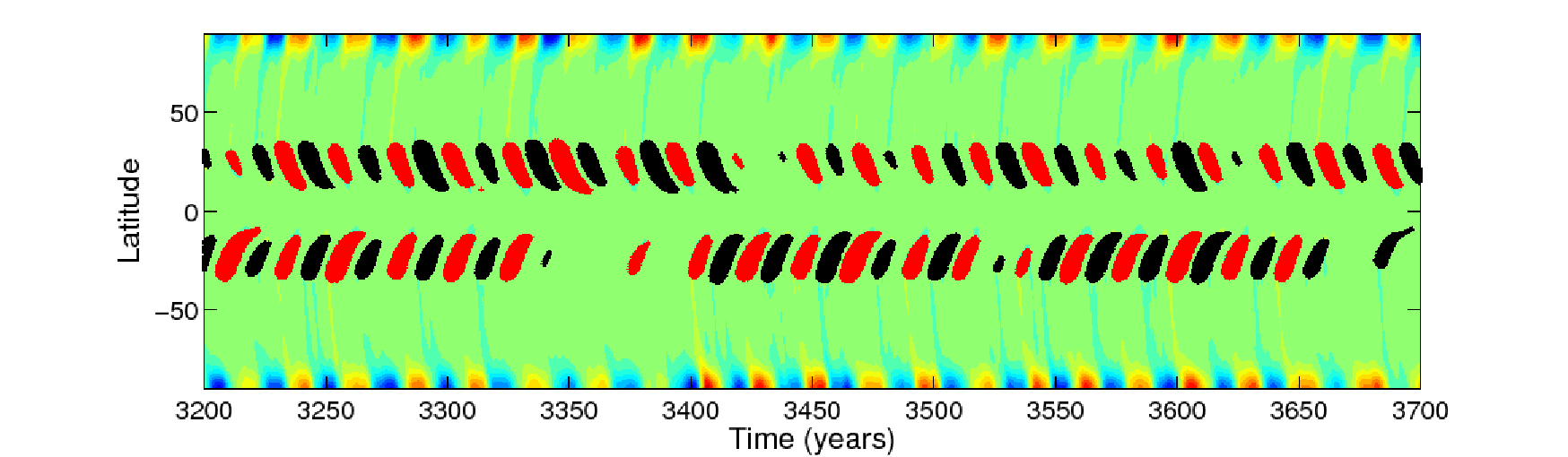}
     \caption{First panel shows the evolution of parity (red colour) and 22 year averaged normalized signed asymmetry (blue color) obtained from our simulations. Second, third, fourth and fifth panels are simulated butterfly diagrams for different time intervals where parity change takes
place. Selected time intervals are shown in top panel by double arrow. These simulations indicate solar cycle parity changes take place when activity in one hemisphere dominates over the other for a sufficiently large period of time. This simulation corresponds to 75\% fluctuation in Babcock-Leighton mechanism and 150 \% fluctuation in mean field $\alpha$.}
        \label{fig:3}
\end{figure*}
We find no north-south asymmetry in the simulated solar cycle if we perform dynamo simulations without stochastic fluctuation. Thus our result confirms the idea that stochastic fluctuation is the cause for hemispheric asymmetry. Interestingly, we find that yearly averaged absolute asymmetry obtained from our simulation is not stochastic, rather it shows a systematic periodic behaviour.  We also find that hemispheric asymmetry obtained from our simulation is stronger at solar cycle maximum but also sometimes significant at solar cycle minima. We note that maximum amplitude of yearly averaged absolute asymmetry time series obtained from our simulation is 200, this is because in our double ring algorithm we allow maximum 200 double ring eruptions in a year.
 
To confirm the systematic periodic behaviour of hemispheric asymmetry, we have performed a wavelet analysis of both observational and simulated asymmetry time series. The Royal Greenwich Observatory provides the monthly mean hemispheric sunspot area data series (in unit of millionths of a hemisphere) from 1874 to present. We have created a yearly data series from this data set by averaging over 12 month to examine hemispheric asymmetry. We take the absolute difference between the areas of northern and southern hemispheres as a definition of observed hemispheric asymmetry. Figs.~3.4 and 3.5 show the results of wavelet analysis of observational absolute asymmetry and simulated absolute asymmetry data series. Third and fourth panels of Fig~3.4 and 3.5 shows the wavelet power spectrum and global wavelet analysis of observational and simulated absolute asymmetry time series respectively. We find a clear signature of 11 year periodicity in both observed and simulated asymmetry time series which is also outside the cone of influence. Distortion of wavelet power spectra occurs inside the region named the cone of influence, thus periodicities in these regions are unreliable. The significant periodicities in both observed and simulated asymmetry time series are already present in the hemispheric sunspot area time series and the simulated toroidal field time series, respectively, which basically corresponds to the underlying magnetic cycle. Thus, we conclude that the periodicity in the asymmetry reflects the underlying periodicity of the parent cycle.

Next we focus our attention on the relationship between parity and hemispheric asymmetry obtained from our dynamo simulations. Figures 3.6 and 3.7 are the representative plots of parity and signed asymmetry relationship with different level of fluctuations. The top panels in Figure 3.6 and 3.7 shows the time evolution of parity and 22 year averaged signed asymmetry. Comparing the time evolutions of parity and signed asymmetry we find that parity changes are always associated with strong dominance of eruptions in one hemisphere for a long period of time. We confirm our findings from the simulated butterfly diagrams. The second, third, fourth and fifth panels of figure 3.6 and 3.7 are the corresponding butterfly diagrams for different time intervals where parity change takes place. We also notice that on some occasions there is a strong dominance of eruptions in one hemisphere, but the parity does not change, however, these occasions are rare.

We perform several numerical experiments with different levels of fluctuations in both poloidal field source terms and find that our model results are robust with different level of fluctuations. 

\section{Conclusions}

Here, we first perform solar dynamo simulations considering only the Babcock-Leighton mechanism for poloidal field generation. We find that stochastic fluctuations in the Babcock-Leighton mechanism is a possible candidate for triggering entry into grand minima like episodes. We also confirm that an additional mean-field like $\alpha$-effect capable of working on weak fields is necessary for recovery of the solar cycle. Next we perform simulations where the poloidal field generation takes place through the combined effect of both the Babcock-Leighton mechanism and mean field $\alpha$-effect. By introducing stochastic fluctuations in the poloidal field source terms we find dynamo solutions of changing parity. Earlier results in a different context (without any consideration of stochastic fluctuations in the dynamo source terms) has been indicative that the parity issue may be related to the coupling between hemispheres (Chatterjee \& Choudhuri 2006). We demonstrate that presence of stochastic fluctuations makes hemispheric coupling weak. Thus there may be a possible relationship between hemispheric asymmetry and parity change. A closer investigations reveals that parity changes are likely to occur only when one hemisphere strongly dominates over the other hemisphere for a long period of time persisting over several solar cycles.

Systematic observations over the past century indicates that the solar magnetic field has always been in the dipolar parity state. However, it has been noted that there was large asymmetry at the recovery phase of the Maunder minimum, wherein, appearance of sunspots were almost confined at the southern hemisphere (Ribes \& Nesme-Ribes 1993). At this point, it is unclear whether this was related to any parity change in the Sun before or after the Maunder minimum. Independent simulations using low order dynamo models also predict the possibility of parity flipping in the Sun (Beer et al. 1998; Knobloch et al. 1998). Thus, our results, taken together with other investigations point out that hemispheric coupling, parity shifts and the occurrence of grand minima episodes may be related. These interrelationship needs to be investigated further and may provide a pathway for predicting parity shifts and the onset of grand minima episodes.
\chapter{A New Paradigm of Magnetic Field Dynamics at the Basis of the Sunspot Cycle Based on Turbulent Pumping}

Four hundred years of sunspot observations show the existence of a 11-year periodicity in the appearance of sunspots. At the turn of the 20{$^{th}$} Century, George Ellery Hale discovered that sunspots are strongly magnetized and subsequently the sunspot cycle came to be recognized as the underlying magnetic cycle of the Sun. Solar magnetism is thought to originate via a magnetohydrodynamic dynamo mechanism relying on plasma flows in the Sun's interior. While a fully self-consistent theoretical model of the solar cycle remains elusive, significant progress has been made based on convection simulations and kinematic, flux transport dynamo modelling of the solar interior. The success of these flux transport dynamo models are largely dependent upon a single-cell meridional circulation with a deep equatorward component at the base of the Sun's convection zone. However, recent observations suggest that the meridional flow may in fact be very shallow (confined to the top 10\% of the Sun) and more complex than previously thought. Taken together these observations cast serious doubts on the validity of flux transport dynamo models of the solar cycle. By accounting for the turbulent pumping of magnetic flux as evidenced in magnetohydrodynamic simulations of solar convection, we demonstrate that flux transport dynamo models can generate solar-like magnetic cycles even if the meridional flow is shallow, or altogether absent. These results imply that substantial revisions may be necessary to our current understanding of magnetic flux transport processes within the Sun and by extension, the interior of other solar-like stars.

\section{Introduction}

Despite early, pioneering attempts to self-consistently model the interactions of turbulent plasma flows and magnetic fields in the context of the solar cycle (Gilman 1983; Glatzmaier 1985) such full MHD simulations are still not successful in yielding solutions that can match solar cycle observations. This task is indeed difficult, for the range of density and pressure scale heights, scale of turbulence and high Reynolds number that characterize the SCZ is difficult to capture even in the most powerful supercomputers. An alternative approach to modelling the solar cycle is based on solving the magnetic induction equation in the SCZ with observed plasma flows as inputs and with additional physics gleaned from simulations of convection and flux tube dynamics. These so called flux transport dynamo models have shown great promise in recent years in addressing a wide variety of solar cycle problems (Charbonneau 2010; Ossendrijver 2003).

\begin{figure}[!htb]
  \begin{center}
\begin{tabular}{cc}
\includegraphics[scale=0.43]{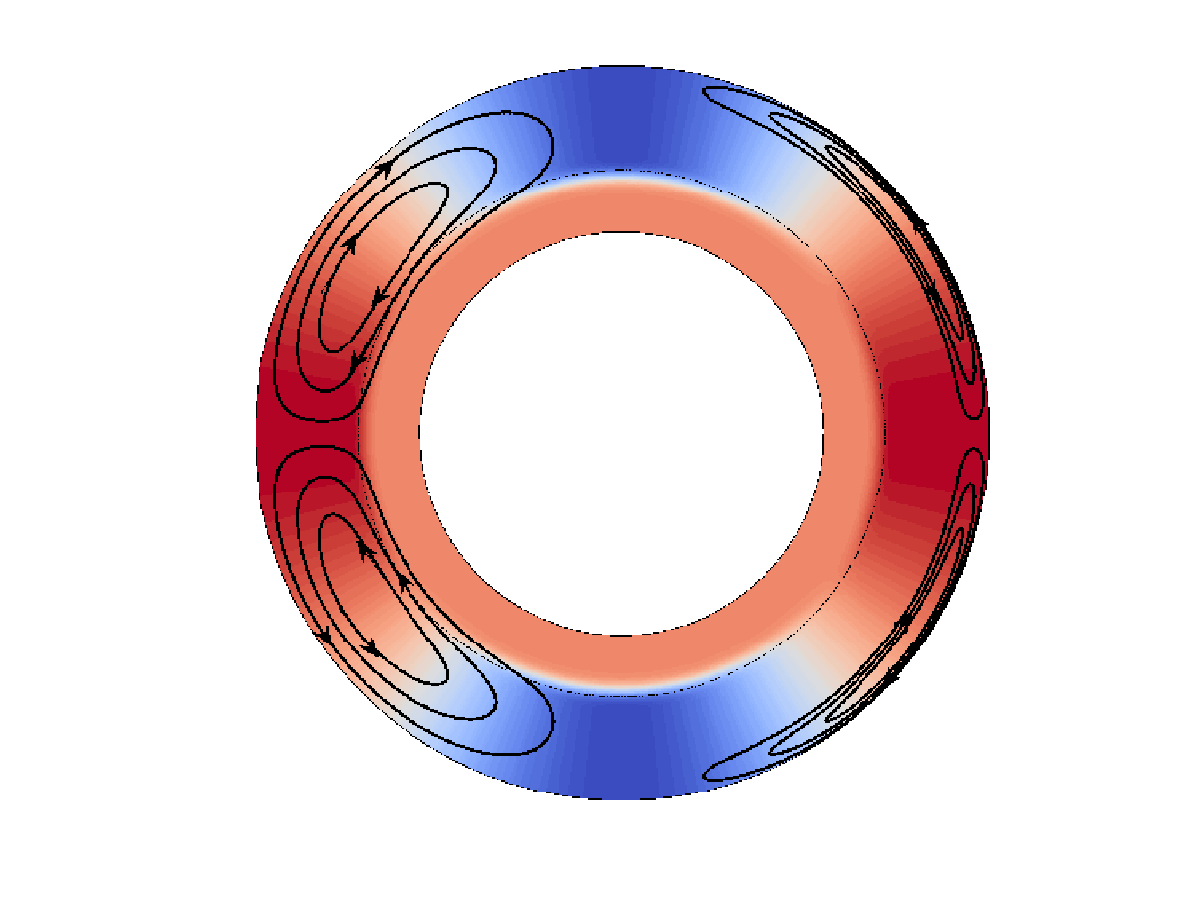}
\end{tabular}
\end{center}
\caption{The outer 45\% of the Sun depicting the internal rotation profile in color. Faster rotation is denoted in deep red and slower rotation in blue. The equator of the Sun rotates faster than the polar regions and there is a strong shear layer in the rotation near the base of the convection zone (denoted by the dotted line). Streamlines of a deep meridional flow (solid black curves) reaching below the base of the solar convection zone (dashed line) is shown on the left hemisphere, while streamlines of a shallow meridional flow confined to the top 10\% of the Sun is shown on the right hemispheres (arrows indicate direction of flow). Recent observations indicate that the meridional flow is much shallower and more complex than traditionally assumed, calling in to question a fundamental premise of flux transport dynamo models of the solar cycle.}
\end{figure}

In particular, solar dynamo models based on the Babcock-Leighton mechanism for poloidal field generation have been more successful in explaining diverse observational features of the solar cycle (Dikpati \& Charbonneau 1999; Nandy \& Choudhuri 2002; Chatterjee et al. 2004;Choudhuri et al. 2004; Nandy et al. 2011; Choudhuri \& Karak 2012, Hazra et al. 2014; Passos et al. 2014). Recent observations also strongly favor the Babcock-Leighton mechanism as a major source for poloidal field generation (Dasi-Espuig et al. 2010; Mu\~noz-Jaramillo et al. 2013). In this scenario, the poloidal field generation is essentially predominantly confined to near-surface layers. For the dynamo to function efficiently, the toroidal field that presumably resides deep in the interior has to reach the near-surface layers for the Babcock-Leighton poloidal source to be effective. This is achieved by the buoyant transport of magnetic flux from the Sun's interior to its surface (through sunspot eruptions). Subsequent to this the poloidal field so generated at near-surface layers must be transported back to the solar interior, where differential rotation can generate the toroidal field. The deep meridional flow assumed in such models (See Fig.~4.1, left-hemisphere) plays a significant role in this flux transport process and is thought to govern the period of the sunspot cycle (Charbonneau \& Dikpati 2000; Hathaway et al. 2003, Yeates et al. 2008, Hazra et al. 2014). Moreover, a fundamentally crucial role attributed to the deep equatorward meridional flow is that it allows the Parker-Yoshimura sign rule (Parker 1955; Yoshimura 1975) to be overcome, which would otherwise result in poleward propagating dynamo waves in contradiction to observations that the sunspot belt migrates equatorwards with the progress of the cycle (Choudhuri et al. 1995; G.Hazra et al. 2014; Passos et al. 2015; Belusz et al. 2015).

While the poleward meridional flow at the solar surface is well observed (Hathaway \& Rightmire 2010, 2011) the internal meridional flow profile has remained largely unconstrained. A recent study utilizing solar supergranules (Hathaway 2012) indicates that the meridional flow is confined to within the top 10\% of the Sun (Fig.~4.1, right-hemisphere) -- much shallower than previously thought. Independent studies utilizing helioseismic inversions is also indicative that the equatorward meridional counterflow may be located at shallow depths (Mitra-Kraev \& Thompson 2007; Zhao et al. 2013). The latter also infer the flow to be multi-cellular and more complex. Here, utilizing a newly developed state-of-the-art Babcock-Leighton flux transport dynamo model, we explore the impact of a shallow meridional flow on our current understanding of the solar cycle and provide a new paradigm which resolves the significant challenges posed by these new observations.

\section{Results}
Our flux transport solar dynamo model (see chapter 3) solves for the coupled, evolution equation for the axisymmetric toroidal and poloidal components of the solar magnetic field with an analytic fit to the observed solar differential rotation, a two-step turbulent diffusivity profile (which ensures a smooth transition to low levels of diffusivity beneath the base of the convection zone), and a new implementation of a double-ring algorithm for buoyant sunspot eruptions that best captures the Babcock-Leighton mechanism for poloidal field generation (Mu\~noz-Jaramillo et al. 2013; Hazra \& Nandy 2013) and which has been tested thoroughly in other contexts. To bring out the significance of the recent observations, we first consider a single cell, shallow meridional flow, confined only to the top 10\% of the convection zone (Fig.~4.1, right-hemisphere). In the first scenario we seek to answer the following question: Can solar-like cycles be sustained through magnetic field dynamics completely confined to the top 10\% of the Sun?
\begin{figure}[!htb]
 \begin{center}
\begin{tabular}{cc}
\includegraphics[scale=0.7]{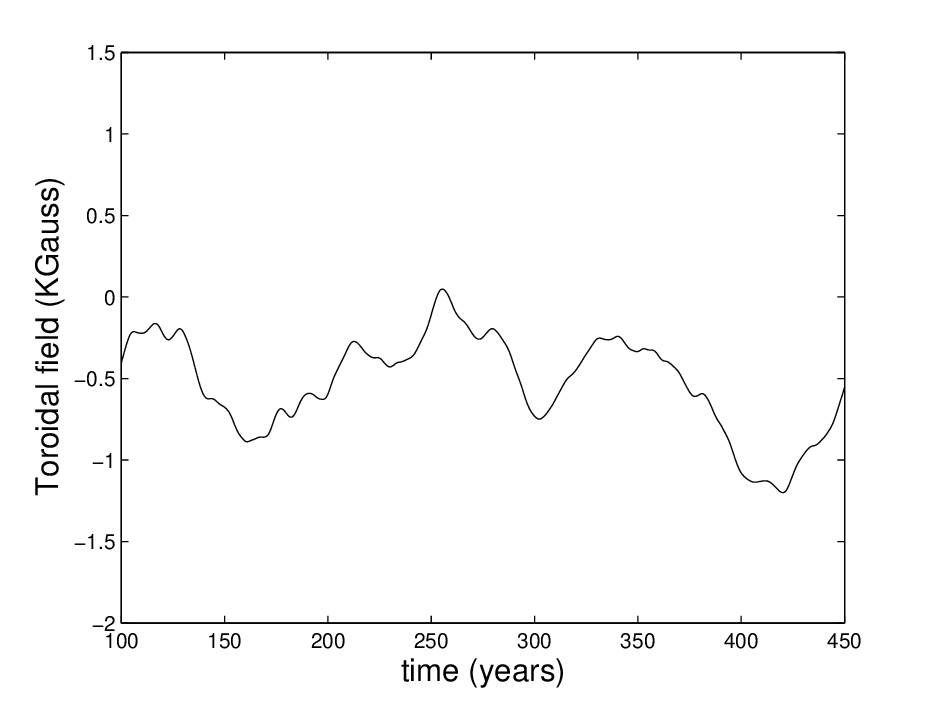} 
 \end{tabular}
\end{center}               
\caption{Evolution of the toroidal field when we allow magnetic flux tubes to buoyantly erupt from near surface layer i.e. 0.90 $R_{\odot}$) above a critical buoyancy threshold of $10^4$ Gauss.}
\end{figure}

\begin{figure}[!htb]
  \begin{center}
\begin{tabular}{cc}
\includegraphics[scale=0.7]{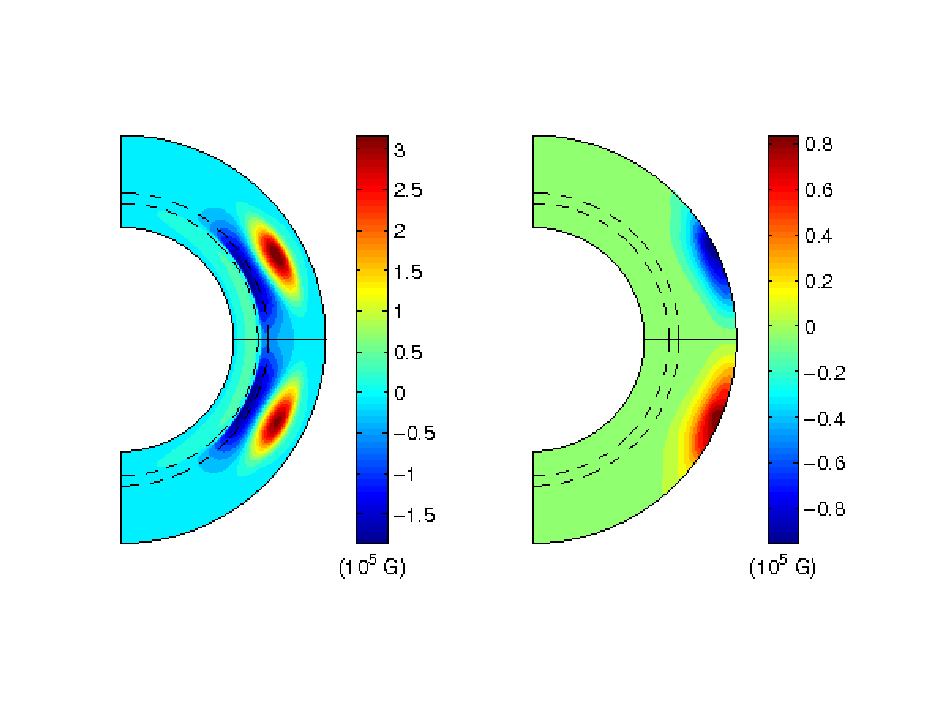} \\
(a)~~~~~~~~~~~~~~~~~~~~~~~~~~~~~~~~~~~~~~~~~~(b)\\
\includegraphics[scale=0.7]{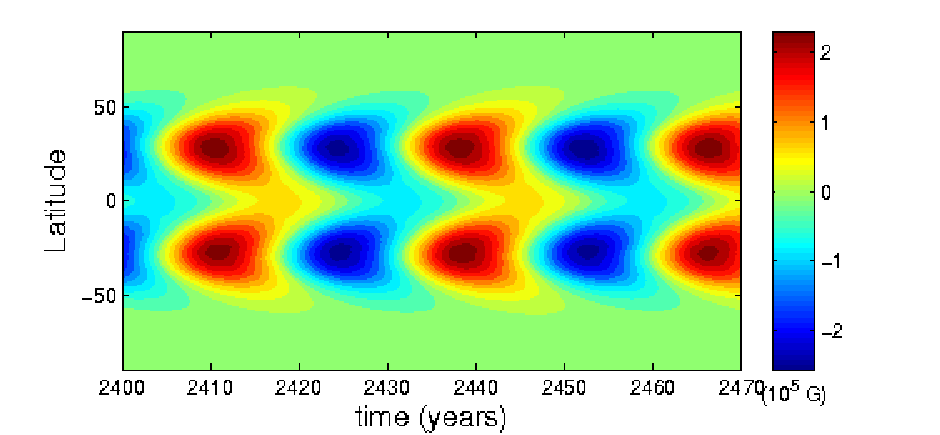} \\
~~~~~~(c)
 \end{tabular}
\end{center}
\caption{Solar cycle simulations with a shallow meridional flow. The toroidal (a) and poloidal (b) components of the magnetic field is depicted within the computational domain at a phase corresponding to cycle maxima. The solar interior shows the existence of two toroidal field belts, one at the base of the convection zone and the other at near-surface layers where the shallow equatorward meridional counterflow is located. Region between two dashed circular arcs indicates the tachocline. (c) A butterfly diagram generated at the base of convection zone showing the spatiotemporal evolution of the toroidal field. Clearly, there is no dominant equatorward propagation of the toroidal field belt and the solution displays quadrupolar parity (i.e., symmetric toroidal field across the equator) which do not agree with observations.}
\end{figure}

 In these simulations, first we allow magnetic flux tubes to buoyantly erupt from 0.90 $R_{\odot}$ (i.e., the depth to which the shallow flow is confined) when they exceed a buoyancy threshold of $10^4$ Gauss (G). In this case we find that the dynamo remains sub-critical (see Fig.~4.2)   with no sunspot eruptions, implying that a solar-like cycle cannot be produced in this case. Given that the upper layers of the SCZ is highly turbulent (characterized by a high turbulent diffusivity), storage and amplification of strong magnetic flux tubes may not be possible in these layers (Parker 1975; Moreno-Insertis 1983) and therefore this result is not unexpected. In the second scenario with a shallow meridional flow, we allow magnetic flux tubes to buoyantly erupt from 0.71 $R_{\odot}$, i.e. from base of the convection zone. In this case we get periodic solutions but analysis of the butterfly diagrams (taken both at the base of SCZ and near solar surface) shows that the toroidal field belts have almost symmetrical poleward and equatorward branches with no significant equatorward migration (see Fig.~4.3). Moreover the solutions always display quadrupolar parity in contradiction with solar cycle observations. Clearly, a shallow flow poses a serious problem for solar cycle models.

\begin{figure}[!htb]
\begin{center}
\begin{tabular}{cc}
\includegraphics[scale=0.45]{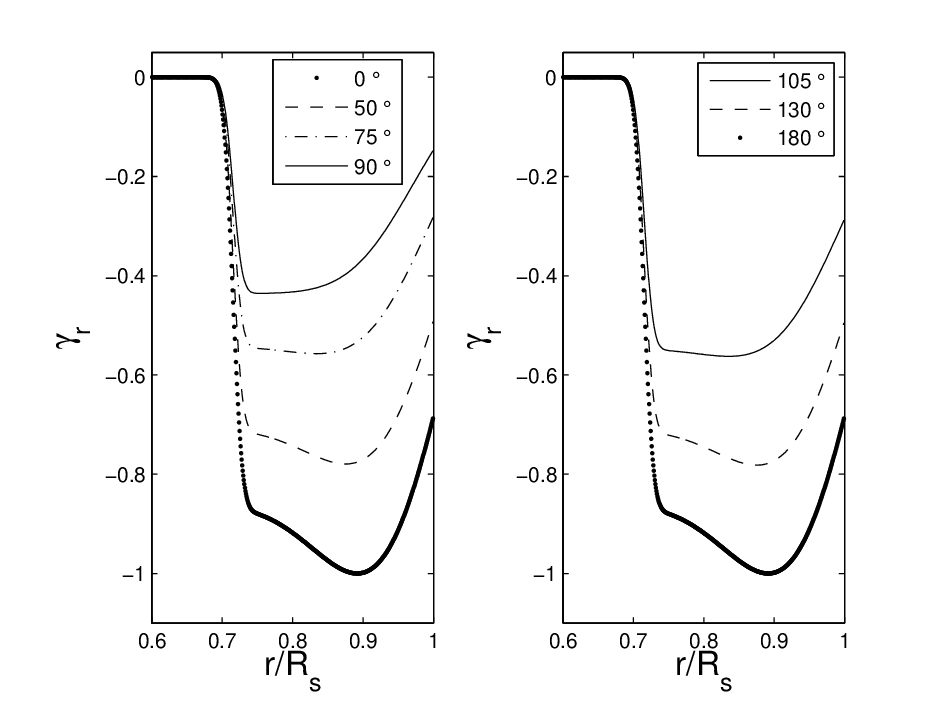}  \includegraphics[scale=0.45]{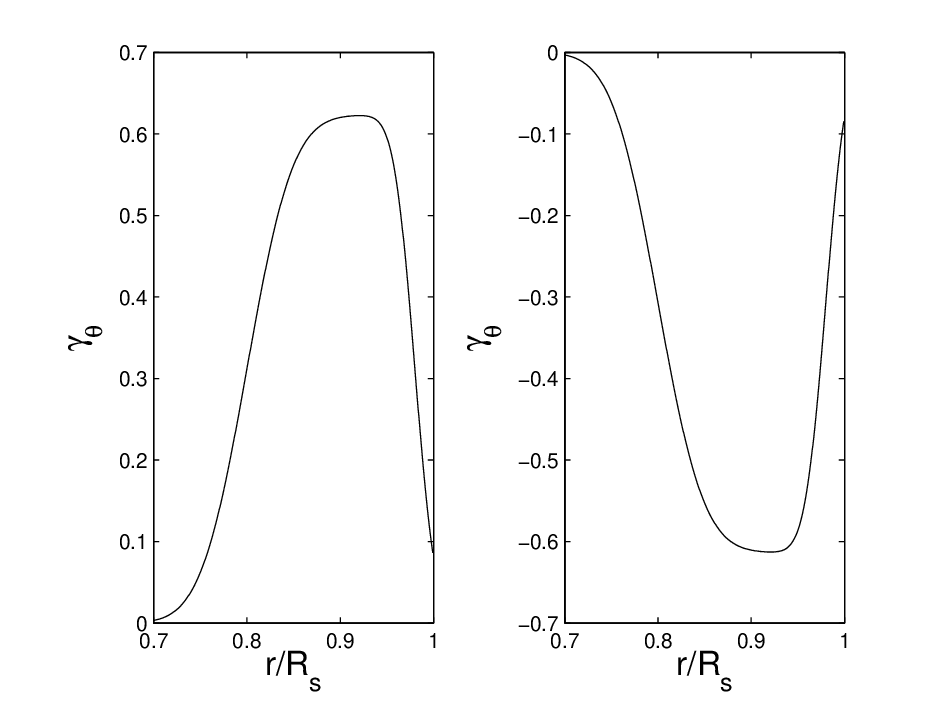}\\
~~(a)~~~~~~~~~~~~~~~~~~~~~~(b)~~~~~~~~~~~~~~~~~~~~~~~~~~~~~~~~~~(c)~~~~~~~~~~~~~~~~~~~~~~~~~(d)~~\\
 \end{tabular}
\end{center}               
\caption{First two plots show the variation of radial pumping $\gamma_r$ (in ms$^{-1}$) at co-latitudes of northern hemisphere and southern hemisphere respectively, with fractional solar radius from the solar surface to the solar interior. Radial turbulent pumping is negative at both hemisphere. Next two plots show the variation of latitudinal pumping $\gamma_\theta$ (in ms$^{-1}$) at $45^\circ$ mid latitudes of both northern hemisphere and southern hemisphere respectively, with fractional solar radius from the solar surface to the solar interior. It is positive in northern hemisphere and negative in southern hemisphere.}
 \end{figure}
 \begin{figure}[!htb]
\begin{center}
\begin{tabular}{cc}
\includegraphics[scale=0.45]{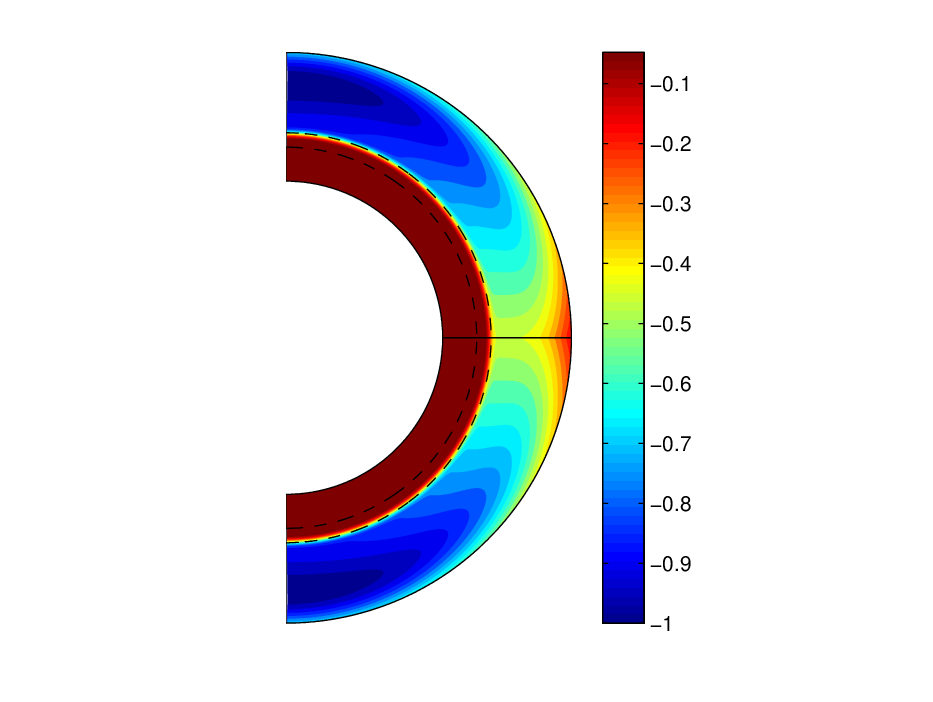}  \includegraphics[scale=0.45]{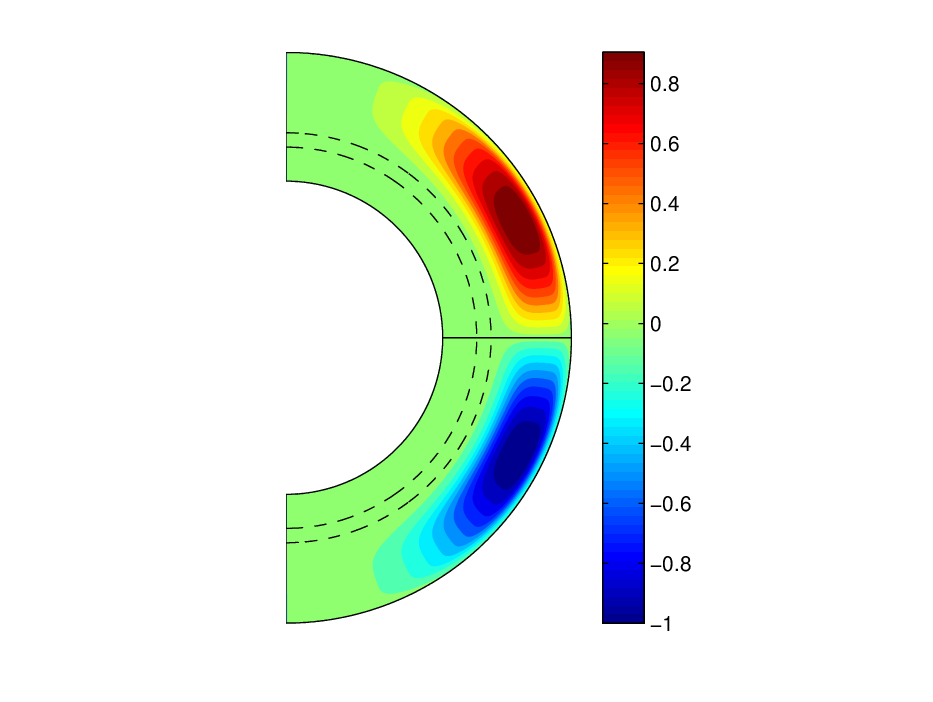}\\
 \end{tabular}
\end{center}               
\caption{ Plot of left side represents the contour plot of radial pumping profile and plot of right side represents the contour plot of latitudinal pumping profile. In this case, peak value of $\gamma_r$ and $\gamma_\theta$ is 0.4 ms$^{-1}$ and 1 ms$^{-1}$ respectively. Region between two dashed
circular arcs indicates the tachocline.}
 \end{figure} 
We note that most flux transport solar dynamo models do not include the process of turbulent pumping of magnetic flux. Magnetoconvection simulations supported by theoretical considerations (Brandenburg et al. 1996; Tobias et al. 2001; Dorch \& Nordlund 2001; K\"apyl\"a et al. 2006; Pipin \& Seehafer 2009; Rogachevskii et al. 2011) have established that turbulent pumping preferentially transports magnetic fields vertically downwards -- likely mediated via strong downward convective plumes which are particularly effective on weak magnetic fields (such as the poloidal component). In strong rotation regimes, there is also a significant latitudinal component of turbulent pumping. The few studies that exist on the impact of turbulent pumping in the context of flux transport dynamo models show it to be dynamically important in the maintenance of solar-like parity and solar-cycle memory (Guerrero \& de Gouveia Dal Pino 2008; Karak \& Nandy 2012; Jiang et al. 2013). Motivated by these considerations, we introduce both radial and latitudinal turbulent pumping in the dynamo model with shallow meridional flow. The turbulent pumping profile is determined from independent MHD simulations of solar magnetoconvection (Ossendrijver et al. 2002; K\"apyl\"a et al. 2006). Profiles for radial and latitudinal turbulent pumping ($\gamma_r$ and $\gamma_\theta$) are:
\begin{eqnarray}
\gamma_r = -  \gamma_{0r} \left[ 1 + \rm{erf}\left( \frac{r - 0.715}{0.015}\right) \right] \left[ 1 - \rm{erf} \left( \frac{r-0.97}{0.1}\right) \right] \nonumber \\
\times \left[ \rm{exp}\left( \frac{r-0.715}{0.25}\right) ^2 \rm{cos}\theta +1\right] ~~~~
\end{eqnarray}
\begin{eqnarray}
\gamma_\theta = \gamma_{0\theta} \left[1+\mathrm{erf}\left(\frac{r-0.8}{0.55}\right)\right]
\left[1-\mathrm{erf}\left(\frac{r-0.98}{0.025}\right)\right] 
\times \cos \theta \sin^4 \theta ~~~~
\end{eqnarray}
The value of $\gamma_{0r}$ and $\gamma_{0\theta}$ determines the amplitude of $\gamma_r$ and $\gamma_\theta$ respectively. Fig.~4.4(a) and (b) shows that radial pumping speed is negative throughout the convection zone corresponds to downward movement of magnetized plasma and vanishes below $0.7R_\odot$ , the radial pumping speed is maximum near the poles and decreases towards the equator. Fig.~4.4(c) and (d) shows that latitudinal pumping speed is positive (negative) in the convection zone of northern (southern) hemisphere and vanishes below the overshoot layer. Contour plots of turbulent pumping profiles are shown in Fig.~4.5.

\begin{figure}[!htb]
\begin{center}
\begin{tabular}{cc}
\includegraphics[scale=0.7]{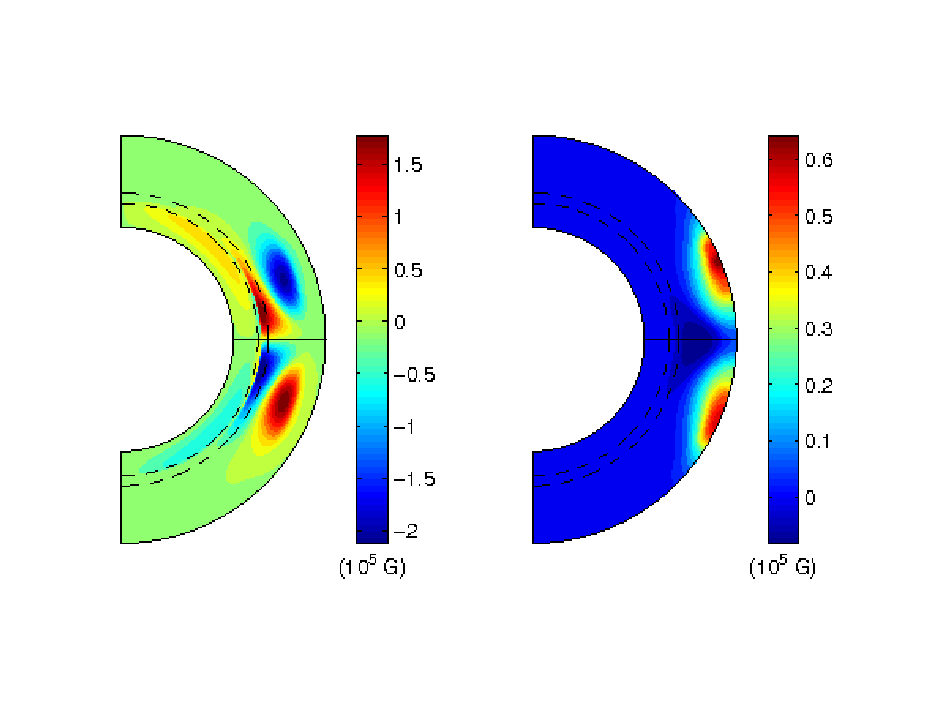} \\
(a)~~~~~~~~~~~~~~~~~~~~~~~~~~~~~~~~~~~~~~~~~~~~~~~(b)\\
\includegraphics[scale=0.7]{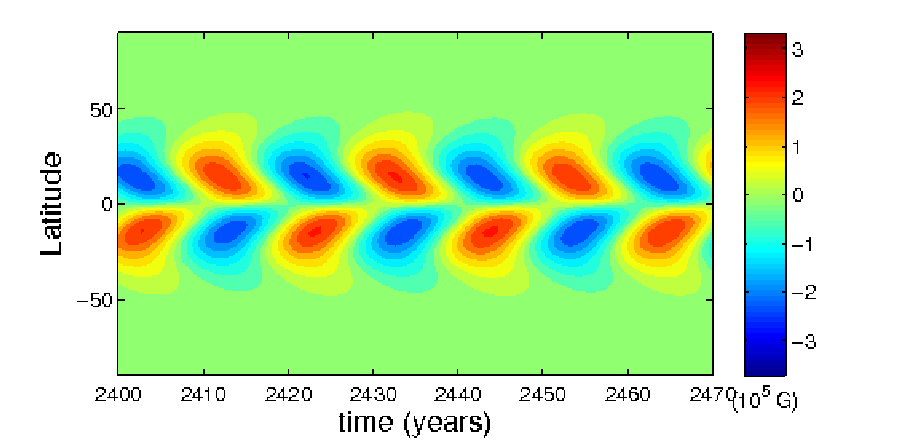} \\
~~~~(c)
 \end{tabular}
\end{center}
\caption{Dynamo simulations with shallow meridional flow but with radial and latitudinal turbulent pumping included (same convention is followed as in Fig.~4.3). The toroidal (a) and poloidal field (b) plots show the dipolar nature of the solutions, and the butterfly diagram at the base of the convection zone clearly indicates the equatorward propagation of the toroidal field that forms sunspots.}
\end{figure}

 Dynamo simulations with turbulent pumping generate solar-like magnetic cycles (Fig.~4.6 and Fig.~4.7). Now the toroidal field belt migrates equatorward, the solution exhibits solar-like parity and the correct phase relationship between the toroidal and poloidal components of the magnetic field (see Fig.~4.7). Evidently, the coupling between the poloidal source at the near-surface layers with the deeper layers of the convection zone where the toroidal field is stored and amplified, the equatorward migration of the sunspot-forming toroidal field belt and correct solar-like parity is due to the important role played by turbulent pumping. We note if the speed of the latitudinal pumping in on order of 1.0 ms$^{-1}$ the solutions are always of dipolar parity irrespective of whether one initializes the model with dipolar or quadrupolar parity. Interestingly, the latitudinal migration rate of the sunspot belt as observed is of the same order.

 \begin{figure}[!htb]
\begin{center}
\begin{tabular}{cc}
\includegraphics[scale=0.75]{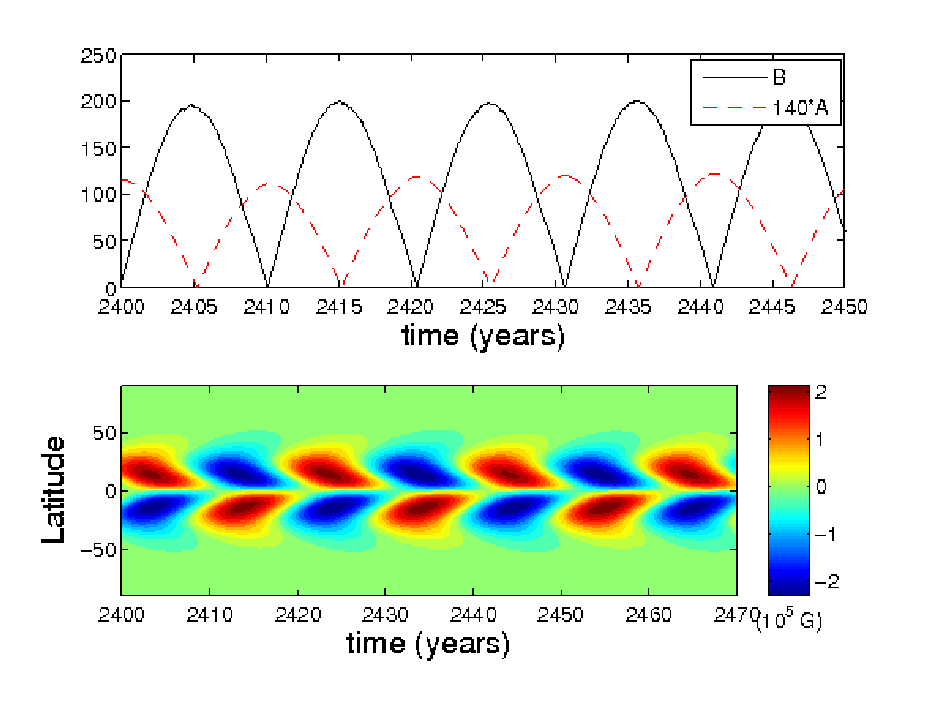} 
 \end{tabular}
\end{center}               
\caption{Dynamo simulations considering both shallow meridional flow and turbulent pumping but started with symmetric initial condition. Plot of top pannel shows the correct phase relationship between toroidal and poloidal field while bottom pannel shows the butterfly diagram taken at the base of the convection zone.}
 \end{figure} 
 
 \begin{figure}[!htb]
\begin{center}
\begin{tabular}{cc}
\includegraphics[scale=0.7]{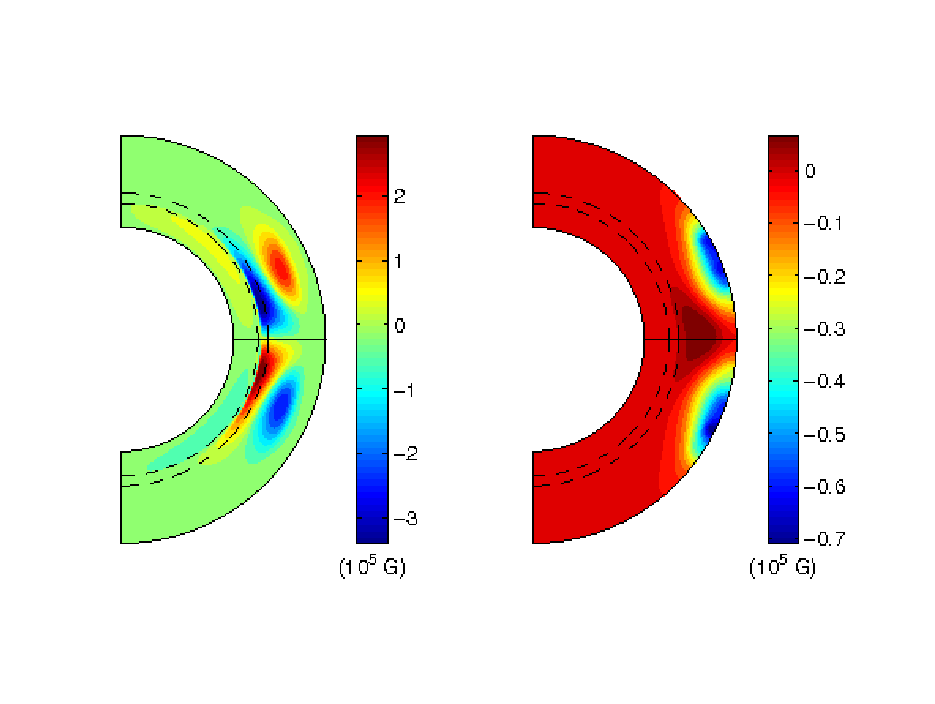} \\
(a)~~~~~~~~~~~~~~~~~~~~~~~~~~~~~~~~~~~~~~~~~~~~~~(b)\\
\includegraphics[scale=0.7]{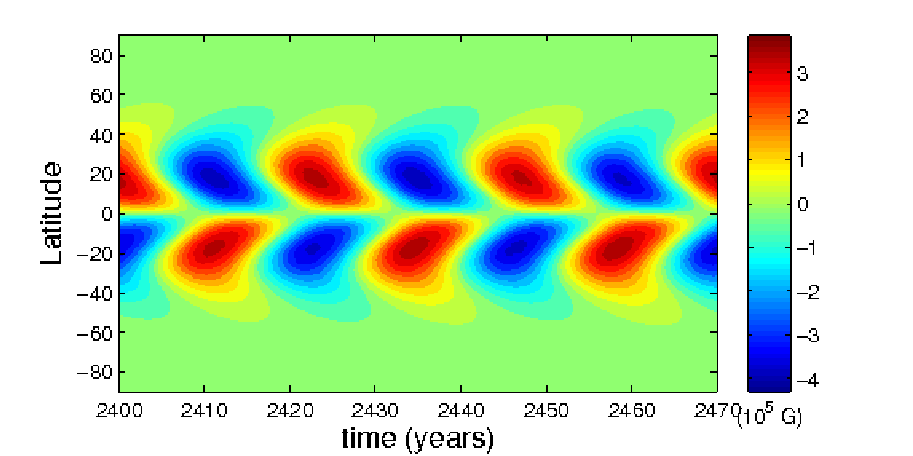} \\
~~~~(c)
 \end{tabular}
\end{center}
\caption{Results of solar dynamo simulations with turbulent pumping and without any meridional circulation. The convention is the same as in Fig.~4.3. The simulations show that solar-like sunspot cycles can be generated even without any meridional plasma flow in the solar interior.}
\end{figure}
The above result begs the question whether flux transport solar dynamo models based on the Babcock-Leighton mechanism that include turbulent pumping can operate without meridional plasma flows. To test this, we remove meridional circulation completely from our model and perform simulations with turbulent pumping included. We find that this model generates solar-like sunspot cycles (see Fig.~4.8) which is qualitatively similar to the earlier solution with both pumping and shallow meridional flow indicating that it is turbulent pumping which predominantly determines the dynamics. This surprising and un-anticipated result suggests a new scenario for magnetic field transport leading to the generation of the solar cycle which we discuss in the next section.

\section{Discussions}

In summary, we have demonstrated that flux transport dynamo models of the solar cycle based on the Babcock-Leighton mechanism for poloidal field generation does not require a deep equatorward meridional plasma flow to function effectively. In fact, our results indicate that when turbulent pumping of magnetic flux is taken in to consideration, dynamo models can generate solar-like magnetic cycles even without any meridional circulation. These findings have significant implications for our understanding of the solar cycle. First of all, the serious challenges that was apparently posed by observations of a shallow (and perhaps complex, multi-cellular) meridional flow on the very premise of flux transport dynamo models stands resolved. Turbulent pumping essentially takes over the role of meridional circulation by transporting magnetic fields from the near-surface solar layers to the deep interior, ensuring that efficient recycling of toroidal and poloidal fields components across the SCZ is not compromised. While these findings augur well for dynamo models of the solar cycle, they also imply that we need to revisit many aspects of our current understanding.

It has been argued earlier that the interplay between competing flux transport processes determine the dynamical memory of the solar cycle governing solar cycle predictability (Yeates et al. 2008). If turbulent pumping is the dominant flux transport process as seems plausible based on the simulations presented herein, the cycle memory would be short and this is indeed supported by independent studies (Karak \& Nandy 2012) and solar cycle observations (Mu\~noz-Jaramillo et al. 2013). It is noteworthy that on the other hand, if meridional circulation were to be the dominant flux transport process, the solar cycle memory would be relatively longer and last over several cycles. This is not borne out by observations.

Previous results in the context of the maintenance of solar-like dipolar parity have relied on a strong turbulent diffusion to couple the Northern and Southern hemispheres of the Sun (Chatterjee et al. 2004), or a dynamo $\alpha$-effect which is co-spatial with the deep equatorward counterflow in the meridional circulation assumed in most flux transport dynamo models (Dikpati \& Gilman 2001). However, our results indicate that turbulent pumping is equally capable of coupling the Northern and Southern solar hemispheres and aid in the maintenance of solar-like dipolar parity. This is in keeping with earlier, independent simulations based on a somewhat different dynamo model (Guerrero \& de Gouveia Dal Pino 2008).

Most importantly, our results point out a completely new alternative to circumventing the Parker-Yoshimura sign rule (Parker 1955; Yoshimura 1975) that would otherwise imply poleward propagating sunspot belts in conflict with observations of equatorward propagation of sunspot belt with the progress of the solar cycle. While a deep meridional counterflow is currently thought to circumvent this constraint and force the toroidal field belt equatorward, our results show that the latitudinal component of turbulent pumping provides a viable alternative to overcoming the Parker-Yoshimura sign rule. 

We note however that our theoretical results should not be taken as support for the existence of a shallow meridional flow, rather we point out that flux transport dynamo models of the solar cycle are equally capable for working with a shallow or non-existent meridional flow, as long as the turbulent pumping of magnetic flux is accounted for. Taken together, these insights suggest a plausible new paradigm for dynamo models of the solar cycle, wherein, turbulent pumping of magnetic flux effectively replaces the important roles that are currently thought to be mediated via a deep meridional circulation within the Sun's interior. Since the dynamical memory and thus predictability of the solar cycle depends on the dominant mode of magnetic flux transport in the Sun's interior, this would also imply that physics-based prediction models of long-term space weather need to adequately include the physics of turbulent pumping of magnetic fields.

\chapter{Observational Studies of Magnetic Field Dynamics in the Solar Atmosphere}

After discussing the generation process of large scale magnetic fields inside the Sun,  in the second part of this thesis we will concentrate on  observations of the photospheric magnetic field of the Sun and constrain the role of photospheric magnetic fields in atmospheric dynamics. Two major explosive events in the solar atmosphere, namely, solar flare and coronal mass ejection, release huge amounts of plasma to outer space. One of the major outstanding problem in astrophysics is the coronal heating problem: the solar corona is much hotter compared to the photosphere. In this chapter, we will discuss some observational techniques and the current theoretical understanding for explaining the observations of the solar atmosphere.
\section{Measuring large scale solar magnetic field} 
  Zeeman effect is a widely used method to measure solar magnetic fields. In this effect energy levels of atom split into several levels in presence of static magnetic field (based on degeneracy of energy level). From the atomic absorption lines interspersed in the black-body continuum spectra, one can easily choose certain magnetically sensitive lines which show larger splitting due to their large Lande-g factor. One can measure different states of polarization from these magnetically sensitive lines, in terms of Stokes parameters I, Q, U, V. The Stokes parameter are used to describe the polarization state of electromagnetic radiation. These parameters are easily measurable by techniques of remote sensing. Stokes parameters of a partially polarized light are defined as (Stokes, 1852):\\

    $\mathbf{I}~= ~~\updownarrow + ~\leftrightarrow ~= ~I_{lin}(0^\circ) + ~I_{lin}(90^\circ)~= ~I_{x}~+~I_{y}$ \\
    
    $~\mathbf{Q}~= ~~\updownarrow - ~\leftrightarrow ~= ~I_{lin}(0^\circ) - ~I_{lin}(90^\circ)~=~I_{x}~-~I_{y}$\\
    
    $\mathbf{U}~= ~~\nearrow - ~\searrow ~= ~I_{lin}(45^\circ) - ~I_{lin}(135^\circ)$ \\
    
    $\mathbf{V}~= ~~\circlearrowleft - ~\circlearrowright ~=~ I_{circ}(left) - ~I_{circ}(right)$\\
    Stokes I vector represents integrated unpolarized light. Stokes Q parameter represents the difference between the x and y axis intensities transmitted through a polaroid. Parameter U is the difference between intensities transmitted through a polaroid having axis at $45^\circ$ and $-45^\circ$ to the x-axis and parameter V is the difference between between the amount of right handed circularly polarized light and left handed circularly polarized light present in the light.
    
 It is possible to determine the strength and direction of magnetic field from the measurement of such polarized signals, using Stokes inversion techniques (Skumanich \& Lites, 1987).
\section{Effects of large scale solar magnetic field}
 \subsection{Null Points and Current Sheet}
Null point is the point where all three components of a magnetic field vanishes. It is a general feature of magnetic fields containing multiple sources, e.g., a field produced by two bar magnets. Current sheets can form only when the medium is conducting plasma, unlike neutral points which can form irrespective of the background medium whether it is conducting plasma or neutral gas.

 A thin current carrying layer across which the magnetic field changes in magnitude or direction or both, is a current sheet. It has been shown that null points typically give rise to current sheets in conducting plasma. R. G. Giovanelli (1946) and F. Hoyle (1949) first suggested that magnetic X-type null points are preferred locations for plasma heating and initiation of solar flares. T. G. Cowling (1953) pointed out that to power a solar flare, a current sheet of only few meters thickness is needed. Around almost the same time, J. W. Dungey (1953) suggested that such current sheets can form by collapsing magnetic field lines near an X-type null point.
\subsection{Magnetic Reconnection}
The Magnetic induction equation is given by,
\begin{equation}
\frac{\partial \mathbf{B}}{\partial t} = \nabla \times (\mathbf{v} \times \mathbf{B}) + \eta \nabla^2 \mathbf{B}
\end{equation}
Now the diffusive term is negligible compared to time derivative term if 
 \begin{equation}
t << T_{decay}= \frac{L^2}{\eta}
\end{equation}
\begin{figure}[t!]
\centering
\includegraphics*[width=0.85\linewidth]{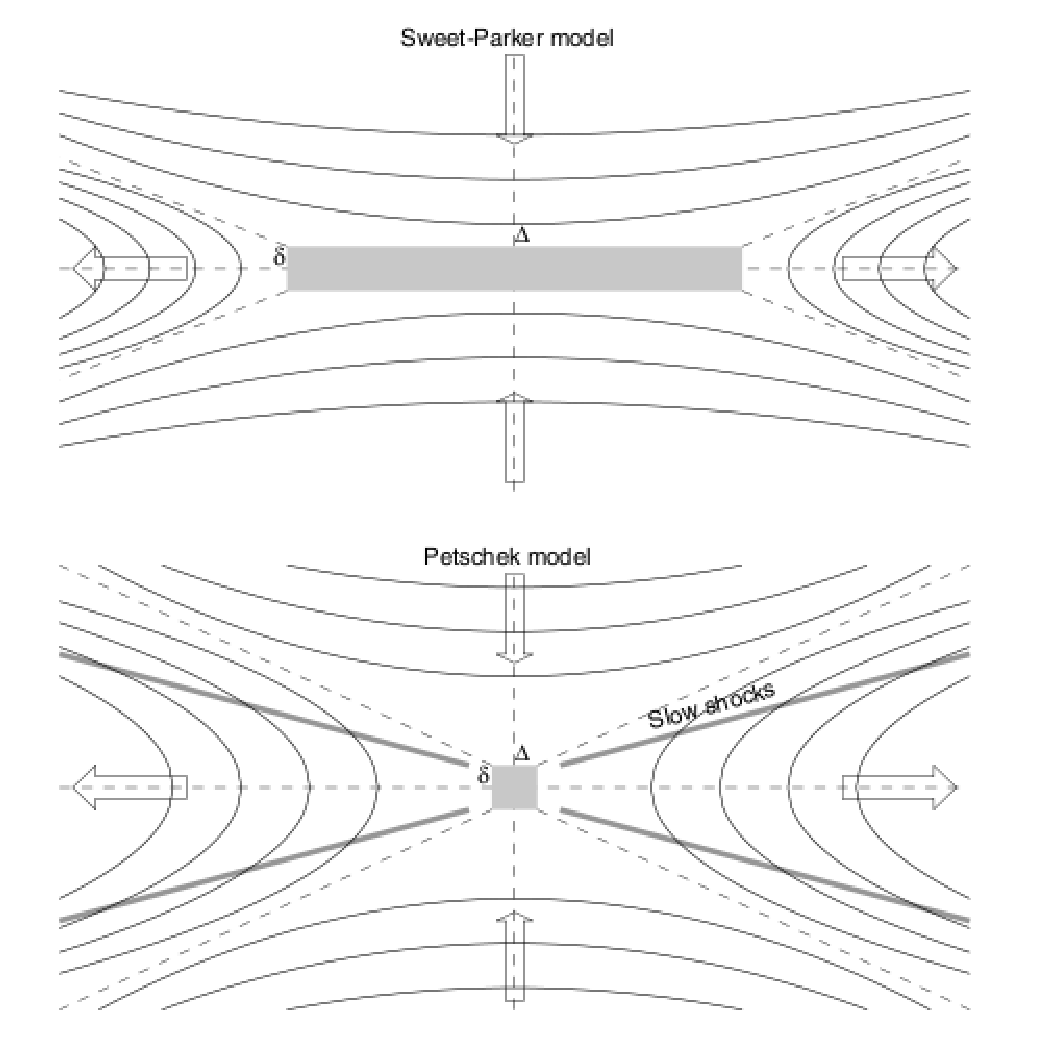}
  \caption{Top figure represents the geometry of Sweet-Parker reconnection model while bottom figure represents Petschek reconnection model. In case of Sweet-Parker reconnection model, diffusion region is a long thin sheet ($\Delta >> \delta$) while for Petschek model, diffusion region is narrow ($\Delta \simeq \delta$). As diffusion region is very narrow, inflow speed abruptly changes to outflow speed, thus Petschek considers slow mode shocks in outward flow region. Image credit: M J Aschwanden.}
\label{fig1}
\end{figure}
In that case flux conservation arises because of the balance between the terms $\frac{\partial \mathbf{B}}{\partial t}$ and $\nabla \times (\mathbf{v} \times \mathbf{B})$ i.e., magnetic field is coupled within the plasma. This mechanism is easily applicable for high Reynold's number plasma system i.e., for astrophysical plasma. So we can safely assume flux freezing condition as long as evolution time scale (t) is small compared to diffusion time scale. This flux freezing condition introduces strong constraints on the dynamics of astrophysical magnetic fields as it implies that connectivity between field lines can not break and their topology is preserved. We can say, two magnetic configurations are topologically equivalent if one can be deformed into the other by continuous motion (without cutting or pasting). But if it is required to cut and paste field lines, then they are of different magnetic topology. Now, we can state the flux freezing statement as: {\it In a time t, magnetic lines of force can slip through the plasma up to the distance $\textit{l}= \sqrt{\eta t}$}. If our scale of interest (say, $\delta$) is very large compared to distance \textit{l} then flux freezing condition is a good approximation. However, there are also some situations, when our scale of interest ($\delta$) is small compared to \textit{l} in which case this flux freezing condition is no longer valid. In that case there might be breaking and reattachment between pair of field lines, which is widely known as magnetic reconnection. Bringing opposing field lines together leads to such large current densities that there is always break in field lines causing a change in topology even if the resistivity is very small. These intense thin current layers (also known as current sheet) are the location for magnetic reconnection.
 
 As topology controls the equilibrium state of the plasma, change in topology corresponds to a change in equilibrium from one configuration to another. On the other hand, change in equilibrium configuration corresponds to conversion of magnetic energy to other forms of energy like heat. If this change is sudden then there is a release of large amounts of energy. This may be the basis for explosive events in solar atmosphere like solar flares and coronal mass ejections.
 
 Sweet (1958) and Parker (1957) first quantitatively modelled magnetic reconnection in two dimension. In their model, they considered the presence of a thin magnetic diffusion layer along the whole boundary between the opposite magnetic fields (thickness of diffusion layer ($\delta$) $\ll$ length of diffusion layer ($\bigtriangleup$)) and they calculate the rate for magnetic reconnection ($v_i$):
  \begin{equation}
v_i=\frac{v_A}{\sqrt{S}}
\end{equation}           
 where $v_A$ is the Alfv\'en velocity and S ($\frac{v_A L}{\eta}$) is known as Lundquist number. It was found that the resulting reconnection rate is very slow and insufficient to explain the energy release in solar flares (Parker, 1963).
 
 Petschek (1964) came up with a brilliant solution of this problem. As breaking and reconnection of field lines is a topological process, thus it is only necessary to break and reconnect field lines near one point. For magnetic reconnection the resistivity of the rest of the flux system is not important. Thus he reduced the size of the diffusion layer to a narrow area ($\bigtriangleup \thicksim \delta$), much less than the length of Sweet-Parker current sheet. As the size of diffusion region is very narrow, thus plasma inflow speed (let say, $v_1$) abruptly changes to outflow speed (let say, $v_2$) thus slow-mode shock arises in the Petschek model. Keeping this in mind, the reconnection rate turns out to be:
  \begin{equation}
v_i=\frac{v_A}{log (S)}
\end{equation} 
This rate is much faster than Sweet-Parker reconnection rate and is able to explain the energy released in solar flare like events.
\subsection{Magnetic Nonpotentiality}
The Navier-Stokes equation for incompressible flow can be written as:
\begin{equation}
 \rho \frac{\partial \mathbf{v}}{\partial t} + \rho (\mathbf{v} \cdot \nabla)\mathbf{v}= -\nabla p + \mathbf{J} \times \mathbf{B}  + \rho \mathbf{g} + \gamma \nabla^2 \mathbf{v}.
\end{equation}
when plasma velocity ($\mathbf{v}$) is small, then the Navier-Stokes equation reduces to:
\begin{equation}
 -\nabla p + \mathbf{J} \times \mathbf{B}  + \rho \mathbf{g} =0.
\end{equation}
In case of solar atmosphere, both the pressure gradient ($-\nabla p$) and gravitational term ($\rho \mathbf{g}$) are important in the dynamical equilibrium. We can subsume the gravity term in the pressure term by expressing it as the gradient of a potential; thus total effective pressure is the term which adjusts itself to equilibrate the Lorentz force. Thus we can write:
\begin{equation}
 -\nabla p_{eff} + \mathbf{J} \times \mathbf{B} =0.
\end{equation}
Using the vector identity, $\nabla(\mathbf{A} \cdot \mathbf{B})=\mathbf{A} \times (\nabla \times \mathbf{B}) + \mathbf{B} \times (\nabla \times \mathbf{A}) + (\mathbf{A} \cdot \nabla)\mathbf{B} +\mathbf{B}(\nabla \cdot \mathbf{A})$; we get:

\begin{equation}
 \frac{1}{\mu_0}(\mathbf{B} \cdot \nabla)\mathbf{B}= \nabla(\frac{\mathbf{B}^2}{2 \mu_0} + p_{eff}).
\end{equation}
The ratio of gas to magnetic pressure is defined as the plasma-$\beta$ parameter i.e., $\beta=\frac{2 \mu_0 p}{\mathbf{B}^2}$. In case of the solar corona, due to low density the plasma-$\beta$ is small and one can consider:
\begin{equation}
  \mathbf{J} \times \mathbf{B} = 0.
\end{equation}
A magnetic field satisfying this condition is known as a force-free field. Now we can satisfy this condition in three possible ways.
First, $\mathbf{B}=0$ everywhere but we are not interested in this situation.
Second, $\mathbf{J}=0$ i.e., current vanishes.  
 Now $\mathbf{J}=0$ implies $\nabla \times \mathbf{B}=0$; thus we can write:
\begin{equation} 
  \mathbf{B}=\nabla \phi,
\end{equation}
where $\phi$ is any scalar potential. This magnetic field configuration is known as potential magnetic field or current-free field.
Another possibility is:
\begin{equation} 
  \nabla \times \mathbf{B}= \mu_0 \mathbf{J}= \alpha(r) \mathbf{B},
\end{equation}
where $\alpha(r)$ is a scalar which is a function of position. If $\alpha(r)=0$ then current vanishes and magnetic field corresponds to potential field. But if $\alpha(r) \neq 0$ then current does not vanish; in this case the corresponding magnetic field is known as nonpotential field. So one can take current as a measure of magnetic nonpotentiality.
Using vector identity $\nabla \cdot (\nabla \times \mathbf{B})=0$, we can show that:
\begin{equation} 
  \nabla \cdot (\nabla \times \mathbf{B})= \nabla \cdot (\alpha \mathbf{B})= \alpha(\nabla \cdot \mathbf{B}) + \mathbf{B} \cdot \nabla \alpha =\mathbf{B} \cdot \nabla \alpha =0.
\end{equation}
It indicates that $\alpha$ does not change along field line. So $\alpha (r)$ is not a scalar function but a constant. Magnetic field satisfying condition (5.11) with constant $\alpha$ is known as linear force free field equation.

If we write the integral form of equation (5.11) with constant $\alpha$, then we get:
\begin{equation} 
 \int_S (\nabla \times \mathbf{B}) \cdot dS = \int_S \alpha \mathbf{B} \cdot dS~~~So, ~~~\oint_C \mathbf{B} \cdot dS = \int_S \alpha \mathbf{B} \cdot dS
\end{equation}
The right hand side of the equation is $\alpha$ times the amount of magnetic field passing through the surface bounded by C while left hand side represents the amount of magnetic field integrated around the circumference of the surface S. Thus relative sizes of these two components are indicated by the force-free parameter $\alpha$, i.e., the amount of twist in the field. This is the physical significance of the force-free parameter $\alpha$. The nonpoteniality parameter represents the departure of the force-free field (observed field in corona) from the potential field.

To model shear arcades Priest (1982) and Sturrock (1994) considered a loop arcade which consists of a sequence of loop with a common axis of curvature and generated a shearing motion by shifting the footpoints parallel to the neutral line on one side along the solar surface. The shear angle of this arcade is higher when the shearing motion is applied for longer time. It was shown that: 
\begin{equation} 
 \tan \theta=\frac{\alpha}{l},
\end{equation}
i.e., shear angle ($\theta$) is proportional to $\alpha$. So if there is an increase in shear angle, it also indicates the increase in the nonpotentiality parameter $\alpha$. Thus shear angle is one of the major indicator of the magnetic field nonpotentiality.

Now from the study of uniformly twisted, cylindrical, force-free flux tubes (Priest, 1982); it was shown that both force free parameter $\alpha$ and geometric shear angle (i.e., angle between twisted and untwisted field lines) is proportional to the  number of twist (say, $N_{twist}$). Thus one can also use geometric shear angle ($\theta$) as an estimate of force-free parameter $\alpha$. Pevtsov et al. (1997) performed an observational study about the relationship between geometric shear angle ($\theta$) and force-free parameter $\alpha$ and found a strong linear correlation between them. Thus one can measure the nonpotentiality of magnetic field by measuring shear angle or the twist parameter ($\alpha$).
\subsection{Magnetic Helicity}
Moreau (1961) and Moffatt (1969) have pointed out that a pseudo scalar quantity called helicity of the form $\int \mathbf{X} \cdot (\nabla \times \mathbf{X}) d^3x$ can be related with topological properties of field lines of $\nabla \times \mathbf{X}$. Similarly, for magnetic field, one can define magnetic helicity as:
\begin{equation} 
 H=\int \mathbf{B} \cdot \mathbf{A}~d^3x,
\end{equation}
where $\mathbf{B}=\nabla \times \mathbf{A}$; $\mathbf{A}$ is the vector potential. This quantity is the measure of the linkage between magnetic field lines. 

Woltjer (1958) showed that for a perfectly conducting plasma (ideal MHD, where field lines are frozen inside the plasma); magnetic helicity is a conserved quantity. He also proved that, if magnetic helicity is invariant, minimum magnetic energy field configuration always satisfies the force-free condition i.e.,
\begin{equation} 
 \nabla \times \mathbf{B}= \alpha \mathbf{B}.
\end{equation}
Taylor (1974) applied this idea to the process of plasma relaxation. He suggested that reconnection can remove all topological constraints except total helicity conservation, making the constant $\alpha$ force-free field accessible for plasma relaxation process. Thus final state after plasma relaxation would be a linear, force-free magnetic field configuration. Note that these two theorems are only valid in case of low-$\beta$ plasma. Magnetic helicity is also an invariant quantity during the evolution of coronal structures, like active region loops (Kusano et al. 2002); flare loops (Pevtsov et al. 1996); filaments (Pevtsov 2002) etc. So in summary magnetic helicity is a globally invariant quantity even under resistive processes like magnetic reconnection.

It has also been shown that magnetic helicity follows a hemispheric sign rule - negative in northern hemisphere and positive in southern hemisphere. However this rule is not very strong, there are also indications about the reversal of helicity sign rule at activity minimum periods. As vector potential $\mathbf{A}$ is not unique, thus it is not possible to calculate a unique value of helicity. Also lack of observations at different heights of solar atmosphere makes the direct calculation of magnetic helicity impossible. Thus to search for hemispheric sign rule, people proposed proxies for magnetic helicity. Seehafer (1990) first used the constant force-free $\alpha$ as a proxy for magnetic twist (i.e., fraction of magnetic helicity) and found that sign of $\alpha$ follows the hemispheric rule - negative in northern hemisphere and positive in southern hemisphere.  Later Pevtsov et al. (1995) studied the helicity sign rule using $\alpha_{best}$ as a proxy for magnetic twist and found that 75 \% of active regions in northern hemisphere and 69 \% of active regions in southern hemisphere follow similar sign rule. $\alpha_{best}$ is the value of $\alpha$ for which computed transverse field best matches with the observed transverse magnetic field. But all of this above method uses force-free field equation, but in reality the photosphere is not force free. So use of these methods for calculating twist is questionable (Leka et al. 2005).
 \subsection{The Coronal Heating Problem}
\begin{figure}[t!]
\centering
\includegraphics*[width=0.75\linewidth]{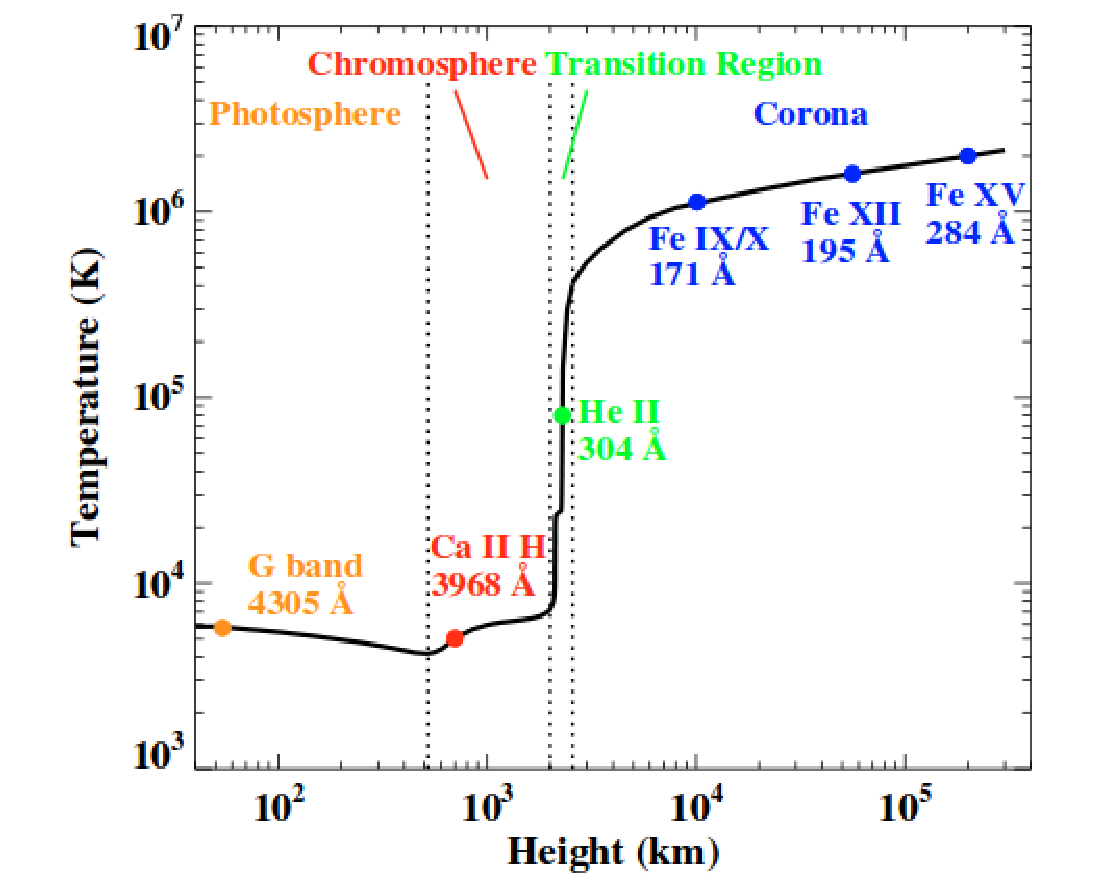}
  \caption{Variation of temperature from solar photosphere to corona. Spectral lines used for observing different regions of solar atmosphere are marked at respective locations. Image credit: Yang et al. (2009).}
\label{fig1}
\end{figure}

One of the major unsolved issues in solar physics is why the corona is so hot (million degree Kelvin), while the temperature of underlying photosphere is only 6000 degrees. According to the second law of thermodynamics it is expected that temperature will drop down steadily above the photosphere. Temperature variation along the different layers of solar atmosphere are shown in Fig. 5.2. Although significant progress has been made in addresing the coronal heating problem this problem remains hotly debated. But there is a general agreement that the heating energy comes from photospheric magnetic field. Solar coronal phenomenona are generally classified into three categories: active region corona(AR), quiet Sun corona (QS) and coronal holes (CH). Temperature trends in these regions are: coronal holes$~$1 MK (lowest temperature); quiet sun corona$~$2 MK; flaring active regions (AR)$~$2-6 MK (hottest) (Aschwanden 2004). The magnetic field structure in the coronal hole is dominated by open magnetic field lines, while in quiet sun and active region corona are mostly closed. In recent time, theories of coronal heating have been broadly classified into two subcategories-- one is DC heating model i.e., nanoflare heating model, and another is AC heating model i.e., wave heating theory.

Photospheric drivers, say random motion of magnetic field line footpoints, plausibly provides the source of energy required for coronal heating. Magnetic disturbances created by photospheric changes propagate towards corona with the Alfv\'en speed ($v_A$). In AC heating model (Wave heating theory), photospheric forcing changes magnetic footpoints rapidly thereby generating waves. These waves propagate up into the corona, get dissipated there and heat the coronal plasma locally.  On the other hand, in the DC heating model, photospheric forcing changes magnetic loop footpoints at time scales much longer than the Alfv\'en transit time thus it allows magnetic stresses to build up over time and dissipate in the corona due to magnetic reconnection, producing heating (Parker 1988). So the basic problem is to find the exact mechanism which is responsible for coronal heating. In chapter 6 of this thesis, we study the relationship between coronal X-ray intensity and active region magnetic fields using high resolution Hinode data and discuss the consequences of our results in the context of coronal heating theory.
\subsection{Solar Flares}
A solar flare is an energetic explosive event high up in the solar atmosphere without any visible energy source. This mystery was resolved when it was understood that it may be produced by an instability of the underlying magnetic field configuration. Due to this event, the magnetic field configuration evolves into a more stable state by changing and reconnecting the magnetic topology.

Due to the random motions of the magnetic footpoints in the solar photosphere, magnetic field lines are twisted. Because of this twisted motion, there is a increase in the magnetic energy much above than the energy of Sun's vacuum dipole field. Now if a rising bubble of the magnetic field from the photosphere collides with the twisted magnetic field, then it produces a reverse field configuration on a very short scale. Then reconnection takes place in this thin region and lowers the energy of the previous field by releasing the twist. In this way, we can explain solar flare like events. Recent observational studies suggest a close relation between photospheric magnetic non-potentiality and flare productivity (Jing et al. 2006; Falconer et al. 2008); which may be helpful for predicting solar flares. But a method of calculating magnetic non-potentiality in terms of twist by using force-free field equation is questionable as the photosphere is not force free. In chapter 7 of this thesis, we use a new flux-tube fitting technique to calculate the best-fit twist ($Q_{fit}$; one of the measures of magnetic non-potentiality) and critical twist threshold necessary for determining the susceptibility of magnetic flux tubes to kink instability mechanism ($Q_{kink}$). We also discuss the importance of the relationship between $Q_{fit}$ and $Q_{kink}$ for predicting solar flare like events.
 
\chapter{The Relationship between Solar Coronal X-Ray Brightness and Active Region Magnetic Fields}

In this chapter, using high-resolution observations of nearly co-temporal and co-spatial {\it Solar Optical Telescope} spectropolarimeter
and {\it X-Ray Telescope} coronal X-ray data onboard {\it Hinode}, we revisit the problematic relationship between global magnetic quantities and coronal X-ray brightness. Co-aligned vector magnetogram and X-ray data were used for this study. The total X-ray brightness over active regions is well correlated with integrated magnetic quantities such as the total unsigned magnetic flux, the total unsigned vertical current and the area-integrated square of the vertical and horizontal magnetic fields. On accounting for the inter-dependence of the magnetic quantities, we inferred that the total magnetic flux is the primary determinant of the observed integrated X-ray brightness. Our observations indicate that a stronger coronal X-ray flux is not related to a higher non-potentiality of active-region magnetic fields. The data even suggest a slight negative correlation between X-ray brightness and a proxy of active-region non-potentiality. Although there are small numerical differences in the established correlations, the main results are qualitatively consistent over two different X-ray filters, the Al-poly and Ti-Poly filters, which confirms the strength of our conclusions and validate and extend earlier studies that used low-resolution data. We discuss the implications of our results and the constraints they set on theories of solar coronal heating.

\section{Introduction}
Solar active-region coronal loops appear bright in EUV and X-ray wavelengths, which is indicative of
very high temperatures of the order of a million degrees Kelvin. The origins
behind these high-temperature coronal structures remain elusive. An energy flux of about 
10$^{7}$~ergs~cm$^{2}$~s$^{-1}$ is required to maintain this high
temperature of the coronal plasma (Withbroe \& Noyes 1977). It has been suggested
that there is a one-to-one correspondence between the location of the magnetic fields
in the photosphere and bright coronal structures in the corona (Vaiana et al. 1973) and we know that
most of the coronal X-ray luminosity is concentrated within active-region magnetic-flux systems.

Several theories have been proposed to explain the heating of coronal structures (Zirker 1993; Narain \& Ulmschneider 1996;
Aschwanden 2004; Klimchuk 2006). These theories are broadly classified into two
subcategories: DC heating model, {\it i.e.} the nano-flare heating model (Parker 1988), and
the AC heating model, {\it i.e.} the wave-heating theory ({\it e.g.} see the review by Aschwanden, 2004).
In the AC heating model, high-frequency MHD waves are generated in the magnetic foot points
of active regions and propagate through magnetic loops in the corona. These
waves dissipate their energy in the corona (Narain \& Ulmschneider 1996). Although recent observations reveal
that MHD waves propagate into the quiet solar corona (Tomczyk et al. 2007), it is unclear
whether these MHD waves alone can heat the corona to such
a high temperature (Mandrini et al. 2000; Cirtain et al. 2013). Alternatively, DC-heating models are proposed
to explain the heating of active regions where nanoflare-like small bursts
(each of energy $10^{24}$ erg) can liberate energy by magnetic reconnection -- driven by the constant shuffling of magnetic foot points by turbulent convective motions just beneath the photosphere (Parker 1988; Cirtain et al. 2013). It has been suggested recently that waves can play a major role in heating the quiet-Sun corona (McIntosh et al. 2011; Wedemeyer-B{\"o}hm et al. 2012), while for coronal active regions the additional DC heating mechanism must play a role (Parker 1988; Klimchuk 2006).

To examine the relative roles of diverse physical mechanisms in the context of coronal heating,
it is essential to have information on the coronal magnetic and velocity field. Current instrumentation is still at a nascent stage, however, and is inadequate for such coronal diagnostics (Lin, Kuhn \& Coulter 2004). Since coronal field lines are linked to the photosphere, another approach is possible: exploring the relationship between photospheric
magnetic-field parameters and brightness of the coronal loops.
In earlier studies, Fisher et al. (1998) and Tan et al. (2007) investigated the
relationship between the X-ray luminosity and photospheric magnetic-field parameters.
These two studies reported a strong correlation between the X-ray luminosity and the total unsigned
magnetic flux. Tan et al. (2007) also found a good correlation between the
average X-ray brightness and average Poynting flux, but ruled out any correlation between the velocity of footpoint motions and total X-ray brightness. Their computed Poynting flux had a range between 10$^{6.7}$ and
10$^{7.6}$~ergs~cm$^{-2}$~s$^{-1}$, which is enough to heat the corona (Withbroe \& Noyes 1977).
Using data from other wavelengths (UV/EUV channels), Chandrasekhar et al. (2013) also found
a good correlation between total emission from bright points and total unsigned photospheric magnetic flux.
Forward-modeling of active regions also suggests a direct correlation between magnetic flux and X-ray luminosity (Lundquist et al. 2008).

The net current is a measure of non-potentiality of the magnetic field in the active region.
In active-region flare and coronal-mass-ejection processes, the non-potentiality of the magnetic field
can play a significant role (Schrijver et al. 2006; Jing et al. 2006; Wang et al. 2008). As a result of the low resistivity, large-scale ($10^3$ km)
currents cannot dissipate sufficiently in the corona (Hagyard 1988); thus these currents may have no contribution to
coronal heating.  Earlier observations do not find a strong relationship between the total X-ray luminosity and total vertical current (Metcalf et al. 1994; Fisher et al. 1998). Note that Wang et al. (2008) showed the existence of 3D current structures over active regions. Another traditionally used measure for magnetic non-potentiality is the parameter $\alpha_\textrm{best}$ which appears in the force-free field equation and which is thought to be related to the wrapping of magnetic-field lines along the axis of an active-region flux tube ({\it i.e.} the twist of magnetic-field lines). Observation shows that there is no significant correlation between X-ray brightness and $\alpha_\textrm{best}$ (Fisher et al. 1998; Nandy 2008). While many studies have used $\alpha_\textrm{best}$ as the measure of the twist in the solar active region, this is questionable because the photosphere is not deemed to be force-free (Leka et al. 2005). Other studies have shown that a polarity-inversion line near coronal-loop foot points and strong magnetic shear may also result in enhanced coronal emission (Falconer 1997; Falconer et al. 1997, 2000).

Longcope (1996) proposed the minimum current corona (MCC) model where coronal heating was described as a
series of small reconnection events punctuating the quasi-static evolution of coronal field.
This model qualitatively predicts the variation of the X-ray luminosity with the total flux that closely
matches observations (Fisher et al. 1998). Wang et al. (2000) have observed bright coronal loops and
diffused coronal loops that are associated with the quasi-separatrix layers (QSLs). Since QSLs
are the places where energy release occurs through 3D magnetic reconnection, they
concluded that QSLs are important for heating the active-region corona and chromosphere.
By analysing the X-ray images taken from {\it Hinode/X-Ray Telescope} (XRT) and corresponding {\it Michelson Doppler Imager} (MDI) line-of-sight
magnetograms, Lee et al. (2010) found a relationship between coronal-loop brightness and
magnetic topologies in AR 10963. They also found that frequent transient brightenings in coronal
loops are related to separators that have a large amount of free energy.

Here we revisit the coronal-heating problem with space-based vector-magnetogram data, which are free from atmospheric seeing effects, which can produce cross talk between various Stokes parameters. Such space-based magnetic field measurements have also reduced atmospheric scattered light contribution. The obtained vector-field data are of very high resolution, thereby reducing the effect of filling factor. In this chapter, we use X-ray images taken from two filters (Ti-poly and thin Al-poly) of the XRT telescope onboard the {\it Hinode} spacecraft and vector magnetic-field measurements taken from the {\it Spectro-Polarimeter} (SP) of {\it Solar Optical Telescope} (SOT) to study the relationship between the X-ray brightness and magnetic-field parameters in active-region flux systems. This study extends previous work that used lower resolution \textit{Yohkoh} data (Fisher et al. 1998). We also, incidentally, explore the effect of the filter response (which is mainly affected by deposition of unknown materials on CCD cameras) on the relationship between X-ray brightness and magnetic-field parameters. In Section 6.2 we provide the details of the data used in this study. In Section 6.3, we detail our results. In Section 6.4, we discuss the implication of our results for the heating of the solar corona.

\section{Data Analysis}
\subsection{Data Selection}
The {\it X-ray telescope} (XRT: Golub et al. 2007) onboard the {\it Hinode} spacecraft (Kosugi et al. 2007)
takes images of the solar corona at a spatial resolution of one arcsec per pixel using different filters.
XRT images are of the size 2k$\times$2k pixel, which covers a 34$\times$34 square arcmin
field of view (FOV) of the solar corona. XRT observes coronal
plasma emission in the temperature range $5.5 <\textrm{log} T < 8$, which is realized by different X-ray
filters, that have their own passband, corresponding to different responses to plasma temperature.
Within a few months of the launch of the {\it Hinode} spacecraft, contaminating materials were deposited on the CCD,
which significantly impacted the filter response, specifically for observations of longer
wavelengths. Regular CCD bakeouts were unable to completely remove this contamination.
As the effect of the contamination is mainly wavelength-dependent (the long-wavelength observations are affected more strongly), the observations from the thin Al-poly/Al-mesh
filter are more heavily affected than the other filters such as Ti-poly, and Be-med. For the
present study, we have used data taken from Ti-poly and the thin Al-poly filter,
which observe the solar coronal plasma at temperatures higher than 2 MK and 0.5 MK,
respectively. Therefore, we have a point of comparison to establish whether filter degradation may play a role in
the inconsistencies of the results.\\
\~~~~~The {\it Spectro-Polarimeter} (SP: Ichimoto et al. 2008) is a separate back-end instrument of
the {\it Solar Optical Telescope} (SOT: Tsuneta et al. 2008) onboard the {\it Hinode} spacecraft. The SP provides 
Stokes signals with high polarimetric accuracy in the 6301 and 6302~\AA~photospheric
lines. The primary product of the Stokes polarimeter are the Stokes-{\it IQUV} profiles, which are suitable for deriving the vector
magnetic field in the photosphere. The spatial resolution along the slit direction is
0.295$^{\prime\prime}$~pixel$^{-1}$; in the scanning direction it is 0.317$^{\prime\prime}$~pixel$^{-1}$.
The Stokes vector was inverted using the MERLIN code, which is based on the Milne--Eddington
inversion method. The inverted data provide the field strength, inclination and azimuth
along with the Doppler velocity, continuum images, and many other parameters. The processed
data were obtained from the Community Spectropolarimetric Analysis Center (\href{http://www.csac.hao.ucar.edu/}{CSAC}).
We corrected for the ambiguity in the transverse component of the magnetic field using the minimum-energy algorithm (Metcalf 1994; Leka et al. 2009). The resulting
magnetic-field vectors were transformed into heliographic co-ordinates
(Venkatkrishnan and Gary 1989). We selected 40 different NOAA active regions observed at different times of the year. We also excluded active regions whose central meridional
distance was greater than $30^{\circ}$. We took the vector magnetogram data close to the
timings of soft X-ray data obtained from both the Ti-poly and the Al-poly filter of the XRT.
In Table 6.1 and 6.2 we list the different active regions used in this study, the date and time of
the observations of the vector magnetogram, and the corresponding soft X-ray data. After these two sets of data, we also obtained the G-band data taken by the {\it X-Ray Telescope}. These data were used for to co-align each of the data sets. For each selected vector magnetogram, we simultaneously took
XRT X-ray (Ti-poly and Al-poly) data and G-band data. Throughout this chapter,
we use the term Ti-poly dataset to represent the X-ray image obtained from the Ti-poly filter of
XRT onboard {\it Hinode}.  Similarly, we use the term Al-poly for the data taken from the Al-poly filter.
There is always a corresponding vector magnetogram associated with these data sets.
The X-ray data were calibrated using the xrt$\_$prep.pro available in the Solarsoft
routines. The calibrated data were normalized to a one-second exposure time.
\begin{table}[]
	\caption{NOAA active regions and time of corresponding XRT X-ray Ti-poly filter and SP magnetogram data}
	\centering
	\begin{tabular}{r  c  c  c}
		Date & NOAA  & Magnetogram scan &  XRT X-ray (Ti-Poly)\\
		& active region & start time [UT]& observation time [UT]\\ \hline
		1 May 2007 & 10953 & 05:00:04 & 05:00:57  \\ 
		1 Jul. 2007 & 10962 & 13:32:05 & 13:31:51 \\ 
		15 Jul. 2010 & 11087 & 16:31:19 & 16:30:53 \\ 
		10 Aug. 2010 & 11093 & 09:15:04 & 09:14:18  \\ 
		31 Aug. 2010 & 11102 & 02:30:04 & 02:30:42  \\ 
		23 Sep. 2010 & 11108 & 07:21:05 & 07:21:12  \\ 
		26 Oct. 2010 & 11117 & 10:45:46 & 10:51:55  \\ 
		22 Jan. 2011 & 11149 & 09:31:28 & 09:43:22  \\ 
		14 Feb. 2011 & 11158 & 06:30:04 & 06:30:02  \\ 
		4 Mar. 2011 & 11164 & 06:15:06 & 06:15:04  \\ 
		31 Jan. 2012 & 11411 & 04:56:32 & 04:57:24  \\ 
		18 Feb. 2012 & 11419 & 11:08:53 & 11:10:10  \\ 
		8 Mar. 2012 & 11429 & 21:30:05 & 21:32:22 \\ 
		22 Apr. 2012 & 11463 & 04:43:05 & 04:48:31  \\ 
		12 May 2012 & 11476 & 12:30:50 & 12:30:41  \\ 
		18 May 2012 & 11479 & 04:47:05 & 04:48:38 \\ 
		5 Jul. 2012 & 11517 & 03:45:35 & 04:18:11  \\ 
		12 Jul. 2012  & 11520 & 11:12:28 & 11:12:45  \\ 
		14 Aug. 2012  & 11543 & 14:35:05 & 14:35:30 \\ 
		25 Sep. 2012  & 11575 & 12:49:06 & 12:50:08 \\ 
	\end{tabular}
\end{table}

\begin{table}[]
	\caption{NOAA active regions and time of corresponding XRT X-ray Al-poly filter and SP magnetogram data}
	\centering
	\begin{tabular}{r  c  c  c}
		Date & NOAA & Magnetogram scan  &  XRT X-ray (Al-Poly) \\
		& active region & start time [UT]& observation time [UT]\\ \hline
		30 Aug. 2011 & 11280 & 07:35:23 & 07:35:36  \\ 
		13 Sep. 2011 & 11289 & 10:34:05 & 10:34:24 \\ 
		28 Sep. 2011 & 11302 & 18:38:05 & 18:38:16 \\ 
		28 Nov. 2011 & 11360 & 00:05:20 & 00:03:04  \\
		31 Jan. 2012 & 11410 & 04:56:32 & 05:25:10  \\ 
		1 Feb. 2012 & 11413 & 08:51:31 & 09:03:35  \\ 
		8 Mar. 2012 & 11429 & 01:20:05 & 01:23:50 \\
		22 Apr. 2012 & 11463 & 04:43:05 & 04:55:22  \\
		16 Aug. 2012 & 11543 & 13:35:05 & 13:35:19  \\ 
		25 Sep. 2012 & 11575 & 12:49:06 & 12:50:37  \\ 
		2 Oct. 2012 & 11582 & 09:53:06 & 09:54:39  \\ 
		17 Oct. 2012 & 11589 & 09:06:01 & 09:06:22  \\
		28 Oct. 2012 & 11594 & 01:40:05 & 01:42:23  \\ 
		17 Nov. 2012 & 11613 & 10:25:06 & 10:25:37 \\
		17 Nov. 2012 & 11619 & 12:49:06 & 12:50:37  \\ 
		10 Feb. 2013 & 11667 & 14:30:04 & 14:31:22  \\ 
		15 Mar. 2013 & 11695 & 09:30:51 & 09:33:22 \\ 
		31 Aug. 2013 & 11836 & 18:14:36 & 18:15:25 \\ 
		27 Sep. 2013 & 11850 & 09:30:05 & 09:30:06 \\ 
	\end{tabular}
\end{table}

\par
\subsection{Data Coalignment}
To overlay the XRT X-ray data with vector magnetograms, we first co-aligned the G-band
data taken by XRT telescope with the continuum image. The continuum image was obtained by inverting the Stokes data set. 
To do this, we first identified the dark center of the
sunspot in the G-band and continuum images. Later, we interpolated the continuum image data
to the XRT image resolution. In the next step, we choose the same field of view (FOV) in the two data sets. 
By using the maximum-correlation method, we then co-aligned the continuum images
with the G-band images. A similar shift was applied to the vector field data to
co-align the entire dataset with X-ray images of the XRT dataset.

\section{Integrated Quantities}
We derived various integrated quantities and compared them with the X-ray
brightness. We computed the individual as well as integrated quantities such as total
magnetic flux, and total magnetic energy {\it etc} and compared them with the
X-ray brightness. Below, we describe each of these quantities.

\subsection{Active-Region Coronal X-Ray Brightness}
The integrated X-ray brightness [Lx] was computed by summing the values of each
bright pixel in the image and then multiplying by the pixel area. The bright pixels were
selected by using the threshold values. We found the rms value in the X-ray image and selected
only those pixels whose value were higher than the 1-$\sigma$ level (the rms value) of the image.
\begin{figure}[!h]
	\begin{center}
		\includegraphics[angle=90,width=1.00\textwidth]{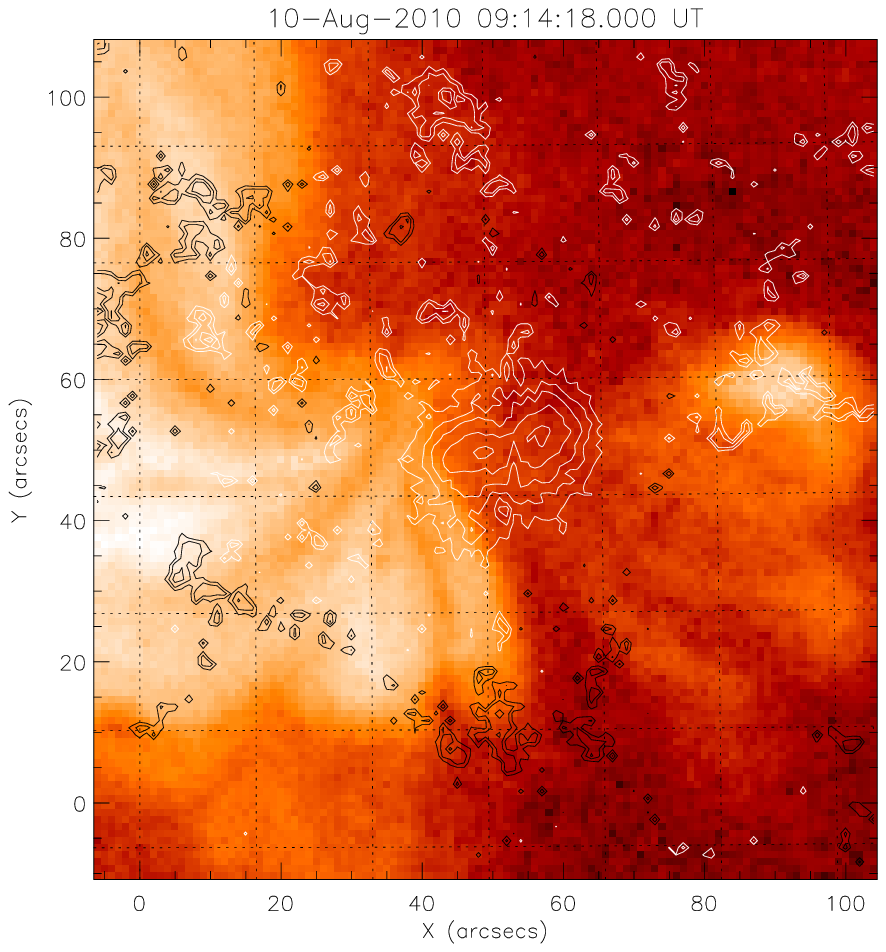}
	\end{center}
	\caption{The contours of vertical magnetic field overlaid upon the X-ray
		image of Active Region NOAA 11093 taken in Ti-poly filter by the XRT telescope. Contours with thick solid lines (white) represent the
		positive magnetic fields with a field strength level of 500, 1000, 1500, 2000, and 3000 G; thin solid
		lines (black) represent the negative vertical magnetic field at the same level.}
	\label{fig:1}
\end{figure}
\subsection{Global Magnetic-Field Quantities}
Since our selected active regions are close to the disk center, the magnetic-field
vectors are horizontal and vertical to the solar surface.  Using the $B_{x}$, $B_{y}$, and
$B_{z}$ components, it is possible to define the integrated quantities, which can be correlated with the
X-ray brightness to find the relationship between the two (for detailed information about
integrated quantities, see Fisher et al. 1998; Leka et al. 2007). We selected pixels in
$B_{x}$, $B_{y}$ and $B_{z}$ whose values are greater than the 1-$\sigma$ level of these images. The
following integrated quantities were computed from magnetic field components:

\begin{equation}
	\phi_\textrm{tot}=\sum |B_{z}| \textrm{d} A
\end{equation}
\begin{equation}
	B_{z,\textrm {tot}}^2=\sum B^{2}_{z}\textrm{d} A
\end{equation}
\begin{equation}
	B^2_{h,\textrm{tot}}=\sum B^2_{h}\textrm{d} A
\end{equation}
\begin{equation}
	J_\textrm{tot}=\sum |J_{z}|\textrm{d} A
\end{equation}

Here $B_{z}$ and $B_{h}$ represent the vertical and horizontal magnetic field,  $J_{z}$ is the vertical
current density, and $\displaystyle\sum \textrm{d}A$ is the effective area on the solar surface. Since the ratio of the vertical current density and magnetic field is related to the handedness or chirality (twist) of the underlying flux tube (Longcope et al. 1998), we also introduced a quantity $\mu_0 J_{\textrm{tot}}/\phi_{\textrm{tot}}$ (ratio of unsigned total current and unsigned total magnetic flux), which has the same units as the twist and can thus be taken as a proxy for it.  Highly twisted flux tubes are strongly non-potential, and thus the quantity above is a measure of the non-potentiality of active region flux systems.

We computed all of the magnetic quantities from the vector magnetogram for all active regions. The average estimated errors of the magnetic variables: $B_z$, $J_z$, $B^2_z$, and $B^2_h$ are 8 G, 45 mA, 64 G$^2$, and 800 G$^2$.

\begin{figure*}[!htb]
	\begin{center}
		\centering
		\Large\Huge\includegraphics[angle=90,width=1.00\textwidth]{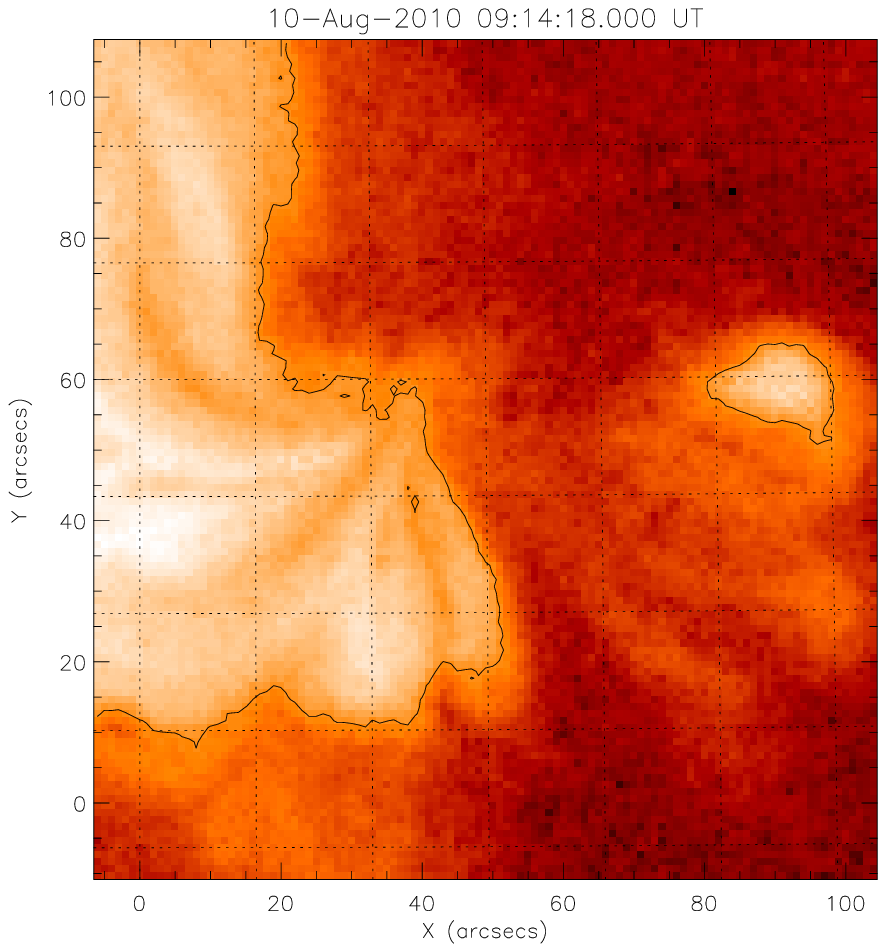}
	\end{center}
	\caption{Contour map of the 1$\sigma$ level of X-ray brightness
		overlaid on the X-ray image of the Active Region NOAA 11093. }
	\label{fig:2}
\end{figure*}
\begin{figure}[!h]
	\begin{center}
		\includegraphics[width=1.0\textwidth]{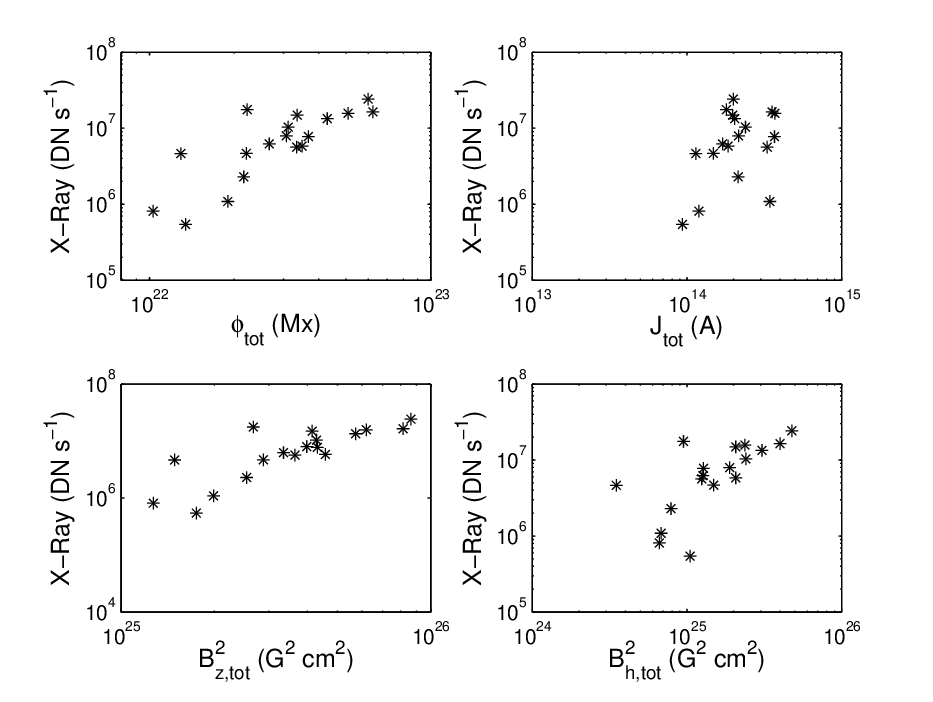}
	\end{center}
	\caption{Relationship between X-ray brightness and global magnetic-field quantities $\phi_{\textrm{tot}}$, $J_{\textrm{tot}}$, $ B^2_{z,\textrm{tot}}$, and $ B^2_{h,\textrm{tot}}$  (using the data set of Table 6.1, {\it i.e.} the Ti-poly filter). Correlation coefficients are listed in Table 6.3.}
	\label{fig:3}
\end{figure}
\begin{figure}[!h]
	\begin{center}
		\includegraphics[width=1.0\textwidth]{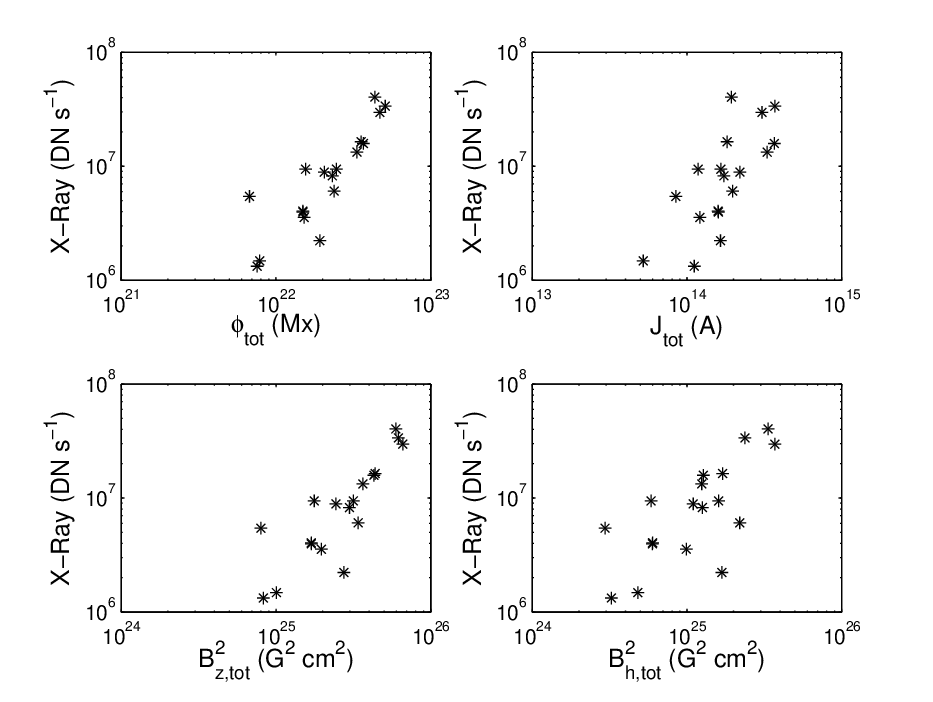}
	\end{center}
	\caption{Relationship between X-ray brightness and global magnetic-field quantities $\phi_{\textrm{tot}}$, $J_{\textrm{tot}}$, $ B^2_{z,\textrm{tot}}$, and $ B^2_{h,\textrm{tot}}$  (using the data set of Table 6.2, {\it i.e.} the Al-poly filter). Correlation coefficients are listed in Table 6.3.}
	\label{fig:4}
\end{figure}
\section{Results}
Figure \ref{fig:1} shows the contours of the $B_{z}$-component of the magnetic
field overlaid on the X-ray image of active region NOAA 11093 after co-aligning the images. 
The contour map shows that the X-ray brightness in the corona overlying the umbral part of the
sunspot is lower than that of the loops emanating from the penumbral part of the active region. The bright loops
are associated with the plage regions as has been observed before (Pallavicini et al. 1979).
The loops are still not fully resolved in the XRT images, but the
cluster of loops clearly turn in a clockwise direction. On the west side of the sunspot, the loop
structures are absent. At the same location in the photosphere, large-scale plage structures
are also absent. This may indicate that large-scale plage regions are essential for
the loops to appear in X-rays. Thus we note that a visual spatial correlation exists between the
location of the plages and the bright loops in X-rays.\\

\subsection{Correlation Between Global Magnetic Field Quantities and X-ray Brightness}

We explored the relationship between total (area-integrated) magnetic quantities and X-ray brightness in active regions.
We used the XRT data for 20 active regions each in the Ti-poly and
Al-poly data sets (all data are listed in Tables 6.1 and 6.2). We
only selected those pixels whose intensity values exceeded a 1-$\sigma$ threshold in the
X-ray and magnetic images. Figure~\ref{fig:2} shows the contour map of 1-$\sigma$ level threshold of X-ray brightness
overlaid upon the X-ray image of Active Region NOAA 11093. The 1-$\sigma$ level threshold line of the contour map clearly indicates the borders of the bright loops.

\begin{figure}[!h]
	\begin{center}
		\includegraphics[width=1.0\textwidth]{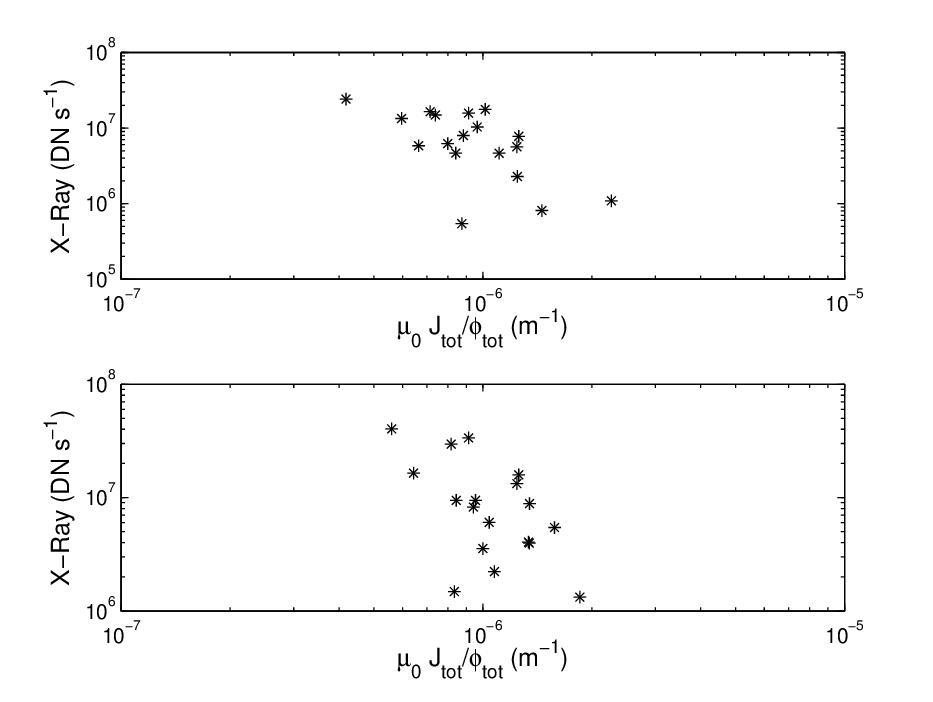}
	\end{center}
	\caption{Scatter plots of X-ray brightness with $\mu_0 J_{tot}/\phi_{tot}$ (top plot is for the data set of Table 6.1, {\it i.e.} the Ti-poly filter, the bottom plot is for the data set of Table 6.2, {\it i.e.} the Al-poly filter). Correlation coefficients are listed in Table 6.3}
	\label{fig:5}
\end{figure}
\begin{table}[!h]
	\caption{Correlation coefficients between different parameters}
	\centering
	\begin{tabular}{c c c c}
		Figure &  Correlated quantities  &   Pearson correlation & Spearman correlation  \\
		number    &   & coefficients with & coefficients with\\
		&   & confidence levels &  confidence levels\\
		
		\hline
		Figure 3 & X-ray brightness {\it vs.} $\phi_{\textrm{tot}}$ & 0.81 ($99.99 \%$) & 0.83 ($99.99 \%$)\\
		(Ti-Poly)       & X-ray brightness {\it vs.} $J_{\textrm{tot}}$ & 0.31 ($98.52 \%$) & 0.50 ($96.63 \%$)\\
		& X-ray brightness {\it vs.} $B^{2}_{z,\textrm{tot}}$ & 0.81 ($99.99 \%$) & 0.75 ($99.99 \%$)\\
		& X-ray brightness {\it vs.} $B^{2}_{h,\textrm{tot}}$ & 0.79 ($99.98 \%$) & 0.79 ($99.98 \%$)\\ \hline
		Figure 4 & X-ray brightness {\it vs.} $\phi_{\textrm{tot}}$ & 0.90 ($99.99 \%$) & 0.89 ($100 \%$)\\
		(Al-Poly)       & X-ray brightness {\it vs.} $J_{\textrm{tot}}$ & 0.62 ($99.99 \%$) & 0.76 ($100 \%$)\\
		& X-ray brightness {\it vs.} $B^{2}_{z,\textrm{tot}}$ & 0.91 ($99.99 \%$) & 0.71 ($99.99 \%$)\\
		& X-ray brightness {\it vs.} $B^{2}_{h,\textrm{tot}}$ & 0.81 ($99.67 \%$) & 0.85 ($99.99 \%$)\\ \hline
		
		Figure 5 & X-ray brightness {\it vs.}  & -0.59 ($96.91 \%$) & -0.54 ($82.16 \%$)\\
		& $\mu_0 J_{\textrm{tot}}/\phi_{\textrm{tot}}$ (Ti-poly)& &\\
		& X-ray brightness {\it vs.}  & -0.55 ($97.1 \%$) & -0.54 ($97.92 \%$)\\
		& $\mu_0 J_{\textrm{tot}}/\phi_{\textrm{tot}}$ (Al-poly)& &\\ \hline
		
		Figure 6 & $J_{\textrm{tot}}$ {\it vs.} $\phi_{\textrm{tot}}$ & 0.57 ($99.99 \%$) & 0.63 ($100 \%$)\\
		& $B^{2}_{z,\textrm{tot}}$ {\it vs.} $\phi_{\textrm{tot}}$ & 0.99 ($99.99 \%$) & 0.98 ($99.99 \%$)\\
		& $B^{2}_{h,\textrm{tot}}$ {\it vs.} $\phi_{\textrm{tot}}$ & 0.91 ($99.99 \%$) & 0.88 ($99.99 \%$)\\
		& $\mu_0 J_{\textrm{tot}}/\phi_{\textrm{tot}}$ {\it vs.} $\phi_{\textrm{tot}}$ & -0.54 ($98.86 \%$) & -0.61 ($99.94 \%$)\\
	\end{tabular}
	\label{tab:3}
\end{table}

Figures~\ref{fig:3} and \ref{fig:4} depict the relationship between the X-ray brightness and total unsigned magnetic flux (top-left), $B^{2}_{z,\textrm{tot}}$ (top-right), $B^{2}_{h,\textrm{tot}}$ (bottom-left), and unsigned $J_{\textrm{tot}}$ (bottom-right)
in logarithmic scale.  Figure~\ref{fig:3} is for the data sets of Table 6.1, {\it i.e.} Ti-poly filter and Figure~\ref{fig:4}
is for the data sets of Table 6.2, {\it i.e.} the Al-poly filter. The coronal X-ray brightness and
the global magnetic-field parameters in both data sets are clearly correlated; although the correlation coefficients are
numerically somewhat different, they are qualitatively similar (for quantitative correlation coefficients see Table 6.3).

Non-potential flux systems are known to be storehouses of free energy, and it is often assumed that therefore, a coronal energy release in X-rays should be positively correlated with measures of non-potentiality. Figure \ref{fig:5} depicts the relationship between X-ray brightness and the non-potentiality measure $\mu_{0}J_{\textrm{tot}}/\phi_{\textrm{tot}}$. The top plots are for data sets of Table 6.1 (Ti-poly filter data), the bottom plots are for data sets of Table 6.2 (Al-poly filter data). The X-ray brightness is anti-correlated with $\mu_{0}J_{\textrm{tot}}/\phi_{\textrm{tot}}$ in both cases.

To determine which of the magnetic quantities contributes predominantly to the X-ray brightness, we need to  examine
whether there is any inter-dependence between the global magnetic quantities. In subsection 6.4.2, we follow Fisher et al. (1998) in this analysis and establish the correlation between each of the magnetic parameters with the total unsigned flux first and also perform a partial correlation analysis to extract the true underlying dependencies.

\subsection{Correlations Among Global Magnetic-Field Quantities and Partial Correlation Analysis}
\begin{figure}[!h]
	\begin{center}
		\includegraphics[width=1.0\textwidth]{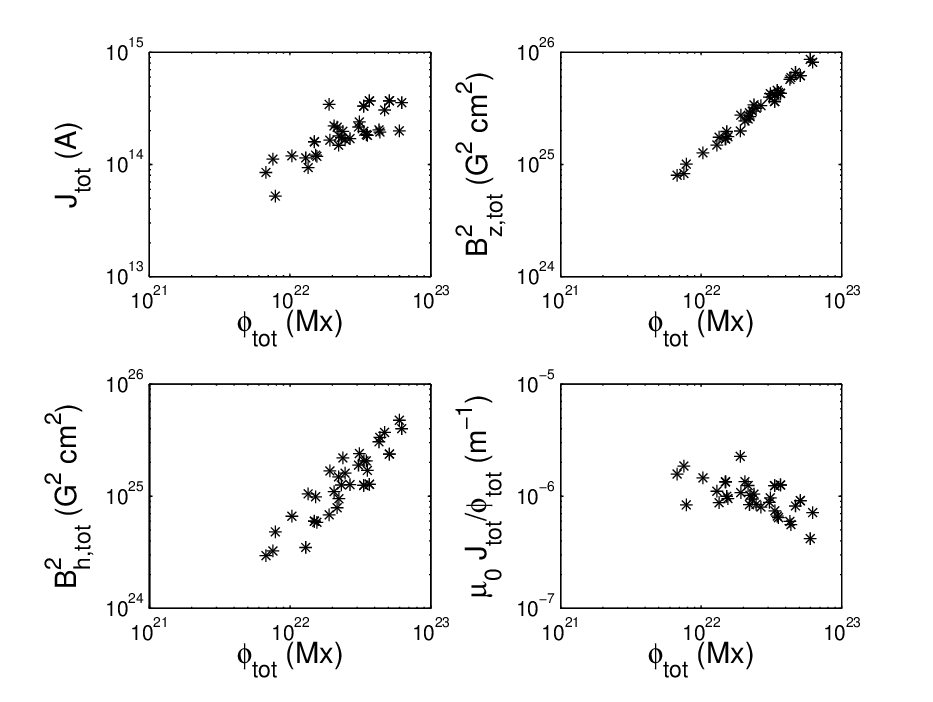}
	\end{center}
	\caption{Relationship of global magnetic quantities $J_{\textrm{tot}}$, $B^2_{z,\textrm{tot}}$, $B^2_{h,\textrm{tot}}$, and $\mu_0 J_{\textrm{tot}}/\phi_{\textrm{tot}}$ with $\phi_{\textrm{tot}}$. Correlation coefficients are listed in Table 6.3.}
	\label{fig:6}
\end{figure}

Figure~\ref{fig:6} (see also Table 6.3) shows the inter-dependence of the total unsigned magnetic flux and
all other magnetic variables, such as the total absolute current, $B^{2}_{z,\textrm{tot}}$, $B^{2}_{h,\textrm{tot}}$, and
$\mu_0 J_{\textrm{tot}}/\phi_{\textrm{tot}}$.  Each of these magnetic parameters shows a good correlation with
the total unsigned magnetic flux, which means that they are related to each other through size (area integration).
To find the relationships between the different magnetic parameters, we carried out a partial-correlation analysis. In the partial-correlation technique, the correlation between the two dependent variables is
examined after removing the effects of other variables.
\begin{table}[!h]
	\caption{Partial correlation coefficients between different quantities for different filters}
	\centering
	\begin{tabular}{c c c } 
		\hline
		Correlated quantities&\multicolumn{2}{c}{Partial Correlation Coefficient} \\
		(controlling $\phi_{\textrm{tot}}$ ) &  Ti-Poly  &   Al-Poly \\
		&  (Table 1 data set) & (Table 2 data set)\\
		\hline
		X-ray brightness {\it vs.} $J_{\textrm{tot}}$   & -0.45  & -0.64 \\ 
		X-ray brightness {\it vs.} B$^{2}_{z,\textrm{tot}}$   & 0.37  & 0.29\\ 
		X-ray brightness {\it vs.} B$^{2}_{h,\textrm{tot}}$   & 0.32  & 0.25\\ 
	\end{tabular}
	\label{tab:4}
\end{table}

Table~\ref{tab:4} shows the partial correlation coefficients between the X-ray
brightness and integrated magnetic quantities (except for the magnetic flux) after removing the
effect of magnetic flux. Again, we find that although the correlation coefficients are numerically somewhat different, they are qualitatively similar across the two filters. We do not find any significant correlation between X-ray brightness and other magnetic quantities (except for a slightly negative correlation for $J_{\textrm{tot}}$, which is lower for the Ti-poly filter). Thus, it appears that the total magnetic flux is the primary positive contributor to the total coronal X-ray flux over solar active regions.

\subsection{Filter Issues in the X-ray Data}

Our analysis shows that there are minor differences in the established relationships gleaned from the Ti-poly and Al-poly data sets. We suggest that this small difference in results can be explained as a consequence of contamination in CCDs that could have altered the filter response. The Ti-poly X-ray data and Al-poly X-ray data have strong linear correlation (linear correlation coefficient 0.99) which indicates that there are no calibration problems with the XRT data (see Figure 6.7). Taken together with the fact that the results are qualitatively similar from both filters, this lends strong credence to the data and our conclusions.

\begin{figure}[!h]
	\begin{center}
		\includegraphics[angle=90, width=1.0\textwidth]{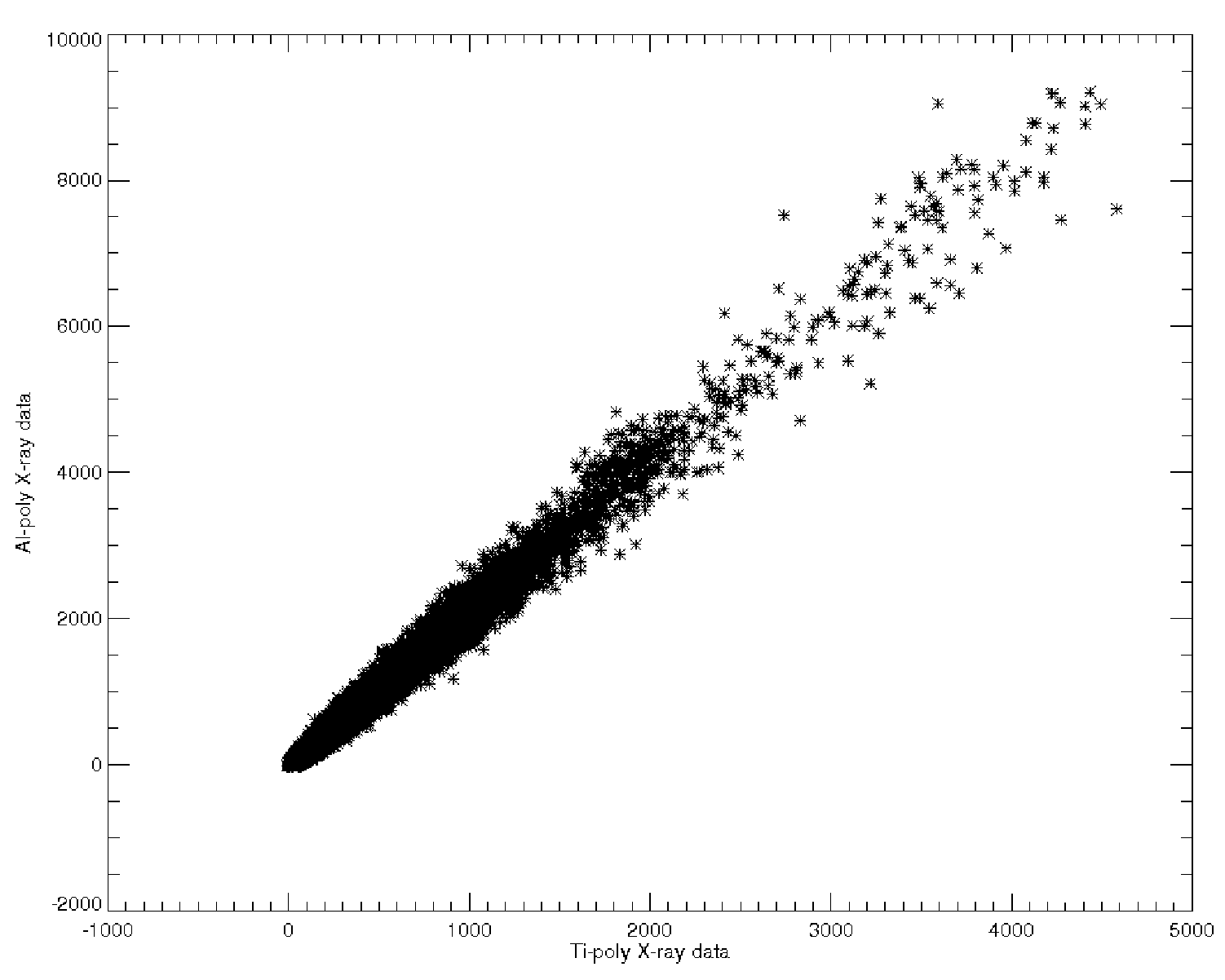}
	\end{center}
	\caption{X-ray data obtained from the Ti-poly and Al-poly filters. The linear correlation coefficient is 0.99.}
	\label{fig:7}
\end{figure}

\section{Summary and Discussion}

A dominant fraction of coronal X-ray emission is known to originate within strongly magnetized active-region structures. To establish which of the magnetic-field quantities within these active regions contributes to the observed X-ray brightness, we have analyzed the X-ray data from the XRT instrument and vector magnetic field measurements from the SP instrument onboard the {\it Hinode} spacecraft. We observed a good correlation between the total area-integrated magnetic field parameters and the
X-ray brightness. A strong correlation is observed  with the total unsigned magnetic flux and closer inspection indicates that other magnetic parameters are correlated with the X-ray brightness through their dependence on magnetic flux. This establishes that the magnetic flux (and thus size) of the system matters. It is generally observed that larger active regions have higher magnetic flux than smaller active regions, which suggests that larger active regions are brighter in X-rays than the small active regions. This result reconfirms the earlier result of Fisher et al. (1998), which was based on lower resolution data and is valid across a range of orders of magnitudes across stars and other astrophysical objects (Pevtsov et al. 2003).

A large amount of total current is indicative of a highly non-potential active region with a large reservoir of energy. Does this larger energy reserve due to non-potentiality directly translate into stronger coronal X-ray emission?. Large-scale current systems are known to produce large-scale flares (Schrijver et al. 2008). It has previously been shown by Nandy et al. (2003) that the variance in the distribution of the local twist within active-region flux systems is also an indicator of the flare productivity of active regions. However, even if this background is suggestive of the role of active region non-potentiality in the release of energy and one might surmise also in coronal heating, we did not find this to be the case here. In fact, we found a (weak) negative correlation between X-ray flux and a measure of non-potentiality, namely $\mu_0 J_{\textrm{tot}}/\phi_{\textrm{tot}}$. If it does really exist, this correlation has no obvious explanation (at least at this time). Previous studies have shown that active-region non-potentiality has a stronger correlation with flare productivity than magnetic flux (Song et al. 2006; Jing et al. 2006). On the other hand, we found a stronger correlation between X-ray brightness and unsigned magnetic flux. Thus, one can argue that while non-potentiality may be an important determinant of localized heating related to flare productivity, the total unsigned magnetic-flux content is the primary factor governing large-scale coronal heating over the active regions.

For the Alfv\'en wave-heating model ({\it e.g.} see the review by Aschwanden, 2004), magnetic flux is related to the power dissipated at the active region through the square of the Alfv\'en velocity, whereas the X-ray brightness would be some fraction of this power -- which also indicates that there should be a relationship between total X-ray brightness and the total magnetic flux. However, based on a detailed analysis, Fisher et al. (1998) showed that the energy in these waves is not sufficient to explain the observed level of coronal heating. The MCC model (Longcope 1996) also predicts a strong correlation between total X-ray brightness and total magnetic flux. On the other hand, in the nano-flare heating  model (Parker 1988) the power dissipated in active-region coronae is related to $B_{z,\textrm{tot}}^2$, suggesting that the total X-ray brightness would be strongly correlated with $B_{z,\textrm{tot}}^2$ rather than $\phi_{\textrm{tot}}$.

Our observations and analysis suggest that the MCC model is a viable contender as a physical theory for the heating of solar and stellar coronae. Nevertheless, we note that it is very likely that a variety of physical processes may contribute to coronal heating to different extents; there are numerous other subtleties in the coronal-heating problem that are far from being settled and need further investigations.

 \chapter{Kink Instability, Coronal Sigmoids and Solar Eruptive Events}

The scientific community is divided on whether the magnetohydrodynamic kink instability mechanism in highly twisted sunspot magnetic structures can generate solar storms such as flares and CMEs and form coronal sigmoidal structures. To explore this issue we utilize high resolution vector magnetograms from the Hinode satellite and a new observational technique for measuring the twist of photospheric magnetic fields. Following this, we perform a comparative study of a subset of solar active region magnetic structures, associated flares and overlying sigmoids or lack thereof, to determine whether the kink instability mechanism can lead to coronal X-ray sigmoids and solar flares. We find that on the rare occasions that the twist in magnetic structures exceed the kink instability criterion, the active regions always had a flare associated with them but not necessarily a coronal X-ray sigmoid. Our results and analysis indicate that kink instability is a plausible source for solar eruptive events and provides a viable methodology for forecasting solar storms based on this mechanism.

\section{Introduction}
Solar flares and Coronal Mass Ejections (CME) are explosive events responsible for major disturbances in space weather. It has been well established that solar flares are associated with sunspots and an analysis of the magnetic structures of active regions (sunspots grouped together) is necessary to shed more light on the origin of these solar storms.

When a magnetic flux tube rises up through the convection zone it acquires twist due to a variety of physical processes including helical turbulence, differential rotation and the Coriolis force. When this twisted flux tube emerges through the surface of the Sun -- the photosphere, the magnetic field expands rapidly as background gas pressure decreases, and overlying coronal loops associated with this active region are formed and observed in high energy radiation. The emerged flux tube can be further twisted by shearing photospheric motions such as foot-point motions and magnetic reconnection. The more a flux tube and associated magnetic loop is twisted the more it deviates from the potential field condition (i.e., a current free state) and currents are developed in the system driving it to a higher energy state. This stressed structure can release the excess energy in the form of solar flares via magnetic reconnection (Parker 1957, 1963; Petschek 1964; Svestka 1976; Priest \& Forbes 2000; Yokoyama et al. 2001; Takaso et al. 2012; Su et al. 2013; Dudik et al. 2014). However, other alternatives exists for triggering solar eruptive events.

Several observational studies have been performed to explore the relationship between solar flares and non-potentiality of magnetic fields. These studies find that non-potentiality of magnetic field is closely linked with the release of  solar flares (Tian et al. 2002, Hahn et al. 2005; Falconer et al. 2008; Jing et al. 2010; Tiwari et al. 2010). One important parameter used to quantify magnetic non-potentiality is magnetic helicity (which accounts for both twist and writhe of an isolated flux tube). Theoretical considerations and numerical simulations show that when the twist of a flux tube exceeds a certain threshold, the flux tube becomes kink unstable and this instability suddenly converts the twist of the flux tube to writhe or axial deformation (Linton et al. 1996, 1999). Traditionally, the force free parameter $\alpha$ -- calculated from the force-free field equation $ \nabla \times \bf{B} = \alpha \bf{B}$ -- has been used as a measure of the twist of solar active regions. The parameter $\alpha$ is roughly equivalent to twice the magnitude of the twist \textit{q}. Several methods have been proposed to calculate  $\alpha$, e.g.,  $\alpha_{best}$ (Pevtsov et al. 1995), $\alpha_{av}$ (Hagino \& Sakurai 2004) and $\alpha_{peak}$ (Leka et al. 2005) from vector magnetograms. The force free field equation arises from an assumption that the gas pressure is significantly lesser than the magnetic pressure which holds true in the corona.  However the usage of the force free field equation on vector magnetograms is questionable since the gas pressure in the photosphere is comparable to the magnetic pressure (Metcalf et al. 1995). To take care of this Nandy et al. (in preparation) proposed a new method for calculating the twist of photospheric magnetic structures, namely the flux tube fitting technique method, which does not rely on the force-free field  assumption.

Leamon et al. (2002) utilized the force-free parameter $\alpha_{best}$ in an earlier comparative analysis in a subset of solar active regions to emphatically state that kink instability does not play a role in solar eruptions. However, Fan (2005) used MHD simulations to demonstrate that a flux tubes can become kink unstable and erupt through the overlying fields if the twist in the flux tube exceeds a certain value. Leka et al. (2005) performed a blind test on a single magnetogram using the  $\alpha_{peak}$ method proposed in the same paper and suggested that kink instability can be a possible trigger mechanism for solar flares. They argued that the $\alpha_{best}$ method underestimates the value of the twist by an order of magnitude as it samples the whole active region structure. However the $\alpha_{peak}$ method too suffers from a drawback -- it takes only the peak value of the spatial distribution of $\alpha$ gleaned from one pixel which makes it susceptible to errors.

In this chapter, we explicitly test the idea that kink unstable flux tubes are prone to flaring using the twist calculation method suggested by Nandy et al. (in preparation) which is free from the shortcomings mentioned above. We provide details about the data used for our study in Section~7.2. In Section~7.3, we briefly discuss the flux tube fitting technique developed by Nandy et al. (in preparation) for calculating best-fit twist and critical twist threshold for kink instability. In Sections~7.4 and 7.5 we proceed to analyse several vector magnetograms with the said method and present our results and conclusion.
  
\section{Data Selection and Analysis} 
For this study we have used data from the spectro-polarimeter (Ichimoto et al. 2008, SP) back-end instrument of the solar optical telescope (Tsuneta et al. 2008; SOT) onboard the Hinode satellite. This instrument measures the stokes signal for spectral lines Fe I 6301 and 6302~\AA~respectively. It scans the active region in fast scan mode with spatial resolution $0.295''/$pixel$^{-1}$ (along the slit direction) and $0.317''/$pixel$^{-1}$ (along the scanning direction) with a integration time of 1.6 s. Physical magnetic parameters from stokes signal are obtained by inverting the data sets using MERLIN code (based on Milne-Eddington inversion method) and this inverted data sets are available at the Community Spectro-Polarimeter Center (http://sot.lmsal.com/data/sot/level2d and http://www.csac.hao.ucar.edu/). The inherent $180^{\circ}$ ambiguity in the transverse field of the vector magnetogram are resolved using the minimum energy algorithm (Metcalf 1994; Leka et al. 2009). Then the magnetic field vectors are transformed into heliographic coordinates (Venkatkrishnan \& Gary 1989).

For this study, we used 14 different NOAA active regions observed at different times. To minimize the impact of projection effect on magnetic parameter estimation we select active regions which lie within a central meridional distance of 30 degrees from the disc center.

For this active region data set we subsequently determine whether they had any associated flares or coronal X-ray sigmoids. Flare information, including timing and class was taken from the NOAA Flare Catalogue and sigmoidal data was provided by the Hinode X-Ray Telescope (XRT) Team (see Savcheva et al. 2014). The resulting database of active regions is detailed in Table~7.1.

\section{Methods}
\subsection{Measuring Twist by Cylindrical-Flux-Tube-Fitting Technique}
\begin{figure}
 \centering
\includegraphics*[width=\linewidth]{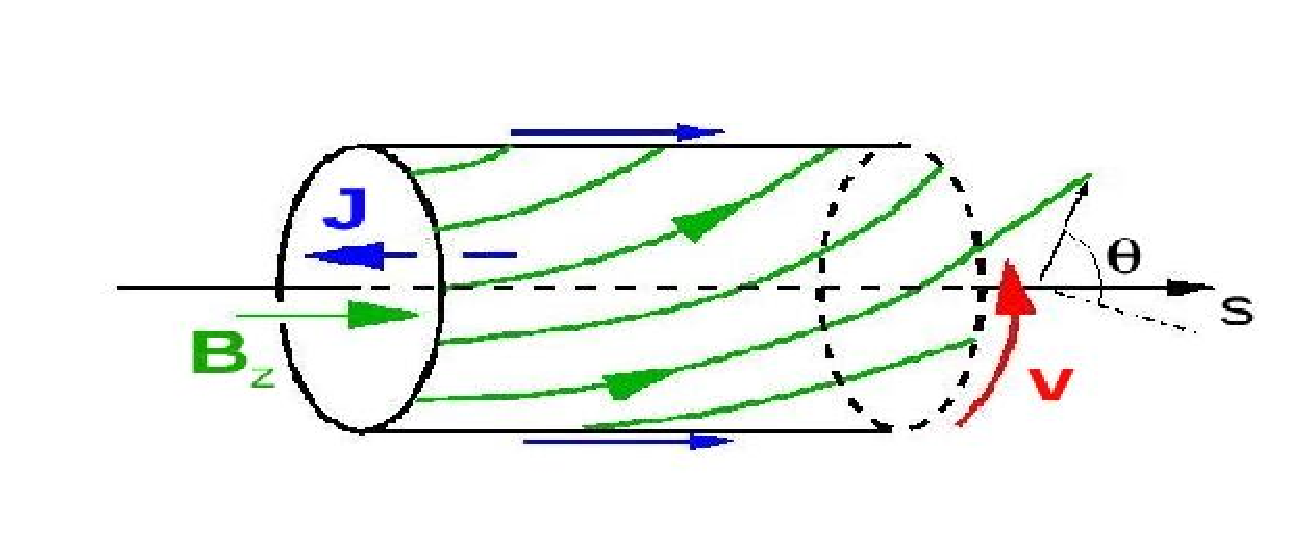}
  \caption{A cartoon image of a twisted flux tube. This figure depicts the conversion of $B_z$ in to the azimuthal $B_\theta$ component. Image Credit: Dana Longcope.}
\end{figure}
To measure twist and the kink instability criterion self consistently from vector magnetograms we follow the flux-tube-fitting technique developed by Nandy et al. (in preparation). In this technique, they assume that solar active regions form photospheric cross-sections of vertical legs of cylindrical flux tubes. The magnetic field structure in the flux tube when represented in cylindrical co-ordinates has three components $B_r$, $B_\theta$ and $B_z$. Considering z-axis as the axis of the flux tube, if we twist the flux tube by an amount q(r) then some part of  $B_z$ component is converted in to the azimuthal $B_\theta$ component, and $B_\theta$ is given by $B_\theta= q(r) r B_z$ (see Figure 7.1 for schematic illustration). This implies that if a flux tube  has no twist it will have no $B_\theta$  component. Assuming flux tubes are uniformly twisted, we can calculate the best fit twist $Q_{fit}$ following the equation $Q_{fit}= B_\theta/r B_z$. To do this, we follow the series of steps described below.

We choose our substructure by manually selecting four pixels from a vector magnetogram which form the edges of the rectangle which encloses our region of interest. We assume that the umbral part of these substructures form a coherent flux tube. To ensure that only the umbral part of the substructures is used in our calculation, we select those pixels whose transverse $B_z$ value lies above a certain critical threshold value (1600 Gauss) ensuring that weak, outlying structures are excluded. More often than not, the isolated coherent structure has a circular cross-section that one expects to be associated with a cylindrical flux tube. Once we have identified this coherent flux tube we have to express the magnetic field structure in cylindrical co-ordinates from the $B_x$, $B_y$ and $B_z$ components in local heliographic coordinate system. We use the flux-weighted-centroid (FWC) as the origin of this local cylindrical co-ordinate system. To calculate the position of the flux-weighted-centroid, we take the value of vertical $B_z$ component in a pixel as the weighted flux of this pixel (say $P_i$). The position of the flux-weighted-centroid is then given by
\begin{eqnarray}
  x_c= \frac{\sum{P_i x_i}}{\sum{P_i}},
 \end{eqnarray}
\begin{eqnarray}                    
y_c= \frac{\sum{P_i y_i}}{\sum{P_i}},
\end{eqnarray}                      
where ($x_c$,$y_c$) is the position of flux-weighted centroid, $P_i$ is the weighted flux in a pixel and $x_i$, $y_i$ are the coordinates of each pixel. After finding the origin of the cylindrical flux tube, a coordinate transformation is performed to obtain the values of magnetic field components in cylindrical coordinate system.\\
\begin{eqnarray}
 B_\theta(r)= -B_x sin (\theta) + B_y cos (\theta)
\end{eqnarray}
\begin{eqnarray}
 B_z(r)=B_z
\end{eqnarray}
Determining the values, we finally plot $B_\theta/B_z$ versus \textit{r} and evaluate the best-fit (linear) slope which gives the best-fit twist of the identified flux tube:\\
\begin{equation}
Q_{fit}= \frac{B_\theta}{r B_z}.
\end{equation}
Thus $Q_{fit}$ represents the average twist per unit length over the identified flux tube. We note that $Q_{fit}$ is related to the twist component of total helicity given by $\Phi^2~(T/2\pi)$ -- where $\Phi$ is the magnetic flux and $T$ is the total twist of a flux tube of axial length L ($T = Q_{fit}L$).
 \subsection{Establishing the Kink Instability Criterion}
 To evaluate the critical twist threshold ($Q_{kink}$) above which flux tubes become kink unstable we utilize the theoretical foundations outlined in Linton et al. (1999); this study, backed by numerical simulations show that the twist threshold for kink instability is $\mu^{1/2}$, where $\mu$ is the $r^2$ coefficient in the Taylor series expansion of the axial magnetic field profile of a cylindrical magnetic flux tube expressed as:
\begin{eqnarray}
B_z(r)=B_0(1- \mu r^2 + ....) .
\end{eqnarray}
Here $B_0$ is the strength of magnetic field along the axis of the flux tube. We plot $B_z$ versus radial distance (r), and perform a fit corresponding to the above equation to determine the $r^2$ coefficient thereby generating the critical twist threshold:
\begin{equation}
Q_{kink}= \mu^{1/2}.
\end{equation}
If the flux tube twist $Q_{fit}$ exceeds $Q_{kink}$ the flux tube will be susceptible to the kink instability mechanism.

  \begin{figure}
 \centering
\includegraphics*[width=\linewidth]{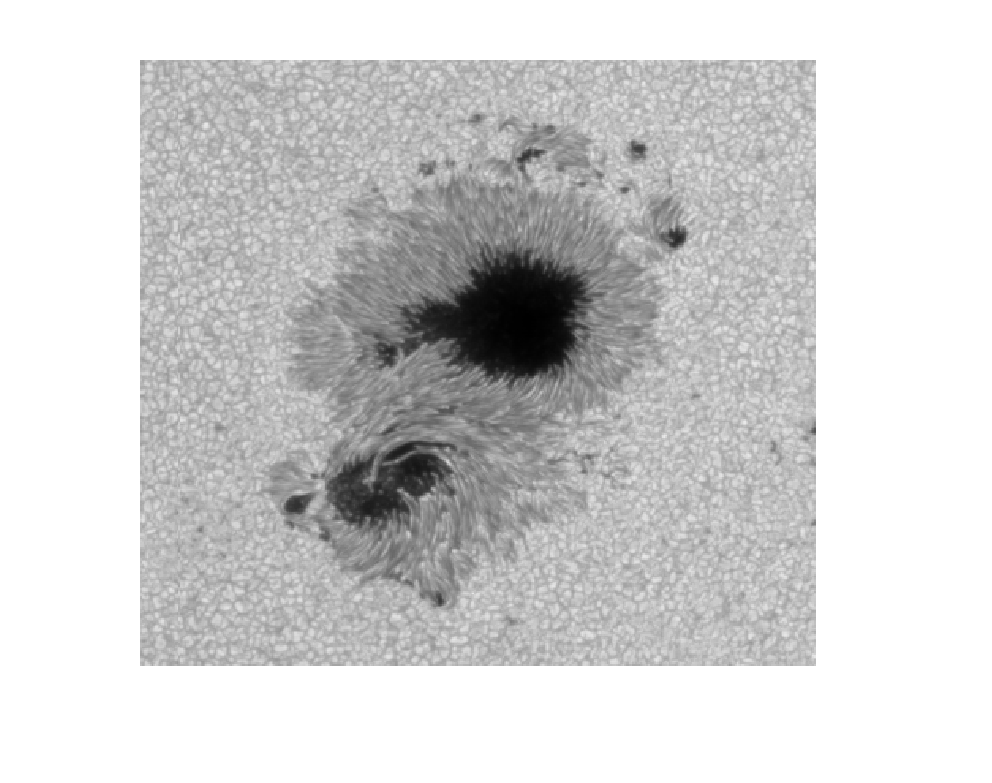}
  \caption{Image of AR 10930. Top spot is the negative spot and bottom spot is the positive spot.}
\end{figure}
 \section{Results}
 
 \begin{figure}
\begin{center}
\begin{tabular}{cc}
\includegraphics[scale=0.45]{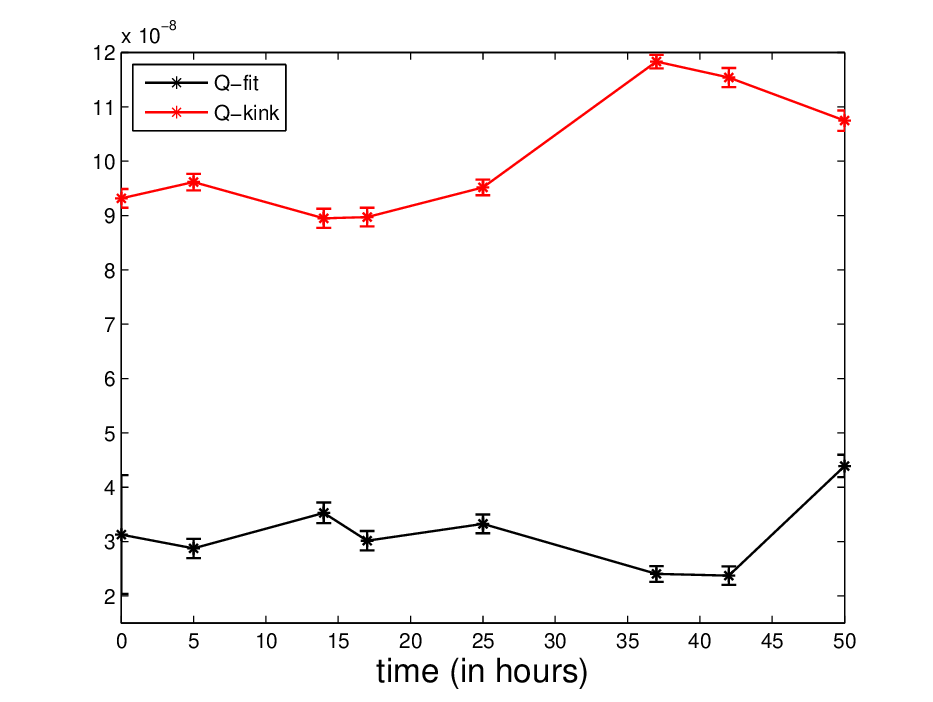} & \includegraphics[scale=0.45]{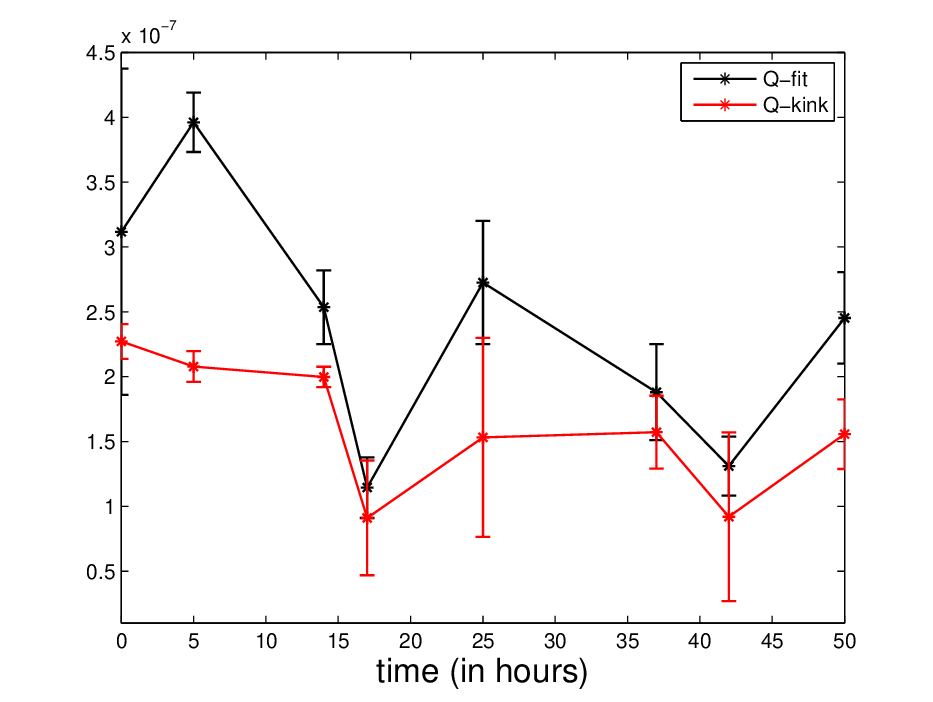}\\
                (a)                      &                   (b)             \\
 \end{tabular}
\end{center}               
\caption{ (a) Temporal evolution of $Q_{fit}$ and $Q_{kink}$ for negative spot of AR 10930
 (b) Temporal evolution of $Q_{fit}$ and $Q_{kink}$ for positive spot of AR 10930. Note that $Q_{fit}$ and $Q_{kink}$ are denoted as Q-fit and Q-kink inside the figure. Error bars refer to 95 \% confidence bound of both $Q_{fit}$ and $Q_{kink}$.}
\end{figure}
 Table 7.1 lists 14 NOAA active regions obtained from the spectro-polarimeter onboard Hinode. Active regions which contain single or multiple loops and appear as either S or inverse S shaped in the soft X-ray image of corona are known as sigmoids (Rust \& Kumar 1996; McKenzie \& Canfield 1999). It has been shown by Canfield et al. (1999) that active regions associated with sigmoids are more flare productive. Thus we chose to include some active regions associated with sigmoid structures in our study.  For this work we first divided active regions into coherent substructures (with a well-formed umbral part) and then calculated $Q_{fit}$ and $Q_{kink}$ for all of these substructures at different times depending on the availability of data. Note that $Q_{fit}$ and $Q_{kink}$ are denoted as Q-fit and Q-kink inside the figures.

Fig.~7.2 is an image of AR 10930, which released an X-class flare. This is a bipolar active region with both positive and negative spots. We have  studied the temporal evolution of $Q_{fit}$ and $Q_{kink}$ for active region 10930 over a time period of three days. In the case of the negative spot, $Q_{fit}$ never exceeds $Q_{kink}$ (see Fig.~7.3(a)); whereas in case of the positive spot, $Q_{fit}$ consistently exceeds $Q_{kink}$ (see Fig.~7.3(b)). We find no relation between the flare release time and the time when $Q_{fit}$ is above the threshold value. 

We note that positive and negative spots associated with the observed vector magnetogram of an active region often have different values of twist; this is thought to be because all magnetic field lines emanating from one of the polarities do not necessarily close on the other spot within the vector magnetogram. Some may connect to regions out of the field of view of the magnetogram or converge on weaker regions which are difficult to associate with the primary spots. In a previous study Inou et al. (2011) utilized non-linear-force-free reconstruction to study AR 10930. They found that only a small isolated region of the flux system had high twist exceeding a full turn and could not clearly establish whether kink instability had a role to play in the flaring dynamics of AR 10930. Clearly, statistical studies are important in this context, whereby a larger sample of active regions and their analysis may generate more compelling constraints. This motivates our study with a larger sample of active regions.

 \begin{figure}
 \centering
\includegraphics*[width=\linewidth]{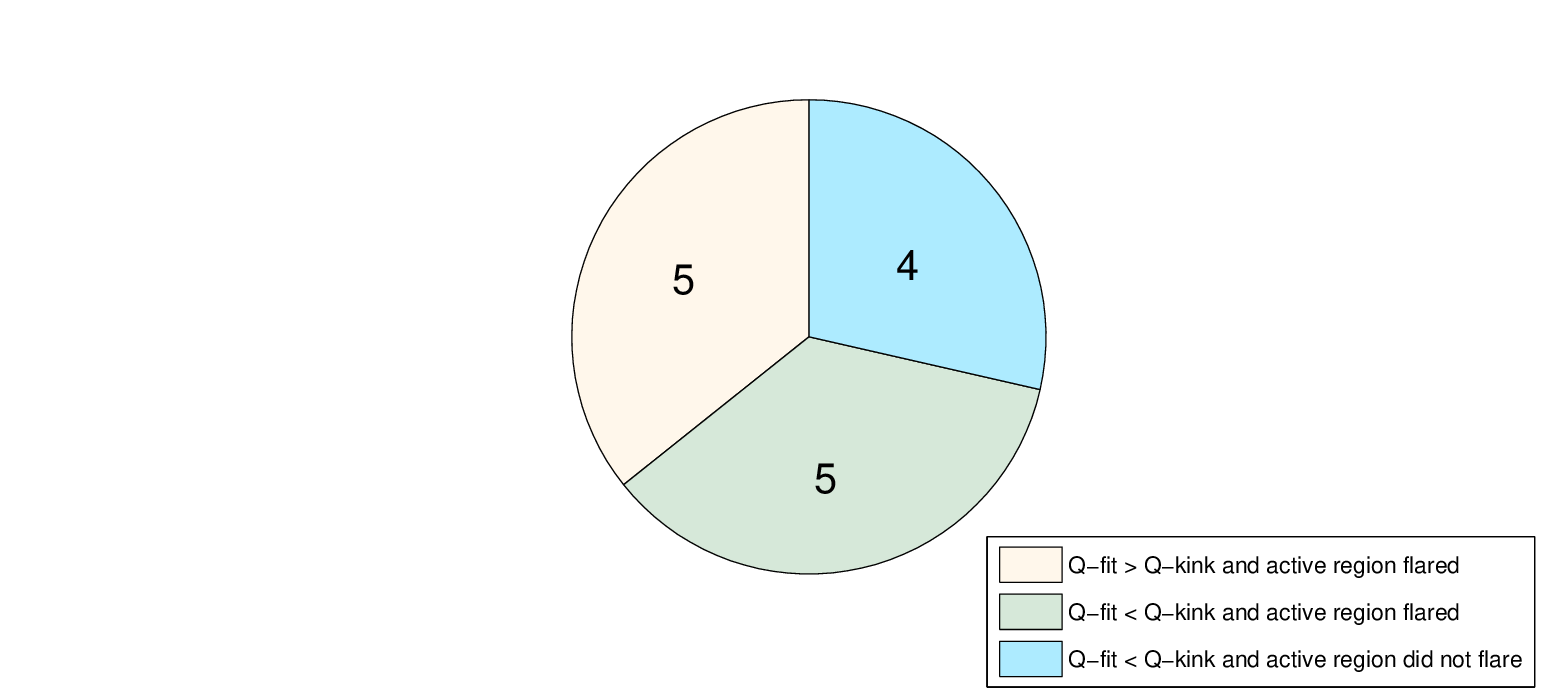}
  \caption{$Q_{fit}$, $Q_{kink}$, flaring and non flaring active regions.}
\end{figure}

\begin{figure}
 \centering
\includegraphics*[width=\linewidth]{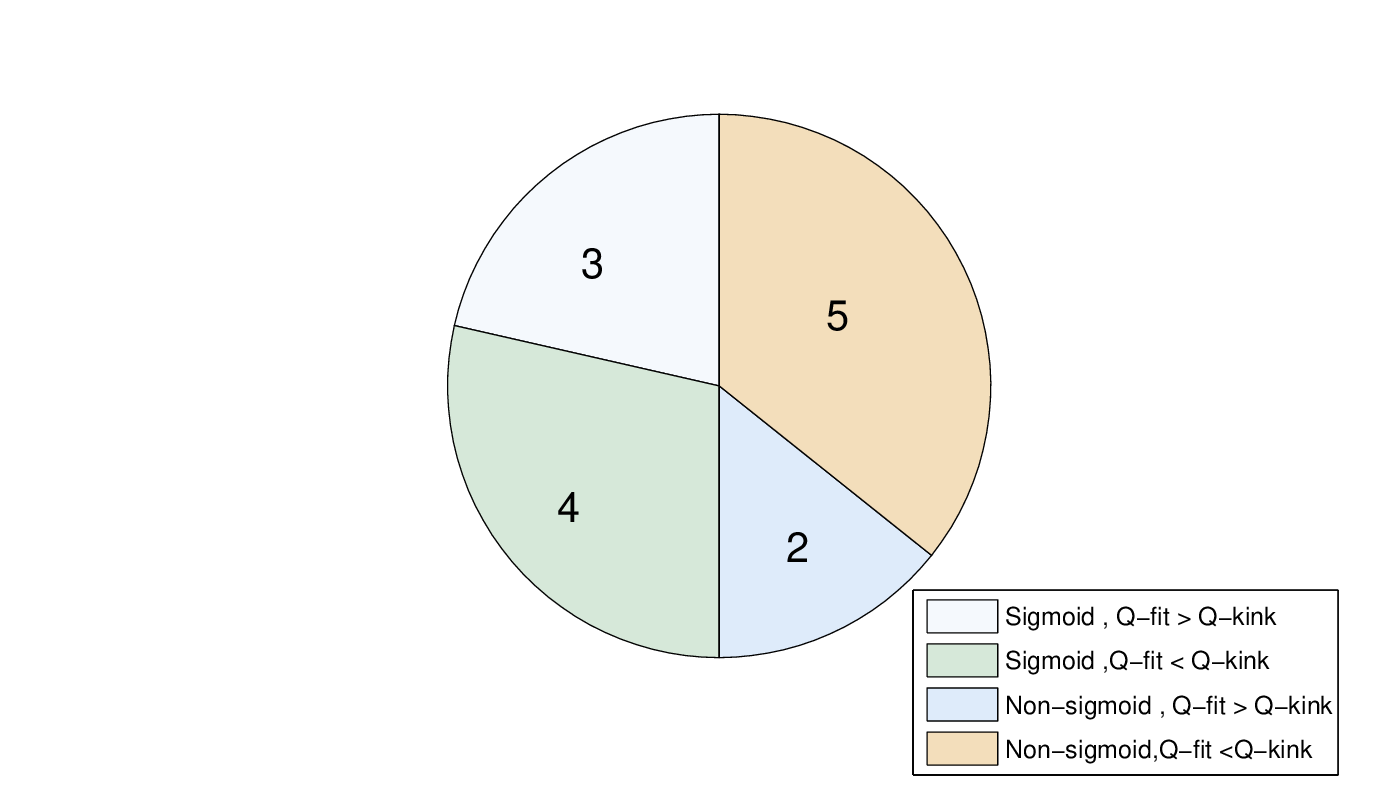}
  \caption{Comparison between sigmoid and non-sigmoid active regions in terms of twist and flares}
\end{figure}
 From Fig.~7.4, we see that out of the analyzed 14 active regions, in five cases $Q_{fit}$ exceeds $Q_{kink}$ at least one time. All of these 5 active regions had at the least C-class flares associated with them. In case of other 9 active regions (where $Q_{fit}$ does not exceed $Q_{kink}$), we find that five of them are associated with flares but the rest are not. Our analysis suggests that if $Q_{fit}$ exceeds $Q_{kink}$, the flux structure is very likely to be associated with a flare but if $Q_{fit}$ does not exceed $Q_{kink}$, there may be or may not be a flare associated with it depending on other factors.

In our study (see Fig.~7.5), 7 active regions were chosen which were associated with sigmoids. The purpose of doing so was to investigate whether the flux tube kinking has anything to do with it being of sigmoidal shape. Out of the 7 sigmoids studied, in only three cases do we find $Q_{fit}$ exceeding $Q_{kink}$. As mentioned before these 3 active regions were also associated with flares. For the 4 other sigmoids not associated with flares, $Q_{fit}$ does not exceed $Q_{kink}$. We conclude from this that a high value of twist and kink unstable flux tubes does not necessarily generate sigmoidal shaped flux loops in all cases.

Our study also shows that eight active regions which contain both positive and negative spots (such as that depicted in Fig.~7.2) had associated flares with them; indicating that complexity in active regions makes them more flare prone.

  \begin{table}[]
 \centering
\caption{List of values of $Q_{fit}$ and $Q_{kink}$ for active regions (in units of $10^{-7} m^{-1} $) }
 \resizebox{\columnwidth}{!}{
\begin{tabular}{ | l | l | l | l | l | l | l | l | l |}
\hline\hline
NOAA & Date & Time  &  $Q_{fit}$ for  & $Q_{fit}$ for& $Q_{kink}$ for & $Q_{kink}$ for & Flare & Kink  \\
       region&  &   & positive spot & negative spot & positive spot & negative spot & & Instability\\
      & & & (with 95 \% & (with 95 \%  &(with 95 \%  &(with 95 \%  & &\\
      & & & confidence bound)& confidence bound) & confidence bound) & confidence bound) & &
        \\ \hline
                &11.12.2006 & 031004 & -3.12 (-4.37, -1.85) & & 2.27 (2.13, 2.39)& & &\\
            & & & & -0.313 (-0.422, -0.203) & &  0.932 (0.914, 0.949) & &\\
       10930  & 13.12.2006 & 125104 & -1.7 (-1.87, -1.53)& & 1.92 (1.79, 2.04) &  & Yes   & \checkmark \\
            & & & & -0.383 (-0.393, -0.372) & & 1.1 (1.08, 1.11)& & \\
                & & 162104 & -1.8 (-2.07, -1.52)& & 1.23 (0.71, 1.58) & & & \\
            & & & &  -0.53 (-0.557, -0.502)& & 1.01 (0.99, 1.03)& &  \\           \hline
   10960 & 5.6.2007 & 65225 & & 0.235 (0.161, 0.308)& & 1.71 (1.58, 1.83) & Yes &  \\ 
      & & & & 0.199 (0.152, 0.246)& & 1.53 (1.43, 1.61) & & \\ \hline  
 11092 & 1.8.2010 & 203051 & &-0.219 (-0.248, -0.189) & & 1.32 (1.28, 1.36) & No  & \\ \hline  
 11093 & 7.8.2010 & 223051 & & -0.235 (-0.291, -0.179) & & 1.45 (1.36, 1.53) & No & \\ \hline
 11158 & 13.02.2011 & 160004 & 0.503 (0.299, 0.706)& & 1.74 (1.55, 1.91) & &  &\\ 
             & & & & -0.333 (-0.485, -0.179) & & 2.06 (1.93, 2.17)& & \\ 
          & 14.02.2011 & 063004 & 4.09  (3.51, 4.68)& & 1.18 (0.647, 1.78)& & Yes & \checkmark  \\
            & & & &  -1.74 (-1.95, -1.53)& & 1.44 (1.36, 1.50)& & \\  \hline 
  11166 & 10.03.2011 & 094136 & 0.951 (0.868, 1.034)& & 1.6 (1.49, 1.69)&  & &  \\
           & & & & 0.582 (0.533, 0.630) & & 1.74 (1.63, 1.85) & Yes & \checkmark \\
            & & & & 5.43 (2.27, 8.62) & & 0.314 (0.922, 0.809) & &\\ 
           & & & & -2.08 (-3.61, -0.541)& & 3.4 (2.15, 4.29) & &\\ \hline  
  11196 & 24.04.2011 & 31546 & 1.32 (1.15, 1.49)&-& 1.11 (1.01, 1.19) &- & Yes & \checkmark \\ \hline  
 11216 & 22.05.2011 & 04605 & 0.794 (0.717, 0.870) &-& 1.87 (1.69, 2.01)&-&  No &\\
         & & 40305 & 0.851 (0.792,0.909)&-& 1.56 (1.34, 1.77)&-& &\\ \hline 
  11263 & 3.8.2011 & 195005& 0.273 (0.179, 0.367) & & 2.15 (1.97, 2.30) & & Yes & \\ 
           & & & &  -0.214 (-0.247, -0.180)& & 1.02 (-0.979, 1.067) & & \\ \hline
 11289 & 13.9.2011& 103405 & & -1.49 (-1.93, -1.05) & & 1.66 (1.31, 1.95)  & No &  \\ 
        & & & & -0.553 (-0.579, -0.525) & & 1.27 (1.20, 1.33) & & \\ \hline
 11302 & 26.9.2011 & 102831 & -0.252 (-0.293, -0.211)&-& 1.48 (1.41, 1.54) & & Yes & \\ 
            & 27.9.2011 & 181806 & & -0.883 (-0.935, -0.830)& &1.03 (0.978, 1.07)& & \\ 
           & & & & -0.258 (-0.312, -0.202) & & 1.29 (1.24, 1.34)&  & \\ \hline  
11339 & 6.11.2011 & 200005 & &-0.516 (-0.563, -0.469) & & 1.4 (1.36, 1.44)& Yes & \\
             & & & & -0.596 (-0.674, -0.516) & & 1.86 (1.78, 1.93)& & \\
            & & & & 0.143 (0.128, 0.158) & & 1.48 (0.279, 2.069) & & \\
            & & & & -0.516 (-0.672, -0.404) & & 1.36 (0.741, 1.77) & & \\
            & & & & -0.955 (-1.123, -0.787) & & 2.06 (1.63, 2.41)& & \\
           & & & -0.518 (-0.870, -0.166) & & 2.08 (1.30, 2.63) & & & \\ \hline
  11471 & 2.5.2012 & 100005 & 0.87 (0.68, 1.05) &-& 0.489 (0.068, 0.910)& &  Yes & \checkmark \\ \hline
 11515 & 4.7.2012 & 224035 & -0.446 (-0.548, -0.342)& & 1.6 (1.39, 1.77) & & Yes & \\
            & & & &  -0.619 (-0.713, -0.525) & & 1.88 (1.70, 2.03) & &  \\
            & 5.7.2012 & 034535 & -0.627 (-0.719, -0.533)& & 1.7 (1.59, 1.79) & & & \\ 
            & & & &  -0.153 (-0.230, -0.069) & & 1.93 (1.85, 2.01)& & \\ \hline
     \hline       
     \end{tabular}
     }
\end{table} 

\section{Summary and Discussions}
 The scientific community has not been able to reach a consensus on whether kink instability is a cause for releasing solar eruptive events. Numerical flux tube simulations indicate kink instability as a possible mechanism for triggering solar flares (Linton et al. 1996, 1999; Fan 2005). However one observational study -- using the force-free parameter $\alpha_{best}$ as a measure of twist (Leamon et al. 2002) -- claimed that kink instability is not responsible for solar flares. But Leka et al. (2005) showed that the calculation of $\alpha_{best}$ for an entire active region is not able to account for highly twisted sub-structures and the above calculation is strongly influenced by horizontal field distribution. Also, the $\alpha_{best}= 2q$ assumption is only valid for a thin flux tube. The proposed flux tube fitting technique for calculating twist does not rely on the assumptions that the entire active region is uniformly twisted and the flux tube is thin. Thus we believe this technique is a better approach for calculating twist.

Here we have studied the susceptibility of active region flux tubes to the kink instability mechanism using this new flux tube fitting technique for calculating twist. Using this technique, we find that if the best fit twist ($Q_{fit}$) of any active region exceeds the threshold necessary for kink instability a flare is always associated with this active region. On the other hand, we also find some active regions where best-fit twist does not exceed the kink instability threshold which had flares associated with them. On another front, Rust \& Kumar (1996) suggested that sigmoids are kink unstable twisted flux tubes. We do not find this to be the case. Within our seven sigmoid data set, we find only three sigmoids had kink unstable (photospheric) flux tubes associated with them.

In summary, we conclude that kink instability is the cause for at least a subset of solar flares but the corresponding active regions need not have coronal sigmoidal structures associated with them. This result and the analysis outlined here may form the basis of forecasting solar flares in advance when accurate vector  magnetogram measurements are available.




\begin{thebibliography}{}
\bibitem[{Alfv\'en (1942)}]{alf42} Alfv\'en, H. \ 1942, Ark. f. Mat. Astr. o. Fysik., 29B, No. 2
\bibitem[{{Antiochos et al.}(1994)}]{antio94} Antiochos, S.K., Dahlburg, R.B. \& Klimchuk, J.A. \ 1994, ApJL, 420, L41
\bibitem[{Arlt \& Abdolvand (2010)}]{arlt10} Arlt, R. \& Abdolvand, A. 2010, in Physics of Sun and Star Spots, Proceedings IAU Symposium No. 273, 
ed. D.P. Choudhary \& K.G. Strassmeier.
\bibitem[{Aschwanden (2004)}]{ascn04} Aschwanden, M.: 2004, {\it Physics of the Solar Corona}, An Introduction, Springer, Praxis Publishing, Chichester, UK.
\bibitem[{Babcock \& Babcock (1955)}]{bab55} Babcock, H. W. \& Babcock, H. D. \ 1955, ApJ, 121, 349
\bibitem[{Babcock (1959)}]{bab59} Babcock, H. W. \ 1959, ApJ, 130, 364
\bibitem[{Babcock (1961)}]{bab61} Babcock, H. W. \ 1961, ApJ, 133, 572
\bibitem[{Barker \textit{et al.} (2012)}]{bark12} Barker, A.J., Silvers, L.J., Proctor, M.R.E. \& Weiss, N.O. 2012, MNRAS, 424,115
\bibitem[{{Barnes et al.}(2005)}]{barn05} Barnes, G., Longcope, D. W. \& Leka, K. D. \ 2005, ApJ, 629, 561
\bibitem[{Beer \textit{et al.} (1998)}]{beer98} Beer, J., Tobias, S.M. \& Weiss, N. \ 1998, SoPh, 181, 23
\bibitem[{Belucz \textit{et al.} (2015)}]{belus15} Belucz, B., Dikpati, M. \& Forgacs-Dajka, N. \ 2015, ApJ, 806, 169
\bibitem[{Bonanno \textit{et al.} (2002)}]{bona02}Bonanno, A., Elstner, D., R\"udiger, G., \& Belvedere, G. 2002, A\&A, 390, 673
\bibitem[{Bonanno \textit{et~al.} (2006)}]{bona06} {Bonanno}, A., {Elstner}, D., {Belvedere}, G., \& {R{\"u}diger}, G. 2006,
  Memorie della Societa Astronomica Italiana Supplementi, 9, 71
\bibitem[{Braun \& Fan (1998)}]{bra98} Braun, D. C. \& Fan, Y. \ 1998, ApJ, 508, L105
\bibitem[{Brandenburg \textit{et al.} (1992)}]{brand92} Brandenburg, A., Moss, D., \& Tuominen, I. 1992, in Astronomical Society
  of the Pacific Conference Series, Vol.~27, The Solar Cycle, ed. K.~L. {Harvey}, 536--+
\bibitem[{Brandenburg \textit{et al.} (1996)}]{brand96} Brandenburg, A., Jennings, R.L., Nordlund, A., Rieutord, M., Stein, R.M., Tuominen, I. 1996,   JFM, 306, 325
\bibitem[{Brown \textit{et al.} (2010)}]{brow10} Brown, B.P., Browning, M.K., Brun, A.S., Miesch, M.S. \& Toomre, J. 2010, ApJ, 711, 424  
\bibitem[{Bushby \& Tobias (2007)}]{bush07} Bushby, P. J., \& Tobias, S. M. \ 2007, ApJ, 661, 1289
\bibitem[{Caligari \textit{et al.} (1995)}]{calig95} Caligari, P., Moreno-Insertis, F. \& \ Sch\"ussler, M. \ 1995 , ApJ , 441, 886.
\bibitem[{Caligari \textit{et al.} (1998)}]{calig98} Caligari, P.,Sch\"ussler, M. \& Moreno-Insertis, F. \ 1998 , ApJ , 502, 481
\bibitem[{Cameron \textit{et al.} (2012)}]{cam12}Cameron, R., Schmitt, D., Jiang, J., \& Isik, E. 2012, A\&A, 542, 127.
\bibitem[{{Canfield et al.}(1999)}]{canifield99} Canfield, R. C., Hudson, H. S. \& McKenzie, D. E. \ 1996, GRL, 26, 627
\bibitem[{{Canfield et al.}(2000)}]{canifield20} Canfield, R., Hudson, H. S. \&  Pevtsov, A. A. \ 2000, IEEE Trans. Plasma Science, 28, 1786
\bibitem[{{Canfield \& Russel}(2007)}]{cani07} Canfield, R. C. \& Russel, A. J. B. \ 2007, ApJL, 662, L39
\bibitem[{Carrington (1858)}]{carr58} Carrington, R. C. \ 1858, MNRAS, 19, 1
\bibitem[{Cattaneo \& Hughes (2009)}]{catta09} Cattaneo, F., \& Hughes, D. W. \ 2009, MNRAS, 395, L48
\bibitem[{Cattaneo (1999)}]{catta99} Cattaneo, F. \ 1999, ApJL, 515, L39
\bibitem[{Chandrasekhar (1952)}]{chand52} Chandrasekhar, S. \ 1952, Phil. Mag. (7), 43, 501
\bibitem[{Chandrashekhar \textit{et al.} (2013)}]{chand13} Chandrashekhar, K., Krishna Prasad, S., Banerjee, D., Ravindra, B. \& Seaton, D.B. 2013, SoPh, 286, 125
\bibitem[{Charbonneau (2010)}]{charb10} Charbonneau, P. \ 2010, Living Rev. Sol. Phys., 7, 3
\bibitem[{Charbonneau \textit{et al.} (1999)}]{charb99} Charbonneau, P. {\it et al.} \ 1999, ApJ, 527, 445
\bibitem[{Charbonneau \& Dikpati(2000)}]{charb20} Charbonneau, P., \& Dikpati, M.\ 2000, ApJ, 543, 1027
\bibitem[{Charbonneau \textit{et al.} (2004)}]{charb04} Charbonneau, P., Blais-Laurier, G., \& St-Jean, C. \ 2004, ApJ, 616, L183
\bibitem[{Charbonneau \textit{et al.} (2005)}]{charb05} Charbonneau, P., St-Jean, C., \& Zacharias, P. \ 2005, ApJ, 619, 613
\bibitem[{Chatterjee \textit{et al.} (2004)}]{chat04} Chatterjee, P., Nandy, D., \& Choudhuri, A. R. \ 2004, A\&A, 427, 1019
\bibitem[{Chatterjee \& Choudhuri (2006)}]{chat06} Chatterjee, P., \& Choudhuri, A. R. \ 2006, SoPh, 239, 29
\bibitem[{Choudhuri \& Gilman (1987)}]{chou87} Choudhuri, A.R., \& Gilman, P. A. \ 1987, ApJ, 316, 788
\bibitem[{Choudhuri (1990)}]{chou90} Choudhuri, A.R. \ 1990, ApJ, 355, 733
\bibitem[{Choudhuri (1992)}]{chou92} Choudhuri, A.R.\ 1992, A\&A, 253, 277
\bibitem[{Choudhuri \& Karak (2009)}]{chou09} Choudhuri, A.R. \& Karak, B.B. 2009, RAA, 9, 953
\bibitem[{Choudhuri \& Karak (2012)}]{chou12} Choudhuri, A.R., \& Karak, B.B.\ 2012, PRL \, 109, 171103
\bibitem[{Choudhuri \textit{et al.} (1995)}]{chou95}Choudhuri, A. R., Sch\"ussler, M., \& Dikpati, M. \ 1995, A\&A, 303, L29
\bibitem[{Choudhuri \textit{et al.} (2004)}]{chou04}Choudhuri, A.R., Chatterjee, P., \& Nandy, D. 2004, ApJL, 615, L57
\bibitem[{Choudhuri \textit{et al.} (2007)}]{chou07}Choudhuri, A. R., Chatterjee, P., \& Jiang, J. \ 2007, PRL, 98, 13
\bibitem[{Chowdhury \textit{et al.} (2013)}]{chou13}Chowdhury, P., Choudhary, D.P., \& Gosain, S. \ 2013, ApJ, 768, 188
\bibitem[{Christensen-Dalsgaard \textit{et al.} (1996)}]{dals96} Christensen-Dalsgaard, J., Dappen, W., Ajukov, S.V., Anderson, E.R., Antia, H.M., Basu, S. et al. 1996, Science, 272, 1286
\bibitem[{Cirtain \textit{et al.} (2013)}]{cirt13} Cirtain, J.W., Golub, L., Winebarger, A.R., De Pontieu, B. {\it et al.} 2013, Nature, 493, 501.
\bibitem[{Cliver \textit{et al.} (1998)}]{cliv98}	Cliver, E.W., Boriakoff, V. \& Bounar, K.H. 1998, GRL, 25, 897
\bibitem[{Corbard \& Thompson (2002)}]{coba02} Corbard, T. \& Thompson, M. 2002, SoPh, 205, 211 
\bibitem[{Cowling (1953)}]{cowl53} Cowling, T.G. \ 1953, Solar Electrodynamics, in the Sun, ed. G.P. Kuiper (Univ. Chicago Press, Chicago)
\bibitem[{Danilovic \textit{et al.} (2010)}]{Dan10}Danilovic, S., Sch\"ussler, M. \& Solanki, S. K. \ 2010, A\&A, 509, A76
\bibitem[{Dasi-Espuig \textit{et al.} (2010)}]{dasi10}Dasi-Espuig, M. {\it et al.} \ 2010, A\&A,  518, 7
\bibitem[{DeLuca \& Gilman (1987)}]{del87}DeLuca, E. E. \& Gilman, P. A. \ 1987, GAFD,  37, 85 
\bibitem[{DeRosa \textit{et al.} (2012)}]{der12}DeRosa, M.L., Brun, A.S. \& Hokesema, J.T. 2012, ApJ, 757, 96
\bibitem[{DeVore \& Sheeley (1987)}]{dev87} DeVore, C. R., \& Sheeley, N. R., Jr. 1987, SoPh, \ 108, 47
\bibitem[{{DeVore \& Antiochos}(2000)}]{devo20}  DeVore, C.R. \& Antiochos, S.K. \ 2000, ApJ, 539, 954
\bibitem[{Dikpati \& Charbonneau (1999)}]{dikp99} Dikpati, M., \& Charbonneau, P. \ 1999, ApJ, 518, 508
\bibitem[{Dikpati \& Gilman (2001)}]{dikp01}Dikpati, M., \& Gilman, P.A. 2001, ApJ, 559, 428
\bibitem[{Dikpati \textit{et al.} (2002)}]{dikp02} Dikpati, M., Corbard, M., Thompson, M. J. \& Gilman, P. A. \ 2002, ApJ, 575, L41
\bibitem[{Dorch \& Nordlund (2001)}]{dorch01} Dorch, S.\ B.\ F., \& Nordlund, \AA. 2001, A\&A, 365, 562.
\bibitem[{D'Silva \& Choudhuri (1993)}]{silva93} D'Silva, S. \& Choudhuri, A. R. \ 1993, A\&A, 272, 621.
\bibitem[{D'Silva (1993)}]{dsilva93} D'Silva, S. \ 1993, ApJ, 407, 385.
\bibitem[{Dube \& Charbonneau (2013)}]{dube13} Dube, C., \& Charbonneau, P. \ 2013, ApJ, 775, 69
\bibitem[{{Dudik et al.}(2014)}]{dudik14} Dudik, J. {\it et al.} \ 2014, ApJ, 784, 144
\bibitem[{Dungey (1953)}]{dung53} Dungey, J. W. \ 1953, Phil. Mag., 44, 725
\bibitem[{Durney (1995)}]{durn95} Durney, B.~R. 1995, SoPh, 106, 213
\bibitem[{Durney (1997)}]{durn97} Durney, B. R. \ 1997, ApJ, 486, 1065
\bibitem[{Eddy (1988)}]{eddy88} Eddy, J.A\ 1988, \ in Secular Solar and Geomagnetic Variations in the Last 10,000 Years, ed. F. R. Stephenson \& A. W. Wolfendale (NATO ASI Ser. C, 236; Dordrecht: Kluwer),1
\bibitem[{Falconer \textit{et al.} (1997)}]{falc97} Falconer, D.A., Moore, R.L., Porter, J.G., Gary, G.A. \& Shimizu, T. 1997, ApJ, 482, 519. 
\bibitem[{Falconer (1997)}]{falc97a} Falconer, D.A. 1997, SoPh, 176, 123. 
\bibitem[{Falconer \textit{et al.} (2000)}]{falc20} Falconer, D.A., Gary, G.A., Moore, R.L. \& Porter J.G. 2000, ApJ, 528, 1004.
\bibitem[{Falconer \textit{et al.} (2008)}]{falc08} Falconer, D.A., Moore, R. L. \& Gary G. A. \ 2008, ApJ, 689, 1433 
\bibitem[{Fan \textit{et al.} (1993)}]{fan93} Fan, Y., Fisher, G. H. \& DeLuca, E. E. \ 1993, ApJ , 405, 390.
\bibitem[{Fan \textit{et al.} (1994)}]{fan94} Fan, Y., Fisher, G. H. \& McClymont, A. N. \ 1994, ApJ, 436, 907
\bibitem[{Fan \& Gong (2000)}]{fan20} Fan, Y. \& Gong, D. \ 2000, SoPh, 192, 141
\bibitem[{Fan (2005)}]{fan05} Fan, Y. \ 2005, ApJ, 630, 543
\bibitem[{Fan \& Fang (2014)}]{fan14} Fan Y. \& Fang F. 2014, ApJ, 789, 11.
\bibitem[{Fisher {\it et al.} (1998)}]{fish98} Fisher, G.H., Longcope, D.W., Metcalf, T.R. \& Pevtsov, A.A. 1998, ApJ, 508, 885.
\bibitem[{Giles \textit{et al.} (1997)}]{gil97} Giles, P. M. {\it et al.} \ 1997, Nature, 390, 52
\bibitem[{Gilman (1983)}]{gil83} Gilman, P. A. \ 1983, ApJ, 53, 243
\bibitem[{Giovanelli (1946)}]{gio46} Giovanelli, R.G. \ 1946, Nature, 158, 81
\bibitem[{Gizon \& Rempel (2008)}]{giz08} Gizon, L. \& Rempel, M. \ 2008, SoPh, 251, 241
\bibitem[{Glatzmaier (1985)}]{glat85} Glatzmaier, G. A. \ 1985, ApJ, 291, 300
\bibitem[{{Glover et al.}(2000)}]{glov20} Glover, A., Ranns, N., Harra, L. K. \& Culhane, J. L. \ 2000, GRL, 27, 2161
\bibitem[{Goel \& Choudhuri (2009)}]{goel09} Goel, A. \& Choudhuri, A.R. 2009, RAA, 9, 115
\bibitem[{Golub {\it et al.} (2007)}]{golu07} Golub, L., DeLuca, E., Austin, G., Bookbinder, J., Caldwell, D. {\it et al.} 2007, SoPh, 243, 63.
\bibitem[{Gonz\'alez Hern\'andez \textit{et al.} (1999)}]{gon99} Gonz\'alez Hern\'andez, I. {\it et al.} \ 1999, ApJ, 510, L153
\bibitem[{{Gopalswamy et al.}(2000)}]{gopa20} Gopalswamy, N. et al. \ 2000, GRL, 27, 145
\bibitem[{Guerrero \& de Gouveia Dal Pino (2007)}]{guer07} Guerrero, G. \& de Gouveia Dal Pino, E.~M.  2007, A\&A, 464,341
\bibitem[{Guerrero \& de Gouveia Dal Pino (2008)}]{guer08} Guerrero, G. \& de Gouveia Dal Pino, E.M. 2008, A\&A, 485, 267
\bibitem[{{Guerrero} {\it et~al.}(2013)}]{guer2013} {Guerrero}, G., {Smolarkiewicz}, P.~K., {Kosovichev}, A., \& {Mansour}, N.
  2013, in Solar and Astrophysical Dynamos and Magnetic Activity, IAUS 294
\bibitem[{{Hagino \& Sakurai}(2004)}]{hagi04} Hagino, M. \& Sakurai, T. \ 2004, PASJ, 56, 831 
\bibitem[{Hagyard (1988)}]{hagya88} Hagyard, M.J. 1988, SoPh, 115, 107
\bibitem[{{Haygard et al.}(1984)}]{hagy84} Hagyard, M. J., Teuber, D., West, E. A. \& Smith, J. B. \ 1984, SoPh, 91, 115
\bibitem[{{Hahn et al.}(2005)}]{hahn05} Hahn, M., Gaard, S., Jibben, P., Canfield R. C. \& Nandy, D. \ 2005, ApJ, 629, 113
\bibitem[{Hale (1908)}]{hale19} Hale, G. E. \ 1908, ApJ, 28, 315
\bibitem[{Hale et al. (1919)}]{hale919} Hale, G. E., Ellerman, F, Nicholson, F. B., Joy, A. R. \ 1919, ApJ, 49, 153
\bibitem[{Hanasoge et al. (2012)}]{hana12} Hanasoge, S.M., Duvall, T.L., Sreenivason, K.R. \ 2012, PNAS, 109, 11928
\bibitem[{Hathaway (1996)}]{hath96} Hathaway, D. H. \ 1996, ApJ, 460, 1027
\bibitem[{Hathaway \textit{et al.} (1996)}]{hatha96} Hathaway, D. H. {\it et al.} 1996, Science 272, 1306
\bibitem[{Hathaway \textit{et al.} (2003)}]{hath03} Hathaway, D. H., Nandy, D., Wilson, R. M., \& Reichmann, E. J. 2003, ApJ, 589, 665
\bibitem[{Hathaway \& Rightmire (2010)}]{hath10} Hathaway, D. H. \& Rightmire, L. \ 2010, Science 327, 1350
\bibitem[{Hathaway \& Rightmire (2011)}]{hath11} Hathaway, D. H. \& Rightmire, L. \ 2011, ApJ, 729, 80
\bibitem[{Hathaway (2012)}]{hath12} Hathaway, D.H. , 2012, arXiv:1210.3343
\bibitem[{Hazra \& Nandy (2013)}]{haz13} Hazra, S. \& Nandy, D. 2013, In ASI Conf. Ser. 10, International Symposium on Solar Terrestrial Physics, ed. N. Gopalswamy, S.S. Hasan, P.B. Rao \& P. Subramanian, 115
\bibitem[{Hazra \textit{et al.} (2014)}]{ghaz14} Hazra, G., Karak, B.B. \& Choudhuri, A.R. 2014, ApJ, 782, 9.
\bibitem[{Hazra \textit{et al.} (2014)}]{haz14} Hazra, S., Passos, D. \& Nandy, D. 2014, ApJ, 789, 5
\bibitem[{Hazra \textit{et al.} (2015)}]{haz15} Hazra, S., Nandy, D. \& Ravindra, B. 2015, SoPh, 290, 771
\bibitem[{Hotta \& Yokoyama (2010)}]{hotta10} Hotta H., \& Yokoyama T., 2010, ApJ, 709, 1009
 \bibitem[{Howe (2009)}]{howe09} Howe, R. \ 2009, Living Rev. Sol. Phys, 6, 1
 \bibitem[{Hoyle (1949)}]{hoyle49} Hoyle, F. \ 1949, Some Recent Researches in Solar Physics (Cambridge Univ. Press, Cambridge)
\bibitem[{Hoyng (1988)}]{hoyng88} Hoyng, P. \ 1988, ApJ, 332, 857
\bibitem[{Hoyt \& Schatten (1996)}]{hoyt96} Hoyt, D. V., \& Schatten, K. H. \ 1996, SoPh, 165, 181
\bibitem[{Hurlburt, Toomre \& Massaguer (1984)}]{hurl84} Hurlburt, N. E., Toomre, J. \& Massaguer, J. M. \ 1984, ApJ, 282, 557
\bibitem[{Ichimoto \textit{et al.} (2008)}]{ichim08}
Ichimoto, K., Lites, B., Elmore, D., Suematsu, Y., Tsuneta, S. {\it et al.} 2008, SoPh, 249, 233
\bibitem[{Jiang \textit{et al.} (2007)}]{jiang07} Jiang, J., Chatterjee, P. \& Choudhuri, A. R. 2007, MNRAS, 381, 1527
\bibitem[{Jiang \& Wang (2007)}]{jiang07}Jiang, J. \& Wang, J.X. 2007, MNRAS, 377, 711
\bibitem[{Jiang {\it et al.} (2013)}]{jiang13} Jiang, J., Cameron, R. H., Schmitt, D. \& Isik, E. 2013, A\&A, 553, A128
\bibitem[{Jing {\it et al.} (2006)}]{jing06} Jing, J., Song, H., Abramenko, V., Tan, C. \& Wang, H. 2006, ApJ, 644, 1273
\bibitem[{{Jing et al.}(2010)}]{jing10} Jing, J., Tan, C., Yuan, Y., Wang, B., Wiegelmann, T., Xu, Y. \& Wang, H. \  2010, ApJ, 713, 440
\bibitem[{Jouve \& Brun (2007)}]{jouve07} Jouve, L. \& Brun, A.~S. 2007, A\&A, 474, 239
\bibitem[{Jouve \textit{et al.} (2008)}]{jouv08}
{Jouve}, L., {Brun}, A.~S., {Arlt}, R., {Brandenburg}, A. {\it et al.} 2008, A\&A, 483, 949
\bibitem[{Jouve \textit{et al.} (2010)}]{jouve10} Jouve, L., Proctor, M.R.E. \& Lesur, G. 2010, A\&A , 519, A68.
\bibitem[{Kitchatinov \& Olemskoy (2011a)}]{kit11} Kitchatinov, L. L., \& Olemskoy, S. V. \ 2011, Ast. Letters, 37, 655
\bibitem[{Kitchatinov \& R\"udiger (2012)}]{kitch12}Kitchatinov, L.\ L., \& R\"udiger, G. 2012, SoPh,  276, 3
\bibitem[{Klimchuk (2006)}]{klim06} Klimchuk, J.A. 2006, SoPh, 234, 41
\bibitem[{Karak (2010)}]{kara10} Karak, B.B. 2010, ApJ, 724, 1021
\bibitem[{Karak \& Nandy (2012)}]{kara12} Karak, B.B., \& Nandy, D. \ 2012, ApJL, 761, L13
\bibitem[{K\"apyl\"a \textit{et al.} (2006a)}]{kapy06}K\"apyl\"a,P.\ J., Korpi, M.\ J., Ossendrijver, M., \& Stix, M. 2006a, A\&A, 455, 401.
\bibitem[{K\"apyl\"a \textit{et al.} (2006b)}]{kapy06b}K\"apyl\"a, P.\ J., Korpi, M.\ J., \& Tuominen, I. 2006b, AN, 327, 884
\bibitem[{Knobloch \textit{et al.} (1998)}]{knob98} Knobloch, E., Tobias, S.M. \& Weiss, N.O. MNRAS, 297, 1123
\bibitem[{Komm, Howard \& Harvey (1993)}]{kom93} Komm, R. W., Howard, R. F. \& Harvey, J. W. \ 1993, SoPh., 147, 207
\bibitem[{Komm \textit{et al.} (2011)}]{kom11} Komm, R.W., Howe, R., Hill, F. \textit{et al.}  \ 2011, IOP Journal of Phys., 271, 012077
\bibitem[{Kosovichev \textit{et al.} (1997)}]{koso97} Kosovichev, A. G. et al. \ 1997, SoPh, 170, 43
\bibitem[{Kosugi \textit{et al.} (2007)}]{kosu07}
Kosugi, T., Matsuzaki, K., Sakao, T., Shimizu, T. {\it et al.} 2007, SoPh, 243, 3.
\bibitem[{Krause \& R\"adler (1980)}]{kra80} Krause, F., \& R\"adler K. -H. \ 1980, Mean-field magnetohydrodynamics and dynamo theory (Oxford, Pergamon Press, Ltd.)
\bibitem[{Kusano \textit{et al.} (2002)}]{kusa02} Kusano, K., Maeshiro, T., Yokoyama, T. {\it et al.} \ 2002, ApJ, 577, 501
\bibitem[{Latushko (1993)}]{lat93} Latushko, S. M.  \ 1993, SoPh, 146, 401
\bibitem[{Larmor (1919)}]{lar19} Larmor, J. \ 1919, Brit. Assn. Adv. Sci. Rep., 159-160
\bibitem[{{Leamon et al.}(2002)}]{leamon02} Leamon, R. J., Canfield, R. C. \& Pevtsov, A. A. \ 2002, JGR, 102, 1234
\bibitem[{Lee {\it et al.} (2010)}]{lee10}
Lee, J.-Y., Barnes, G., Leka, K.D., Reeves, K.K. {\it et al.} 2010, ApJ, 723, 149.
\bibitem[{Leighton (1969)}]{leigh69} Leighton, R. B. \ 1969, ApJ, 156, 1
\bibitem[{Leka \textit{et al.} (2005)}]{leka05} Leka, K.D., Fan, Y. \& Barnes, G. 2005, ApJ, 626, 1091.
\bibitem[{{Leka et al.}(2009)}]{leka09} Leka, K. D. et al. \ 2009, SoPh., 260, 83
\bibitem[{Leka \textit{et al.} (2009)}]{Leka2009}Leka, K.D., Barnes, G. \& Crouch, A. 2009, In: Lites, B., Cheung, M., Magara, T., Reeves, K. (eds.) The Second Hinode Science Meeting: Beyond Discovery-Toward Understanding. {\bf CS-415}, Astron. Soc. Pac., San Francisco, 365.
\bibitem[{Leka \& Barnes (2007)}]{leka07}Leka, K.D. \& Barnes, G.D. 2007, ApJ, 656, 1173.
\bibitem[{Lin, Kuhn \& Coulter (2004)}]{lin04} Lin, H., Kuhn, J.R. \& Coulter, R. 2004, ApJL, 613, L177.
\bibitem[{{Linton et al.}(1996)}]{lint96} Linton, M. G., Longcope, D. W. \& Fisher, G. H. \ 1996, ApJ, 469, 964
\bibitem[{{Linton et al.}(1998)}]{lint98} Linton, M. G.,  Dahlburg, R. B., Fisher, G. H. \& Fan, Y. \ 1998, ApJ, 507, 404
\bibitem[{{Linton et al.}(1999)}]{lint99} Linton, M. G., Fisher, G. H., Dahlburg, R. B., \& Fan, Y. \ 1999, ApJ, 522, 1190
\bibitem[{Lites (2011)}]{lites11} Lites, B. W. \ 2011, ApJ, 737, 52
\bibitem[{Lockwood \& Owens (2011)}]{lock11}
Lockwood, M. \& Owens, M. J. \ 2011, JGR, 116, 4109
\bibitem[{Longcope (1996)}]{long96} Longcope, D.W. 1996, SoPh, 169, 91.
\bibitem[{Longcope \textit{et al.} (1998)}]{long98} Longcope, D.W., Fisher, G.H. \& Pevtsov, A.A. 1998, ApJ, 507, 871.
\bibitem[{Longcope \& Choudhuri (2002)}]{long02} Longcope, D., \& Choudhuri, A.R. 2002, SoPh, 205, 63
\bibitem[{Lopes \& Passos (2009)}]{lop09} Lopes, I. \& Passos, D. 2009, SoPh, 257, 1
\bibitem[{Lundquist {\it et al.} (2008)}]{lund08} Lundquist, L.L., Fisher, G.H., Metcalf, T.R. \textit{et al.} 2008, ApJ, 689, 1388.
\bibitem[{Mandrini \textit{et al.} (2000)}]{mandr20} Mandrini, C.H., Demoulin, P. \& Klimchuk, J.A. 2000, ApJ, 530, 999.
\bibitem[{McClintock \& Norton (2013)}]{mcc13} McClintock, B.H. \& Norton, A.A. 2009, SoPh, 287, 215
\bibitem[{McIntosh {\it et al.} (2011)}]{mcint11} 	
McIntosh, S.W., de Pontieu, B., Carlsson, M. \textit{et al.} 2011, Nature, 475, 477.
\bibitem[{{McKenzie \& Canifield}(2008)}]{mcke08} McKenzie, D.E. \& Canfield, R.C. \ 2008, A \& A, 481, L65
\bibitem[{Metcalf (1994)}]{metc94} Metcalf, T.R. 1994, SoPh, 155, 235.
\bibitem[{Metcalf {\it et al.} (1994)}]{metc94b}
Metcalf, T.R., Canifield, R.C., Hudson, H.H., Mickey, D.L., Wulser, J.-P., Martens, P.C.H {\it et al.} 1994, ApJ, 428, 860.
\bibitem[{{Metcalf et al.}(1995)}]{metc95} Metcalf, T. R., Jiao, L., McClymont, A. N. \& Canfield, R. C. \ 1995, ApJ, 439, 474
\bibitem[{Miesch \textit{et al.} (2006)}]{mies06} Miesch, M.~S., Brun, A.~S., \&  Toomre, J. 2006, ApJ, 641, 618 
\bibitem[{Mitra-Kraev \& Thompson (2007)}]{mitrak07} Mitra-Kraev, U., \& Thompson, M.J. 2007, AN, 328,1009
\bibitem[{Miyahara \textit{et al.} (2004)}]{miya04}Miyahara, H. et al.\ 2004, SoPh, 224, 317
\bibitem[{Miyahara \textit{et al.} (2010)}]{miya10}	Miyahara, H., Kitazawa, K., Nagaya, K. \textit{et al.} 2010, J. Cosmol., 8, 1970
\bibitem[{Moffatt (1969)}]{moff69} Moffatt, H.K. \ 1969, JFM, 35, 117
\bibitem[{Moreau (1961)}]{more61} Moreau, J.J. \ 1961, C. R. Acad. Sci. Paris, 252, 2810
\bibitem[{Moreno-Insertis, F. (1983)}]{moren83}Moreno-Insertis, F. \ 1983, A\&A, 122, 241
\bibitem[{Moreno-Insertis, F.  et al. (1992)}]{moren92}Moreno-Insertis, F., Sch\"ussler, M., Ferriz-Mas A \ 1992, A\&A, 264, 686
\bibitem[{Mu\~noz-Jaramillo \textit{et al.} (2009)}]{munoz09}Mu\~noz-Jaramillo, A., Nandy, D., \& Martens, P. C. H. \ 2009, ApJ, 698,461
\bibitem[{Mu\~noz-Jaramillo \textit{et al.} (2010)}]{munoz10} Mu\~noz-Jaramillo, A., Nandy, D., Martens, P.C.H. \& Yeates, A.R. 2010, ApJL, 720, L20
\bibitem[{Mu\~noz-Jaramillo \textit{et al.} (2013)}]{munoz13}Mu\~noz-Jaramillo, A., Dasi-Espuig, M., Balmaceda, L. A. \& DeLuca, E. E. \ 2013, ApJ, 767, L25
\bibitem[{Nandy \& Choudhuri (2001)}]{nandy01} Nandy, D., \& Choudhuri, A. R. \ 2001, ApJ, 551, 576
\bibitem[{Nandy (2002)}]{nand02} Nandy, D. \ 2002, Ap \&SS, 282, 209
\bibitem[{Nandy \& Choudhuri (2002)}]{nandy02} Nandy, D., \& Choudhuri, A. R. \ 2002, Science, 296, 1671
\bibitem[{Nandy \it{et al.} (2003)}]{nand03} Nandy, D., Hahn, M., Canifield, R.C.\& Longcope, D.W. 2003, ApJL, 597, L73.
\bibitem[{Nandy (2008)}]{nandy08} Nandy, D. 2008, {\it Subsurface and Atmospheric Influences on Solar Activity, ASP
Conference Series}, 383, 201.
\bibitem[{Nandy \textit{et al.} (2011)}]{nandy11} Nandy, D., Mu{\~n}oz-Jaramillo, A., \& {Martens}, P.~C.~H. 2011, Nature, 471, 80
\bibitem[{Nandy (2012)}]{nandy12} Nandy, D. \ 2012, American Astron. Soc. Meeting Abstracts, 220, 300.01 (http:$//$spd.aas.org/docs/prizetalks/2012/nandy$\_$harvey/)
\bibitem[{{Nandy et al.}(2012)}]{nandys12} Nandy, D., Calhoun, A., Windschitl, J. \& Linton, M. G. \ 2012, (in preparation)
\bibitem[{Narain \& Ulmschneider (1996)}]{nar96} Narain, U. \& Ulmschneider, P. 1996, Space Sci. Rev., 75, 453.
\bibitem[{Nelson \& Miesch (2014)}]{nel14} Nelson, N.J. \& Miesch, M.S. 2014, Plasma Phys. Control. Fusion, 56, 064004
\bibitem[{Newton \& Milsom (1955)}]{newt55} Newton, H.W. \& Milsom, A.S. 1955, MNRAS, 115, 398
\bibitem[{Norton \textit{et al.} (2014)}]{nort14} Norton, A.A., Charbonneau, P. \& Passos, D. 2014, Space Sci. Rev., 186, 251
\bibitem[{Olemskoy \& Kitchatinov (2013)}]{olem13} Olemskoy \& Kitchatinov 2013, ApJ, 777, 71
\bibitem[{Ossendrijver \textit{et al.} (2002)}]{ossen02} Ossendrijver, M., Stix, M., Brandenburg, A., \& R{\"u}diger, G. 2002, A\&A, 394, 
\bibitem[{Ossendrijver (2003)}]{ossen03} Ossendrijver, M. \ 2003, A\&A Rev, 11, 287
\bibitem[{Pallavicini {\it et al.} (1979)}]{palla79}
Pallavicini, R., Vaiana, G.S., Tofani, G. \& Felli, M. 1979, ApJ, 229, 375
\bibitem[{Parker (1955a)}]{park55a} Parker, E. N.\ 1955, ApJ , 121, 491
\bibitem[{Parker (1955b)}]{park55b} Parker, E. N.\ 1955, ApJ , 122, 293
\bibitem[{Parker (1957)}]{park57} Parker, E. N.\ 1957, JGR, 62, 509
\bibitem[{Parker (1963)}]{park63} Parker, E. N.\ 1963, ApJS, 8, 177
\bibitem[{Parker (1975)}]{park75} Parker, E. N.\ 1975, ApJ , 198, 205
\bibitem[{Parker (1988)}]{park88}Parker, E.N. 1988, ApJ, 330, 474.
\bibitem[{Parker (1993)}]{park93} Parker, E. N.\ 1993, ApJ , 408, 707
\bibitem[{Passos \& Lopes (2011)}]{pass11} Passos, D. \& Lopes, I.P. \ 2011 , JASTP \textbf{73,} 191
\bibitem[{Passos \textit{et al.} (2012)}]{pass12} Passos, D., Charbonneau, P. \& Beaudoin, P. \ 2012 SoPh, 279, 1
\bibitem[{Passos \textit{et al.} (2014)}]{pass14} Passos, D., Nandy, D., Hazra, S. \& Lopes, I.P., \ 2014, A\&A, 563, A18
\bibitem[{Passos \textit{et al.} (2015)}]{pass15} Passos, D., Charbonneau, P. \& Miesch, M. \ 2015 ApJL, 800, L18
\bibitem[{Petschek (1964)}]{pets64} Petschek, H. E.\ 1964, Magnetic field annhilation, in Physics of Solar Flares, ed. W.N. Hess (NASA SP-50, Washington, DC), 425
\bibitem[{Petrovay \& Szakaly (1993)}]{pet93} Petrovay, K. \& Szakaly, G. \ 1993, A\&A, 274, 543
\bibitem[{Pevtsov \textit{et al.} (1995)}]{pevt95} Pevtsov, A.A., Canfield, R.C., \& Metcalf, T.R., \ 1995, ApJL, 440, L109
\bibitem[{Pevtsov \textit{et al.} (1996)}]{pevt96} Pevtsov, A.A., Canfield, R.C., \& Zirin, H., \ 1996, ApJ, 473, 533
\bibitem[{Pevtsov \textit{et al.} (1997)}]{pevt97} Pevtsov, A.A., Canfield, R.C., \& McClymont, A.N., \ 1997, ApJ, 481, 973
\bibitem[{Pevtsov \& Acton (2001)}]{pevt01} Pevtsov, A.A. \& Acton, L.W. 2001, ApJ, 554, 416.
\bibitem[{Pevtsov (2002)}]{pevt02} Pevtsov, A.A. \ 2002, SoPh, 207, 111.
\bibitem[{Pevtsov {\it et al.} (2003)}]{pevt03}
Pevtsov, A.A., Fisher, G.H., Acton, L.W., Longcope, D.W. {\it et al.} 2003, ApJ, 598, 387.
\bibitem[{Pipin \& Seehafer (2009)}]{pip09}Pipin, V. V. \& Seehafer, N. 2009, ApJ, 493, 819
\bibitem[{Pipin \& Kosovichev (2011)}]{pip11}Pipin, V. V. \& Kosovichev A.G. 2011, ApJ, 738, 104
\bibitem[{Pipin \& Kosovicev (2013)}]{pip13}Pipin, V. V. \& Kosovichev A.G. 2013, arXiv:1302.0943
\bibitem[{Poluianov \textit{et al.} (2014)}]{polu14} Poluianov, S.V., Usoskin, I.G. \& Kovaltsov, G.A. 2014, SoPh, 289, 4701
\bibitem[{Priest (1982)}]{prist82} Priest, E.R. \ 1982, Solar Magnetohydrodynamics (Reidel, Dordrecht).
\bibitem[{{Priest \& Forbes}(2000)}]{pris20} Priest, E. \& Forbes, T. \ 2000, Magnetic Reconnection: MHD theory and Applications (Cambridge: Cambridge University Press)
\bibitem[{Racine \textit{et al.} (2011)}]{rac11} Racine, E., Charbonneau, P., Ghizaru, M., Bouchat, A. \& Smolarkiewicz, P.K. 2011, ApJ, 735, 46
\bibitem[{Rempel (2006)}]{remp06} Rempel, M. 2006, ApJ, 647, 662
\bibitem[{Ribes \& Nesme-Ribes (1993)}]{rib93}Ribes, J.C. \& Nesme-Ribes, E. 1993, A\&A 276, 549
\bibitem[{Rogachevskii \textit{et al.} (2011)}]{rog11} Rogachevskii, I., Kleeorin, N., K\"apyl\"a, P. \& Brandenburg, A. 2011, PRE, 84, 056314
\bibitem[{{Rust \& Kumar}(1996)}]{rust96} Rust, D. M. \& Kumar, A. \ 1996, ApJL, 464, L199
\bibitem[{{Rust}(1999)}]{rust99}Rust, D. M. \ 1999, in M. R. Brown, R. C. Canfield, \& A. A. Pevtsov (eds.), Magnetic Helicity in
Space and Laboratory Plasmas, American Geophysical Union, Washington, D.C., p. 221
\bibitem[{Savcheva \textit{et al.} (2014)}]{savc14} Savcheva, A.S., McKillop, S.C., McCauley, P.I., Hanson, E.M. \& DeLuca, E.E. 2014, SoPh, 289, 3297
\bibitem[{{Sawyer et al.}(1986)}]{sawy86} Sawyer, C., Warwick, J. W. \& Dennett, J. T.  \ 1986, Solar Flare Presiction (Colorado: Colorado Association University Press)
\bibitem[{Schaten \textit{et al.} (1978)}]{scha78} Schatten, K.H., Scherrer, P.H., Svalgaard, L. \& Wilcox, J.M. 1978, GRL, 5, 411
\bibitem[{Schaten (2005)}]{scha05} Schatten, K.H., GRL, 32, L21106
\bibitem[{Schou \textit{et al.} (1998)}]{scho98} Schou, J. et al. \ 1998, ApJ, 505, 390
\bibitem[{Schou \& Bogart (1998)}]{schou98} Schou, J. \& Bogart, R. S. \ 1998, ApJ, 504, L131
\bibitem[{Schrijver {\it et al.} (2006)}]{sch06}
Schrijver, C.J., De Rosa, M.L., Metcalf, T.R. {\it et al.} 2006, SoPh, 235, 16
\bibitem[{Schrijver {\it et al.} (2008)}]{sch08}
Schrijver, C.J., De Rosa, M.L., Metcalf, T. {\it et al.} 2008, ApJ, 675, 1637
\bibitem[{Schwabe (1844)}]{sch44} Schwabe, S. H. \ 1844, AN, 21,2
\bibitem[{Seehafer (1990)}]{seeh90} Seehafer, N.\ 1990, SoPh, 125, 219
\bibitem[{{Shibata \& Magara}(2011)}]{shibata11} Shibata, K. \& Magara, M. \ 2011, LRSP, 8, 6
\bibitem[{Simard \textit{et al.} (2013)}]{sima13} Simard, C., Charbonneau, P. \& Bouchat, A., \ 2013, ApJ,768,16
\bibitem[{Sivaraman \textit{et al.} (2010)}]{siva10} Sivaraman, K. R., Sivaraman, H., Gupta, S. S. \& Howard, R. F. 2010, SoPh, 266, 247
\bibitem[{Skumanich \& Lites (1987)}]{skum87} Skumanich, A. \& Lites, B.W. \ 1987, ApJ, 322, 473
\bibitem[{Snodgrass \& Dailey (1996)}]{snod96} Snodgrass, H. B. \& Dailey, S. B. \ 1996, SoPh, 163, 21
\bibitem[{Sokoloff \& Nesme-Ribes (1994)}]{sok94} Sokoloff, D. \& Nesme-Ribes, E. 1994, A\&A 288, 293
\bibitem[{Song {\it et al.} (2006)}]{song06}
Song, H., Jing, J., Tan, C. \& Wang, H. 2006, AAS/Solar Physics Division Meeting, 37, 09.0
\bibitem[{Spiegel \& Weiss (1980)}]{spi80}Spiegel, E. A. \& Weiss, N. W. \ 1980, Nature, 287, 616
\bibitem[{Steinhilber \textit{et al.} (2010)}]{stein10}
Steinhilber, F., Abreu, J. A., Beer, J. \& McCracken, K. G.\ 2010, JGR , 115, 1104
\bibitem[{Steenbeck, Krause \& R\"adler (1966)}]{steen66}Steenbeck, M., Krause, F. \& R\"adler K. H. \ 1966, Z. Natur. Teil A, 21, 369
\bibitem[{Stokes (1852)}]{stok52} Stokes, G.G. \ 1852, Philosophical Transactions of the Royal Society of London, 142, 463
\bibitem[{Sturrock (1994)}]{sturr94} Sturrock, P.A. \ 1994, Plasma Physics (Cambridge Univ. Press, Cambridge).
\bibitem[{{Su et al.}(2013)}]{su13} Su, Y. {\it et al.}  \ 2013, Nature, 9, 489
\bibitem[{Svalgaard \& Kamide (2013)}]{sva13} Svalgaard, L. \& Kamide, Y. 2013, ApJ, 763, 23
\bibitem[{{Svestka}(1976)}]{sves76} Svestka, Z. \ 1976, Solar Flares (Dordrecht:Reidel)
\bibitem[{Sweet (1958)}]{sweet58} Sweet, P.A. \ 1958, in Electromagnetic Phenomena in Cosmical Physics, IAU Symp. 6. ed. B. Lehnert (Cambridge Univ. Press, London), 123
\bibitem[{{Takasao et al.}(2012)}]{takasa12} Takasao, S., Asai, A., Isobe, H. \& Shibata, K. \ 2012, ApJL, 745, L6
\bibitem[{Taylor (1974)}]{tayl74} Taylor, J.B. \ 1974, PRL, 33, 1139
\bibitem[{Tan {\it et al.} (2007)}]{tan07} Tan, C., Jing, J., Abramenko, V.I. {\it et al.} 2007, ApJ, 665, 1460.
\bibitem[{Temmer \textit{et al.} (2006)}]{tem06} Temmer, M., Rybak, J., Bendík, P., Veronig, A. {\it et al.} 2006, A\&A, 447, 73
\bibitem[{Thompson (1951)}]{thomp51} Thompson, W. \ 1951, Phil. Mag., 42, 1417
\bibitem[{Thompson \textit{et al.} (1996)}]{thomp96} Thompson, M. J. et al. \ 1996, Science, 272, 1300
\bibitem[{{Tian et al.}(2002)}]{tian02} Tian, L., Liu, Y. \& Wang, J. \ 2002, SoPh, 209, 361
\bibitem[{{Tiwari et al.}(2010)}]{tiwa10} Tiwari, S. K., Venkatakrishnan, P. \& Gosain, S. \ 2010, ApJ, 721, 622
\bibitem[{Tobias \textit{et al.} (2001)}]{tobias01}Tobias SM, Brummell NH, Clune TL \& Toomre J. 2001, ApJ, 549, 1183
\bibitem[{Tobias \textit{et al.} (2006)}]{tobi06} Tobias, S. M., Hughes, D. W. \& Weiss N., \ 2006 , Nature , 442, 26.
\bibitem[{Tomczyk {\it et al.} (2007)}]{tomc07} Tomczyk, S., McIntosh, S.W., Keil, S.L. {\it et al.} 2007, Science, 317, 1192.
\bibitem[{Tsuneta {\it et al.} (2008)}]{tsun08}
Tsuneta, S., Ichimoto, K., Katsukawa, Y., Nagata, S {\it et al.} 2008, SoPh, 249, 167.
\bibitem[{Tsuneta \textit{et~al.} (2008)}]{tsuns08}
{Tsuneta}, S., {Ichimoto}, K., {Katsukawa}, Y., {Lites}, B.~W. {\it et al.} 2008, ApJ, 688, 1374
\bibitem[{{US National Academy of Sciences}(2008)}]{usn08} US National Academy of Sciences-National Research Council 2008 Report on “Severe Space Weather Events - Understanding Societal and Economic Impacts”(Washngton, D.C., National Academies Press)
\bibitem[{Usoskin \textit{et al.}(2003)}]{usosk03}
Usoskin, I.G. {\it et al.} \ 2003, PRL, 91, 21
\bibitem[{Usoskin \textit{et al.}(2000)}]{usosk20}
Usoskin, I.G., Mursula, K. \& Kovaltsov, G.A. 2000, A\&A, 354, L33
\bibitem[{Usoskin \textit{et al.}(2005)}]{usosk05}
Usoskin, I.G., Sch\"ussler, M., Solanki, S.K. \& Mursula, K.\ 2005, JGR, 110, A10102
\bibitem[{Usoskin \textit{et al.}(2007)}]{usosk07}
Usoskin, I.G., Solanki, S.K. \& Kovaltsov, G.A. 2007, A\&A 471, 301
\bibitem[{Usoskin \textit{et al.} (2014)}]{usos14} Usoskin, I.G., Hulot, G., Gallet, Y. \textit{et al.} 2014, A\&A 562, L10
\bibitem[{Vaiana {\it et al.} (1973)}]{vaia73}
Vaiana, G.S., Davis, J.M., Giacconi, R. {\it et al.} 1973, ApJL, 185, L47
\bibitem[{van Ballegooijen (1982)}]{van82}  van Ballegooijen, A. A. \ 1982, A\&A, 113, 99
\bibitem[{van Ballegooijen \textit{et al.} (1998)}]{van98}  van Ballegooijen, A. A., Cartledge, N. P., \& Priest, E. R., \ 1998, ApJ, 501, 866.
\bibitem[{van Ballegooijen \& Mackay (2007)}]{van07} van Ballegooijen, A.~A. \& Mackay, D.~H. \ 2007, ApJ, 659, 1713
\bibitem[{Vaquero \textit{et al.}(2011)}]{vaqu11}
Vaquero, J.M., Gallego, M.C., Usoskin, I.G. \& Kovaltsov, G.A. 2011, ApJL, 731, L24
\bibitem[{Venkatakrishnan \& Gary (1989)}]{venkat89} Venkatakrishnan, P. \& Gary, G.A. 1989, SoPh, 120, 235.
\bibitem[{Waldmeier (1955)}]{wald55} Waldmeier M., 1955, "Ergebnisse und Probleme der Sonnenforschung", Leipzig, Geest, \& Portig, 2nd Edition.
\bibitem[{Waldmeier (1971)}]{wald71} Waldmeier M., 1971, SoPh, 20, 332
\bibitem[{Wang \textit{et al.} (1989)}]{wang89}Wang, Y.-M., Nash, A. G. \& Sheeley, N. R., \ 1989, Science, 245, 712
\bibitem[{Wang \& Sheeley (2003)}]{wang03}Wang, Y.-M., Nash \& Sheeley, N.R., Jr. 2003, ApJ, 593, 1241
\bibitem[{Wang {\it et al.} (2000)}]{wan20}
 Wang, H., Yan, Y., Sakurai, T. \& Zhang, M. 2000, SoPh, 197, 263.
 \bibitem[{{Wang et al.}(2007)}]{wang07} Wang, T. J., Sui, L. H. \& Qiu, J. D \ 2007, ApJL, 661, L207
 \bibitem[{Wang {\it et al.} (2008)}]{wan08}
 Wang, H., Jing, J., Changyi, T., Wiegelmann, T. \& Kubo, M. 2008, ApJ, 687, 658.
 \bibitem[{Wedemeyer-B{\"o}hm {\it et al.} (2012)}]{wedem12} 
Wedemeyer-B{\"o}hm, S., Scullion, E., Steiner, O., van der Voort, L.R. {\it et al.} 2012, Nature, 486, 505.
\bibitem[{Weiss (1981)}]{wess81}Weiss, N. \ 1981, JFM, 108, 247
\bibitem[{Wilmot-Smith \textit{et al.}(2005)}]{wilm05} Wilmot-Smith, A. L., Martens, P. C. H., Nandy, D., Priest, E. R., \& Tobias, S. M.\ 2005, MNRAS, 363, 1167
\bibitem[{Wilmot-Smith \textit{et al.}(2006)}]{wilm06} Wilmot-Smith, A. L., Nandy, D., Hornig, G., \& Martens, P. C. H. \ 2006, ApJ, 652, 696
\bibitem[{Withbroe \& Noyes (1977)}]{withbr77}Withbroe, G.L. \& Noyes, R.W. 1977, Annu. Rev. Astron. Astrophys., 15, 363.
\bibitem[{Woltjer (1958)}]{wolt58} Woltjer, L. \ 1958, PNAS, 44, 489
\bibitem[{Yang \textit{et al.} (2009)}]{yang09} Yang, S.H., Zhang, J., Jin, C.L. \textit{et al.} \ 2009, A\&A, 501, 745
\bibitem[{Yeates \textit{et al.} (2008)}]{yeat08} Yeates, A.R., Nandy, D. \& Mackay, D.H.\ 2008, ApJ, 673, 544
\bibitem[{{Yokoyama et al.}(2001)}]{yoko01} Yokoyama, T., Akita, K., Morimoto, T., Inoue, K. \& Newmark, J. \ 2001, ApJL, 546, L6
\bibitem[{Yoshimura (1975)}]{yosh75} Yoshimura, H. \ 1975, ApJS, 29, 467
\bibitem[{{Yurchyshyn et al.}(2001)}]{yurc01} Yurchyshyn, V. B., Wang, H., Goode, P. R. \& Deng, Y. \ 2001, ApJ, 563, 381
\bibitem[{Zhao \textit{et al.} (2013)}]{zhao13} Zhao, J., {Bogart}, R.S., Kosovichev, A.G., Duvall, Jr., T.L. \& Hartlep, T. 2013, ApJL, 774, L29
\bibitem[{Zirker (1993)}]{zirk93} Zirker, J.B. 1993, SoPh, 148, 43.
\end{thebibliography}

\begin{appendices} 

\chapter{Numerical Methods}

	We provide a overview of numerical methods which is used to solve dynamo equations in a two dimensional geometry. The evolution equation for the poloidal and toroidal field is given by:
\begin{equation}
   \frac{\partial A}{\partial t} + \frac{1}{s}\left[ \mathbf{v_p} \cdot \nabla (sA) \right] = \eta\left( \nabla^2 - \frac{1}{s^2}  \right)A + S (r,\theta,B),
\end{equation}
\begin{equation}
   \frac{\partial B}{\partial t}  + s\left[ \mathbf{v_p} \cdot \nabla\left(\frac{B}{s} \right) \right] + (\nabla \cdot \mathbf{v_p})B = \eta\left( \nabla^2 - \frac{1}{s^2}  \right)B + s\left(\left[ \nabla \times (A\bf \hat{e}_\phi) \right]\cdot \nabla \Omega\right)   + \frac{1}{s}\frac{\partial (sB)}{\partial r}\frac{\partial \eta}{\partial r},
\end{equation}
where $s= r \sin (\theta)$. These two equations are coupled, nonlinear partial differential equation. We solve these equations as a intial value problem using Alternating Direction Implicit (ADI) method (Press et al. 1988). In this method, we write these equations in operator splitting method:
\begin{equation}
   \frac{\partial A}{\partial t} = [L_r +L_{\theta}] A + S_p,
\end{equation}
\begin{equation}
   \frac{\partial B}{\partial t} = [N_r +N_{\theta}] B + S_t,
\end{equation}
where $L_r$ and $N_r$ are the operators involving r-derivative; similarly $L_{\theta}$ and $N_{\theta}$ are the operators involving $\theta$ derivatives. $S_p$ and $S_t$ (rotational shear term) are the source terms for the poloidal and toroidal field respectively. In our model, $S_t=s\left(\left[ \nabla \times (A\bf \hat{e}_\phi) \right]\cdot \nabla \Omega\right)$ i.e., source term for toroidal field and $S_p = S (r, \theta, B)$, i.e., source term for poloidal field due to the Babcock-Leighton mechanism which is modelled by double ring algorithm.
 
 From equation (A.1), the operators $L_r$ and $L_{\theta}$ is given by:
  \begin{equation}
  L_r A= - \frac{v_r}{r} \frac{\partial}{\partial r} (rA) + \eta \left[ \frac{\partial^2 A}{\partial r^2} +\frac{2}{r} \frac{\partial A}{\partial r} - \frac{A}{2 r^2 sin^2{\theta}}\right],
\end{equation}

 \begin{equation}
  L_\theta A= - \frac{v_\theta}{r  sin \theta} \frac{\partial}{\partial \theta} (A sin \theta) + \eta \left[ \frac{1}{r} \frac{\partial^2 A}{\partial \theta^2} +\frac{cot \theta}{r^2} \frac{\partial A}{\partial \theta} - \frac{A}{2 r^2 sin^2{\theta}}\right].
\end{equation}

Here we describe the basic theme of Alternating Direction Implicit (ADI) method in case of two dimensional system. In Alternating Direction Implicit (ADI) method, each time step is divided into two half time steps for two dimensional geometry. In the first half time step, one direction (say r), is advanced implicitly then other direction (say $\theta$) is advanced explicitly. In the second half step, $\theta$ direction is advanced implicitly and r direction is advanced explicitly. 
Let the value of A at grid point (i,j) at time step m is $A^m_{i,j}$. Then the ADI scheme consists of following two time steps:

 \begin{equation}
  A^{m+\frac{1}{2}}_{i,j} - A^{m}_{i,j}= (\overline{L}_r A^{m+\frac{1}{2}}_{i,j} + \overline{L}_\theta A^{m}_{i,j}) \frac{\Delta t}{2} 
\end{equation}

\begin{equation}
 A^{m+1}_{i,j}- A^{m+\frac{1}{2}}_{i,j} = (\overline{L}_r A^{m+\frac{1}{2}}_{i,j} + \overline{L}_\theta A^{m+1}_{i,j}) \frac{\Delta t}{2} 
\end{equation}
Here $\overline{L}_r$ and $\overline{L}_\theta$ represent the difference forms of the operators $L_r$ and $L_\theta$. \\
Here we treat the diffusion term by Crank-Nicholson scheme. We use an first order accurate upwind scheme to treat the terms $\frac{\partial A}{\partial r}$ and $\frac{\partial A}{\partial \theta}$. The hyperbolic advective terms [$- \frac{v_r}{r} \frac{\partial}{\partial r} (rA)$] is handled by Lax-Wendroff scheme, which is second order accurate in time and avoids mesh drifting, large numerical dissipation. 

Treating various terms in this way, we get the following forms of $\overline{L}_r$ and $\overline{L}_\theta$:
\begin{equation}
 \overline{L}_r A^{m}_{i,j} = d(i,j) A^{m}_{i-1,j} + e(i,j) A^{m}_{i,j} + f(i,j) A^{m}_{i+1,j}
 \end{equation}
 
\begin{equation}
\overline{L}_\theta A^{m}_{i,j} = a(i,j) A^{m}_{i,j-1} + b(i,j) A^{m}_{i,j} + c(i,j) A^{m}_{i,j+1} 
\end{equation}
The matrix coefficients can be calculated with straight forward algebra and take the form:\\

\begin{multline} 
a(i,j)= \frac{\eta}{(r \Delta \theta)^2} + \frac{U_\theta(i,j)}{\Delta \theta} sin(\theta - \frac{\Delta \theta}{2})
 \left[ \frac{1}{2} + \frac{\Delta t}{4 \Delta \theta} U_\theta(i, j-\frac{1}{2})sin(\theta- \Delta \theta) \right]
\end{multline}

\begin{multline} 
 b(i,j)= - \frac{2 \eta}{(r \Delta \theta)^2}- \frac{\eta cot{\theta}}{r^2 \Delta \theta}-\frac{\eta}{2 r^2 sin^2{\theta}} 
 -\frac{U_\theta(i,j)}{\Delta \theta} \left[ sin(\theta +\frac{ \Delta \theta}{2})  \right. \\
  \left.      \left\lbrace \frac{1}{2} + \frac{\Delta t}{4 \Delta \theta} U_\theta(i, j+\frac{1}{2}) sin \theta \right\rbrace 
    - sin(\theta -\frac{\Delta \theta}{2}) \left\lbrace \frac{1}{2} - \frac{\Delta t}{4 \Delta \theta} U_\theta(i, j-\frac{1}{2}) sin \theta \right\rbrace \right]
\end{multline}

\begin{multline} 
c(i,j)= \frac{\eta}{(r \Delta \theta)^2} + \frac{\eta cot{\theta}}{r^2 \Delta \theta} -\frac{U_\theta(i,j)}{\Delta \theta} sin(\theta + \frac{\Delta \theta}{2}) \left[\frac{1}{2} - \frac{\Delta t}{4 \Delta \theta} U_\theta(i, j+\frac{1}{2})sin(\theta + \Delta \theta) \right] 
\end{multline}

\begin{multline}
d(i,j) = \frac{\eta}{(\Delta r)^2} +\frac{U_r (i,j)}{\Delta r} (r-\frac{\Delta r}{2}) \left [\frac{1}{2} + \frac{\Delta t}{4 \Delta r} U_r(i-\frac{1}{2}, j) (r- \Delta r) \right]
\end{multline}

\begin{multline} 
e(i,j) = - \frac{2 \eta}{(\Delta r)^2} - \frac{2 \eta}{ r \Delta r}- \frac{\eta}{2 r^2 sin^2{\theta}} - \frac{U_r(i,j)}{\Delta r}  \\ 
  \left[\frac{\Delta r}{2}+ \frac{\Delta t}{4 \Delta r} r  \left\lbrace U_r(i+\frac{1}{2},j)(r+ \frac{\Delta r}{2}) 
 + U_r(i-\frac{1}{2}, j) (r- \frac{\Delta r}{2})\right\rbrace  \right]
\end{multline}

\begin{multline} 
f(i,j)= \frac{\eta}{(\Delta r)^2} + \frac{2 \eta}{r \Delta r} - \frac{U_r(i,j)}{\Delta r} (r+\frac{\Delta r}{2}) 
 \left[ \frac{1}{2} - \frac{\Delta t}{4 \Delta r} U_r(i+\frac{1}{2}, j)(r +\Delta r) \right]
\end{multline}
Here $U_r(i,j)$ and $U_\theta(i,j)$ give the values of $\frac{v_r}{r}$ and $\frac{v_\theta}{r sin \theta}$ at the grid point (i,j). \\

Using (A.9) and (A.10), we can write the equation (A.7) in the form:

\begin{equation}
 a(i,j)  A^{m+\frac{1}{2}}_{i,j-1} + [1 + b(i,j)] A^{m+\frac{1}{2}}_{i,j} + c(i,j) A^{m+\frac{1}{2}}_{i, j+1} = \Psi(i,j),
\end{equation}
 
 where,
 \begin{equation}
 \Psi(i,j)= -d(i,j) A^{m}_{i-1,j} +[1- e(i,j)]A^{m}_{i,j} -f(i,j) A^{m}_{i+1, j}.
  \end{equation}
  
 Similarly in second step of ADI method, we can write the equation (A.8) in the form:
\begin{equation}
d(i,j) A^{m+1}_{i-1,j} +[1+ e(i,j)]A^{m+1}_{i,j} +f(i,j) A^{m+1}_{i+1, j} = \Phi (i,j),
\end{equation}
 where,
\begin{eqnarray}
\Phi (i,j)= -a(i,j) A^{m+ \frac{1}{2}}_{i,j-1} +[1- b(i,j)]A^{m+\frac{1}{2}}_{i,j} -c(i,j) A^{m +\frac{1}{2}}_{i, j+1} \nonumber \\
           = -\Psi(i,j) + 2 A^{m+\frac{1}{2}}_{i,j}.
\end{eqnarray}

The coefficients of all these equations form a tridiagonal matrix. Source term ($S_p$) is treated explicitly in the code. In the first half step of ADI method, we solve the equation (A.17) and get the values of $A^{m+\frac{1}{2}}_{i,j}$ at all grid points.  In the second half step, we solve the equation (A.19), and get the values of $A^{m+1}_{i,j}$ at all grid points. 

Similarly, we can solve the toroidal field evolution equation. In case of the toroidal field evolution equation, 
For first half step of ADI method,
\begin{equation}
 ab(i,j)  B^{m+\frac{1}{2}}_{i,j-1} + [1 + bb(i,j)] B^{m+\frac{1}{2}}_{i,j} + cb(i,j) B^{m+\frac{1}{2}}_{i, j+1} = \chi(i,j),
\end{equation}
 where,
 \begin{equation}
 \chi(i,j)= -db(i,j) B^{m}_{i-1,j} +[1- eb(i,j)]B^{m}_{i,j} -fb(i,j) B^{m}_{i+1, j}.
  \end{equation}
  and for second half step of ADI method:
  \begin{equation}
db(i,j) B^{m+1}_{i-1,j} +[1+ eb(i,j)]B^{m+1}_{i,j} +fb(i,j) B^{m+1}_{i+1, j} = \Gamma(i,j),
\end{equation}
 where,
\begin{eqnarray}
\Gamma(i,j)= -ab(i,j) B	^{m+ \frac{1}{2}}_{i,j-1} +[1- bb(i,j)]B^{m+\frac{1}{2}}_{i,j} -cb(i,j) B^{m +\frac{1}{2}}_{i, j+1} \nonumber \\
           = -\chi(i,j) + 2 B^{m+\frac{1}{2}}_{i,j}.
\end{eqnarray}

The expression for the matrix coefficients necessary for the toroidal field evolution equation are given by:

\begin{multline} 
ab(i,j)= \frac{\eta}{(r \Delta \theta)^2} + \frac{U_\theta(i,j-\frac{1}{2})}{\Delta \theta} sin(\theta - \frac{\Delta \theta}{2})
 \left[ \frac{1}{2} + \frac{\Delta t}{4 \Delta \theta} U_\theta(i, j-1)sin(\theta- \Delta \theta) \right]
\end{multline}

\begin{multline} 
bb(i,j)= - \frac{2 \eta}{(r \Delta \theta)^2}- \frac{\eta cot{\theta}}{r^2 \Delta \theta}-\frac{\eta}{2 r^2 sin^2{\theta}} -
 \left[ \frac{U_\theta(i,j+\frac{1}{2})}{\Delta \theta} sin(\theta +\frac{ \Delta \theta}{2}) \right.\\
   \left.     \left\lbrace \frac{1}{2} + \frac{\Delta t}{4 \Delta \theta} U_\theta(i, j) sin \theta \right\rbrace 
    - \frac{U_\theta(i,j-\frac{1}{2})}{\Delta \theta} sin(\theta -\frac{\Delta \theta}{2}) \left\lbrace \frac{1}{2} - \frac{\Delta t}{4 \Delta \theta} U_\theta(i, j) sin \theta \right\rbrace \right]
\end{multline}

\begin{multline} 
cb(i,j)= \frac{\eta}{(r \Delta \theta)^2} + \frac{\eta cot{\theta}}{r^2 \Delta \theta} -\frac{U_\theta(i,j+\frac{1}{2})}{\Delta \theta} 
  sin(\theta + \frac{\Delta \theta}{2}) \left[\frac{1}{2} - \frac{\Delta t}{4 \Delta \theta} U_\theta(i, j+1)sin(\theta + 
\frac{\Delta \theta}{2}) \right] 
\end{multline}

\begin{multline}
db(i,j) = \frac{\eta}{(\Delta r)^2} +\frac{\overline{U}_r (i,j)}{\Delta r} (r- \frac{\Delta r}{2}) \left [\frac{1}{2} + \frac{\Delta t}{4 \Delta r} \overline{U}_r(i-\frac{1}{2}, j) (r- \Delta r) \right]
\end{multline}

\begin{multline} 
 eb(i,j) = - \frac{2 \eta}{(\Delta r)^2} - \frac{2 \eta}{ r \Delta r}- \frac{\eta}{2 r^2 sin^2{\theta}} - \overline{U}'_r(i,j) -\frac{\overline{U}_r(i,j)}{\Delta r} \\
   \left[\frac{\Delta r}{2}+ \frac{r \Delta t}{4 \Delta r} \left\lbrace \overline{U}_r(i+\frac{1}{2},j)(r+ \frac{\Delta r}{2}) 
+ \overline{U}_r(i-\frac{1}{2}, j) (r- \frac{\Delta r}{2})\right\rbrace  \right]
\end{multline}

\begin{multline} 
fb(i,j)= \frac{\eta}{(\Delta r)^2} + \frac{\eta}{r \Delta r} - \frac{\overline{U}_r(i,j)}{\Delta r} (r+\frac{\Delta r}{2}) 
 \left[ \frac{1}{2} - \frac{\Delta t}{4 \Delta r} \overline{U}_r(i+\frac{1}{2}, j)(r +\frac{\Delta r}{2}) \right]
\end{multline}
where, $\overline{U}_r= \frac{U_r - \frac{d \eta}{dr}}{r}$.

We solve these equations in a $n \times n$ grid with the appropriate initial and boundary conditions.

\end{appendices}

\printthesisindex
\end{document}